\documentclass[10pt,a4paper]{article}

\usepackage{lmodern}
\usepackage{epsf}
\usepackage[latin1]{inputenc}
\usepackage[english]{babel}
\usepackage{amsfonts}
\usepackage{amssymb}
\usepackage{amsmath}
\usepackage{bm}
\usepackage{graphicx}

\usepackage{multirow}
\usepackage{multicol}
\usepackage{array}
\usepackage{version}
\usepackage{url}
\usepackage{lscape}
\usepackage{mathrsfs}
\usepackage[nottoc]{tocbibind} %no number biblio
\usepackage{layout} 
\usepackage[textwidth=15.5cm,textheight=22cm]{geometry}
\usepackage{makecell}
\usepackage{lipsum}
\usepackage{cite}
\usepackage{float}
\usepackage{quoting}
\usepackage{authblk}

\newlength{\bibitemsep}
\newlength{\bibparskip}
\let\oldthebibliography\thebibliography
\renewcommand\thebibliography[1]{%
  \oldthebibliography{#1}%
  \setlength{\parskip}{\bibitemsep}%
  \setlength{\itemsep}{\bibparskip}%
}
\title{Gravitational waves from inflation}

\author[1,2]{M.C.~Guzzetti\thanks{mariachiara.guzzetti@pd.infn.it}}
\author[1,2,3]{N.~Bartolo\thanks{nicola.bartolo@pd.infn.it}}
\author[1,2,3]{M.~Liguori\thanks{michele.liguori@pd.infn.it}}
\author[1,2,3,4]{S.~Matarrese\thanks{sabino.matarrese@pd.infn.it}}

\affil[1]{Dipartimento di Fisica e Astronomia ``G. Galilei'', Universit\`a degli Studi di Padova, via Marzolo 8, I-35131, Padova, Italy}
\affil[2]{INFN, Sezione di Padova, via Marzolo 8, I-35131, Padova, Italy}
\affil[3]{INAF-Osservatorio Astronomico di Padova, Vicolo dell'Osservatorio 5, I-35122 Padova, Italy}
\affil[4]{Gran Sasso Science Institute, INFN, Viale F. Crispi 7, I-67100 L'Aquila, Italy}

\date{\today\endgraf\bigskip {\small Prepared as invited review for La Rivista del Nuovo Cimento.}}

\begin{document}

\maketitle

\begin{abstract}
The production of a stochastic background of gravitational waves is a fundamental prediction of {\it any} cosmological inflationary model. The features of such a signal encode unique information about the physics of the Early Universe and beyond, thus representing an exciting, powerful window on the origin and evolution of the Universe. 
We review the main mechanisms of gravitational-wave production, ranging from quantum fluctuations of the gravitational field to other mechanisms that can take place during or after inflation. 
These include e.g. gravitational waves generated as a consequence of extra particle production during inflation, or during the (p)reheating phase. Gravitational waves produced in inflation scenarios based on modified gravity theories and second-order gravitational waves are also considered. For each analyzed case, the expected power-spectrum is given.
We discuss the discriminating power among different models, associated with the validity/violation of the standard consistency relation between tensor-to-scalar
ratio $r$ and tensor spectral index $n_{\rm T}$.
In light of the prospects for (directly/indirectly) detecting primordial gravitational waves, we give the expected present-day gravitational radiation spectral energy-density, highlighting the main characteristics 
imprinted by the cosmic thermal history, and we outline the signatures left by gravitational waves on the Cosmic Microwave Background and some imprints in the Large-Scale Structure of the Universe.
Finally, current bounds and prospects of detection for inflationary gravitational waves are summarized.
\end{abstract}

\tableofcontents 

\newpage

\section{Introduction}

A general, and extremely important, prediction of cosmological Inflation is the generation of a stochastic background of primordial gravitational waves (GW), the detection of which would be of massive importance for Cosmology. Primordial GW are in fact {\em not} expected in the framework of non-inflationary Early-Universe models, making them a {\em smoking-gun} probe of Inflation.\footnote{During the early stages of the Universe, other processes, over than inflation, can act as sources of GW, such as the electroweak phase transition \cite{Huber:2015znp}, the first-order phase transitions \cite{Caprini:2015zlo,Dev:2016feu} (and refs. therein) and the topological defects \cite{Sanidas:2012ee,Figueroa:2012kw}.} In the standard single-field, slow-roll inflationary scenario, tensor fluctuations of the metric (\textit{i.e.} primordial GW) are characterized by a nearly scale-invariant power-spectrum on super-horizon scales. The amplitude of the GW signal is usually described by the tensor-to-scalar ratio $r$, defined as the ratio between the tensor and scalar power-spectrum amplitudes, at a given pivot scale $k_{\ast}$. 
The current best bound on $r$ comes from the joint analysis of Planck, BICEP2, Keck Array and other data, which yields $r<0.07$ at $95\%$ C.L. for $k_{\ast}=0.05 \mbox{Mpc}^{-1}$ \cite{Array:2015xqh} assuming the consistency relation $r=-8n_{\rm T}$, where $n_{\rm T}$ is the tensor spectral index. Excluding temperature data and assuming a scale-invariant GW power-spectrum, the bound becomes $r<0.09$ at $95\%$ C.L. for $k_{\ast}=0.05 \mbox{Mpc}^{-1}$ \cite{Array:2015xqh}, corresponding to a present time spectral energy-density $\Omega_{\rm GW}\left(f\right)\simeq 10^{-15}$ for $f\simeq10^{-17}$ Hz. 
A crucial point is that, even in the simplest, single-field framework, different inflationary scenarios predict different values of $r$. 
The study of observational signatures of primordial GW thus provide not only a way to probe the general inflationary paradigm, but also to discriminate in detail among specific models. If we move beyond single-field Inflation, even more specific signatures can be generated. For example, in presence of additional fields besides the Inflaton, an extra GW background, not due to vacuum oscillations, can be produced \cite{Sorbo:2011rz,Barnaby:2012xt,Senatore:2011sp,Binetruy:2012ze,Biagetti:2013kwa}. Interestingly, classical generation of GW is also possible during the inflationary reheating phase \cite{Khlebnikov:1996mc}; the primordial GW background therefore also opens a potential window on the study of reheating mechanisms, and related parameters. On top of this, a primordial GW detection would not only be of paramount importance for Cosmology, as discussed so far, but also have far reaching consequences for High Energy and Fundamental Physics. 
The inflationary GW background is in fact generated at energy scales which are many orders of magnitude {\em above} those achievable in collisions by present-day particle accelerators. The energy scale of Inflation is moreover directly linked to the value of the tensor-to-scalar-ratio. Therefore, a detection of $r$ not only would provide strong evidence for Physics beyond the Standard Model of Particle Physics, but also give a precise indication of the energy regime of such new Physics. It is also very important to stress that inflationary tensor fluctuations of the background Friedmann-Robertson-Walker metric arise from quantum fluctuations in the gravitational field itself, via a mechanism that is similar to that leading to their scalar counterparts. Their observation would thus also produce the first experimental evidence of a quantum gravity phenomenon. In light of all this, it is not at all surprising that primordial GW are the object of a growing experimental effort, and that their detection will be a major goal for Cosmology in the forthcoming decades. The main observational signature of the inflationary GW background is a curl-like pattern (``B-mode'') in the polarization of the Cosmic Microwave Background (CMB). A number of, present or forthcoming, ground-based or balloon-borne experiments, such as ACTPol \cite{Calabrese:2014gwa}, Polarbear \cite{Ade:2014afa}, CLASS \cite{Eimer:2012ny}, Piper \cite{piper} and Spider \cite{Crill:2008rd}, are specifically aimed at B-mode detection. In addition, CMB satellites such as WMAP and {\it Planck} have, in recent years, provided bounds on $r$, such as the one reported above. Finally, several next-generation CMB space missions have been proposed in recent years, with the specific goal of B-mode detection in mind, like COrE \cite{Bouchet:2011ck}, PRISM \cite{Andre:2013nfa}, LiteBIRD \cite{Matsumura:2013aja} and PIXIE \cite{Kogut:2011xw}. In addition to the B-mode, evidence of primordial GW could come from galaxy and CMB curl-like lensing signatures, induced by tensor modes \cite{Dodelson:2003bv,Masui:2010cz,Jeong:2012nu}, or from parameters related to the small modification in the expansion history of the Universe, due to the GW contribution to the overall energy budget \cite{Smith:2006nka}. Finally, the possibility of a future direct detection, by experiments such as aLIGO \cite{TheLIGOScientific:2014jea} or eLISA \cite{elisaweb,Klein:2015hvg}, cannot be ruled out, especially if some specific inflationary mechanism produced a blue-tilted primordial tensor spectrum. 
This point holds even more true in these days, in light of the recent, exciting discovery of a gravitational wave signal, interpreted as the gravitational radiation emitted by a black-hole merger,\footnote{A different source for the detected GW is not completely excluded; in \cite{Bird:2016dcv,Sasaki:2016jop,Clesse:2016vqa} the interpretation of the signal as due to a merger of a black-hole binary of \textit{primordial} origin is discussed.} by the LIGO and Virgo collaborations \cite{Abbott:2016blz}. Whatever the origin of the signal is, this very important result does provide the first direct experimental confirmation of GW, and increase our confidence that, as they just did for Astronomy and Astrophysics (see, for example, \cite{Blair:2016idv} for a recent review), GW might soon open a new observational window and a new era in Cosmology.\\
Armed with this - reasonable - hope, and given all the important scientific premises above, we feel it is a proper time to review the current theoretical and observational status of primordial GW from Inflation, with the following plan: in section \ref{cap1} we overview predictions about GW related to the standard single-field slow-roll inflationary scenario, and we illustrate the main properties that make them a significant physical observable. Then, in section \ref{secondo} we explain the possible mechanisms, different from vacuum oscillations of the gravitational field, by which primordial GW can be produced during inflation. GW generation during the reheating stage of the Universe is reviewed in section \ref{analisi}, followed by GW predictions related to a few inflationary models built in the framework of modified gravity. In section \ref{overview}, an overview of the analyzed models is provided.
In section \ref{quantum} we rapidly outline the issue of the quantum to classical transition of inflationary fluctuations. The predictions about the validity and the violation of the standard inflationary consistency relation between the tensor-to-scalar ratio $r$ and the tensor spectral index $n_{\rm T}$ are presented in section \ref{sezioneconsistency}. In section \ref{sezstoria}, signatures of the thermal history of the Universe on the present GW spectral energy-density are shown. Afterward, significant imprints on CMB and LSS of primordial GW are outlined. In section \ref{exp} we summarize current results and observational prospects about primordial GW. We conclude in section \ref{conclusioni}.

\section{Gravitational waves from single-field slow-roll inflation}\label{cap1}

The inflationary scenario provides an elegant solution to some internal inconsistencies of the Big Bang Theory, such as the \textit{horizon} and \textit{flatness problems} \cite{Brout:1977ix,Starobinsky:1980te,Kazanas:1980tx,Sato:1980yn,Guth:1980zm,Linde:1981mu,Albrecht:1982mp}. It consists in a sufficiently long period of accelerated expansion of the Universe at early times \cite{Abbott:1982hn}. Besides solving the mentioned problems, considering its quantum aspects 
reveals that it provides an elegant mechanism for generating the initial seeds of all observed structures in the Universe and the anisotropies of the CMB radiation, which otherwise have to be implemented {\it by hand}, without a physical motivation \cite{Mukhanov:1981xt,Hawking:1982cz,Guth:1982ec,Starobinsky:1982ee,Abbott:1984fp,Mukhanov:1985rz}. This result is achieved by considering quantum fluctuations of the fields that describe the dynamics of the Universe in such an epoch: usually a neutral scalar field and the metric tensor. Developing a perturbation theory within General Relativity, one finds that, besides a set of perturbations coupled to the energy density of the Universe, tensor perturbations are produced. The latter are due to fluctuations of the metric tensor and constitute the so called Gravitational Wave background.\\
We start this section with a summary of the classical aspect of the basic inflationary paradigm and then we move to considering quantum aspects, in order to show how the primordial GW should have been produced.

\subsection{The Physics of inflation}

Standard cosmology is built starting from an isotropic and homogeneous Universe described by Friedman-Robertson-Walker (FRW) metric:
\begin{equation}\label{frw}
	{\rm d}s^{2}=-{\rm d}t^{2}+a^{2}(t)\left[\frac{{\rm d}r^{2}}{1-K r^{2}}+r^{2}({\rm d}\theta^{2}+\sin^{2}\theta {\rm d}\varphi^{2})\right]\:,
\end{equation}
where $t$ is the cosmic time, $r, \theta, \varphi$ are the comoving spherical coordinates and $K$ the curvature of the three-dimensional spatial hyper-surfaces. The metric is identified by the evolution of the ``scale-factor" $a(t)$ and the spatial curvature 
parameter $K$. To get the evolution of the scale-factor via Einstein's equations we need to specify the energy form of the cosmic medium. Under the hypothesis of homogeneous and isotropic Universe $T_{\mu\nu}$ can be that of a perfect fluid:
\begin{equation}\label{fluid}
	T_{\mu\nu}=(\rho +P){\rm u}_{\mu}{\rm u}_{\nu}+Pg_{\mu\nu}\:,
\end{equation}
where $\rho$ is the density, $P$ the pressure, ${\rm u}_{\mu}$ the 4-velocity of fluid elements and $g_{\mu\nu}$ the metric tensor.
Then, using this expression for the stress-energy tensor and the metric \eqref{frw} in the Einstein's equations, the Friedman equations are obtained:
\begin{equation}\label{friedman1}
	H^{2}=\frac{8\pi G}{3}\rho -\frac{K}{a^{2}}\,,\qquad
\frac{\ddot{a}}{a}=-\frac{4\pi G}{3}(\rho+3P)\:.
\end{equation}
where $H$ is the Hubble parameter defined as $H\equiv \dot{a}/a$.
Hereafter we will set $K=0$, in agreement with observational constraints which imply negligible spatial curvature \cite{Ade:2015xua}.
The Friedman equations reveal which kind of perfect fluid can drive the dynamics. 
The basic requirement of the inflationary mechanism consists in $\ddot{a}>0$, that is an accelerated expansion of the Universe. From eq.\eqref{friedman1} this corresponds to $P<-\frac{\rho}{3}$.
One guesses immediately that an inflationary period cannot be driven neither by ordinary radiation nor matter.
The simplest way to obtain such a kind of dynamics is the case in which $P\simeq-\rho$, that, in the equality case, means an evolution of the scale-factor like
\begin{equation}\label{aevolution}
	a(t)=a_{\rm I}e^{H_{\rm i}(t-t_{\rm i})}\:,
\end{equation}
with Hubble parameter nearly constant in time $H=H_{i}\simeq {\rm const}$; here 
the subscript $i$ indicates the beginning of the inflationary period.
A period characterized by this evolution of the scale-factor is called \textit{de Sitter stage}. It is now useful to introduce a quantity called Hubble radius (or Hubble horizon), defined as $R_{\rm H}\equiv 1/H\left(t\right)$, which effectively sets the size of causally connected regions at each time. For a non-exotic content of the Universe we have $R_{\rm H}\propto c t$. In a de Sitter model, instead, the physical Hubble radius is constant in time, while physical lengths continue to grow, thus being able to exit the 
Hubble radius at some ``horizon-crossing" time. 
The requirement of sufficiently long inflation corresponds to the requirement that all scales relevant for cosmological observations were able to exceed the Hubble radius during inflation.\\
The simplest way of implementing a source of stress-energy which provides $P\simeq-\rho$, consists in introducing a scalar field $\varphi$ with suitable potential energy $V(\varphi)$. 
Therefore, we introduce a minimally coupled scalar field described by the following Lagrangian: 
\begin{equation}\label{action}
	\mathscr{L}=-\frac{1}{2}\partial^{\mu}\varphi\:\partial_{\mu}\varphi-V(\varphi)\,.
\end{equation}
By varying the action with respect to $\varphi$, the equation of motion for the field, the Klein-Gordon equation $\Box\varphi=\partial V/\partial\varphi$, is obtained. With the FRW background it reads
\begin{equation}\label{friz}
	\ddot{\varphi}+3H\dot{\varphi}-\frac{1}{a^{2}}\nabla^{2}\varphi+V'(\varphi)=0\:,
\end{equation}
where $V'(\varphi)=dV(\varphi)/d\varphi$. 
On the other hand varying the action \eqref{action} with respect to the metric tensor the expression for the stress-energy tensor for the minimally coupled scalar field is obtained,
\begin{equation}
	T_{\mu\nu}=-2\frac{\partial\mathscr{L}}{\partial g^{\mu\nu}}+g_{\mu\nu}\mathscr{L}=\partial_{\mu}\varphi\partial_{\nu}\varphi+g_{\mu\nu}\left[-\frac{1}{2}g^{\alpha\beta}\partial_{\alpha}\varphi\partial_{\beta}\varphi-V(\varphi)\right]\, .
\end{equation}
Comparing this expression with eq.\eqref{fluid}, one finds that a homogeneous scalar field $\varphi(t)$ behaves like a perfect fluid with energy-density background and pressure given by
\begin{equation}\label{rho}
	\rho_{\varphi}=\frac{\dot{\varphi}^{2}}{2}+V(\varphi)\,,\qquad P_{\varphi}=\frac{\dot{\varphi}^{2}}{2}-V(\varphi)\,.
	\end{equation}
Therefore the quantity that establishes the sign of the acceleration of the Universe, from the second equation of \eqref{friedman1}, reads
\begin{equation}\label{pressione}
	\rho_{\varphi}+3P_{\varphi}=2\left[\dot{\varphi}^{2}-V\left(\varphi\right)\right]\:,
\end{equation}
so one concludes that
$V(\varphi)>\dot{\varphi}^{2}$
suffices to obtain accelerated expansion. 
In particular to obtain a quasi-de Sitter stage, from eqs.\eqref{pressione} we need
\begin{equation}\label{slow}
	V(\varphi)\gg\dot{\varphi}^{2}\,.
\end{equation}
From this simple calculation, we realize that a scalar field whose energy density is dominant in the Universe and whose potential energy dominates over the kinetic one gives rise to an inflationary period. 
The simplest way to satisfy eq.\eqref{slow} is to introduce a scalar field slowly rolling towards the minimum of its potential.

\subsubsection{Slow-roll conditions}
\begin{figure}[ht]
    \centering
    \includegraphics[width=0.7\textwidth]{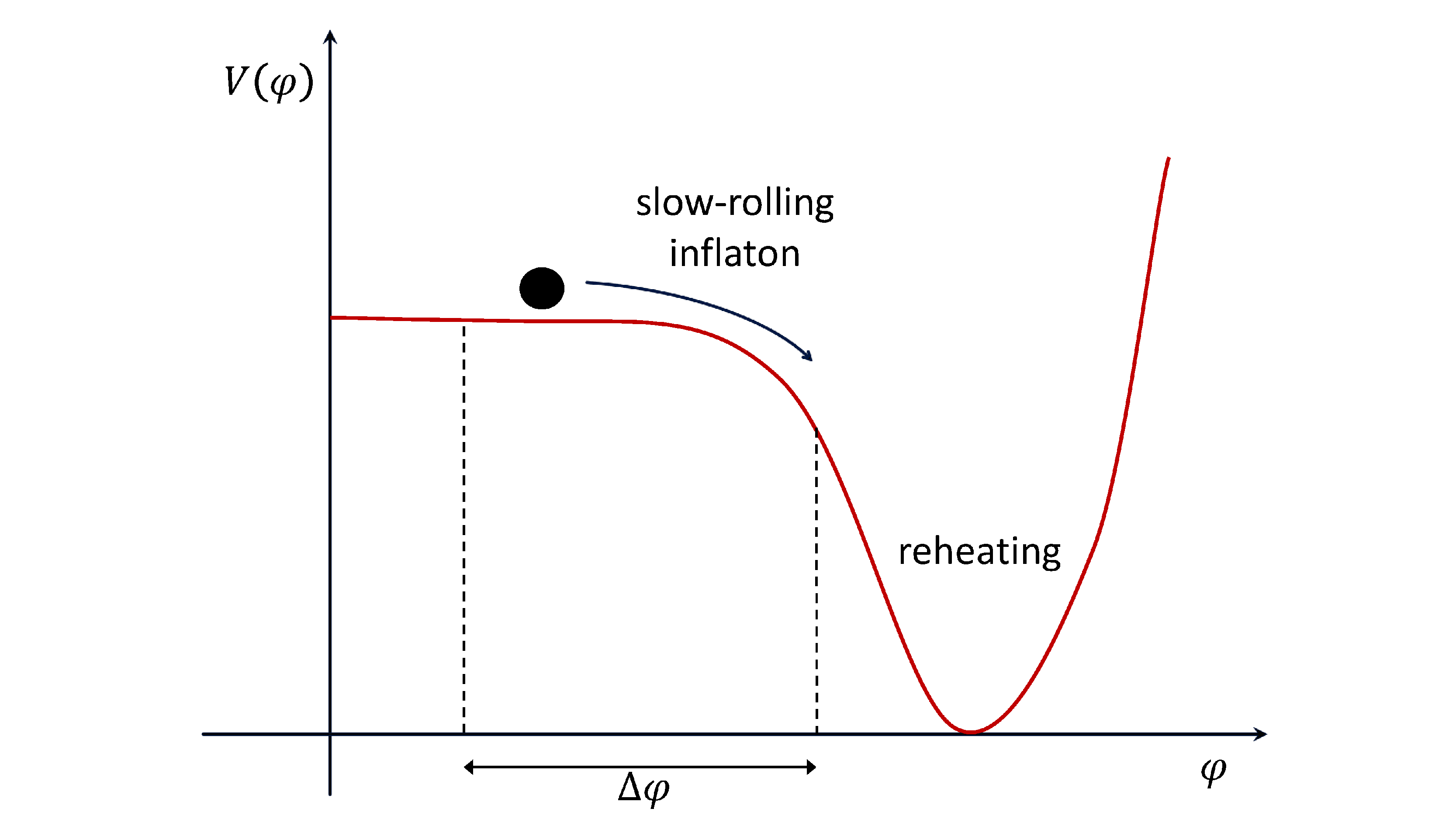}
    \caption{Example of inflationary potential with a ``flat" region. After the slow-roll of the inflaton field $\varphi$, the reheating phase starts, the field is supposed to oscillate around the minimum of the potential and to decay in other particles; see section \ref{sezionere}. $\Delta\varphi$ indicates the inflaton excursion between the horizon exit of a given comoving scale and the end of inflation; see section \ref{escursione}.}
  
    \label{potenziale}
\end{figure}

Let us now better quantify under which circumstances a scalar field and its potential may give rise to a period of inflation.\\
The simplest way to satisfy eq.\eqref{slow} is to require that there exist regions of field-configuration space where the potential is sufficiently flat, see fig.\ref{potenziale}. 

\noindent
In such a situation, for sufficiently late times the evolution of the scalar field is driven by the friction term, that is we can consider $\ddot{\varphi}\ll 3H\dot{\varphi}$. Exploiting the Friedman equations, these conditions can be summarized by restrictions on the form of the {\it inflaton} potential $V(\varphi)$ and its derivatives. Employing eq.\eqref{slow} and the second condition just mentioned, in eqs.\eqref{friedman1}-\eqref{friz} the equations become
\begin{equation}
	H^{2}\simeq \frac{8\pi G}{3}V(\varphi)\:, \qquad
	3H\dot{\varphi}+V_{\varphi}=0\:,
\end{equation}
where we have assumed that the inflaton is a homogeneous field that dominates the energy density of the Universe and the subscript $\varphi$ means the derivation w.r.t. such a field. The second expression gives $\dot{\varphi}$ as a function of $V'\left(\varphi\right)$, then the slow-roll condition \eqref{slow} is satisfied provided that
\begin{align}\label{parametri}
	\frac{\left(V_{\varphi}\right)^{2}}{V}\ll H^{2} \qquad &\Longrightarrow \qquad   \textbf{$\epsilon$}\equiv\frac{M^{2}_{\rm pl}}{2}\left(\frac{V_{\varphi}}{V}\right)^{2}\ll1\,,\\
	V_{\varphi\varphi}\ll H^{2}\:  \qquad &\Longrightarrow  \qquad \textbf{$\eta$}\equiv M^{2}_{\rm pl}\,\frac{V_{\varphi\varphi}}{V}\ll1\,,\qquad \qquad
\end{align}
where $\epsilon$ and $\eta$ are the so-called \textit{slow-roll parameters} \cite{Lidsey:1995np,Liddle:1994dx,Lyth:1998xn} and $M_{\rm pl}\equiv\left(8\pi G\right)^{-1/2}$ is the reduced Planck mass. Notice that we can also write the first parameter in terms of the Hubble parameter and its derivative as $\epsilon=-\dot{H}/H^{2}$. Then, the slow-roll conditions can be expressed by a hierarchy of slow-roll parameters involving higher-order derivatives of the potential $V\left(\varphi\right)$ \cite{Liddle:1994dx}; for example we can define a second-order slow-roll parameter $\zeta^{2}=1/\left(8\pi G\right)\left(V_{\varphi}V_{\varphi\varphi\varphi}/V_{\varphi\varphi}\right)$.
Once these constraints are satisfied, the inflationary process happens generically for a wide class of models $V\left(\varphi\right)$. As soon as these conditions fail, inflation ends. 
During inflation, the slow-roll parameters can be considered constant in time at first order, since, as it is easy to show, $\dot{\epsilon},\dot{\eta}=O\left(\epsilon^{2},\eta^{2}\right)$. 

\subsubsection{Duration and end of inflation}
\begin{figure}[ht]
    \centering
    \includegraphics[width=0.9\textwidth]{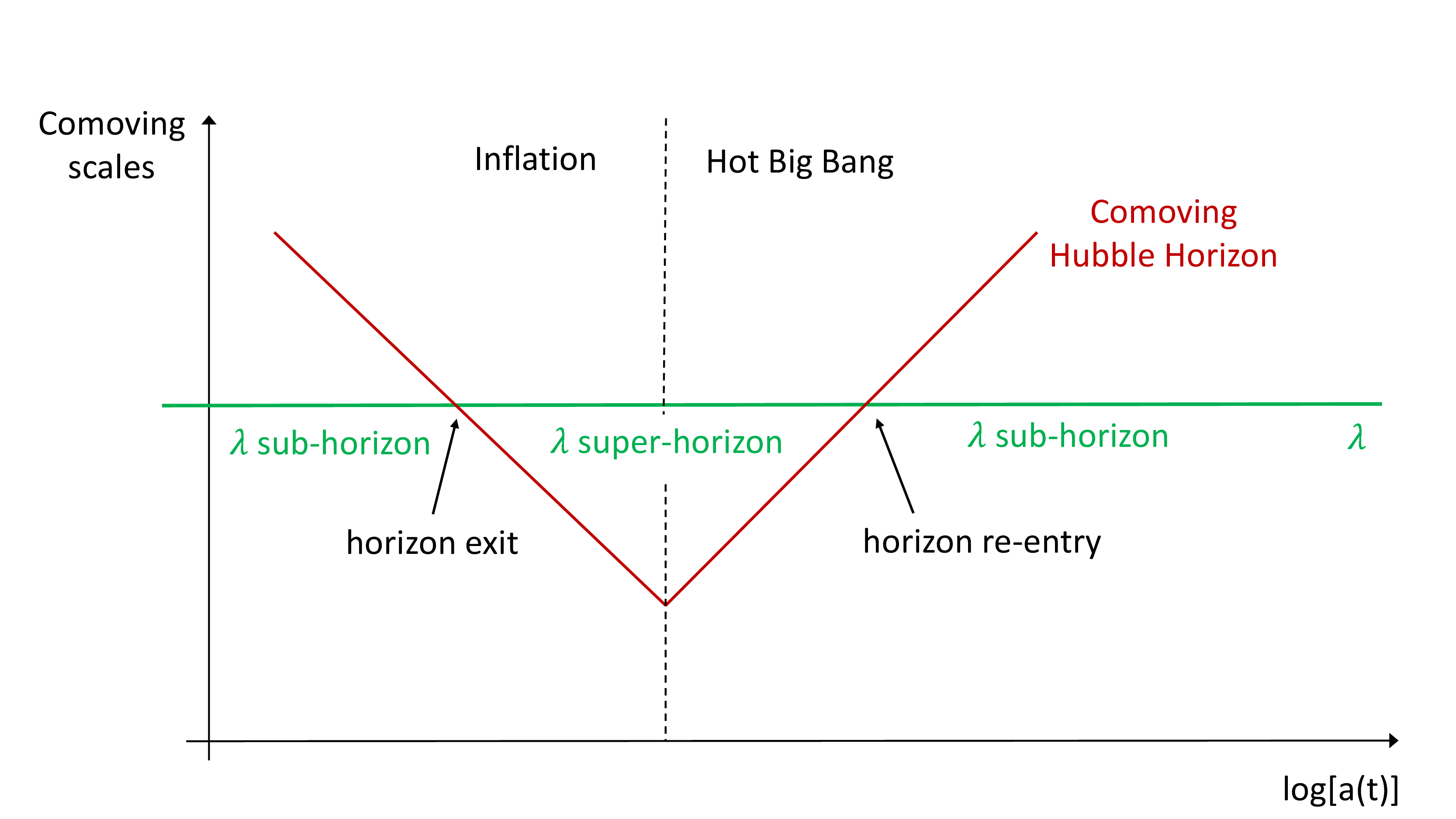}
    \caption{Time evolution of the comoving Hubble horizon during inflation and the following epoch, compared to the evolution of a comoving scale $\lambda$ \cite{Lucchin:1985wy}. During the accelerated expansion the comoving Hubble horizon decreases in time, while it grows during the radiation and matter dominated epochs. At a certain time during inflation, the comoving scale $\lambda$ exits the comoving Hubble horizon and then re-enters after inflation is over. The behavior of the comoving Hubble horizon shown in this figure, provides a solution to the \textit{horizon problem}.}
  
    \label{lunghezze}
\end{figure}

Successful inflation must last for a long enough period in order to solve the horizon and flatness problems, which means that, at least, all what is now inside the Hubble horizon, in particular those regions which entered the Hubble horizon during radiation and matter dominance, was inside a causally connected region at some time in the past. 
Therefore, we need a primordial period of accelerated expansion long enough that a smooth region smaller than the Hubble horizon at that time, can grow up to encompass at least the entire observable Universe; see fig.\ref{lunghezze}. Typically, this feature is expressed in terms of the number of e-foldings \cite{Liddle:1994dx}, defined as:
\begin{equation}
	N_{\rm tot}\equiv\int^{t_{\rm f}}_{t_{\rm i}}H\:{\rm d}t\:\:,
\end{equation}
where $t_{\rm i}$ and $t_{\rm f}$ are the starting and ending time of inflation, that, in case the scale-factor evolution is described by \eqref{aevolution}, reads $N=\mathrm{ln}\left(a_{\rm f}/a_{\rm i}\right)$, where $a_{\lambda}=a\left(t\left(\lambda\right)\right)$. The lower bound required to solve the horizon problem number is $N\gtrsim \mathrm{ln} 10^{26}\thicksim 60$ \cite{Ade:2015lrj}.
%\vspace{5cm}

\subsubsection{Reheating phase}\label{sezionere}
Inflation cannot proceed forever: the greatest successes of the Standard Big Bang model, such as primordial nucleosynthesis and the origin of the CMB, require the standard evolutionary progression from radiation to a matter domination era.\\
In the single-field slow-roll scenario, inflation ends when the inflaton field starts rolling fast along its potential, it reaches the minimum and then oscillates around it. 
Anyhow, we know that the Universe must be repopulated by hot radiation in order to initiate the hot Big Bang phase. 
The process by which the Universe moves from the inflationary dynamics to the hot Big Bang is called \textit{reheating} \cite{Albrecht:1982mp,Linde:1981mu,Abbott:1982hn,Dolgov:1982th}.\\
By investigating primordial GW, we cannot neglect this stage, for several reasons. First, there are many models for the reheating period which provide further GW production, besides that of the inflationary phase. 
Moreover, it can be shown that reheating parameters are related to inflationary power-spectra ones, so that the constraints on tensor perturbations are related to those on the reheating period of the Universe.\\
%Here we present the main features of the reheating models and later we will examine in some detail the role of GW.\\
The main requirement for the developing of the hot Big Bang is a radiation dominated Universe at $T\simeq 1{\rm MeV}$. However at the end of inflation most of the energy density of the Universe is stored in the scalar field(s), as the other components have been diluted by the accelerated expansion. The reheating process so consists in the conversion of such an energy into other forms, which ultimately lead to a radiation dominated scenario in thermal equilibrium. 
The temperature of the Universe at the time this process is substantially completed is called \textit{reheat temperature}. 
Many models have been proposed to describe this transition, some of which include the perturbative decay of the inflaton field while others involve non-perturbative mechanisms, such as parametric resonance decay.
If the fluctuations are sufficiently small, inflaton quanta could decay into relativistic products. This happens as soon as the inflaton decay rate $\Gamma$ becomes comparable to the Hubble constant. 
If the decay is slow, only fermionic decays are available.
Usually each decay product is supposed to thermalize quickly so that their energy distribution can be described by a black-body function and the reheating temperature for a sudden process is $T_{\rm reh}\sim\sqrt{M_{\rm pl}\Gamma}$.
Then a mechanism is supposed to take place that leads to energy transfer of the decay products into radiation.
Otherwise, if the scalar field decays into bosonic particles, we can have a rapid decay through a parametric resonance mechanism. The process may be so fast that it ends after a few oscillations of the inflaton field. 
This phase is called \textit{preheating phase} \cite{Kofman:1994rk}.

\subsection{Quantum fluctuations: origin of cosmological perturbations}
We can now move to consider quantum aspects of the inflationary paradigm. The current understanding of structure formation and generation of CMB anisotropies requires the existence of small fluctuations that entered the horizon during the radiation and matter era. Employing only the standard cosmology we cannot explain the presence of perturbations. On the other hand, the quantum aspects of the inflationary mechanism constitute a natural way to explain the presence of such small seeds.\\
According to quantum field theory, each physical field involved in a theory is characterized by quantum fluctuations: they oscillate with all possible wavelengths maintaining zero average on a sufficient macroscopic time. 
The inflationary accelerated expansion can stretch the wavelength of these fluctuations to scales greater than the Hubble horizon $k\gg aH$, where $k$ is the comoving wave number of a given fluctuation, so that they become classical \cite{Mukhanov:1981xt,Guth:1982ec,Hawking:1982cz,Linde:1982uu,Starobinsky:1982ee}.
Here the fluctuation amplitudes approximately do not change in time in contrast with their wavelengths that go on increasing exponentially. When inflation ends and the radiation and matter dominated eras develop, these perturbations are encompassed a second time by the Hubble horizon starting from the smallest ones. When a perturbation returns to be embedded by a causally connected region we have fluctuations on sufficient large scales and with non-zero amplitude so that the action of gravity leads to the present  large-scale structure (LSS) and CMB anisotropy pattern.\\
In virtue of such a mechanism, the quantities we are interested in are the perturbations left over by the accelerated expansion on super-horizon scales.\\
In the basic inflationary scenario the fields involved in the dynamics of the Universe are two: the inflaton and the metric tensor which describes the gravitational degrees of freedom. In what follows we will consider in such a scenario 
the fluctuations of these fields and study their dynamics.
We will find out that the inflaton fluctuations are coupled to scalar perturbations of the metric while tensor perturbations constitute the real degrees of freedom of the gravitational field, {\em i.e.} gravitational waves\footnote{See \cite{Ashoorioon:2012kh,Krauss:2013pha} for a discussion about the quantum/classical origin of inflationary GW.}.

\subsubsection{Perturbed tensors}\label{sezionetens}
To get the dynamical equations for the perturbations, we have to perturb tensor objects, the metric and the stress-energy tensor. The most useful way to do it consists in decomposing perturbations in parts which have well-defined transformation properties with respect to the underlying three-dimensional space.

\paragraph{Perturbations of the metric tensor}
Defining the conformal time $\tau\equiv \int {\rm d}t/a\left(t\right)$, the perturbed FRW metric can be decomposed in the following way \cite{Bruni:1996im,Matarrese:1997ay,Acquaviva:2002ud}: 
\begin{align}\label{metric}	
&g_{00}=-a\left(\tau\right)\left(1+2\sum^{+\infty}_{r=1}\frac{1}{r!}\Psi^{\left(r\right)}\right)\\
	&g_{0i}=a^{2}\left(\tau\right)\sum^{+\infty}_{r=1}\frac{1}{r!}\omega^{\left(r\right)}_{i}\\	&g_{ij}=a^{2}\left(\tau\right)\left\{\left[1-2\left(\sum^{+\infty}_{r=1}\frac{1}{r!}\Phi^{\left(r\right)}\right)\right]\delta_{ij}+\sum^{+\infty}_{r=1}\frac{1}{r!}h^{\left(r\right)}_{ij}\right\}\,,
\end{align}
where we can recognize the background metric \eqref{frw} and where the functions $\Phi^{\left(r\right)}$, $\omega_{i}^{\left(r\right)}$, $\Psi^{\left(r\right)}$, $h_{ij}^{\left(r\right)}$ represent the \textit{r}th-order perturbations of the metric and $h_{ij}^{\left(r\right)}$ is a transverse ($\partial^{i}h_{ij}^{T}=0$) and traceless ($h_{i}^{\left(r\right)i}=0$) tensor.\\
It is useful to decompose these quantities in objects with well-defined transformation under spatial rotations \cite{Bardeen:1980kt,Kodama:1985bj}, since their dynamics is uncoupled at first order. 
Exploiting Helmholtz theorem, we can decompose each vector object into a solenoidal and a longitudinal part, respectively called \textit{vector part} and \textit{scalar part}:
\begin{equation}	
\omega_{i}=\partial_{i}\omega^{\parallel}+\omega_{i}^{\bot}\,,
\end{equation}
where $\omega_{i}^{\bot}$ is a solenoidal vector, {\em i.e.} $\partial^{i}\omega_{i}^{\bot}=0$ and $\omega^{\parallel}$ the longitudinal object. Similarly, the traceless part of the spatial metric can be written as
\begin{equation}
	h_{ij}=D_{ij}h^{\parallel}+\partial_{i}h_{j}^{\bot}+\partial_{j}h_{i}^{\bot}+h_{ij}^{\rm T}\,,
\end{equation}
where $h^{\parallel}$ is a suitable scalar function, $h_{i}^{\bot}$ is a solenoidal vector field, and the \textit{tensor part} $h^{T}_{ij}$ is symmetric, solenoidal and trace-free. 
We have used the trace-free operator $D_{ij}:=\partial_{i}\partial_{j}-\delta_{ij}\nabla^{2}/3$, and we omitted the apex $\left(r\right)$ for simplicity. Hereafter, where we neglect such an apex we mean a perturbation of first order.

\paragraph{Perturbations of the stress-energy tensor}
The stress-energy tensor for a fluid can be written as
\begin{equation}\label{stress}
	T_{\mu\nu}=\left(\rho+P_{0}\right){\rm u}_{\mu}{\rm u}_{\nu}+P_{0}g_{\mu\nu}+\pi_{\mu\nu}\,,
\end{equation}
with $\rho$ the energy density, $P_{0}$ the pressure, ${\rm u}^{\mu}$ the four-velocity and $\pi_{\mu\nu}$ the anisotropic stress tensor. 
The latter tensor is subject to the constraints $\pi_{\mu\nu}{\rm u}^{\nu}=0$, $\pi^{\mu}_{\nu}=0$, and vanishes for a perfect fluid or a minimally coupled scalar field.
Perturbing eq.\eqref{stress} and decomposing each physical quantity according to its transformation properties, the first-order components of the stress-energy tensor can be written as:
\begin{align}
	&T^{0}_{\,0}=-\rho_{0}+\delta\rho\,,\\
	&T^{i}_{\,i}=3\left(P_{0}+\delta P\right)=3P_{0}\left(1+\pi_{\rm L}\right)\,,\label{pi} \\
	&T^{i}_{\,0}=T^{0}_{\,i}=0\,,\\
	&T^{i}_{\,j}=P_{0}\left[\left(1+\pi_{\rm L}\right)\delta^{i}_{\:j}+\pi_{{\rm T},\,\,j}^{\,\,\,\,i}\right]\,,
\end{align}
where we have neglected vector perturbations. $\pi_{\rm L}$ is interpreted as the amplitude of an isotropic pressure perturbation and correspondingly $\pi_{\rm T}$ is interpreted as the amplitude of an 
anisotropic stress perturbation, practically imperfections of the fluid.

\paragraph{The gauge problem}
Because of the Einstein's equations, we have to study at the same time inflaton and metric perturbations. We have then to manipulate perturbations of objects which live on a manifold, such as the stress-energy tensor, and at the same time consider the perturbation of the manifold itself, as described by the metric tensor. This situation determines the so-called gauge problem \cite{Bardeen:1980kt}: a generic perturbation $\Delta T$ of a tensor field $T$ is usually defined as the difference between the value $T$ has in the physical (that is perturbed) space-time and the value $T_{0}$the same quantity has in the given background space-time. The two considered tensors are so defined on two different varieties, the physical and the background space-times. However, in order to make the comparison of tensors meaningful, one has to consider them at the same point. Therefore they can be compared only after a prescription for identifying points of these varieties is given. This is a \textit{gauge choice}, that is a one-to-one correspondence between the background and the physical space-time. A change in the correspondence between physical and background points, keeping the background coordinates fixed, is called a \textit{gauge transformation}.\\
The standard procedure to address the issue consists in finding the relation between quantities defined in several gauges and then constructing variables that do not change under gauge transformations and which describe the physical quantities, that is \textit{gauge-invariant} objects \cite{Kodama:1985bj,Mukhanov:1990me}. Tensor perturbations $h_{ij}$ are gauge-invariant objects at linear order.

\subsubsection{The Dynamics}
Here we present the dynamics of the inflationary perturbations at linear order.\\
The system of interest can be described by the action of a scalar field minimally coupled to gravity, {\em i.e.}
\begin{equation}\label{azione}
	S=\int {\rm d}^{4}x\sqrt{-g}\left[\frac{1}{2}M_{\rm pl}^{2}R-\frac{1}{2}g^{\mu\nu}\partial_{\mu}\varphi\partial_{\nu}\varphi-V\left(\varphi\right)\right]\,,
\end{equation}
where $R$ is the Ricci scalar. 
We know that with the energy density dominated by a scalar field, the Universe metric is of the form \eqref{frw}, where the evolution of the scale-factor depends on the relation between the kinetic and the potential energy of the scalar field. 
Then, the background metric that follows from the previous action is described by a FRW metric.\\
In the next sections we will examine separately each kind of perturbation. As anticipated, at first order tensor perturbations will result uncoupled with the other ones. 
Moreover, we will find that the scalar (when expressed in terms of a suitable gauge-invariant potential) and the tensor perturbations remain almost frozen until their wavelengths correspond to super-horizon scales, 
so that the amplitude at the time they re-enter into the causally connected region is the same as the first horizon crossing during inflation.

\paragraph{The power-spectrum}
An efficient way to characterize the properties of a field perturbations is given by the power-spectrum. For a generic random field $g(\textbf{x},t)$, which can be expanded in Fourier space as
\begin{equation}	
g(\textbf{x},t)=\int{\frac{{\rm d}^{3}\textbf{k}}{(2\pi)^{3/2}}\,e^{i\textbf{k}\cdot\textbf{x}}\,g_{\textbf{k}}(t)}\, ,
\end{equation}
the dimensionless power-spectrum $P_{\rm g}(k)$ is defined as
\begin{equation}	
\langle g_{\textbf{$k_{1}$}},g^{\ast}_{\textbf{$k_{2}$}}\rangle\equiv\frac{2\pi^{2}}{k^{3}}\,P_{\rm g}(k)\,\delta^{(3)}(\textbf{k}_{1}-\textbf{k}_{2})\, ,
\end{equation}
where angle brackets denote ensemble average. The power-spectrum measures the amplitude of the fluctuation at a given mode $k$. This definition leads to the usual relation
\begin{equation}
	\langle g^{2}(\textbf{x},t)\rangle=\int\frac{{\rm d}k}{k}P_{\rm g}(k)\, ,
\end{equation}
which tells that $P_{\rm g}$ is the contribution to the variance per unit logarithmic interval in wave-number $k$.
To describe the slope of the power-spectrum a \textit{spectral index} is also defined in the following manner
\begin{equation}\label{spectralindex}
	n_{\rm g}(\textit{k})-1\equiv\frac{{\rm d\,ln}\,P_{\rm g}}{{\rm d\,ln}\,\textit{k}}
\end{equation}
Let us specify the form that the power-spectrum gets when the random field is a canonically quantized scalar field $\chi$.\\
We split the scalar field as $\chi(\textbf{x},\tau)=\chi(\tau)+\delta\chi(\textbf{x},\tau)$, where $\chi(\tau)$ denotes the homogeneous classical value of the scalar field and $\delta\chi(\tau,\textbf{x})$ the fluctuation.
Before performing the quantization, it is useful to perform the redefinition
$\widetilde{\delta\chi}=a\,\delta\chi$.
We promote $\widetilde{\delta\chi}$ to an operator and we decompose it defining two operators $a_{\textbf{k}}$ and $a_{\textbf{k}}^{\dag}$:
\begin{equation}\label{tilde}	
\widetilde{\delta\chi}(\textbf{x},\tau)=\int\frac{{\rm d}^{3}\textbf{k}}{(2\pi)^{3/2}}\left[u_{k}(\tau)a_{\textbf{k}}e^{i\textbf{k}\cdot \textbf{x}}+u_{k}^{\ast}(\tau)a_{\textbf{k}}^{\dag}e^{-i\textbf{k}\cdot \textbf{x}}\right]\,,
\end{equation}
where $u_{\textbf{k}}$ and $u^{\ast}_{\textbf{k}}$ satisfies the canonical commutation relations $u_{k}^{\ast}u_{k}'-u_{k}u_{k}'^{\ast}=-i$ by
\begin{equation}\label{comm}
	\left[a_{\textbf{k}},a_{\textbf{k}'}\right]=0\, , \qquad [a_{\textbf{k}},a_{\textbf{k}'}^{\dag}]=\delta^{3}(\textbf{k}-\textbf{k}')\,.
\end{equation}
From the redefinition of $\widetilde{\delta\chi}$ and eqs.\eqref{tilde}-\eqref{comm} we get
\begin{equation}
\langle\delta\chi_{{\textbf{k}_{1}}}\delta\chi_{{\textbf{k}_{2}}}^{\ast}\rangle=\frac{\left|\textit{u}_{\textit{k}}\right|^{2}}{a^{2}}\delta^{(3)}(\textbf{k}_{1}-\textbf{k}_{2})\, ,
\end{equation}
which leads to the power-spectrum
\begin{equation}\label{spettro}	
P_{\delta\chi}(\textit{k})=\frac{\textit{k}^{3}}{2\pi^{2}}\left|\delta\chi_{{\textit{k}}}\right|^{2}\, ,
\end{equation}
with $\delta\chi_{{\textit{k}}}\equiv\textit{u}_{\textit{k}}/a$.

\subsubsection{Scalar perturbations}\label{scalaristandard}
In the present section we only deal with scalar perturbations. Later, we will concentrate on the tensor perturbations in a dedicated section. For a detailed analysis of scalar perturbations see \cite{liddle2000cosmological} and refs. therein.

\paragraph{Gauge-invariant curvature perturbation of the uniform energy-density hypersurfaces}
We need a gauge-invariant quantity which univocally describes scalar perturbations. Let us work with a space-time described by the perturbed metric \eqref{metric} at first order. Consider the intrinsic spatial curvature on hyper-surfaces of constant 
conformal time at linear order,
\begin{equation}
	^{\left(3\right)}R=\frac{4}{a^{2}}\nabla^{2}\hat{\Phi}\,\qquad\mbox{where}\qquad
	\hat{\Phi}\equiv\Phi+\frac{1}{6}\nabla^{2}\chi^{\parallel}\,.
\end{equation}
$\hat{\Phi}$ is usually referred to as the \textit{curvature perturbation}, however it is not a gauge-invariant quantity, since under a transformation on constant time hyper-surfaces $\tau\rightarrow\tau+\alpha$ we have: $\hat{\Phi}\rightarrow\tilde{\hat{\Phi}}=\hat{\Phi}-\mathcal{H}\alpha$,
where $\mathcal{H}\equiv a'/a$ is the Hubble parameter in conformal time and the prime denote differentiation w.r.t. it.
What we need is a gauge-invariant combination that reduces to the curvature perturbation choosing a particular gauge. Consider the following expression:
\begin{equation}\label{zeta}
	-\zeta\equiv\hat{\Phi}+\mathcal{H}\frac{\delta\rho}{\rho'}\,.
\end{equation}
Considering the $\hat{\Phi}$ transformation and the gauge transformation for scalars, the quantity \eqref{zeta} results gauge-invariant and it is referred to a \textit{gauge-invariant curvature perturbation of the uniform energy-density hyper-surfaces}. 

\paragraph{Power-spectrum of curvature perturbations}
A way to track the evolution of $\zeta$ consists in exploiting the perturbed Klein-Gordon equation for the field $\varphi$ from the action \eqref{azione}:
\begin{equation}\label{eqmotodelta}
\delta\varphi''+2\mathcal{H}\delta\varphi'-\nabla^{2}\delta\varphi+a^{2}\delta\varphi\frac{\partial^{2}V}{\partial\varphi^{2}}a^{2}+2\Psi\frac{\partial V}{\partial\varphi}-\varphi'_{0}\left(\Psi'+3\Phi'+\nabla^{2}\omega^{\parallel}\right)=0\,.
\end{equation}
To get a simpler equation of motion we introduce the so-called Sasaki-Mukhanov gauge-invariant variable \cite{Sasaki:1986hm}
\begin{equation}\label{sasaki}
	Q_{\varphi}\equiv\delta\varphi+\frac{\varphi'}{\mathcal{H}}\Phi\,.
\end{equation}
This quantity is linked to $\zeta$, so if we are able to solve \eqref{eqmotodelta} for this variable we are also able to get the power-spectrum for $\zeta$.\\
Let us introduce the field $\tilde{Q}_{\varphi}=aQ_{\varphi}$, so that the Klein-Gordon equation reads \cite{Taruya:1997iv}
\begin{equation}\label{moto1}
	\tilde{Q_{\varphi}}''+\left(k^{2}-\frac{a''}{a}+\mathscr{M}^{2}_{\varphi}a^{2}\right)\tilde{Q}_{\varphi}=0\,,
\quad
\mbox{where}
\quad
\mathscr{M}^{2}_{\varphi}=\frac{\partial^{2} V}{\partial\varphi^{2}}-\frac{8\pi G}{a^{3}}\left(\frac{a^{3}}{H}\varphi^{2}\right)\,.
\end{equation}
In the slow-roll approximation the latter expression reduces to
$\mathscr{M}_{\varphi}^{2}/H^{2}=3\eta-6\epsilon$,
where $\eta$ and $\epsilon$ are the slow-roll parameters defined in eqs.\eqref{parametri}.
Moving to Fourier space, the solution of \eqref{moto1} is a combination of the Hankel functions of the first and second order, which for super-horizon scales and at lowest order in the slow-roll parameters, are approximated by\footnote{We will specify this procedure in section \ref{sezionegw}.}
\begin{equation}\label{pow}	
\left|Q_{\varphi}\left(k\right)\right|=\frac{H}{\sqrt{2k^{3}}}\left(\frac{k}{aH}\right)^{3/2-\nu_{\varphi}}\,,
\end{equation}
where $\nu_{\varphi}\simeq \frac{3}{2}+3\epsilon-\eta$.
To obtain the $\zeta$ power-spectrum, let us consider the gauge-invariant curvature perturbation on comoving hyper-surfaces, which, in the case of a stress-energy tensor of a single scalar field, reads \cite{Lukash:1980iv,Lyth:1984gv,Lidsey:1995np,Lyth:1998xn}
\begin{equation}\label{curvatura}
	\mathscr{R}\equiv\hat{\Phi}+\frac{\mathcal{H}}{\varphi'}\delta\varphi\,. 
\end{equation}
From eq.\eqref{sasaki} we immediately have $\mathscr{R}=\mathcal{H}Q_{\varphi}/\varphi'$.
On the other hand, $\mathscr{R}$ is related to the curvature perturbation $\zeta$ by
\begin{equation}
	-\zeta=\mathscr{R}+\frac{2\rho}{9\left(\rho+P\right)}\left(\frac{k}{aH}\right)^{2}\Psi\,,
\end{equation}
where $\Psi$ is the perturbation that appears in eq.\eqref{metric}. From this relation on large scales we have $\mathscr{R}\simeq-\zeta$. Then, combining eq.\eqref{pow} and the expression of $\mathcal{R}$ into eq.\eqref{spettro}, we obtain the power-spectrum for $\zeta$ on large scales:
\begin{equation}\label{scal}	
P_{\zeta}=\left(\frac{H^{2}}{2\pi\dot{\varphi}}\right)^{2}\left(\frac{k}{aH}\right)^{3-2\nu_{\varphi}}\simeq\left(\frac{H^{2}}{2\pi\dot{\varphi}}\right)^{2}_{\ast}\,,
\end{equation}
where the asterisk denotes quantities evaluated at the epoch a given perturbation mode leaves the horizon during inflation, 
that is $k=aH$. Equation \eqref{scal} shows that curvature perturbations remain  time-independent on super-horizon scales. 
So, the solution obtained for $\zeta$ is valid throughout the different evolution eras of the Universe until the mode remains super horizon. We will see that the same happens with tensor perturbations.\\
The spectral index at the lowest order in slow-roll reads
\begin{equation}
	n_{\zeta}-1=3-2\nu_{\varphi}=-6\epsilon+2\eta\,.
\end{equation}

\paragraph{Scalar power-spectrum parametrization}
In order to compare these theoretical predictions with observational data, it is useful to introduce a phenomenological parametrization of the power-spectrum \cite{Ade:2015lrj}
\begin{equation}
	P_{\rm S}=A_{\rm S}\left(\frac{k}{k_{\ast}}\right)^{n_{\rm S}-1+\frac{1}{2}\frac{\mathrm{d}n_{\rm S}}{\mathrm{d\,ln}k} {\rm ln}\left(k/k_{\ast}\right)+...}\,,
\end{equation}
where $A_{\rm S}$ is the amplitude of the perturbations to a fixed pivot scale $k_{\ast}$, $n_{\rm S}$ is the spectral index and $\mathrm{d}n_{\rm S}/\mathrm{d\,ln}k$ the running of the spectral index.
These quantities are usually expressed by the so called Hubble flow-functions $\epsilon_{i}$ \cite{Liddle:1994cr,Liddle:1994dx}, which express the conditions of slow-roll in terms of deviations with respect to an exact exponential expansion: 
$\epsilon_{1}=-\dot{H}/H^{2}$, $\epsilon_{i+1}\equiv\dot{\epsilon}_{i}/\left(H\epsilon_{i}\right)$. These parameters are linked to those in eq.\eqref{parametri} by: $\epsilon_{1}\simeq \epsilon$ nad $\epsilon_{2}\simeq -2\eta+4\epsilon$, at the first order in 
slow-roll parameters.
Up to second order the power-spectrum parameters read \cite{Ade:2015ava}:
\begin{align}	\label{tiltscalari}
&n_{\rm S}-1=-2\epsilon_{1}-\epsilon_{2}+-2\epsilon^{2}_{1}-\left(2C+3\right)\epsilon_{1}\epsilon_{2}-C\epsilon_{2}\epsilon_{3}\\
&\frac{\mathrm{d}n_{\rm S}}{\mathrm{d\,ln}k} =-2\epsilon_{1}\epsilon_{2}-\epsilon_{2}\epsilon_{3}
\end{align}
where $C=\mathrm{ln}2+\gamma_{\rm E}-2\approx-0.7296$, with $\gamma_{\rm E}$ the Euler-Mascheroni constant.

\paragraph{First order vector perturbations}
For vector perturbations there is only a constraint equation which relates the gauge-invariant vector metric perturbation to the divergence-free velocity of the fluid, which obviously vanishes in the presence of scalar fields only.

\subsection{Gravitational waves from inflation}\label{sezionegw}
The inflationary scenario predicts also the production of a background of stochastic GW \cite{Grishchuk:1974ny,Starobinsky:1979ty,Rubakov:1982df,Fabbri:1983us,Abbott:1984fp}.
Tensor fluctuations of the metric represent the degrees of freedom of the gravitational sector: there are no constraint equations coming from the stress-energy continuity equation for these modes (in the case of a perfect fluid). 
Their evolution is only regulated by the traceless spatial part of the Einstein equation, which, in the presence of perfect fluids does not contain direct influence from the energy content of the Universe except for the underlying background solution.
We will see later that a coupling between GW and the content of the Universe grows up only in the presence of anisotropic stress tensor.

\subsubsection{Evolution equation and power-spectrum}\label{modistandard}
Perturbing \eqref{azione} at first order leads to the following action for tensor perturbations \cite{Grishchuk:1974jy,Grishchuk:1977zz}:
\begin{equation}\label{azionestandard}
   S_{\rm T}^{\left(2\right)}=\frac{M_{\rm pl}^{2}}{8}\int\,{\rm d}^{4}x\,a^{2}\left(t\right)\left[\dot{h}_{ij}\dot{h}_{ij}-\frac{1}{a^{2}}\left(\nabla h_{ij}\right)^{2}\right]\,;
\end{equation}
as already mentiones $h_{ij}$ is a gauge-invariant object, so varying the action with respect to this quantity, we get the required equation of motion
\begin{equation}\label{onde}
	\nabla^{2}h_{ij}-a^{2}\ddot{h}_{ij}-3a\dot{a}\dot{h}_{ij}=0\,.
\end{equation}
It is now clear that tensor perturbations solve a wave equation, hence the name \textit{gravitational waves}.
Recalling that $h_{ij}$ is symmetric, transverse and trace-free,
the solutions of eq.\eqref{onde} present the following form
\begin{equation}\label{soluzione}
	h_{ij}\left(\textbf{x},t\right)=h\left(t\right)e_{ij}^{\left(+,\times\right)}\left(\textbf{x}\right)\,,
\end{equation}
where $e_{ij}^{\left(+,\times\right)}$ is a polarization tensor satisfying the conditions $e_{ij}=e_{ji}$, $k^{i}e_{ij}=0$, $e_{ii}=0$, with $+,\times$ the two GW polarization states \cite{Misner:1974qy}.
Equation \eqref{soluzione} reflects the fact that tensor modes are left with two physical degrees of freedom: starting from six of the symmetric tensor $h_{ij}$, four constraints are given by the requirement of being trace-free and transverse. 
In summary the most general solution of eq.\eqref{onde} reads
\begin{equation}	
h_{ij}\left(\textbf{x},t\right)=\sum_{\lambda=\left(+,\times\right)}h^{\left(\lambda\right)}\left(t\right)e^{\left(\lambda\right)}_{ij}\left(\textbf{x}\right)\,.
\end{equation}
To get the solution of the equation of motion it is useful to perform the transformation
\begin{equation}
	v_{ij}\equiv \frac{a M_{\rm pl}}{\sqrt{2}}h_{ij}\,.
\end{equation}
In terms of $v_{ij}$ the action \eqref{azionestandard} reads
\begin{equation}
	S_{\rm T}^{\left(2\right)}=\frac{M_{\rm pl}^{2}}{8}\int\,{\rm d}^{4}x \left[v'_{ij}v'_{ij}-\left(\nabla v_{ij}\right)^{2}+\frac{a''}{a}v_{ij}v_{ij}\right]\,,
\end{equation}
which can be interpreted as the action for two scalar fields in Minkowski space-time, with effective mass squared equal to $a''/a$ \footnote{The appearance of this effective mass term indeed follows from the non-invariance under Weyl transformations 
of the tensor mode action \eqref{azionestandard}.}.
Being interested in the power-spectrum, we move to Fourier space and write
\begin{equation}\label{azionev}
v_{ij}\left(\textbf{x},t\right)=\int\frac{{\rm d}^{3}\textbf{k}}{\left(2\pi\right)^{3}}\sum_{\lambda=\left(+,\times\right)}v^{\left(\lambda\right)}_{\textbf{k}}\left(t\right)e^{\left(\lambda\right)}_{ij}\left(k\right)e^{i\textbf{k}\cdot \textbf{x}} \,,
\end{equation}
where $v^{\left(\lambda\right)}_{\textbf{k}}$ is the Fourier transform of the scalar amplitude. From \eqref{azionev}, the equation of motion for each mode $v^{\left(\lambda\right)}_{\textbf{k}}$ then reads:
 \begin{equation}\label{mototens}	
v_{\textbf{k}}^{\left(\lambda\right)}\,''+\left(k^{2}-\frac{a''}{a}\right)v^{\left(\lambda\right)}_{\textbf{k}}=0\,.
\end{equation}
We obtained a wave equation.
Let us study the qualitative behavior of its solutions. We can identify two main regimes depending on the relative magnitude of the second and third term.
First, consider the case in which $a''/a \ll k^2$. Ignoring the second term in parenthesis, the equation for $v_{\textbf{k}}$ becomes that of a free harmonic oscillator, so that tensor perturbations $h_{ij}$
oscillate with a damping factor $1/a$. This approximation corresponds to overlook the effect of the expansion of the Universe. To make explicit the physical condition corresponding to this regime, notice that, since $a''/a\sim (a'/a)^{2}$, $a''/a\ll k^2$ corresponds to $k\gg aH$, {\em i.e.} to the \textit{sub-horizon} behavior (check for example the case of a de Sitter space-time where $a\left(\tau\right)\sim 1/\tau$). Keeping in this regime, the solution of \eqref{mototens} reads
\begin{equation}
	v_{k}\left(\tau\right)=Ae^{ik\tau}\,,
\end{equation}
which means that the amplitude of the modes of the original field $h_{ij}$ decrease in time with the inverse of the scale-factor as an effect of the Universe expansion.
Consider now the regime in which the second term is negligible with respect to the third one: $k^2\ll a''/a$. There are two possible solutions of the equation \eqref{mototens}:
\begin{equation}
	v_{k}\left(\tau\right)\propto a\,,\qquad
\mbox{and}\qquad
	v_{k}\left(\tau\right)\propto 1/a^{2}\,,
\end{equation}
which corresponds to $h\propto {\rm const}$ and a decreasing in time solution, respectively. This situation clearly corresponds to the \textit{super-horizon} regime. In particular we will be interested in the solutions with constant amplitude.\\
Now we calculate more accurately the power-spectrum of tensor perturbations, solving \eqref{mototens}.
We perform the standard quantization of the field writing
\begin{equation}	
v^{\left(\lambda\right)}_{\textbf{k}}=v_{k}\left(\tau\right)\hat{a}_{\textbf{k}}^{\left(\lambda\right)}+v^{\ast}_{k}\left(\tau\right)\hat{a}_{-\textbf{k}}^{\left(\lambda\right)\dag}\,,
\end{equation}
where the modes are normalized so that they satisfy $v_{k}^{\ast}v'_{k}-v_{k}v'^{\ast}_{k}=-i$, and this condition ensures that $\hat{a}_{\textbf{k}}^{\left(\lambda\right)}$ and $\hat{a}_{\textbf{-k}}^{\left(\lambda\right)\dag}$ behave as the canonical creation and annihilation operators.
Following the simplest and most natural hypothesis, as initial condition, we assume that the Universe was in the vacuum state defined as $\hat{a}_{\textbf{k}}^{\left(\lambda\right)}|0>=0$ at past infinity, that is the ``Bunch-Davies vacuum state" \cite{Bunch:1978yq}.\\
Equation \eqref{mototens} is a Bessel equation, which, in case of de Sitter spacetime, has the following exact solution \cite{Abbott:1984fp}:
\begin{equation}\label{soltens}
	v_{\textbf{k}}\left(\tau\right)=\sqrt{-\tau}\left[C_{1}H_{\nu}^{\left(1\right)}\left(-k\tau\right)+C_{2}H_{\nu}^{\left(2\right)}\left(-k\tau\right)\right]\,,
\end{equation}
where $C_{1},C_{2}$ are integration constants, $H_{\nu}^{\left(1\right)},H_{\nu}^{\left(2\right)}$ are Hankel functions of first and second order and
$\nu\simeq 3/2+\epsilon$.
Remember we have negative sign to $\tau$ because, from its definition, it lies in $-\infty<\tau<0$. To determining $C_{1}$ and $C_{2}$, we impose that in the UV regime, that is sub-horizon scales, the solution matches the
plane-wave solution $e^{-ik\tau}/\sqrt{2k}$ found before. This hypothesis is a direct consequence of the Bunch-Davies vacuum condition.
Using the asymptotic form of Hankel functions
\begin{equation}
	H_{\nu}^{\left(1\right)}\left(x\gg 1\right)\sim\sqrt{\frac{2}{\pi x}}e^{i\left(x-\frac{\pi}{2}\nu-\frac{\pi}{4}\right)}\,,\qquad H_{\nu}^{\left(2\right)}\left(x\gg 1\right)\sim\sqrt{\frac{2}{\pi x}}e^{-i\left(x-\frac{\pi}{2}\nu-\frac{\pi}{4}\right)}\,,
\end{equation}
the second term in the solution has negative frequency, so that we have to fit $C_{2}=0$, while matching the asymptotic solution to a plane wave leads to
\begin{equation}
	C_{1}=\frac{\sqrt{\pi}}{2}e^{i\left(\nu+\frac{1}{2}\right)\frac{\pi}{2}}\,.
\end{equation}
Then the exact solution becomes
\begin{equation}\label{solstandard}	
v_{\textbf{k}}=\frac{\sqrt{\pi}}{2}e^{i\left(\nu+\frac{1}{2}\right)\frac{\pi}{2}}\sqrt{-\tau}H_{\nu}^{\left(1\right)}\left(-k\tau\right)\,.
\end{equation}
In particular, for our purpose we are interested in the super-horizon wavelength behaviour, where the Hankel function reads
\begin{equation}
	H_{\nu}^{\left(1\right)}\left(x\ll 1\right)\sim\sqrt{2/\pi}e^{-i\frac{\pi}{2}}2^{\nu-\frac{3}{2}}\left[\Gamma\left(\nu\right)/\Gamma\left(3/2\right)\right]x^{-\nu}\,,
\end{equation}
so that the fluctuations on such scales become
\begin{equation}
	v_{\textbf{k}}=e^{i\left(\nu-\frac{1}{2}\right)\frac{\pi}{2}}2^{\left(\nu-\frac{3}{2}\right)}\frac{\Gamma\left(\nu\right)}{\Gamma\left(3/2\right)}\frac{1}{\sqrt{2k}}\left(-k\tau\right)^{\frac{1}{2}-\nu}\,,
\end{equation}
where $\Gamma$ is the Euler function.\\
With the latter equation we can now write the sought tensor power-spectrum. Employing the expression \eqref{spettro} and considering that here we deal with two polarization states, we have
\begin{equation}\label{spettrogw}
	P_{\rm T}\left(k\right)=\frac{k^{3}}{2\pi^{2}}\sum_{\lambda}\left|h_{\textbf{k}}^{\left(\lambda\right)}\right|^{2}\,,
\end{equation}
so that on super-horizon scales the following power-spectrum holds
\begin{equation}\label{tens}	
P_{\rm T}\left(k\right)=\frac{8}{M_{\rm pl}^{2}}\left(\frac{H}{2\pi}\right)^{2}\left(\frac{k}{aH}\right)^{-2\epsilon}\,.
\end{equation}
Notice that it is almost scale-invariant, which means that all the GW produced, nearly frozen on super-horizon scales, have all the same amplitude. Moreover, from eq.\eqref{parametri} the tensor spectral index, defined in eq.\eqref{spectralindex}, has to be negative in order to 
have $\dot{H}<0$, that is in order to satisfy the Null Energy Condition (NEC) \cite{Lucchin:1985wy}. In this case the power-spectrum is called \textit{red}, while for $n_{\rm T}>0$ it is indicated as \textit{blue} \cite{Mollerach:1993sy}. Later on we will refer to the case in which $n_{\rm T}=0$ as \textit{scale-invariant}.

\paragraph{Tensor power-spectrum parametrization}
In analogy with the scalar perturbation power-spectrum, it is useful to parametrize \eqref{tens} in the following manner \cite{Ade:2015ava}
\begin{equation}\label{spettrotensori}
	P_{\rm T}\left(k\right)=A_{\rm T}\left(\frac{k}{k_{\ast}}\right)^{n_{\rm T}+\frac{1}{2}\frac{\mathrm{d}n_{\rm T}}{\mathrm{d}{\rm ln}k} {\rm ln}\left(k/k_{\ast}\right)+...}\,,
\end{equation}
where $A_{\rm T}$ is the tensor amplitude at some pivot scale $k_{\ast}$, $n_{\rm T}$ is the tensor spectral index, and $\mathrm{d}n_{\rm T}/\mathrm{d}{\rm ln}k$ the running of the spectral index.\\
Again, introducing the Hubble flow-functions, we can rewrite these quantities in terms of the Hubble parameter and its derivatives. Up to second order they read
\begin{align}\label{tilttensori}
&n_{\rm T}=-2\epsilon_{1}-2\epsilon^{2}_{1}-2\left(C+1\right)\epsilon_{1}\epsilon_{2}\\
&\frac{\mathrm{d}n_{\rm T}}{\mathrm{d\,ln}k} =-2\epsilon_{1}\epsilon_{2}
\end{align}
where $C=\mathrm{ln}2+\gamma_{\rm E}-2\approx-0.7296$.

\subsubsection{Consistency relation}\label{sezconsistency}
In the considered inflationary scenario an interesting consistency relation holds between quantities which involve tensor perturbations. To get it, we introduce the \textit{tensor-to-scalar ratio}
\begin{equation}
	r\left(k_{\ast}\right)\equiv \frac{A_{\rm T}\left(k_{\ast}\right)}{A_{\rm S}\left(k_{\ast}\right)}
\end{equation}
that yields the amplitude of the GW with respect to that of the scalar perturbations at some fixed pivot scale $k_{\ast}$.
From eqs.\eqref{scal}-\eqref{tens}, this quantity depends on the time-evolution of the inflaton field, as
\begin{equation}\label{evoluzionephi}
	r=\frac{8}{M^{2}_{\rm pl}}\left(\frac{\dot{\varphi}}{H}\right)^{2}\,,
\end{equation}
that is $r=16\epsilon$.
Furthermore, we have shown that a nearly scale-invariant spectrum of tensor modes is expected, being $n_{\rm T}=-2\epsilon$.
Therefore at the lowest order in slow-roll parameters, one finds the following consistency relation \cite{liddle2000cosmological}:
\begin{equation}\label{consistency}
	r=-8n_{\rm T}\,.
\end{equation}
Clearly, this equality can be checked only with a measure of the tensor power-spectrum, {\em i.e.} not only of its amplitude, but also of its spectral index. 
Furthermore if this relation really holds true it means that it will be very hard to measure any scale dependence of the tensors, since a large spectral index would invalidate the consistency relation. 
At present we have only an upper bound on the tensor-to-scalar ratio: $r_{0.05}<0.07$ at $95\%$ C.L. \cite{Array:2015xqh}, assuming the consistency relation \eqref{consistency}, where the subscript indicates the pivot scale in $\mathrm{Mpc}^{-1}$ units. 

\subsubsection{Second-order gravitational waves}
Up to now we have considered phenomena concerning first-order perturbation theory on a FRW background. At that order, scalar, vector and tensor modes evolve governed by uncoupled equations of motion. This fact does not hold at higher order. In particular, the combination of first-order scalar perturbations represents a source for GW at second order. This means that when curvature perturbations are present we always have generation of GW, even if tensor perturbations of first order are absent. See section \ref{secondo}.

\subsubsection{Post-inflationary evolution of gravitational waves}
Let us have a look at how GW behave at the time of radiation and matter domination, when accelerated expansion has already ended. Inflation stretches tensor perturbations wavelengths to super-horizon scales, making their amplitude almost frozen. During the radiation and subsequent matter eras, tensor perturbation wavelengths re-enter the horizon sequentially. When this happens the decaying solution has substantially disappeared, so what re-enters the causally connected space is the almost scale-invariant power-spectrum at the time of first horizon crossing, which occurred during inflation. Then, modes that are inside the horizon, start oscillating with the amplitude damped by a factor $1/a$. In particular, during radiation and matter dominance the scale-factor evolves as $a\sim\tau$ and $a \sim \tau^{2}$ respectively, so that eq.\eqref{mototens} becomes a Bessel equation with the following solutions respectively, in terms of $h_{ij}$ modes:
\begin{equation}\label{radiation}
	h_{k}\left(\tau\right)=h_{k,{\rm i}}j_{\rm 0}\left(k\tau\right)\,,\qquad h_{k}\left(\tau\right)=h_{k,{\rm i}}\left(\frac{3j_{1}\left(k\tau\right)}{k\tau}\right)\,,
\end{equation}
where $h_{k,{\rm i}}$ is the amplitude at horizon crossing and $j_{0}$ and $j_{1}$ are the Bessel functions. Looking at the dependence on $k$, these solutions tells us that tensor perturbations start oscillating with a damping factor greater for high frequency waves.\\
During an era of pure dominance of the cosmological constant, the space-time assumes a de Sitter metric so that the scale-factor evolves in a exponential way, as during inflation in case of $\epsilon=0$. Then, in such an epoch, the form of the solution of the GW equation of motion \eqref{mototens} is given by eq.\eqref{soltens}.
In a dedicated section we will investigate the features impressed in the present GW energy density due to these different ways of evolving. Recently, the effect due to the presence of scalar fluctuations during the matter dominated era on the GW background has been estimated in terms of a local blue or red shift of the GW spectrum, proportional to the amplitude of scalar perturbations \cite{Alba:2015cms}; see also \cite{Adamek:2015mna}.

\paragraph{Energy-density of gravitational waves}

Let us now introduce some useful definitions, in particular to identify the GW energy-density.
Consider the weak-field limit, where GW can be described as space-time ripples propagating on a \textit{fixed} background.
The vacuum field equations read $G_{\mu\nu}=0$, which is equivalent to $R_{\mu\nu}=0$.
Making explicit the Ricci tensor as a sum of a background term and perturbative terms up to second order, $R_{\mu\nu}=\bar{R}_{\mu\nu}+R_{\mu\nu}^{\left(1\right)}\left(h\right)+R^{\left(2\right)}_{\mu\nu}\left(h\right)+\mathcal{O}\left(h^{3}\right)$, one can deduce from the vacuum equations, how the presence of the GW affects the background $\bar{R}_{\mu\nu}$ (where, for example, $R^{\left(2\right)}_{\mu\nu}\left(h\right)$ indicates the contribution to the Ricci tensor which contains terms as $\sim h\cdot h$). The terms that play this role then can be interpreted as a stress-energy tensor $t_{\mu\nu}$ due to the presence of GW. In this direction it is useful to note that $R_{\mu\nu}$ can be written as a sum of two kinds of terms, those representing a smooth contribution and others which encode the fluctuating part. Each of the two contributions vanishes on its own \cite{Misner:1974qy}. 
The background term $\bar{R}_{\mu\nu}$ varies only on large scales with respect to some coarse-graining scale, therefore we are interested in the equation for the smooth contributions.
The only linear term $R_{\mu\nu}^{\left(1\right)}\left(h\right)$ solves by itself $R_{\mu\nu}^{\left(1\right)}\left(h\right)=0$\footnote{More precisely, the perturbation $h_{\mu\nu}$ may contain non-linear corrections $j_{\mu\nu}$, which lead to a non-linear term, that we call $R_{\mu\nu}^{\left(1\right){\rm NL}}\left(j\right)$. The latter contributes to the fluctuating part of the Ricci tensor, but, being non-linear, is not constrained by the equation just shown in the text. 
%In fact, the only non-vanishing part of $R_{\mu\nu}^{\left(1\right){\rm NL}}$ is the fluctuating one since it contains only terms $\sim j_{\mu\nu}$ which are clearly fluctuating.
In fact, in general, smoothed parts can be obtained only from combinations $\sim h_{\mu\nu}h_{\rho\sigma}$, where the two high frequencies of each perturbation $h_{\mu\nu}$ can cancel each other, leading to a smooth contribution \cite{Maggiore:1900zz}.}.
Then, the remaining equation for the smooth part of the vacuum equation reads \cite{Misner:1974qy,Maggiore:1900zz}:
\begin{equation}
	\bar{R}_{\mu\nu}+\langle R_{\mu\nu}^{\left(2\right)}\rangle=0\,,
\end{equation}
where $\langle...\rangle$ indicates the average over several wavelengths which extracts the smooth contribution with respect to the coarse-graining scale. An analogous reasoning can be enlarged to the Einstein tensor, so that one gets the following Einstein equations, in vacuum:
\begin{equation}
	\bar{G}_{\mu\nu}=\bar{R}_{\mu\nu}-\frac{1}{2}\bar{R}\bar{g}_{\mu\nu}=\langle R_{\mu\nu}^{\left(2\right)}\rangle-\frac{1}{2}\bar{g}_{\mu\nu}\langle R^{\left(2\right)}\rangle\,.
\end{equation}
The terms on the RHS tell how the presence of GW affects the background metric, then they can be interpreted as the GW stress-energy tensor $t_{\mu\nu}$, apart from a factor $8\pi{\rm G}$. In terms of the tensor perturbations of the metric it reads \cite{Misner:1974qy}:
\begin{equation}\label{energygw}
	t_{\mu\nu}=\frac{1}{32\pi G}\langle \partial_{\mu}h_{ij}\partial_{\nu}h^{ij}\rangle\,;
\end{equation}
see also \cite{Maggiore:1900zz,Watanabe:2006qe}. From the previous equation, the GW energy-density, on a FRW background, reads
\begin{equation}\label{densitygw}
\rho_{\rm gw}=\frac{1}{32\pi G a^{2}}\langle h'_{ij}\left(\textbf{x},\tau\right)h'^{ij}\left(\textbf{x},\tau\right)\rangle\,.
\end{equation}
However, more often one makes use of the GW energy-density per logarithmic frequency interval, normalized to the critical density $\rho_{\rm c}\equiv 3H^{2}/8\pi G$,
\begin{equation}\label{definizioneomega1}
	\Omega_{\rm GW}\left(k,\tau\right)\equiv \frac{1}{\rho_{\rm c}}\frac{\mathrm{d}\rho_{\rm gw}}{\mathrm{d\,ln} k}.
\end{equation}

\subsection{Why are primordial gravitational waves so interesting?}
Primordial GW represent a very interesting tool to constrain different aspects of the early Universe and of the underlying fundamental physics theory. 

\subsubsection{Test and constrain single-field slow-roll inflation}
If the single-field slow-roll inflationary paradigm holds, a detection of the GW power-spectrum would provide an estimate of the fundamental scales involved.

\paragraph{Energy scale of inflation}
GW carry direct information about their generation mechanism: a measurement of the amplitude of the tensor power-spectrum would provide the way to fix the energy scale of such a mechanism \cite{Lyth:1984yz}. 
From eq.\eqref{parametri}-\eqref{scal}, the scalar power-spectrum is related to the Hubble parameter, evaluated during inflation at the horizon exit of a pivot scale, and to the slow-roll parameter $\epsilon$ in the following manner:
\begin{equation}	
P_{\rm S}\left(k\right)=\frac{1}{2M_{\rm pl}^{2}\epsilon}\left(\frac{H_{\ast}}{2\pi}\right)^{2}\left(\frac{k}{aH_{\ast}}\right)^{n_{\rm S}-1}\,,
\end{equation}
so that a measurement of the amplitude of scalar perturbations would provide an estimate of $H_{\ast}$ in terms of the slow-roll parameter $\epsilon$. Furthermore, the Friedman equations in the slow-roll limit give a relation between the Hubble parameter and the energy scale of inflation $V$: $H^{2}=V/3M_{\rm pl}^{2}$.
In virtue of the latter equation, we can relate the energy scale of inflation at the time when the pivot scale leaves the horizon, directly to the parameter $\epsilon$: $V=24\pi^{2}M_{\rm pl}^{4}A_{\rm S}\epsilon$.
From the link between $\epsilon$ and the tensor-to-scalar ratio $r$ we have $V=\left(3\pi^{2}A_{\rm S}/2\right)M_{\rm pl}^{4}r$, so that considering the scalar amplitude estimated by the Planck Collaboration \cite{Ade:2015lrj} one gets the following relation between the energy scale of inflation at the time when the pivot scale leaves the Hubble radius, and the tensor-to-scalar ratio:
\begin{equation}
	V=\left(1.88\times 10^{16}{\rm GeV}\right)^{4}\frac{r}{0.10}\,.
\end{equation}
Then $r$ provides the energy scale of inflation.

\begin{figure}[ht]
    \centering
    \includegraphics[width=0.7\textwidth]{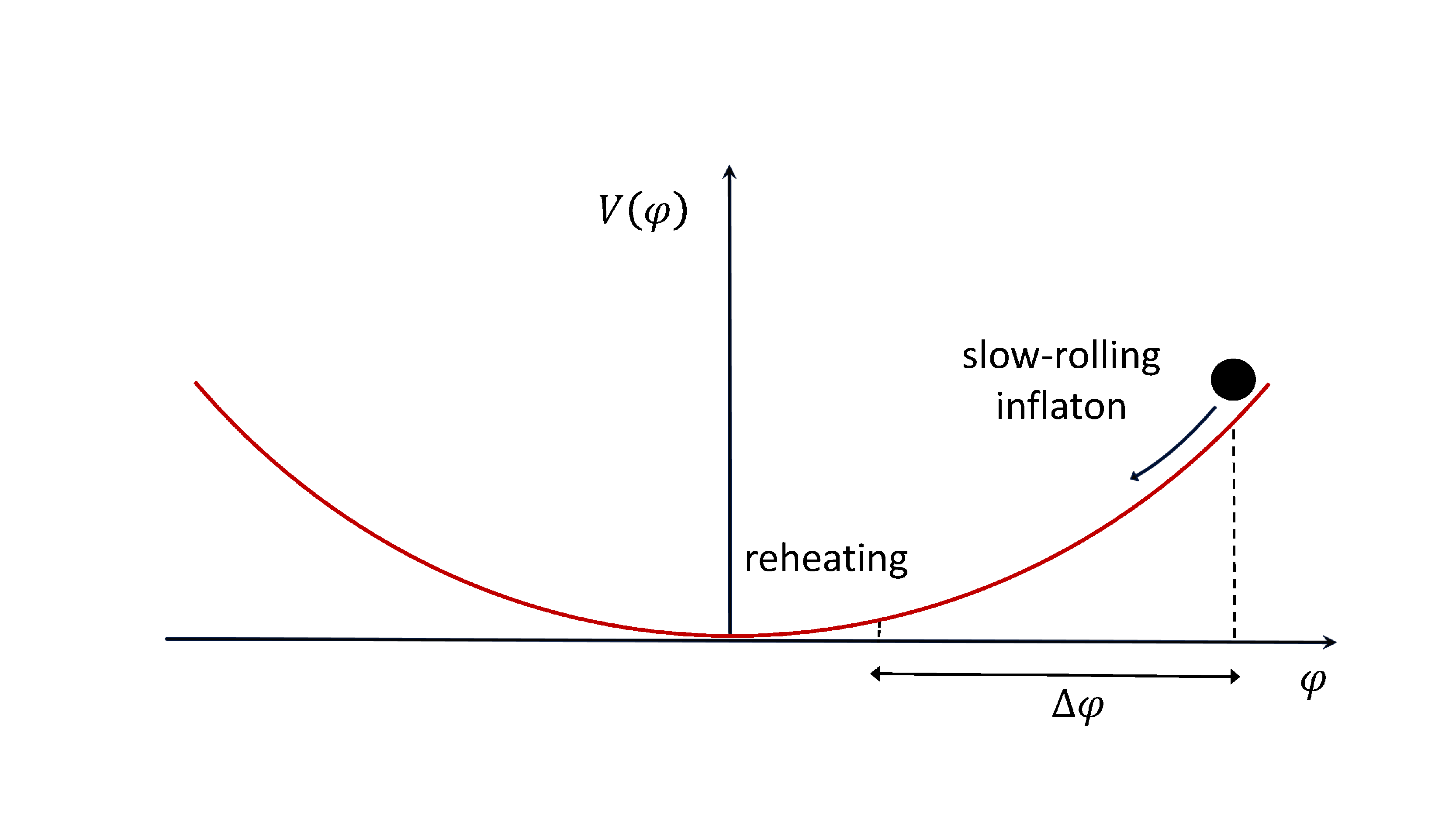}
    \caption{Example of large field inflationary potential. $\Delta\varphi$ indicates the inflaton excursion between the horizon exit of a given comoving scale and the end of inflation.}
  
    \label{largefield}
\end{figure}

\paragraph{Scalar field excursion during inflation}\label{escursione}
An estimation of the tensor-to-scalar ratio might enlighten the variation of the inflaton field expectation value from the horizon crossing of large-scale perturbations to the end of inflation \cite{Lyth:1996im,Boubekeur:2005zm,Baumann:2006cd,Boubekeur:2012xn}.
We have just seen that in the slow-roll model the tensor-to-scalar ratio relates to the slow-roll parameter $\epsilon$ and then to the evolution of the inflaton field as \eqref{evoluzionephi}.
Restoring the definition of the e-foldings $N$, we can express the evolution in time of the field via such a quantity, that is
\begin{equation}
	r=\frac{8}{M_{\rm pl}}\left(\frac{{\rm d}\varphi}{{\rm d}N}\right)^{2}\,.
\end{equation}
Integrating ${\rm d}\varphi$ from the horizon crossing of a pivot scale to the end of inflation and making explicit the dependence of $r$ on $N$ (see figs.\ref{potenziale}-\ref{largefield}), we have
\begin{equation}	
\frac{\Delta\varphi}{M_{\rm pl}}=\left(\frac{r\left(\varphi_{\rm cross}\right)}{8}\right)^{1/2}\int^{N\left(\varphi_{\rm end}\right)}_{N\left(\varphi_{\rm cross}\right)}\left(\frac{r\left(N\right)}{r\left(\varphi_{\rm cross}\right)}\right)^{1/2}{\rm d}N\,.
\end{equation}
One can consider the second factor as the effective number of e-foldings
\begin{equation}	
N_{\rm e}\equiv\int^{N\left(\varphi_{\rm end}\right)}_{N\left(\varphi_{\rm cross}\right)}\left(\frac{r\left(N\right)}{r\left(\varphi_{\rm cross}\right)}\right)^{1/2}{\rm d}N\,,
\end{equation}
so that
\begin{equation}
	\frac{\Delta\varphi}{M_{\rm pl}}=\left(\frac{r\left(\varphi_{\rm cross}\right)}{8}\right)^{1/2}N_{\rm e}\,.
\end{equation}
$N_{\rm e}$ depends on the evolution of the tensor-to-scalar ratio during inflation and so it is model-dependent. In particular for the standard slow-roll model, $r$ is constant up to the second order, 
then the integral becomes simply the number of e-foldings. Keeping this approximation, in agreement with the chosen pivot scale, one can find a lower bound for the field excursion. 
The first evaluation of this bound was given by Lyth \cite{Lyth:1996im} and refers to the scales corresponding to $1<l<100$, and so he considered $N_{\rm e}\simeq 4$. Putting $N_{\rm e}\simeq 30$ we 
obtain \cite{Boubekeur:2005zm,Boubekeur:2012xn}
\begin{equation}
	\frac{\Delta\varphi}{M_{\rm pl}}\gtrsim 1.06\left(\frac{r\left(\varphi_{\rm cross}\right)}{0.01}\right)^{1/2}\,.
\end{equation}
This bound tells that a model producing a large amount of GW would involve a field excursion of the order of the Planck mass. This constraint leads to a classification of inflationary models according to the field excursion: \textit{small field} and \textit{large field} models, where the discriminating value is the Planck mass. Being this inflationary features strictly related to the UV completion of gravity, constraining the inflaton excursion could provide useful information about the correct quantum gravity theory; see \cite{Efstathiou:2005tq,Boubekeur:2012xn,Garcia-Bellido:2014eva,Garcia-Bellido:2014wfa,Easther:2006qu}.

%%%%%%%%%%
\paragraph{Role of the consistency relation}
Another interesting check involving the tensor perturbation amplitude and spectral index is the consistency relation \eqref{consistency}. This would be a strong check to establish if the single-field slow-roll model is that realized by nature, but also could constrain features of the inflationary models which lead to a peculiar deviation from it. We will explain in detail these aspects in the dedicated section \ref{sezioneconsistency}.

\subsubsection{Beyond single-field slow-roll inflation: gravitational waves as a test of the inflationary models}
A measurement of the tensor power-spectrum would represent not only a way to constrain the features of the standard inflationary model, but also a way to discriminate among the many inflationary scenarios and to test the underlying fundamental physical theory.\\
Leaving the standard single-field slow-roll inflationary model, a great variety of scenarios is currently admitted. We can refer to two main categories: models built employing General Relativity (GR) as the theory of gravity and scenarios based on theories of modified gravity (MG). We will specify soon what we mean by this distinction. Many of them can be distinguished in terms of the predictions concerning the tensor power-spectrum, such as its amplitude, spectral index, and the tensor bispectrum, as a measure of their non-Gaussianity.\\
Among the models built on GR we can identify some main phenomena which might lead to unusual GW power-spectra: the production of tensor perturbations due to extra mechanisms, and secondary modifications of the standard inflationary model, such as the presence of spectator scalar fields. Moreover, one can also consider predictions coming from specific scenarios, as for example \textit{solid} \cite{Endlich:2012pz} and \textit{elastic} \cite{Gruzinov:2004ty} inflation, or scenarios of warm inflation \cite{Bartrum:2013fia,Bastero-Gil:2014jsa,Bastero-Gil:2016qru}.\\
As anticipated, one can also consider that GW can be produced during the reheating period too and can be useful to constrain the physics of such a period (and viceversa).\\
Many inflationary models built on MG theories have been proposed as a way to obtain an accelerated expansion stage in the early Universe. 
Considering the action
\begin{equation}
	S=\int\sqrt{-g}\left(\mathscr{L}_{\rm grav}+\mathscr{L}_{\rm mat}\right){\rm d}^{4}x\,,
\end{equation}
where $\mathscr{L}_{\rm grav}$ is the gravitational Lagrangian and $\mathscr{L}_{\rm mat}$ the Lagrangian describing the matter content of the Universe, to obtain the accelerated expansion, one can introduce a suitable $\mathscr{L}_{\rm mat}$ and employ the Einstein-Hilbert Lagrangian as $\mathscr{L}_{\rm grav}$, or one can apply modifications of GR to achieve acceleration.\\
In the latter models the dynamics is usually governed by the gravitational sector without involving any other fields since the equation of motion of the gravitational degrees of freedom alone can lead to an accelerated expansion stage. 
Clearly, as for models based on GR, one has to take care of the duration and of the end of such a period in order to get to the radiation and matter dominated eras.
%We will also examine predictions concerning tensor modes which emerge from an effective field theory approach to inflation {\bf SM refs...}. 

\paragraph{Primordial gravitational waves in the EFT approach of inflation}\label{eft}
In the direction of testing the inflationary physics, the latter has been investigated also implementing the approach of the Effective Field Theory \cite{Cheung:2007st} (see \cite{Tsujikawa:2014mba} for a review). In this way several inflationary models can be taken into account in a unique analysis at the same time. 
The basic idea of this approach is to write the most general action compatible with the symmetries of the theory and given the fields that drive the dynamics, and then calculate the predictions for observables as functions of the general coefficients of the operators included in the action. 
While time diffeomorphism is surely broken, it has been investigated also what happens to tensor modes when also spatial diffeomorphisms are broken for fluctuations \cite{Cannone:2014uqa,Cannone:2015rra,Graef:2015ova,Lin:2015cqa,Abolhasani:2015cve,Bartolo:2015qvr}. In this case the action for tensor perturbations reads \cite{Cannone:2014uqa,Bartolo:2015qvr}:
\begin{equation}\label{left}
	S_{\rm T}^{\left(2\right)}=\frac{M_{\rm pl}^{2}}{4}\int {\rm d}\tau\,d^{3}x\,a^{2}\left(\tau\right)\alpha\left[h_{ij}'h_{ij}'-c_{\rm T}^{2}\left(\nabla h_{ij}\right)^{2}-m^{2}h^{2}_{ij}\right]\,,
	\end{equation}
where the parameters $m$, $c_{\rm T}$ and $\alpha$ are obtained by combinations of the coefficients of the operators appearing in the original action \cite{Cheung:2007st}. GW can get a mass $m$ and a speed $c_{\rm T}$ different form that of light during inflation. From this action, at the leading order in slow roll and with $m/H\ll 1$ and requiring as initial condition the Bunch-Davies vacuum, the tensor mode power-spectrum results of the form:
\begin{equation}
	P_{\rm T}=\frac{2H^{2}}{\pi^{2}M_{\rm pl}^{2}c_{\rm T}}\left(\frac{k}{k_{\ast}}\right)^{n_{\rm T}}\,,\qquad \mbox{with}\qquad
	n_{\rm T}=-2\epsilon+\frac{2}{3}\frac{m^{2}}{\alpha H^{2}}\left(1+\frac{4}{3}\epsilon\right)\,.
\end{equation}
Notice that, given a pivot scale, the amplitude can be enhanced with respect to the standard one by the non-canonical speed of tensor modes. Moreover, if the ratio $m/H$ is sufficiently large the tilt can be blue, while preserving the NEC, leading to a violation of the consistency relation \cite{Cannone:2014uqa}. For a model in EFT approach, in which the speed of sound for tensors is considered time-dependent see \cite{Cai:2016ldn}.

\paragraph{A test for the theory of gravity and high energy physics} As anticipated, constraining inflationary models inevitably constitutes a test for the fundamental theories on the basis of which each model is built, first of all the theory of gravity. If, on one side MG theories naturally allow for a period of accelerated expansion, on the other side they could lead to unusual features of the gravitational degrees of freedom. In particular, we will show in section \ref{capmodified} that the propagation velocity of GW can be different from the speed of light, in contrast with GR, or there can appear modifications of the friction term in their equation of motion. We will see how these special properties affect the tensor power-spectrum.

\subsubsection{Alternatives to inflation}
An accelerated expansion phase in the early Universe is not the only way to solve standard cosmology problems. Early Universe models which provide an alternative to inflation have been proposed. We mention here the String Gas Cosmology \cite{Brandenberger:2006xi}, the Pre-Big Bang Cosmology \cite{Gasperini:2002bn} and the Ekpyrotic Universe \cite{Khoury:2001wf}. For each of these models predictions about GW have been obtained. These are however outside the scope of the present review.

\section{Classical production of primordial gravitational waves during inflation}\label{secondo}

In the next sections we will investigate the main mechanisms of GW production, beyond vacuum oscillations of the gravitational field, that can take place during the inflationary epoch and the preheating stage. At the end, in section \ref{overview}, we will recap in a summarizing table all the models we discuss in the present work.\\
During inflation and preheating, GW can be produced in two ways: from vacuum fluctuations of the gravitational field or by a \textit{classical} mechanism.
The first case is that we have shown in the previous section for the single-field slow-roll inflation. For this kind of production, different predictions for the tensor power-spectrum follow from different theories of gravity underlying the inflationary model. On the other hand, the GW classical production takes place when a source term in the GW equation of motion (eq. \ref{onde}) is present. Such a kind of term can be provided by several situations, such as a particle production or the presence of more than one scalar field during inflation. Clearly, the features of the source term determine the GW power-spectrum produced in such a way. The interesting point then is to investigate what can provide a source term during the inflationary and preheating stages and, for each kind of them, to determine the form of the GW power-spectrum.\\
In the following sections we will examine the classical production of GW and later on their production from vacuum oscillations for several gravity theories different from General Relativity.\\
\\
In the previous section we have considered phenomena concerning first-order perturbation theory on a FRW background. At such order scalar, vector and tensor modes are independent. This fact does not hold at higher perturbative order: already at second order, suitable combinations of scalar modes can give rise to second-order vector or tensor perturbations, while, e.g., linear tensor modes give rise to scalar perturbations \cite{tomita,Matarrese:1997ay,Carrilho:2015cma}. This fact of course also holds at higher order, with the only constraint that scalars, vectors and tensors of the same perturbative order remain uncoupled. 
Indeed, a combination of two first order objects invariant under spatial rotations, that is two scalars, might not be still invariant under such transformation. As a consequence, when second-order perturbations 
of the metric and of the stress-energy tensor are taken into account, the free wave equation that at the first order describes the dynamic of the tensor modes, gets a source term. 
Combination of first-order scalar or vector perturbations represent possible sources for GW \cite{Matarrese:1997ay}. First this means that when curvature perturbations are present we always have 
generation of GW, even if tensor perturbations of the first order are absent.
More precisely, by looking at the traceless and divergence-less spatial components of the second-order of Einstein's equations, one finds that, in contrast to the first-order case, other terms besides 
those involving the usual GW wave equations are left: combinations of scalar and vector perturbations coming from the Einstein tensor and from the anisotropic part of the stress-energy tensor. 
These terms give rise to a source in the GW equation of motion 
\cite{Matarrese:1992rp,Matarrese:1993zf,Matarrese:1997ay,Nakamura:2003wk,Nakamura:2004rm,Carbone:2004iv,Noh:2003yg,tomita,Carbone:2005nm}.
Unlike GW generated by quantum fluctuations of the gravitational field, here we are dealing with {\it classical} mechanisms of GW production.
Concerning the inflationary physics, second-order sources are present during the accelerated expansion stage, given by the quantum fluctuations of the inflaton or provided by other mechanisms,
but also after the inflationary period during the radiation and matter dominated epochs, when scalar and tensor perturbations re-enter the horizon. Here we are specially interested in the first situation.\\
In most standard inflationary scenarios the extra amount of GW produced by this mechanism results negligible compared to the first-order production and to the planned experimental capabilities.
Nonetheless, there are many cases in which second-order GW play a significant role because of the presence of efficient sources.

\subsection{Production of second-order gravitational waves}\label{secondoordine}
Consider the spatial part of second-order Einstein equations and project them into their transverse and traceless parts:
\begin{equation}\label{sopra}	
\hat{\Pi}_{ij}^{\,lm}G^{\left(2\right)}_{lm}=\kappa^{2}\hat{\Pi}_{ij}^{\,lm}T^{\left(2\right)}_{lm}\,,
\end{equation}
where $\hat{\Pi}_{ij}^{lm}$ is the projector operator $\hat{\Pi}_{ij}^{\,\,lm}=\Pi^{i}_{l}\Pi^{j}_{m}-\frac{1}{2}\Pi_{ij}\Pi^{lm}$ with $\Pi_{ij}=\delta_{ij}-\partial_{i}\partial_{j}/\Delta$ and $\kappa^{2}=8\pi G$.
Consider the flat FRW second-order perturbed metric form eq.\eqref{metric} neglecting for simplicity first-order vector and tensor perturbations, and
employ $h_{ij}\equiv h_{ij}^{\left(2\right)}$.
From this expression the Einstein tensor at second order results \cite{Acquaviva:2002ud}:
\begin{align}	
G^{\left(2\right)i}_{j}=&a^{-2}\left[\frac{1}{4}\left(h''^{i}_{j}+2\mathcal{H}h'^{i}_{j}-\nabla^{2}h^{i}_{j}\right)+ 
2\Psi^{\left(1\right)}\partial^{i}\partial_{j}\Psi^{\left(1\right)}-2\Phi^{\left(1\right)}\partial_{i}\partial_{j}\Psi^{\left(1\right)}\right.\nonumber\\
&+4\Phi^{\left(1\right)}\partial_{i}\partial_{j}\Phi^{\left(1\right)}+\partial^{i}\Psi^{\left(1\right)}\partial_{j}\Psi^{\left(1\right)}-\partial^{i}\Psi^{\left(1\right)}\partial_{j}\Phi^{\left(1\right)}-\partial^{i}\Phi^{\left(1\right)}\partial_{j}\Psi^{\left(1\right)}\nonumber\\
&\left.+3\partial^{i}\Phi^{\left(1\right)}\partial_{j}\Phi^{\left(1\right)}+\left(\Psi^{\left(2\right)},\Psi^{\left(2\right)},\omega_{i}^{\left(2\right)} \mathrm{term}\right)+\left(\mathrm{diagonal\,\,part}\right)\delta^{i}_{\,j}\right]\,.
\end{align}
The stress-energy tensor of a perfect fluid perturbed at second order reads \cite{Acquaviva:2002ud}:
\begin{equation}	T^{\left(2\right)i}_{\,\,\,j}=\left(\rho^{\left(0\right)}+P^{\left(0\right)}\right)v^{\left(1\right)i}v^{\left(1\right)}_{j}+P^{\left(0\right)}\pi^{\left(2\right)i}_{\,\,\,j}+ 
P^{\left(1\right)}\pi^{\left(1\right)i}_{\,j}+P^{\left(2\right)}\delta^{i}_{\,j}\,.
\end{equation}
Using the expressions for the first-order perturbations of the energy-momentum tensor in terms of the linear metric perturbations and of the background value of the energy-momentum tensor 
\cite{Baumann:2007zm}, eq.\eqref{sopra} becomes
\begin{equation}\label{ondeseconde}
	h''_{ij}+2\mathcal{H}h'_{ij}-\nabla^{2}h_{ij}=-4\hat{\Pi}_{ij}^{lm}\mathscr{S}_{lm},
\end{equation}
with $\mathscr{S}_{lm}$:
\begin{align}
	\mathscr{S}_{lm}\equiv& \,2\Psi\partial^{l}\partial_{m}\Psi-2\Phi\partial^{l}\partial_{m}\Psi
+4\Phi\partial^{l}\partial_{m}\Phi+4\Psi\partial^{l}\partial_{m}\Psi+\nonumber\\
&+\partial^{l}\Psi\partial_{m}\Psi-\partial^{l}\Psi\partial_{m}\Phi-\partial^{l}\Phi\partial_{m}\Psi+3\partial^{l}\Phi\partial_{m}\Phi+\nonumber\\
&-\frac{4}{3\left(1+\omega\right)\mathcal{H}^{2}}\partial_{l}\left(\Phi'+3\mathcal{H}\Psi\right)\partial_{m}\left(\Phi'+3\mathcal{H}\Psi\right)\nonumber\\
&-\frac{2c_{\rm S}^{2}}{3\omega\mathcal{H}^{2}}\left[3\mathcal{H}\left(\mathcal{H}\Psi-\Phi'\right)+\nabla^{2}\Phi\right]\partial_{l}\partial_{m}\left(\Psi-\Phi\right)\,,
\end{align}
and $\omega\equiv P^{\left(0\right)}/\rho^{\left(0\right)}$, $\Psi\equiv \Psi^{\left(1\right)}$, $\Phi\equiv \Phi^{\left(1\right)}$ and $c_{\rm S}=P^{\left(1\right)}/\rho^{\left(1\right)}$. Notice that 
the source $\mathscr{S}_{ij}$ is composed of terms coming from the Einstein tensor and others coming from the stress-energy tensor.
In order to solve eq.\eqref{ondeseconde} we Fourier transform the tensor perturbations as
\begin{equation}	
h_{ij}\left(\textbf{x},\tau\right)=\int\frac{\mathrm{d}^{3}\textbf{k}}{\left(2\pi\right)^{3/2}}
e^{i\textbf{k}\cdot\textbf{x}}\left[h_{\textbf{k}}\left(\tau\right)\texttt{e}_{ij}\left(\textbf{k}\right)+\bar{h}_{\textbf{k}}\left(\tau\right)\bar{\texttt{e}}_{ij}\left(\textbf{k}\right)\right]\,.
\end{equation}
The two polarization tensors $\texttt{e}_{ij}$, $\bar{\texttt{e}}_{ij}$ can be expressed by the polarization vectors $\texttt{e}_{i}\left(\textbf{k}\right)$, $\bar{\texttt{e}}_{i}\left(\textbf{k}\right)$ 
orthogonal to the propagation vector 
$\textbf{k}$ as
\begin{align}	
&\texttt{e}_{ij}\left(\textbf{k}\right)\equiv\frac{1}{\sqrt{2}}\left[\texttt{e}_{i}\left(\textbf{k}\right)\texttt{e}_{j}\left(\textbf{k}\right)-\bar{\texttt{e}}_{i}\left(\textbf{k}\right)\bar{\texttt{e}}_{j}\left(\textbf{k}\right)\right]\,,\\
&\bar{\texttt{e}}_{ij}\left(\textbf{k}\right)\equiv\frac{1}{\sqrt{2}}\left[\texttt{e}_{i}\left(\textbf{k}\right)\bar{\texttt{e}}_{j}\left(\textbf{k}\right)-\bar{\texttt{e}}_{i}\left(\textbf{k}\right)\texttt{e}_{j}\left(\textbf{k}\right)\right]\,.
\end{align}
In terms of the polarization tensors, then the RHS of eq.\eqref{ondeseconde} is written as
\begin{equation}	
\hat{\Pi}_{ij}^{lm}\mathscr{S}_{lm}\left(\textbf{x}\right)=\int\frac{\mathrm{d}^{3}\textbf{k}}{\left(2\pi\right)^{3/2}}e^{i\textbf{k}\cdot\textbf{x}}
\left[\texttt{e}_{ij}\left(\textbf{k}\right)\texttt{e}^{lm}\left(\textbf{k}\right)+\bar{\texttt{e}}_{ij}\left(\textbf{k}\right)\bar{\texttt{e}}^{lm}\left(\textbf{k}\right)\right]\mathscr{S}_{lm}\left(\textbf{k}\right)\,,
\end{equation}
where $\mathscr{S}_{lm}\left(\textbf{k}\right)$ is the Fourier transform of $\mathscr{S}_{lm}\left(\textbf{x}'\right)$.
Then, the equation of motion of second-order tensor modes in Fourier space, for each polarization state, reads
\begin{equation}\label{fouriersecondo}
	h''_{\textbf{k}}+2\mathcal{H}h'_{\textbf{k}}+k^{2}h_{\textbf{k}}=\mathscr{S}\left(\textbf{k},\tau\right)\,,
\end{equation}
where the quantity
\begin{equation}\label{esse2}
	\mathscr{S}\left(\textbf{k}, \tau\right)=-4\texttt{e}^{lm}\left(\textbf{k}\right)\mathscr{S}_{lm}\left(\textbf{k}\right)
\end{equation}
is the convolution of two linear scalar perturbations.
The equality \eqref{fouriersecondo} is a wave-equation with a source, whose solution reads
\begin{equation}\label{soluzioneseconda1}	
h_{\textbf{k}}\left(\tau\right)=\frac{1}{a\left(\tau\right)}\int\mathrm{d}\tilde{\tau}\,G_{\textbf{k}}\left(\tau;\tilde{\tau}\right)\left[a\left(\tilde{\tau}\right)\mathscr{S}\left(\textbf{k},\tilde{\tau}\right)\right]\,,
\end{equation}
where the Green function $G_{\textbf{k}}$ solves the eq.\eqref{fouriersecondo} with the source given by $\left(1/a\right)\delta\left(\tau-\tilde{\tau}\right)$. $G_{\textbf{k}}$ then 
depends only on the evolution of the scale-factor.
Given eq.\eqref{soluzioneseconda1}, the expression for the GW correlator can be written in terms of that of the source as
\begin{equation}\label{spettroseconda}
\left\langle h_{\textbf{k}}\left(\tau\right)h_{\textbf{k}'}\left(\tau\right)\right\rangle=\frac{1}{a^{2}\left(\tau\right)}\int^{\tau}_{\tau_{0}}{\rm d}\tilde{\tau}_{1}{\rm d}\tilde{\tau}_{2}\,a\left(\tilde{\tau}_{1}\right)a\left(\tilde{\tau}_{2}\right) G_{\textbf{k}}\left(\tau;\tilde{\tau}_{1}\right) G_{\textbf{k}'}\left(\tau;\tilde{\tau}_{2}\right)\left\langle \mathscr{S}\left(\textbf{k},\tilde{\tau}_{1}\right)\mathscr{S}\left(\textbf{k}',\tilde{\tau}_{2}\right)\right\rangle\,,
\end{equation}
where $\tau_{0}$ is the time when the source switches on. Equation \eqref{spettroseconda} represents the general expression for the GW power-spectrum due to tensor modes that solve eq.\eqref{ondeseconde}. Then, now the interesting point is to find out the solution for specific cases of the source term.

\subsubsection{Second-order gravitational waves sourced by inflaton perturbations}\label{secondordinarie}
The immediate application of second-order perturbation theory consists in considering the inflationary scalar perturbations as a source for GW. We have just seen that the very existence of scalar perturbations 
gives rise to tensor modes, independently of how the first-order scalars have been generated. Knowing the scalar power-spectrum during the inflationary period, the sourced-GW power-spectrum can be calculated too.
More precisely, $\mathscr{S}_{lm}\left(\textbf{k}\right)$ can be written highlighting the dependence on the perturbation $\Phi_{\textbf{k}}\left(\tau\right)$ evaluated at early times, so that the correlator 
\eqref{spettroseconda} can be written in terms of the primordial power-spectrum $P_{\Phi}\left(k\right)$:
\begin{equation}
	\left\langle \Phi_{\textbf{k}} \Phi_{\textbf{k}'}\right\rangle=\frac{2\pi^{2}}{k^{3}}P_{\Phi}\left(k\right)\delta\left(\textbf{k}+\textbf{k}'\right)\,,
\end{equation}
which is strongly constrained by CMB and LSS measurements. 
Scalar perturbations play the role of GW source both during inflation and at the end of inflation, when they re-enter the horizon after having been frozen \cite{Matarrese:1997ay}.
Tensor modes generated by curvature perturbations that re-enter the horizon after the end of inflation, can lead to non-negligible contributions. More precisely, scalar perturbations that enter the horizon during the radiation dominated epoch \cite{Ananda:2006af,Baumann:2007zm}, generates tensor perturbations that at the present time results on scales that could be interesting for experiments of direct GW detection \cite{Ananda:2006af}. 
However, assuming an inflationary power law spectrum on all scales for scalar perturbations, in accordance with current constraints coming from the CMB, the GW spectral energy-density at the present time is several orders of magnitude smaller than the sensitivity curve of planned experiments \cite{Ananda:2006af}. 
In particular, for a power-law scalar power-spectrum with a red tilt $n_{\rm S}=0.95$, a GW spectral energy-density of $\Omega_{\rm GW}\simeq 10^{-22}\left(f/\mbox{Hz}\right)^{-0.1}$ is expected \cite{Ananda:2006af}. Notice that in doing these estimates one exploits values extrapolated from the scales of the CMB to constrain the scalar power-spectrum on scales 
smaller by $\sim20$ orders of magnitude.\\
The presence of second-order tensor modes sourced by first-order curvature perturbations that re-enter the horizon after the matter-radiation equality, clearly affects also the CMB polarization predictions \cite{Mollerach:2003nq}.
This effect limits the ability of estimating the inflationary first-order power-spectrum of tensor modes and then it relaxes the constraints on the energy scale of inflation. The amount of B modes due to the presence of second-order GW is estimated as the second contribution after weak lensing. Numerically estimating the B-mode power-spectrum taking into account second-order vector and tensor modes, and physical effects at recombination, the contribution to the GW power-spectrum of second-order vector and tensor modes is found to be comparable to that coming from primordial GW for multipoles lower than $\ell\simeq 100$, for $r\simeq 10^{-7}$ and lower than $\ell\simeq 700$ for $r\simeq 10^{-5}$ \cite{Fidler:2014oda}.\\
The generation of second-order GW has been investigated also with respect to the reheating phase, usually considered as a matter dominated epoch \cite{Assadullahi:2009nf}.\\
On the other hand the second-order contribution plays an interesting and non-negligible role in several inflationary models, as for example scenarios with events of particle production. 
In the next sections we will investigate some of those models.
In order to do so, the main work consists in solving the equation of motion of the extra field, in order to write the source term and then solve the tensor mode equation \eqref{fouriersecondo} by the Green's 
function method. Only the tensor perturbations in the metric are usually considered and the scalar and vector ones neglected. We will start by looking at the form taken by eq.\eqref{sopra} with this assumption.

\subsubsection{Gravitational wave equation neglecting scalar and vector metric perturbations}\label{procedura}
Let us consider the FRW metric perturbed at second order, neglecting scalar and vector perturbations of first and second order. The equation of motion for GW \eqref{sopra} becomes:
\begin{equation}\label{ondeconsorgente}
	h''_{ij}+2\mathcal{H}h'_{ij}-\nabla^{2} h_{ij}=\frac{2}{M_{\rm pl}^{2}}\hat{\Pi}_{ij}^{lm}T_{lm}\,,
\end{equation}
where $T_{lm}$ is a generic stress-energy tensor. Notice that here the source term is given only by the stress-energy tensor, having set to zero scalar and vector perturbations in the Einstein tensor. Equation \eqref{ondeconsorgente} is solved by:
\begin{equation}\label{soluzionesorgente}
	h_{ij}\left(\textbf{k},\tau\right)=\frac{2}{M_{\rm pl}^{2}}\int\mathrm{d}\tilde{\tau} G_{\textbf{k}}\left(\tau,\tilde{\tau}\right)\hat{\Pi}_{ij}^{\,lm}\left(\textbf{k}\right)T_{lm}\left(\textbf{k},\tilde{\tau}\right)\,,
\end{equation}
with $G_{\textbf{k}}$ the Green function.
Proceeding as before, the amplitude of the GW of a fixed polarization states reads
\begin{equation}\label{soluzioneseconda}	
h_{\textbf{k}}\left(\tau\right)=\frac{1}{a\left(\tau\right)}\int\mathrm{d}\tilde{\tau}G_{\textbf{k}}\left(\tau,\tilde{\tau}\right)\left[a\left(\tilde{\tau}\right)\mathscr{T}\left(\textbf{k},\tilde{\tau}\right)\right]\,,
\end{equation}
where $\mathscr{T}\left(k,\tau'\right)$ is defined similarly to eq.\eqref{esse2}. In order to specify the solution we need to fix the evolution of the scale-factor and the projected stress-energy tensor.
An exact solution for the Green function exists for a de Sitter stage, and for a radiation or matter dominated epoch. In most of the following scenarios we will consider a de Sitter background. 
For such a case the Green function takes the following simple form \cite{Cook:2011hg}:
\begin{equation}\label{green}
	G_{\textbf{k}}\left(\tau,\tilde{\tau}\right)=\frac{1}{k^{3}\tilde{\tau}^{2}}\left[\left(1+k^{2}\tau\tilde{\tau}\right)\sin k\left(\tau-\tilde{\tau}\right)+k\left(\tilde{\tau}-\tau\right)\cos k\left(\tau-\tilde{\tau}\right)\right]\Theta\left(\tau-\tilde{\tau}\right)\,,
\end{equation}
with $\Theta$ the Heaviside function.

\subsection{Gravitational waves sourced by scalar perturbations}
We will consider two examples of inflationary models where the source of GW is due to the presence of a further scalar field besides the inflaton. We investigate the amount of GW generated by the perturbations of the extra field, without taking into account the source terms coming from the Einstein tensor (that is we will calculate the Einstein tensor neglecting scalar and vector perturbations). 

\subsubsection{Second-order gravitational waves in the curvaton scenario}\label{curvatone}
In the \textit{curvaton scenario} \cite{Enqvist:2001zp}, a scalar field, the \textit{curvaton}, is added besides the inflaton, requiring that it does not influence the inflationary dynamics. In the simplest curvaton scenario, the curvature perturbations of the inflaton are considered negligible and not the seeds of structure formation. 
This situation is achieved by lowering the energy scale of inflation which leads to a suppression of the amount of GW produced by the standard mechanism. 
More precisely, asking that the inflaton curvature perturbation is much smaller than required to explain CMB anisotropies corresponds to $H_{\ast}\ll 10^{-5}M_{\rm pl}$ \cite{Bartolo:2007vp}, with $H_{\ast}$ the Hubble rate estimated during inflation. At the same time, this requirement means that the amount of GW generated from vacuum oscillations of the metric tensor is very small.
Then, in this scenario second-order GW might be significant with respect to those produced at first order
\cite{Bartolo:2007vp,Enqvist:2008be,Suyama:2011pu,Kawasaki:2013xsa}.\\
In \cite{Kawasaki:2013xsa}, the particular case in which the inflaton contributes significantly in generating curvature perturbations, in a way that allows the curvaton scalar 
perturbations to be blue tilted, is considered.\\
Considering the most standard scenario, let us examine the role of the curvaton: the presence of a second field leads to the generation of isocurvature perturbations. 
After the end of inflation, the curvaton decays and isocurvature perturbations lead to adiabatic perturbations, which give rise to structure formation.
Tensor modes are sourced by isocurvature perturbations that re-enter the horizon between the end of inflation and the time of curvaton decay, and by curvature perturbations after the curvaton decay. 
The latter lead to a GW spectral energy-density of the order of $\Omega_{\rm GW}\simeq 10^{-20}$ for thsoe modes that cross the horizon during the radiation dominated era \cite{Bartolo:2007vp}. 
The most interesting result is that isocurvature perturbations that enter the horizon between the end of inflation and the epoch of curvaton decay source an amount of GW larger than that due to curvature fluctuations \cite{Bartolo:2007vp}. We consider the first contribution.\\
The equation that governs second-order scalar-sourced GW in which the source term is given by isocurvature perturbation reads
\begin{equation}\label{eqcurvatone}	
h''_{ij}+2\mathcal{H}h'_{ij}-\nabla^{2}h_{ij}=-\frac{2}{M_{\rm pl}^{2}}\hat{\Pi}^{lm}_{ij}\partial_{l}\delta\sigma\partial_{m}\delta\sigma\,,
\end{equation}
where $\delta\sigma$ are isocurvature perturbations of the curvaton field. The solution of this equation is given by eq.\eqref{soluzionesorgente}, where the Green function relative to the radiation dominated era reads
$G_{\textbf{k}}\left(\tilde{\tau},\tau\right)=\sin\left[k\left(\tau-\tilde{\tau}\right)\right]/k$,
and the integration over time starts form the horizon entry of the perturbations.
The form of the Green function tells us that the main contribution to each wavelength of the GW spectrum blows up at horizon entry of the scale. 
Then, we approximate the power-spectrum to that computed at horizon entry. 
The source term in the integrand is determined by the curvaton fluctuations $\delta\sigma$ so that the source form depends on the time-dependence of the curvaton fluctuations at horizon crossing, 
more precisely on whether it is already oscillating around the minimum of its potential or not. If the zero mode of the curvaton decay is already oscillating, curvaton perturbations are found to scale as $\delta\sigma_{\textbf{k}}\left(\tau\right)\sim a^{-3/2}$, while those modes which enter the horizon before the zero mode of the curvaton starts oscillating, scale as $\delta\sigma_{\textbf{k}}\left(\tau\right)\sim a^{-1}$ \cite{Bartolo:2007vp}. This leads to two different power-spectra depending on the range of scale considered.\\
One can see that $\Omega_{\rm GW}$, defined in \eqref{definizioneomega1}, which depends on $\delta\sigma$, can be written in terms of, in principle measurable quantities (that is relative to the curvature perturbations $\zeta$ and the non-Gaussianity parameter $f_{\rm NL}^{\rm local}$) and quantities relative to the curvaton physics.
Super-horizon isocurvature perturbations produced during a de Sitter stage are found to be $\delta\sigma_{\textbf{k}}=H_{\ast}/2\pi$ \cite{Enqvist:2001zp}.
These perturbations are then converted into curvature perturbation $\zeta_{\textbf{k}}$ after curvaton decay. In particular, they are linked by \cite{Bartolo:2007vp}:
\begin{equation}
	\zeta_{\textbf{k}}\simeq r_{\rm c}\left(\frac{\delta\sigma_{\textbf{k}}}{\bar{\sigma}_{\ast}}\right)\,,
\end{equation}
where $r_{\rm c}\equiv\left(\rho_{\sigma}/\rho\right)_{\rm D}$ is the ratio between the produced radiation energy-density and the total energy-density at the epoch of the decay, and where $\bar{\sigma}_{\ast}$ is the background value of the curvaton during inflation. 
The connection between the isocurvature perturbation and the adiabatic ones, that are at the origin of structures and CMB anisotropies, allows to write the GW power-spectra in terms of measurable quantities, that is to substitute the dependence on the isocurvature perturbations with the curvature one.
It is interesting to note that $r_{c}$ can be connected with $f_{\rm NL}^{\rm local}$, a parameter which quantifies the level of (a certain type - called ``local" - of) non-Gaussianity of primordial scalar perturbations, which is strongly constrained by Planck measurements \cite{Ade:2015ava}. The isocurvature perturbations power-spectrum, that inevitably appear in the second-order GW power-spectrum, can be written in terms of the $\zeta$ power-spectrum, $f_{\rm NL}^{\rm local}$ and other quantities related to the curvaton physics, such as the curvaton decay rate $\Gamma$, the curvaton mass $m$ and the decay temperature $T_{\rm D}$.\\
We now assume a scale-invariant curvature power-spectrum and, according to the predictions of the curvaton scenario, we take $A_{\rm S}\simeq 2.5\times 10^{-9}$. Defining $k_{D}$ as the scale that enters the horizon at the time of the curvaton decay, the GW spectral energy-density today for each of the introduced range, results \cite{Bartolo:2007vp}
\begin{equation}	
\Omega_{\rm GW}\simeq10^{-15}\left(\frac{f_{\rm NL}^{\rm local}}{10^{2}}\right)^{2}\left(\frac{k}{k_{\rm D}}\right)^{5}\left(\frac{\Gamma}{m}\right)^{7/2}\,,
\end{equation}
for $k_{\rm D}\leq k\leq\left(m/\Gamma\right)^{1/2}k_{\rm D}$, that is for modes which enter the horizon when the curvaton zero mode is already oscillating, and
\begin{equation}	
\Omega_{\rm GW}\simeq10^{-15}\left(\frac{f_{\rm NL}^{\rm local}}{10^{2}}\right)^{2}\left(\frac{\Gamma}{m}\right)\,,
\end{equation}
for $k\geq \left(m/\Gamma\right)^{1/2}k_{\rm D}$.
From these expression, in the perturbative regime $\Gamma\lesssim m$ and maximizing the current constraints on the non-Gaussianity \cite{Ade:2015ava}, the present GW spectral energy-density $\Omega_{\rm GW}$ can be of the order of $\Omega_{\rm GW}\simeq 10^{-19}$. To get such an amplitude in the range of frequencies where planned experiments of direct detection are expected to be more sensitive, a temperature $T_{\rm D}\lesssim 10^{8}\mathrm{GeV}$ is needed.

\subsubsection{Second-order gravitational waves sourced by spectator scalar fields}\label{sezionespectator}
Let us now present another inflationary model in which second-order GW can play a significant role. In this scenario a so-called \textit{spectator} scalar field $\sigma$, different from the inflaton, is assumed to be light and not affecting the dynamics of the background. It plays a crucial role in possibly leading to a significant production of second-order GW \cite{Biagetti:2013kwa,Biagetti:2014asa,Fujita:2014oba}. 
Contrary to the curvaton scenario, here curvature perturbations are generated also by the inflaton field. The intriguing fact is that, for a speed of sound $c_{\rm S}$ of the spectator field smaller than unity, we have a more efficient second-order GW production,
with respect to the case of a spectator scalar field with $c_{\rm S}=1$.
This can be easily obtained, for example, with a general Lagrangian $P\left(X,\sigma\right)$, with $X$ the canonical kinetic term, so that the speed of sound reads $c_{\rm S}=\partial_{X}P/\left(\partial_{X}P+2X\partial^{2}_{XX}P\right)$ (other examples are reported in \cite{Biagetti:2013kwa}).\\
Consider the action for the scalar perturbations of the spectator field \cite{Biagetti:2013kwa}:
\begin{equation}\label{azionespect}
S_{\delta\sigma}^{\left(2\right)}=\int\mathrm{d}\tau\,\mathrm{d}^{3}x\,a^{4}\left\{\frac{1}{2a^{2}}\left[\delta\sigma'^{2}-c_{\rm S}^{2}\left(\nabla\delta\sigma\right)^{2}\right]-V_{\left(2\right)}\right\}\,,
\end{equation}
where $V_{\left(2\right)}$ is the second-order potential.
In general, the presence of the spectator field leads to the production of scalar perturbations, besides those due to the inflaton, whose amplitude is determined by $c_{\rm S}$.
As all scalar perturbations, also those of the spectator field represent a source for second-order GW. Considering the role of $\sigma$ fluctuations, the equation of motion for tensor modes reads:
\begin{equation}	
h''_{ij}+2\mathcal{H}h'_{ij}-\nabla^{2}h_{ij}=-2\frac{c_{\rm S}^{2}}{M_{\rm pl}^{2}}\hat{\Pi}^{lm}_{ij}\partial_{l}\delta\sigma\partial_{m}\delta\sigma\,.
\end{equation}
Notice that a new factor appears in the source: the sound speed of the scalar field. Proceeding as in section \ref{procedura}, one gets the solution \eqref{soluzioneseconda}. Then, numerically integrating over the inflationary period, the second-order GW power-spectrum sourced by the spectator field on super-horizon scales is found to be \cite{Biagetti:2013kwa}:
\begin{equation}\label{tensorspectator}
	P_{\rm T}=c\frac{H^{4}}{c_{\rm S}^{18/5}M_{\rm pl}^{4}}\,,
\end{equation}
where $c$ is a numerical factor $c\simeq 3$ (\cite{Biagetti:2014asa} and \cite{Fujita:2014oba} make the analysis with two different covariant formulations of \eqref{azionespect}, providing the results for the related GW power-spectrum; the significant, and not obvious, point is that, in each case, the amplitude of the sourced GW is found to be inversely proportional to a power of $c_{\rm S}$). In summary, the total scalar and tensor power-spectra are respectively a sum of two terms, one due to vacuum fluctuations and the other due to the presence of the spectator field. Therefore the overall tensor-to-scalar ratio results to be sensitive to the sound speed of the spectator field.
Clearly the dependence of the tensor-to-scalar ratio $r$ on $c_{\rm S}$, introduces a degeneracy between different parameters of the model. As a consequence, in particular, the constraints on $r$ no longer correspond directly to an upper bound on the energy scale of inflation. On the other hand, we have to take into account that CMB data provide a measurement of the amplitude of the scalar perturbations on CMB scales that has to be satisfied. This constraint restricts the admitted range of values for the sound speed of the spectator field, which, at the end, provides a limit on the extra GW production \cite{Biagetti:2014asa,Fujita:2014oba}.
In particular, assuming that the spectator field does not significantly source curvature perturbations, leads to a strict upper bound on the amplitude of the GW sourced by the spectator field, on CMB scales, that results in a negligible contribution. On the other hand, admitting a significant role of the curvature perturbations sourced by the spectator field, and considering the amplitude of scalar perturbations obtained from CMB data, the value of the tensor-to-scalar ratio can be larger than in the previous case. Notice that, the presence of the sourced contribution, which includes a dependence on $c_{\rm S}$, introduces a degeneracy between $r$ and the energy scale of inflation.\\
In the case of a spectator field with a tiny mass $m$ and evolving in a quasi-de Sitter background, the spectral tilt of the power-spectrum for $\delta\sigma$ does not vanish (similarly to what happens for ordinary scalar perturbations). This affects the generated GW power-spectrum, providing a tensor tilt \cite{Biagetti:2013kwa}
\begin{equation}\label{tiltspectator}
	n_{\rm T}\simeq 2\left(\frac{2m^{2}}{3H^{2}}-2\epsilon\right)-\frac{18}{5}\frac{\dot{c}_{\rm S}}{Hc_{\rm S}}\,,
\end{equation}
considering also that the sound speed can slowly vary during the inflationary period. Equation \eqref{tiltspectator} shows that the second-order tensor power-spectrum can be blue, contrary to the first-order GW produced by vacuum oscillations. Notice that, since $\dot{c}_{\rm s}/Hc_{\rm s}$ appears in the scalar spectral index of the power-spectrum due to the spectator field too, in order to avoid strict bounds on such quantity coming from CMB data, curvature perturbations due to the spectator field are required to be negligible.

\subsection{Particle production as a source of gravitational waves}\label{particelle}
Several inflationary models involving the production of quanta of extra fields during inflation, have been proposed. If the inflaton is minimally, or non-minimally, coupled to another scalar or gauge field, its energy can move from its sector to the others and give rise to the production of extra quanta \cite{Chung:1999ve}. The new particles provide a further contribution to the stress-energy tensor with a non vanishing anisotropic component, giving rise to a source of GW \cite{Barnaby:2010vf,Sorbo:2011rz,Barnaby:2011vw,Barnaby:2011qe,Anber:2012du,Namba:2015gja}.
Different inflationary models have been built in this framework\footnote{For a recent work about analogous models in the framework of bouncing cosmologies, see \cite{Ben-Dayan:2016iks}.}, in particular scenarios where the extra quanta production occurs during the slow-roll of the inflaton field, and others where the production occurs during the inflaton oscillations at the end of the accelerated expansion phase. In this section we examine the first case, next, in section \ref{analisi}, we will take into account the second one.\\
Among these models, the most interesting ones are those presenting a coupling between the inflaton and a gauge field $A_{\mu}$ \cite{Garretson:1992vt}: a band of modes of the produced field is subjected to an exponential growth which generates a large anisotropic stress. This fact leads to a large amount of sourced GW, but, at the same time, the presence of the new particles sources the production of curvature perturbations, for which we have strict constraints from CMB observations. Current data concerning scalar perturbations restrict hardly the parameter space of this kind of models but allows also for an amount of GW in the window of sensitivity of some planned experiments.\\
The first models investigated in this framework present a non-minimal coupling between the inflaton and the extra field \cite{Cook:2011hg,Barnaby:2011qe,Barnaby:2010vf,Sorbo:2011rz,Barnaby:2011vw}. In particular the coupling with a massive scalar field and a gauge field has been studied. In the first case a burst of particle production happens while in the latter a continuous production develops during the inflationary stage. 
In both scenarios the coupling between the two fields leads to narrow the regions compatible with current observational constraints related to CMB scales. 
Considering this fact, the case where the coupling which leads to the production of the gauge quanta, is moved from the inflaton to another \textit{auxiliary} field have been investigated \cite{Barnaby:2012xt,Cook:2013xea}. In the latter case, the particle production is described by a hidden sector constituted of a pseudo-scalar field and the gauge field, minimally coupled to the inflaton, so that the production of GW can be efficient, preserving at the same time current bounds on the associated features related to scalar quantities.

\subsubsection{Inflaton coupled to a scalar field}\label{particellescalari}
Consider a system described by the following Lagrangian \cite{Cook:2011hg}
\begin{equation}\label{lagscalar}	
\mathscr{L}=-\frac{1}{2}\partial_{\mu}\varphi\partial^{\mu}\varphi-V\left(\varphi\right)-\frac{1}{2}\partial_{\mu}\chi\partial^{\mu}\chi-\frac{g^{2}}{2}\left(\varphi-\varphi_{0}\right)^{2}\chi^{2}\,,
\end{equation}
where $\varphi$ is the inflaton, $V\left(\varphi\right)$ is the potential that drives the dynamics of the inflationary period, $\chi$ is an extra scalar field, and, for simplicity, the self-interaction of the 
field $\chi$ is neglected. The mass of the secondary field, $m_{\chi}$, depends on time, being related to the value of the inflaton field, which is rolling down its potential. When the inflaton $\varphi$ reaches the value 
$\varphi_{0}$, $m_{\chi}$ vanishes and the production of $\chi$ quanta becomes energetically favored. During the period of time around which the inflaton is equal to $\varphi_{0}$, a non-perturbative production of such particles takes place. After this interval of time the Universe is filled with $\chi$ particles besides the inflaton ones.
The presence of the $\chi$ quanta gives rise to a contribution to the stress-energy tensor of the system, more precisely its spatial part reads $T_{ab}=\partial_{a}\chi \partial_{b}\chi+\delta_{ab}\left(...\right)$, 
where the factor proportional to the Kronecker delta will be projected away by $\hat{\Pi}_{ij}^{lm}$.\\
We promote the scalar field $\chi\left(\textbf{k},\tau\right)$ to an operator $\hat{\chi}\left(\textbf{k},\tau\right)$, and we move to Fourier space
\begin{equation}	
\hat{\chi}\left(\textbf{x},\tau\right)=\frac{1}{a\left(\tau\right)}\int\frac{\mathrm{d}^{3}\textbf{k}}{\left(2\pi\right)^{3/2}}e^{i\textbf{k}\cdot\textbf{x}}\hat{\chi}\left(\textbf{k},\tau\right)\,.
\end{equation}
Substituting this expression into the stress-energy tensor and then in eq.\eqref{soluzionesorgente}, the correlator of the sourced GW reads \cite{Cook:2011hg}
\begin{align}\label{espressione}
	\left\langle h_{ij}\left(\textbf{k},\tau\right)h_{ij}(\textbf{k}',\tau)\right\rangle=&\frac{1}{2\pi^{3}M_{\rm pl}^{4}}\int\frac{\mathrm{d}\tilde{\tau}_{1}}{a\left(\tilde{\tau}_{1}\right)^{2}}G_{k}\left(\tau,\tilde{\tau}_{1}\right)\times\nonumber\\	
	&\times\int\frac{\mathrm{d}\tilde{\tau}_{2}}{a\left(\tilde{\tau}_{2}\right)^{2}}G_{k'}\left(\tau,\tilde{\tau}_{2}\right)\Pi_{ij}^{\,ab}\left(\textbf{k}\right)\Pi_{ij}^{\,cd}\left(\textbf{k}'\right)\times\nonumber\\	&\times\int\mathrm{d}^{3}\textbf{p}\,\mathrm{d}^{3}\textbf{p}'p_{a}\left(\textbf{k}_{\textbf{b}}-\textbf{p}_{\textbf{b}}\right)p'_{c}\left(\textbf{k}'_{\textbf{d}}-\textbf{p}'_{\textbf{d}}\right)\times\nonumber\\
&\times\left\langle \hat{\chi}\left(\textbf{p},\tilde{\tau}_{1}\right)\hat{\chi}\left(\textbf{k}-
\textbf{p},\tilde{\tau}_{1}\right)\hat{\chi}\left(\textbf{p}',\tilde{\tau}_{2}\right)\hat{\chi}\left(\textbf{k}'-\textbf{p}',\tilde{\tau}_{2}\right)\right\rangle\,.
\end{align}
Exploiting Wick's theorem and neglecting the disconnected term, the GW power-spectrum results a function of the two-point correlator of the scalar operators $\left\langle \hat{\chi}\left(\textbf{p},\tilde{\tau}_{1}\right)\hat{\chi}\left(\textbf{q},\tilde{\tau}_{2}\right)\right\rangle$. 
The latter quantity is obtained solving the equation of motion for the scalar field $\chi$ coming from the Lagrangian \eqref{lagscalar}. 
Decomposing $\hat{\chi}\left(\textbf{k},\tau\right)$ in terms of creation 
and annihilation operators, as
\begin{equation}
	\hat{\chi}\left(\textbf{k},\tau\right)=\chi\left(\textbf{k},\tau\right)\hat{a}_{\textbf{k}}+\chi
^{\ast}\left(-\textbf{k},\tau\right)\hat{a}^{\dag}_{-\textbf{k}}\,,
\end{equation}
from the Lagrangian \eqref{lagscalar}, the equation of motion for $\chi$ reads
\begin{equation}\label{eqmotochi}
	\chi''\left(\textbf{k},\tau\right)+\omega\left(\textbf{k},\tau\right)^{2}\chi\left(\textbf{k},\tau\right)=0\,,
\end{equation}
with
\begin{equation}
	\omega\left(\textbf{k},\tau\right)^{2}\equiv k^{2}+g^{2}a\left(\tau\right)^{2}\left[\varphi\left(\tau\right)-\varphi_{0}\right]^{2}-\frac{a''\left(\tau\right)}{a\left(\tau\right)}\,.
\end{equation}
This expression can be approximated in different ways depending on the behavior of the system. In particular, three main periods can be identified:
\begin{itemize}
\item At the beginning of the inflationary stage, the Universe does not contain quanta of the $\chi$ field, and then the source term of tensor modes equation vanishes.
\item When the inflaton approaches $\varphi_{0}$, and then $m_{\chi}\left(t\right)$ goes to zero, the production of the $\chi$ quanta starts and for a period $\Delta t_{\rm nad}$ the evolution of 
$m_{\chi}$ is non-adiabatic, that is $\dot{m}_{\chi}\geq m_{\chi}^{2}$. In order to get efficient production $\Delta t_{\rm nad}$ has to be shorter than the Hubble time. During a de Sitter stage, 
the evolution equation of the inflaton is approximated by a linear relation $\varphi\left(t\right)=\varphi_{0}+\dot{\varphi_{0}}t$. Therefore, from the condition of non-adiabaticity and the 
expression of $m_{\chi}$, one gets $\Delta t_{\rm nad}\simeq \left(g\dot{\varphi}_{0}\right)^{-1/2}$, and then the efficiency request, reads 
$g\gg H^{2}/\left|\dot{\varphi}_{0}\right|$.
At the same time this condition allows us to neglect the expansion of the Universe in that interval of time.
\item When the inflaton leaves the $\varphi_{0}$ value, $m_{\chi}$ comes back to evolve adiabatically and the production of $\chi$ quanta stops, but the Universe is left with a 
significant content of $\chi$ particles which works as a source for GW. 
\end{itemize}
Let us start by considering the latter stage. In order to obtain the power-spectrum of the sourced GW, one needs to find the amount of $\chi$ quanta left by the non-adiabatic stage. 
Under the condition of non-adiabaticity and the approximation of linear evolution of the inflaton, eq.\eqref{eqmotochi} takes the form
\begin{equation}\label{nonadieq}
	\ddot{\chi}+\left(k^{2}H^{2}\tau_{0}^{2}+g^{2}\dot{\varphi}^{2}_{0}t^{2}\right)\chi=0\,,
\end{equation}
where $\varphi\left(t=0\right)=\varphi\left(\tau=\tau_{0}\right)=\varphi_{0}$.
From this equation, following the procedure of \cite{Kofman:1997yn}, 
the amount of $\chi$ quanta produced during the non-adiabatic period can be found. 
During the adiabatic stage, the expression of $\omega$ can be simplified as $\omega\simeq\left|g\left[\varphi\left(\tau\right)-\varphi_{0}\right]/\left(H\tau\right)\right|$ \cite{Cook:2011hg}, where $\varphi\left(\tau\right)-\varphi_{0}\simeq-\left(\dot{\varphi}_{0}/H\right)\mathrm{ln}\left(\tau/\tau_{0}\right)$. 
For $\tau\rightarrow 0$, that is at the end of inflation, the correlator \eqref{espressione} results \cite{Cook:2011hg}
\begin{equation}
	\left\langle h_{ij}\left(\textbf{k}\right) h_{ij}\left(\textbf{k}'\right)\right\rangle=\frac{\delta^{\left(3\right)}\left(\textbf{k}+
	\textbf{k}'\right)}{2\pi^{5}k^{6}\left|\tau_{0}\right|^{3}}\frac{H^{4}}{M_{\rm pl}}\left(1+\frac{1}{4\sqrt{2}}\right)\times\left(\frac{g\dot{\varphi}_{0}}{H^{2}}\right)^{3/2}F_{\left|\Delta\tau_{\rm nad}/\tau_{0}\right|}\left(k\left|\tau_{0}\right|\right)\,,
\end{equation}
where
\begin{equation}
	F_{\epsilon}\left(y\right)\equiv\left|\int^{1-\epsilon}_{0}x\frac{\left(\sin xy-xy\cos xy\right)}{\mathrm{ln}x}\mathrm{d}x\right|^{2}\stackrel{\epsilon\rightarrow 0}{\simeq}\left[\left(y\cos y-\sin y\right)\mathrm{ln}\epsilon\right]^{2}\,.
\end{equation}
To get the total inflationary GW power-spectrum, this correlator has to be added to the contribution from the vacuum oscillations of the metric tensor. Then we have
\begin{align}	
P_{\rm h}\left(k\right)\simeq\frac{2H^{2}}{\pi^{2}M_{\rm pl}^{2}}&\left[1+4.8\times10^{-4}\frac{\left(k\tau_{0}\cos k\tau_{0}\sin k\tau_{0}\right)^{2}}{\left|k\tau_{0}\right|^{3}}\times\right.\nonumber\\
&\left.\times\frac{H^{2}}{M_{\rm pl}^{2}}\left(\frac{g\dot{\varphi}_{0}}{H^{2}}\right)^{3/2}\mathrm{ln}^{2}\left(\frac{\sqrt{g\dot{\varphi}_{0}}}{H}\right)\right]\,.
\end{align}
The effect due to the particle production is represented by a scale-dependent contribution added to the usual scale-invariant power-spectrum.
Observing that $\dot{\varphi}_{0}=\sqrt{2\epsilon}H\,M_{\rm pl}$ and considering reasonable values of $\epsilon$, one can find that the correction to the standard tensor power-spectrum is at most of the order of $10^{-2}\sqrt{H/M_{\rm pl}}$, which is several orders of magnitude smaller than unity.
We thus conclude that the presence of a scalar particle gas during a de Sitter stage, does not give rise to a significant amount of GW able to produce observable features in the power-spectrum.\\
Let us now consider the non-adiabatic period, when $\chi$ quanta are rapidly produced.
Solving eq.\eqref{nonadieq}, one can show that the contribution to the tensor power-spectrum due to this stage is of the same order of magnitude of that of the adiabatic period, modulo a 
logarithmic term \cite{Cook:2011hg}. Therefore, again this term is negligible with respect to the usual one. It has been noted that also if several bursts of scalar particles production develop during the rolling down of the inflaton (\textit{trapped inflation}; see \cite{Pearce:2016qtn} for a recent analysis of this scenario), the amount of generated GW is still several orders of magnitude smaller than the contribution from vacuum oscillations \cite{Cook:2011hg}.\\
In summary, the production of an extra scalar field coupled to the inflaton as described by eq.\eqref{lagscalar}, does not lead to a significant contribution to the GW power-spectrum.\\

\subsubsection{Axion inflation: pseudoscalar inflaton coupled to a gauge field}\label{sezgauge}
Consider a system described by the Lagrangian
\begin{equation}\label{modellogauge}	
\mathscr{L}=-\frac{1}{2}\partial_{\mu}\varphi\partial^{\mu}\varphi-V\left(\varphi\right)-\frac{1}{4}F_{\mu\nu}F^{\mu\nu}-\frac{\varphi}{4f}F_{\mu\nu}\tilde{F}^{\mu\nu}\,,
\end{equation}
where the potential $V\left(\varphi\right)$ drives the slow-roll evolution, $f$ is the measure of the coupling between the pseudo-scalar inflaton $\varphi$ and the gauge field $A_{\mu}$, $F_{\mu\nu}=\partial_{\mu}A_{\nu}-\partial_{\nu}A_{\mu}$ is the field strength associated to the gauge field and $\tilde{F}^{\mu\nu}=\tau^{\mu\nu\beta}F_{\alpha\beta}/\left(2\sqrt{-g}\right)$ its dual.\\
The coupling between the two fields leads to two main phenomena: the production of GW (and of scalar perturbations) which we are interested in, but also the back-reaction on the background dynamics. 
In fact the production of the gauge quanta involves transfer of energy form the inflaton sector to the gauge sector, so that there is a new form of energy that affects the background dynamics (see \cite{Anber:2009ua,Barnaby:2011qe,Barnaby:2011vw,Meerburg:2012id,Anber:2012du,Domcke:2016bkh}). Practically this translates into the presence of an additional friction term in the equation of motion of the scalar field, which slows down inflation.\\
As in the previous case the equation of motion for tensor modes is of the form \eqref{ondeconsorgente}, with the solution \eqref{soluzionesorgente}. Working in the Coulomb gauge, $A_{\mu}$ can 
be described by a vector potential $\textbf{A}\left(\tau,\textbf{x}\right)$, defined by $a^{2}\textbf{B}=\nabla\times \textbf{A}$, $a^{2}\textbf{E}=-\textbf{A}'$, where $\textbf{E}$ and $\textbf{B}$ have the usual relation with $F_{\mu\nu}$, so that the spatial stress-energy takes the form $T_{ab}=-a^{2}\left(E_{a}E_{b}+B_{a}B_{b}\right)+\left(...\right)\delta_{ab}$.\\
We need to solve the equation of motion for the gauge field in order to insert the expression in the stress-energy tensor \cite{Barnaby:2010vf,Barnaby:2011vw}. The equations of motion for the vector potential introduced before, read
\begin{equation}	
\left(\frac{\partial^{2}}{\partial\tau^{2}}-\nabla^{2}-\frac{\varphi'}{f}\nabla\times\right)\textbf{A}=0\,,\qquad \qquad \nabla\cdot\textbf{A}=0\,,
\end{equation}
where $\varphi$ spatial gradients have been neglected. We promote $\textbf{A}\left(\tau,\textbf{x}\right)$ to an operator $\hat{\textbf{A}}\left(\tau,\textbf{x}\right)$ and decompose its modes in terms of creation and annihilation operators
\begin{equation}	
\hat{A}_{i}\left(\textbf{x},\tau\right)=\int\frac{\mathrm{d}^{3}\textbf{k}}{\left(2\pi\right)^{2/3}}e^{i\textbf{k}\cdot\textbf{x}}\hat{A}_{i}\left(\tau,\textbf{k}\right)=\sum_{s=\pm}\int\frac{\mathrm{d}\textbf{k}}{\left(2\pi\right)^{3/2}}\left[\epsilon^{i}_{s}\left(\textbf{k}\right)A_{s}\left(\tau,\textbf{k}\right)\hat{a}_{s}^{\textbf{k}}e^{i\textbf{k}\cdot\textbf{x}}+\mathrm{h.c.}\right]\,,
\end{equation}
where $\epsilon^{i}_{s}\left(\textbf{k}\right)$ are such that $\epsilon^{i}_{\pm}\epsilon^{i}_{\mp}=1$ and $\epsilon^{i}_{\pm}\epsilon^{i}_{\pm}=0$.
Assuming a de Sitter background, and approximating $\varphi'/a=\sqrt{2\epsilon}HM_{\rm pl}\simeq\mathrm{const}$, the equations of motion for the amplitude $A_{\pm}$ read
\begin{equation}\label{gauge}
	\frac{\mathrm{d}A_{\pm}\left(k,\tau\right)}{\mathrm{d}\tau^{2}}+\left[k^{2}\pm 2k\frac{\xi}{\tau}\right]A_{\pm}\left(\tau,k\right)=0\,,\qquad\mbox{where}\qquad
	\xi\equiv\frac{\dot{\varphi}}{2fH}=\sqrt{\frac{\epsilon}{2}}\frac{M_{\rm pl}}{f}\,,
\end{equation}
expresses the strength of the coupling between the gauge field and the inflaton.
Equation \eqref{gauge} shows a different behavior for the two helicity states of the gauge field. Depending on the sign of $\xi$, one polarization mode is subjected to an instability, while the other is 
approximately equal to zero. This fact is a direct consequence of the parity violation of the slowly rolling inflaton, and will generate a parity violating power-spectrum of GW. Assuming, for example, $\xi>0$, 
the solution of eq.\eqref{gauge} for $\left(8\xi\right)^{-1}\lesssim \left|k\tau\right|\lesssim 2\xi$ can be approximated by
\begin{equation}\label{solgauge}
	A_{+}\left(k,\tau\right)\simeq\frac{1}{\sqrt{2k}}\left(\frac{k}{2\zeta a H}\right)^{1/4}e^{\pi\xi-2\sqrt{2\zeta k/aH}}\,,
\end{equation}
and at the same time we can put $A_{-}\simeq 0$, neglecting such modes.\\
In order to reveal the different behavior of the two helicity state of the GW we split the tensor modes in the two contribution. Going to the momentum space and projecting $h_{ij}$ on the two helicity modes, 
GW can be described by the functions $h_{s}\left(\textbf{k},\tau\right)$:
\begin{equation}	
h^{ij}\left(\textbf{k}\right)=\sqrt{2}\sum_{s=\pm}\epsilon^{i}_{s}\left(\textbf{k}\right)\epsilon^{j}_{s}\left(\textbf{k}\right)h_{s}\left(\textbf{k}\right)\,,
\end{equation}
where $h_{s}\left(\textbf{k},\tau\right)$ are defined by $h_{\pm}\left(\textbf{k}\right)=\hat{\Pi}^{ij}_{\pm}\left(\textbf{k}\right)h_{ij}\left(\textbf{k}\right)$
with the polarization tensor\\
$\hat{\Pi}^{ij}_{\pm}\left(\textbf{k}\right)=\epsilon^{i}_{\mp}\left(\textbf{k}\right)\epsilon^{j}_{\mp}\left(\textbf{k}\right)/\sqrt{2}$. Promoting $h_{\pm}$ to an operator $\hat{h}_{\pm}$, 
its expression can be explained in terms of the Green function in an analogous way as \eqref{soluzioneseconda}, \cite{Barnaby:2010vf,Barnaby:2011vw}:
\begin{align}\label{hh}	
\hat{h}_{\pm}\left(\textbf{k}\right)=&-\frac{2H^{2}}{M_{\rm pl}^{2}}\int\mathrm{d}\tilde{\tau}\,G_{k}\left(\tau,\tilde{\tau}\right)\tilde{\tau}^{2}\int\frac{\mathrm{d}^{3}\textbf{q}}{\left(2\pi\right)^{2/3}}\hat{\Pi}^{lm}_{\pm}\left(\textbf{k}\right)\times\nonumber\\	&\times\left[\hat{A}'_{l}\left(\textbf{q},\tilde{\tau}\right)\hat{A}'_{m}\left(\textbf{k}-\textbf{q},\tilde{\tau}\right)-\varepsilon_{lab}q_{a}\hat{A}_{b}\left(\textbf{q},\tilde{\tau}\right)\varepsilon_{mcd}\left(k_{c}-q_{c}\right)\hat{A}_{d}\left(\textbf{k}-\textbf{q},\tilde{\tau}\right)\right]\,,
\end{align}
where the Green function is given by eq.\eqref{green}.\\
Putting \eqref{solgauge} into the last expression and using Wick's theorem, the GW power-spectrum can be written in terms of the Green's functions and the amplitude of the gauge field, in particular of the 
parameter $\xi$. For $\xi>1$ the correlator results:
\begin{align}
	\left\langle h_{s}\left(\textbf{k}\right) h_{s}(\textbf{k}')\right\rangle=&\frac{H^{4}\xi}{4\pi^{3}M_{\rm pl}^{4}}e^{4\pi\xi}\delta\left(\textbf{k}+\textbf{k}'\right)\int\mathrm{d}\tilde{\tau}_{1}\mathrm{d}\tilde{\tau}_{2}\left|\tilde{\tau}_{1}\right|^{3/2}\left|\tilde{\tau}_{2}\right|^{3/2}G_{k}\left(\tau,\tilde{\tau}_{1}\right)G_{k}\left(\tau,\tilde{\tau}_{2}\right)\times\nonumber\\	&\times\int\mathrm{d}^{3}\textbf{q}\left|\epsilon^{i}_{-s}\left(\textbf{k}\right)\epsilon^{i}_{+}\left(\textbf{q}\right)\right|^{2}\left|\epsilon^{j}_{-s}\left(\textbf{k}\right)\epsilon^{j}_{+}\left(\textbf{k}-\textbf{q}\right)\right|^{2}\times\nonumber\\	&\times\sqrt{\left|\textbf{k}-\textbf{q}\right|}\sqrt{q}e^{-2\sqrt{2\zeta}}\left(\sqrt{\left|\tilde{\tau}_{1}\right|}+\sqrt{\left|\tilde{\tau}_{2}\right|}\right)\left(\sqrt{q}+\sqrt{\left|\textbf{k}-\textbf{q}\right|}\right)\,.
\end{align}
The two terms in the second line show the different behavior of the two polarization states. In the limit $k\tau\rightarrow 0$ the above integrals are computed numerically, but can 
be approximated by an analytical expression, with accuracy improving with increasing $\xi$ \cite{Sorbo:2011rz}:
\begin{align}
	&\left\langle h_{+}\left(\textbf{k}\right)h_{+}(\textbf{k}')\right\rangle\simeq8.6\times 10^{-7}\frac{H^{4}}{M_{\rm pl}^{4}}\frac{e^{4\pi\xi}}{\xi^{6}}\frac{\delta^{\left(3\right)}\left(\textbf{k}+\textbf{k}'\right)}{k^{3}}\,,\\
	&\left\langle h_{-}\left(\textbf{k}\right)h_{-}(\textbf{k}')\right\rangle\simeq1.8\times 10^{-9}\frac{H^{4}}{M_{\rm pl}^{4}}\frac{e^{4\pi\xi}}{\xi^{6}}\frac{\delta^{\left(3\right)}\left(\textbf{k}+\textbf{k}'\right)}{k^{3}}\,.
\end{align}
The numerical factor reveals a difference of magnitude of about $3$ between the two scale-invariant correlators. Moving to the power-spectra and adding the contribution of GW coming from vacuum 
fluctuations, we have, respectively,
\begin{align}\label{spettro1}
	&P_{\rm T}^{+}=\frac{H^{2}}{\pi^{2}M_{\rm pl}^{2}}\left(1+8.6\times 10^{-7}\frac{H^{2}}{M_{\rm pl}^{2}}\frac{e^{4\pi\xi}}{\xi^{6}}\right)\,,\\
	&P_{\rm T}^{-}=\frac{H^{2}}{\pi^{2}M_{\rm pl}^{2}}\left(1+1.8\times 10^{-9}\frac{H^{2}}{M_{\rm pl}^{2}}\frac{e^{4\pi\xi}}{\xi^{6}}\right)\,.
\end{align}
The parity violation can be quantified by the chirality parameter \cite{Anber:2006xt}:
\begin{equation}
	\Delta\chi=\frac{P_{\rm T}^{+}-P_{\rm T}^{-}}{P_{\rm T}^{+}+P_{\rm T}^{-}}\,,
\end{equation}
which in our case reads
\begin{equation}	
\Delta\chi=\frac{4.3\times10^{-7}\frac{e^{4\pi\xi}}{\xi^{6}}\frac{H^{2}}{M_{\rm pl}^{2}}}{1+4.3\times10^{-7}\frac{e^{4\pi\xi}}{\xi^{6}}\frac{H^{2}}{M_{\rm pl}^{2}}}\,.
\end{equation}
For small $\xi$, when vacuum oscillations dominate the tensor power-spectrum, $\Delta\chi\rightarrow 0$, while at large $\xi$, when sourced GW constitute the main contribution, $\Delta\chi\rightarrow 1$. 
Thus, the departure of $\Delta\chi$ from zero represents an interesting feature, being a signature of a parity violation mechanism, which is not expected for GW from vacuum fluctuations. In this direction 
cross-correlations could carry significant information \cite{Lue:1998mq,Vachaspati:2001nb,Barnaby:2011vw,Cook:2013xea,Shiraishi:2013kxa,Bartolo:2014hwa}.

\paragraph{Constraints from current data}
Since the gauge field is coupled to the inflaton, its inverse decay leads to the production of scalar perturbations besides those coming form the vacuum oscillations of $\varphi$. More precisely, the presence of the gauge field fluctuations gives rise to a source in the equation of motion of the inflaton perturbations. The gauge-invariant scalar perturbation $\hat{\zeta}$ results proportional to the factor $e^{2\pi\xi}/\xi^{3}$ \cite{Cook:2013xea}, so that a weaker coupling leads to a smaller amplitude of the sourced scalar perturbations (this is interesting only if the inflaton field is the main source of the curvature fluctuations). 
The scalar perturbations power-spectrum is well constrained by measurements of CMB anisotropies, concerning their amplitude, spectral index and level of non-Gaussianity. The main point here is that, sourced curvature perturbations are expected to be highly non-Gaussian \cite{Barnaby:2010vf,Sorbo:2011rz,Barnaby:2011vw,Anber:2012du}. For this reason, a strict constraint for this kind of inflationary model, is provided by the upper bounds on non-Gaussianity. For weak gauge regime, the expected non-Gaussianity level is predicted to be \cite{Barnaby:2010vf,Sorbo:2011rz}
\begin{equation}\label{nngauss}
	f_{\rm NL}^{\rm equil}\simeq 8.9\times 10^{4}\frac{H^{6}}{M_{\rm pl}^{6}}\frac{\epsilon^{3}e^{6\pi\xi}}{\xi^{9}}\,,
\end{equation}
for an equilateral configuration, which is that expected to present the maximal amplitude.
Considering the constraints on $f_{\rm NL}$ obtained in the Planck 2015 release \cite{Ade:2015ava}, results in $\xi<2.5$. 
Imposing this limit, the correction to the power-spectrum due to the sourced GW results negligible on cosmological scales compared to planned experimental sensitivity.

\paragraph{The possibility of a blue spectrum}\label{bluepart}
Actually, constraints coming from CMB measurements are limited to a range of scales. From that bounds, in principle, we cannot extract limits on other scales. This is an interesting fact which leaves open the possibility of observing these GW by interferometer experiments. In fact the model parameter $\xi$ is actually time dependent, more precisely it usually grows during the inflationary period in according to the inflationary potential \cite{Cook:2011hg,Domcke:2016bkh}. 
Larger values of such a parameter, means larger amplitudes of the sourced GW, correspondent to a blue power-spectrum. Limits coming from the CMB put constraints on the value of $\xi$ only at the time of exit the horizon of CMB scales, this means that we could have a higher $\xi$ value corresponding to later times, that is on scales interesting by interferometer experiments. The limits found above then applies only to $\xi$ evaluated on CMB scales.
Moreover, when the gauge field becomes non-negligible with respect to the background evolution, the bound \eqref{nngauss} due to non-Gaussianity of scalar perturbation cannot be applied yet, and then $\xi$ is admitted to grow rapidly \cite{Barnaby:2011qe,Anber:2012du,Domcke:2016bkh}.
The time dependence of $\xi$ is obtained from the potential shape of the inflaton field. 
Fixing the range of scales constrained by CMB experiments, and choosing one of the allowed values of $\xi$ corresponding to a scale in the explored range, one can find the evolution of the parameter $\xi$ and then the GW power-spectrum corresponding to the chosen value of $\xi$ and to the inflationary potential. 
Based on these considerations, for chaotic and Starobinsky inflationary potentials, a large amount of primordial GW can be obtained, in principle detectable by future experiments \cite{Domcke:2016bkh} such as the the Evolved Laser Interferometer Space Antenna (eLISA) \cite{elisaweb,Klein:2015hvg} and the Advanced-Laser Interferometer Gravitational-wave Observatory (aLIGO)-Advanced-Virgo (adVirgo) \cite{TheLIGOScientific:2014jea,Acernese:2004jn} network.

\paragraph{Alternative scenarios}
In order to have positive prospects for GW detection, in particular for CMB experiments, the model \eqref{modellogauge} seems to become interesting also admitting some variations. In this direction, two modifications appear significant 
\cite{Sorbo:2011rz}: the introduction of a second scalar field with the role of a curvaton, or of a great number of coupled gauge fields, which fits in the framework of string theory.\\
Let us briefly consider the curvaton case. The presence of such a scalar field liberates the inflaton from the task of generating all the amount of the observed scalar perturbations. This fact also liberates the slow-roll parameter $\epsilon$ to be so small as in the standard inflationary models. In this case $f_{\rm NL}^{\rm equil}$ continues to depend on the Hubble parameter, $\epsilon$ and on $\xi$, but from the larger freedom in $\epsilon$ we gain the possibility to consider higher values of $\xi$, on CMB scales, with respect to that of the previous model, so that for example a fully chiral system with $r\simeq0.009$ is admitted by current constraints \cite{Ade:2015ava}. It is found that in this case the parity violation would be detected at $95\%$ confidence level by a cosmic-variance-limited CMB experiment \cite{Gluscevic:2010vv}.\\
Another interesting modification seems to be the following \cite{Barnaby:2012xt}: the strict constraints on our model parameters are due to the presence of sourced scalar perturbations, and we have seen that current constraints on the non-Gaussianity of scalar perturbations lead to an upper bound for $\xi$, and then from eq.\eqref{spettro1}, to an upper bound on the generated GW on CMB scales.
Therefore, to get more freedom in the model parameters, the ratio between the sourced scalar and tensor modes has to be minimized. 
A way to relax this bound appeared to be the introduction of a pseudo-scalar field, coupled to the gauge field through $\alpha/f$, and both minimally coupled to the inflaton \cite{Barnaby:2012xt}.\\
However, in general it is found that, for these kinds of scenarios, the tensor-to-scalar ratio between sourced perturbations is of the order $\sim \epsilon^{2}$, so that usually it is not possible to reach a large value of $r$ thanks to these kind of mechanisms, as noted in \cite{Barnaby:2012xt} and then in \cite{Mirbabayi:2014jqa,Ferreira:2014zia}.

\subsubsection{Scalar inflaton and pseudoscalar field coupled to a gauge field}\label{pseudoparticelle}
Considering the mentioned idea, a system described by the following Lagrangian \cite{Barnaby:2012xt} was considered:
\begin{equation}\label{modpseudo}
\mathscr{L}=\frac{1}{2}\partial_{\mu}\varphi\partial^{\mu}\varphi-V\left(\varphi\right)-\frac{1}{2}\partial_{\mu}\psi\partial^{\mu}\psi-U\left(\psi\right)-\frac{1}{4}F_{\mu\nu}F^{\mu\nu}-\frac{\psi}{4f}F_{\mu\nu}\tilde{F}^{\mu\nu}\,,
\end{equation}
with $\varphi$ the inflaton field, and $\psi$ a pseudo-scalar; see also \cite{Mukohyama:2014gba,Namba:2015gja}. The gauge field is now minimally coupled to the main source of curvature perturbations, the inflaton.
The calculations needed to get the tensor power-spectrum sourced by the gauge field are exactly the same as for eq.\eqref{modellogauge}.\\
Contrary to the first intuition, the curvature power spectrum sourced by the perturbations of the scalar field $\delta \psi$ is found to be the same as the previous-analyzed model \cite{Ferreira:2014zia}. 
Also in the case in which the decaying of the pseudo-scalar field is considered \cite{Mukohyama:2014gba}, the constraints coming from the non-Gaussianity of the scalar sector limits the GW production. In particular, in the latter scenario, the amplitude of the sourced scalar perturbations actually results to be proportional to the number of e-foldings during which $\psi$ is rolling. This fact allows to obtain a rather large tensor-to-scalar ratio.
However, curvature perturbations due to the coupling with the pseudo-scalar field, are sourced mainly in correspondence of those modes that cross the horizon when $\dot{\psi}\neq 0$ \cite{Namba:2015gja}. Therefore, choosing a suitable potential shape for the pseudo-scalar field, it is possible to obtain a significant GW production on certain scales, satisfying at the same time the constraints coming from CMB data. In fact, it has to be noted that in this case the primordial non-Gaussianity signal will be mainly non-vanishing up to multipoles $\ell \simeq100$. This would relax the bounds on the non-Gaussianity level since they would be obtained from the CMB temperature anisotropies only up to those scales (see \cite{Namba:2015gja}, for the case of a specific potential and the calculations of CMB signatures). Then, for particular choices of the pseudo-scalar potential, the sourced GW power-spectrum is expected to be widely scale-dependent. Clearly also the current bound on the tensor-to-scalar ratio has to be considered.

\paragraph{Chirality}
An interesting aspect is the capability of planned experiments to capture the parity violation of CMB power-spectra.
Refs. \cite{Gluscevic:2010vv,Cook:2013xea,Namba:2015gja,Shiraishi:2016yun} examined the prospects for a detection of the parity violation for this model in the CMB power-spectra. 
Varying the value of $\xi$ and $\epsilon$, and considering the current upper bound on the $r$ value, it is found that the detection of $\Delta\chi=1$ at $1\sigma$ is possible, in principle, 
by experiments such as Spider \cite{Crill:2008rd} and CMBpol \cite{Baumann:2008aq}. For the $1\sigma$ contours in the $r-\Delta\chi$ plane, see plot 4 of \cite{Barnaby:2012xt}. For further details and references see section \ref{cmbchirali}. 
The capability of a detection of the chirality by laser interferometer experiments is discussed in \cite{Seto:2007tn,Seto:2008sr,Crowder:2012ik}. In particular, in \cite{Crowder:2012ik}, it is shown the significant advantage, in this direction, of collecting data by a network composed of more than two observatories.
In this kind of scenario also the contribution to non-Gaussianity of CMB power-spectra due to tensor modes is a significant aspect to be taken into account too \cite{Cook:2013xea, Namba:2015gja,Shiraishi:2016yun}.
\section{Gravitational wave production during reheating after inflation}\label{analisi}

As seen in section \ref{secondo}, in the presence of large, time-dependent inhomogeneities in the distribution of the energy-density of the Universe, GW are produced in a \textit{classical} way. Such a situation could occur also during the reheating stage, where a source term in the GW equation of motion can be provided by the rapid decay of the inflaton field.\\
The production of gravitational radiation during reheating was first pointed out by Khlebnikov and Tkachev in \cite{Khlebnikov:1996mc}. Since such a stage occurs in most inflationary models, the GW signal generated at that time in principle represents a source of information on the inflationary physics and the subsequent reheating period. GW remain decoupled since the moment of their production and therefore the features of their spectrum represent a very interesting probe of the physics of the reheating period, such as the coupling between the inflaton and other fields; see \cite{Hyde:2015gwa} and refs. therein.\\
At the end of inflation the field that has driven the accelerated expansion starts oscillating around the minimum of its potential. In such a way it produces elementary particles which interact to each other, eventually 
leading to a state of thermal equilibrium. The first stage of this process, in which the inflaton field oscillates was initially described by perturbation theory techniques, considering the oscillating field as a decaying collection of particles \cite{Albrecht:1982mp,lyth2009primordial}. However, if oscillations are large and coherent they lead to a non-perturbative process, in which the inflaton energy is explosively moved to a coupled-energy sector. This rapid mechanism is called \textit{parametric resonance} \cite{Traschen:1990sw,Kofman:1994rk,Kofman:1997yn}. In this case a perturbative description does not work, being the process violent and rapidly efficient. To distinguish such a rapid stage from the whole mechanism, people call it \textit{preheating}. After such an explosive stage the produced particles are not in thermal equilibrium, contrary to the case of the perturbative mechanism; so, another phase is needed to get thermalized radiation.
The preheating is the period we are mainly interest in here, being a scenario of gravitational radiation production.\\
The most studied inflationary scenarios in which GW production during the following preheating phase has been investigated, are chaotic inflation \cite{Linde:1981mu,Albrecht:1982wi,Linde:1983gd}, and hybrid inflation \cite{Linde:1993cn}. In the latter model preheating develops in a slightly different way compared to the first, and the mechanism is called \textit{tachyonic preheating} \cite{Linde:1993cn,GarciaBellido:1997wm,Felder:2000hj,GarciaBellido:2002aj}. In both cases the process of GW production is substantially the same.

\subsection{Preheating mechanisms}

\subsubsection{Preheating with parametric resonance}\label{sezchaotic}
Parametric resonance typically happens when the field that drives inflation is coupled to another field whose mass is negligible during the accelerated 
expansion \cite{Khlebnikov:1996mc,Traschen:1990sw,Kofman:1994rk,Kofman:1997yn,Easther:2006gt}.\\
Consider a system in which the inflaton $\varphi$ is coupled to another light scalar field $\chi$, by the following Lagrangian
\begin{equation}\label{azionepre}
	\mathscr{L}=\frac{1}{2}\partial_{\mu}\varphi\partial^{\mu}\varphi+\frac{1}{2}\partial_{\mu}\chi\partial^{\mu}\chi-\mathcal{V}\left(\varphi,\chi\right)\,,
\end{equation}
with
\begin{equation}
	\mathcal{V}\left(\varphi,\chi\right)=V\left(\varphi\right)+\frac{1}{2}g^{2}\varphi^{2}\chi^{2}-\frac{1}{2}m_{\chi}^{2}\chi^{2}\,,
\end{equation}
with $g$ the coupling between the two scalar fields. During inflation the secondary field is supposed to be light, so that the inflationary dynamics is governed by $\varphi$.
Following \cite{Easther:2006gt}, in the analysis we will neglect the Universe expansion and the mass of the secondary field.
Ignoring the second field, the equation of motion for the background part of the inflaton is eq.\eqref{friz},
where, contrary to what happens during inflation, the field cannot be considered homogeneous and the kinetic energy cannot be neglected.
The dynamics is determined by the potential of the inflaton field. 
Consider the case in which the inflaton potential reads
\begin{equation}
	V\left(\varphi\right)=\frac{1}{2}m_{\varphi}^{2}\varphi^{2}\,.
\end{equation}
As mentioned before, eq.\eqref{friz} approximately reduces to a damped harmonic oscillator, solved by $\varphi\left(t\right)=\Phi\left(t\right)\sin\left(m_{\varphi}t\right)$,
with $\Phi$ the time-dependent amplitude, which varies slowly over a single cycle \cite{Easther:2006gt}.\\
From the action \eqref{azionepre}, the equation of motion for the secondary field results
\begin{equation}
	\ddot{\chi}+3H\dot{\chi}-\frac{1}{a^{2}}\nabla^{2}\chi+g^{2}\varphi^{2}\chi=0\,,
\end{equation}
so that moving to Fourier space and changing variables to \cite{Easther:2006gt}, 
\begin{equation}
	q=\frac{g^{2}\Phi^{2}}{4m_{\varphi}^{2}}\,, \qquad A_{k}=\frac{k^{2}}{m_{\varphi}^{2}}+\frac{g^{2}\Phi^{2}}{2m_{\varphi}^{2}}=\frac{k^{2}}{m_{\varphi}^{2}}+2q\,,\qquad z=m_{\varphi}t\,,
\end{equation}
the equation for a single mode $\chi_{k}$ becomes the Mathieu equation
\begin{equation}
	\frac{{\rm d}^{2}\chi_{k}}{{\rm d}z^{2}}+\left[A_{k}-2q\cos\left(2z\right)\right]\chi_{k}=0\,.
\end{equation}
The solutions of the Mathieu equation are given by the following combination
\begin{equation}
	\chi_{k}\left(z\right)=f_{+}\left(z\right)e^{\mu_{k} z}+f_{-}\left(z\right)e^{-\mu_{k} z}\,,
\end{equation}
where $f_{\pm}$ are periodic functions and $\mu_{k}$ is a complex number, which depends on both the wavenumber and the parameters of the system, included in $A_{k}$ and $q$. If $\mu_{k}$ has an imaginary part, the solution $\chi_{k}$ presents an exponential growth. This is the case which we are interested in, for the production of gravitational radiation to take place. For each mode $k$ one can calculate $A_{k}$ and $q$ and then establish the bands corresponding to stable modes and those for which parametric resonance occurs. The key parameter in distinguishing the two behaviors is $q$. In a rough approximation, broad bands of exponentially growing modes occur for $q>1$. The explosion of the amplitude of those modes can be interpreted as a rapid particle production (at least for bosonic species), being the number density of particles per mode proportional to the mode energy.\\
Preheating ends when the exponential grow becomes energetically disadvantageous, in particular when the energy-density of the created particles becomes comparable with the energy-density of the oscillating field. This could happen after a few oscillations of the inflaton field. The energy distribution resulting from the parametric resonance is clearly highly non-thermal; the pumped modes then dissipate their energy by interaction with other modes, leading to the thermalization of the Universe.\\
If instead the inflationary potential is given by
\begin{equation}\label{vchaotic}
	V\left(\varphi\right)=\frac{\lambda_{\rm c}}{4}\varphi^{4}\,,
\end{equation}
with $\varphi$ massless, the inflaton at the end of the accelerated expansion does not undergo sinusoidal oscillations, but results proportional to the elliptic cosine \cite{Easther:2007vj,Greene:1997fu}.
The equation of motion for $\chi_{k}$, at first order, takes the form of a Mathieu equation also in this scenario. The solutions are of the same kind as the previous case, with oscillatory and exponential behaviors depending on $k$. The band of $k$ which grow exponentially, depends on the parameter $q=g^{2}/\lambda_{\rm c}$.
Then the system evolves as before.\\
For most inflationary models with a secondary field $\chi$ the equation of motion for $\chi_{k}$ can be rewritten as a Mathieu equation, so that the mode band presents an exponential growth. However, the details of the excitation depend on the inflaton potential.\\
Notice that above modeling was performed at the linear level, neglecting the Universe expansion and the possible non-zero mass of the second field. Furthermore, 
we have neglected the effect of the back-reaction of the growing $\chi$ modes. If these features are restored, the situation becomes extremely complicated \cite{Easther:2007vj},
then the only way to study the phenomenon involves numerical simulations.

\paragraph{Bubbly stage} 
Soon after the production of the bumped modes a \textit{bubbly stage} takes place \cite{Felder:2006cc}; see also \cite{Dufaux:2007pt}.
When $\chi$ oscillations become non-linear, inflaton modes are excited too due to the back-reaction of the secondary field, and $\varphi$ oscillations start growing very fast with different and changing frequencies. The profile of $\varphi\left(\textbf{x},t\right)$ becomes a superposition of a still oscillating homogeneous part plus inhomogeneities induced by the coupling with growing modes of the secondary field, with peaks in correspondence with those of the $\chi$ field. These high peaks are called \textit{bubbles}. When the height of these peaks becomes comparable with the background value of the field, the bubbles begin to spread, expanding and colliding with each other.
The collisions among such structures lead to a turbulent phase which eventually brings the system to homogeneization and local thermal equilibrium.
In the short stage of bubble formation and collisions, the main contribution to the production of GW takes place due to the growing inhomogeneities \cite{Felder:2006cc}.\\
In summary preheating after chaotic inflation can be described by four phases \cite{Dufaux:2007pt}: a linear preheating stage with excitations of $\chi$ modes, a non-linear \textit{bubbly} stage, a period of turbulence and a final stage where thermal equilibrium is restored.

\subsubsection{Tachyonic preheating}\label{sezhybrid}
\begin{figure}[ht]
    \centering
    \includegraphics[width=0.7\textwidth]{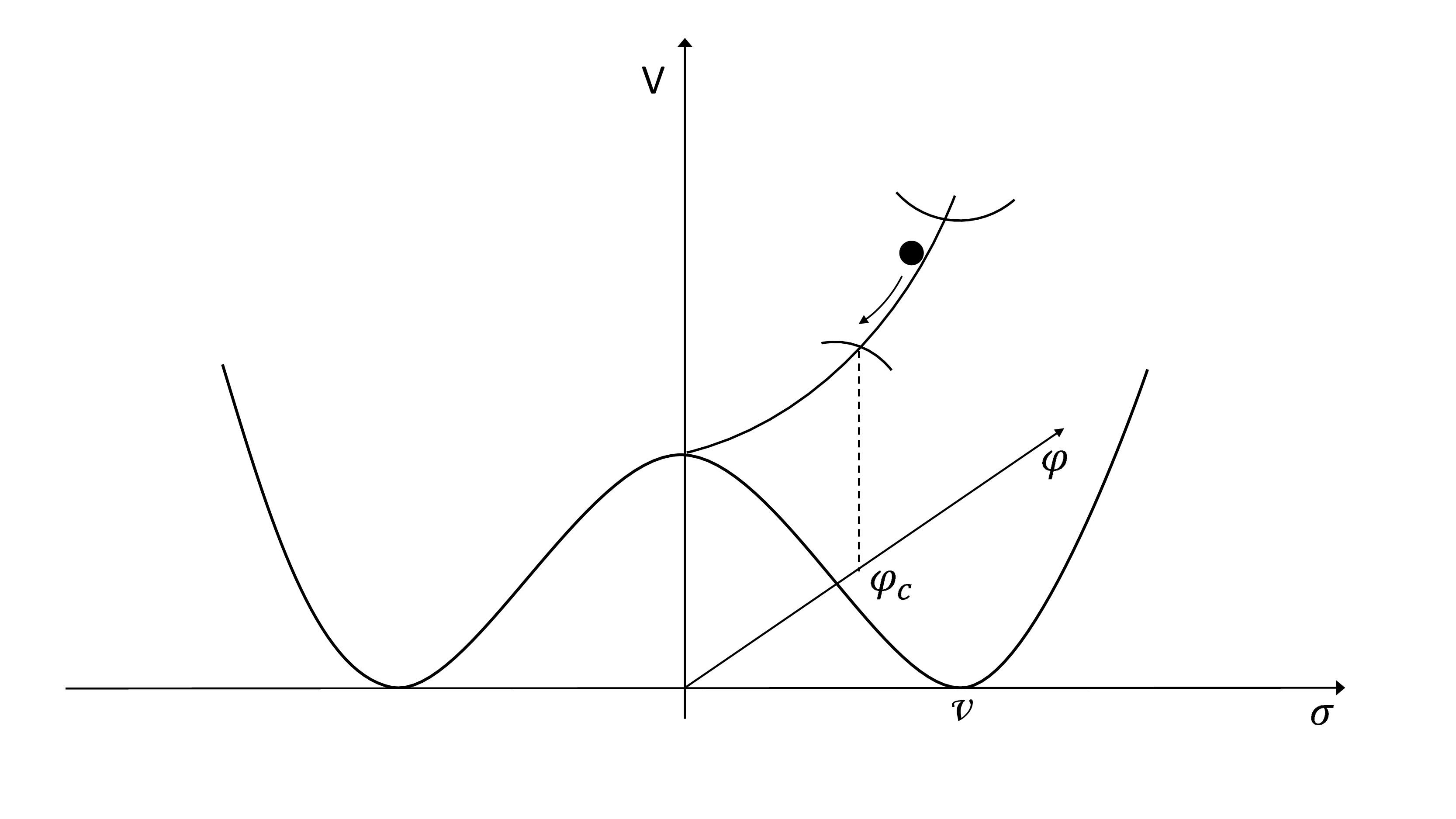}
    \caption{Hybrid inflation potential. The field rolls down the potential up to the critical point $\varphi_{c}$, and then reaches the true minimum of the potential $\varphi=0$, $\sigma=v$.}
  
    \label{ibridofig}
\end{figure}

The main condition for a tachyonic preheating \cite{Dufaux:2008dn,Felder:2000hj,Felder:2001kt,GarciaBellido:2002aj} to take place is a field $\phi$ descending from a maximum of its potential $V\left(\phi\right)$ and then oscillating around the minimum. In fact, around the maximum of the potential there could be a region where the quadratic mass of the field becomes negative and the field fluctuations grow exponentially.\\
In this kind of scenario the equation of motion for such field modes assumes the form
\begin{equation}
	\ddot{\phi}_{k}\left(t\right)+E_{k}^{2}\left(t\right)\phi_{k}\left(t\right)=0\qquad \mathrm{with}\qquad E_{k}^{2}\left(t\right)\equiv k^{2}+m^{2}\left(t\right)\,,
\end{equation}
where $m$ is the mass of the field, that is $m^{2}\left(t\right)=V_{\phi\phi}$. When $\phi$ is around the maximum of the potential, the mass squared becomes negative and $E_{k}^{2}$ 
could pass through zero. This non-adiabatic variation leads to an exponential growth of the field fluctuations. The $\phi$ value in correspondence to the inversion of the potential (that is when the squared mass becomes zero) is called critical point and represents the moment when preheating switches off. The field, after rolling down from the maximum, reaches the minimum and then starts oscillating.  
The whole process of preheating ends when the oscillations around the minimum become too small for the field to get around the maximum.\\
This mechanism is what could happen after Hybrid inflation \cite{Linde:1993cn}.
From our point of view, the advantage of these models consists in the fact that they can occur for a large range of energy scales, from the GUT scale to the GeV scales, contrary to large-field models. Hybrid inflation in fact does not require small couplings to explain the observed CMB anisotropies. 
We will see that this leads to the possibility of production of GW at frequencies and amplitudes more accessible for planned experimental capabilities.\\
In this case the field that descends from the maximum and starts oscillating is a secondary field called \textit{waterfall field} $\sigma$. In these kind of scenario the preheating process is even more violent than in the case of parametric resonance that occurs after chaotic inflation. 
Because of the spinodal instability, some $\sigma$ fluctuations are exponentially amplified leading to a spatial distribution of the field characterized by high peaks, the bubbles \cite{GarciaBellido:2007dg,Price:2008hq}. Then, the field can be considered as a collection of classical waves with a fixed dispersion. When the non-linear regime is reached, collisions and scatterings between bubbles start and the system is driven into a turbulent stage after which local thermal equilibrium is achieved \cite{Dufaux:2008dn}.\\
Consider now models described by the following potential
\begin{equation}\label{ibrido}	
\mathcal{V}\left(\varphi,\sigma\right)=\frac{1}{4}\lambda_{\rm t}\left(\sigma^{2}-v^{2}\right)^{2}+\frac{1}{2}g^{2}\varphi^{2}\sigma^{2}+V\left(\varphi\right)\,,
\end{equation}
where $\left|\sigma\right|^{2}=\sigma_{1}^{2}+\sigma^{2}_{2}$ with $\sigma_{1}$ and $\sigma_{2}$ two real scalar fields, is the \textit{waterfall field}, and the inflaton $\varphi$ a real field too. The vacuum of the system corresponds to $\sigma=\pm v$ and $\varphi=0$.
The critical point at which the curvature changes is given by $\varphi_{\rm c}\equiv v\sqrt{\lambda_{\rm t}}/g$.
For $\varphi>\varphi_{\rm c}$ the inflaton decreases, slowing down in the valley where $\sigma=0$ and the masses are positive. Inflation ends when $\varphi$ reaches the $\varphi_{\rm c}$ value or until the slow-roll conditions do not break up if this happens before. 
When $\varphi_{\rm c}$ is reached, the curvature of the effective potential with respect to $\sigma$ becomes negative, $\sigma$ acquires a tachyonic mass and the fields roll rapidly towards the true minimum, at $\varphi=0$, $\sigma=\pm v$, see fig.\ref{ibridofig}. 

\noindent
When $\sigma$ is around the maximum the homogeneous field energy rapidly decays due to the exponential growth of the field fluctuations.
As in the previous model, we can identify a precise sequence of events \cite{GarciaBellido:2007af}: the exponential growth of $\sigma$ with spinodal instability, the nucleation and collision of bubble-like structures associated with the peaks of a Gaussian random field, a turbulent regime and the final thermalization.\\
As we already mentioned, the bubble stage is not a homogeneous process, during this phase $\sigma$ spherically symmetric bubbles arise and the collision among them leads to short-wavelength inhomogeneities which constitute the source for GW \cite{GarciaBellido:2007af}. The amplitude of GW produced after hybrid inflation was first estimated by \cite{GarciaBellido:1998gm}, employing the formalism of parametric resonance. Then, in \cite{GarciaBellido:2007dg,Price:2008hq} a more accurate calculations have been performed.

\subsection{Gravitational wave production}
We consider here the most investigated models in the literature, namely preheating after chaotic and hybrid inflation.

\subsubsection{Mechanism of production}
The amplification of a band of field modes makes the Universe inhomogeneous. In particular, highly pumped modes lead to large and time-dependent inhomogeneities in the energy-density of the Universe, generating a non-trivial quadrupole moment. On the other hand also asymmetric collisions of the bubbles lead to quadrupolar inhomogeneities. In both situations, when the quadrupolar moment changes significantly fast, GW are produced. The GW production has been analyzed in several works \cite{Felder:2006cc,Easther:2006gt,Easther:2006vd,Dufaux:2007pt,GarciaBellido:2007dg,GarciaBellido:2007af,Easther:2007vj,Price:2008hq,Dufaux:2008dn,Figueroa:2016ojl}.\\
Notice the difference between the production mechanism during inflation and this one: here the production is \textit{classical}, GW are sourced by inhomogeneities which cannot be neglected and leads to a source term in the GW equation of motion. On the contrary, during inflation the production occurs due to a \textit{quantum} mechanism, the inhomogeneities of the fields correspond to small perturbations which can be neglected and lead to a vanishing source term in the equation of motion, then quantum fluctuations are amplified and stretched giving rise to super-horizon tensor modes.\\
At first order in the perturbations the equation of motion for $h_{ij}$ results:
\begin{equation}\label{gweq}	
\ddot{h}_{ij}+3H\dot{h}_{ij}-\frac{1}{a^{2}}\nabla^{2}h_{ij}=\frac{16\pi}{M_{\rm pl}^{2}}\Pi_{ij}\left(\varphi,\chi\right)\,,
\end{equation}
where by $\Pi_{ij}$ is the transverse and traceless part ($TT$) of the total stress-energy tensor \cite{GarciaBellido:2007af}
\begin{equation}\label{energiamomento}	
T_{\mu\nu}=\frac{1}{a^{2}}\left[\partial_{\mu}\chi\partial_{\nu}\chi+\partial_{\mu}\varphi\partial_{\nu}\varphi+g_{\mu\nu}\left(\mathscr{L}-\langle p\rangle\right)\right]\,,
\end{equation}
with $\langle p\rangle$ the background homogeneous pressure.
Let us now quantify $\Pi_{ij}$: being the second term of eq.\eqref{energiamomento} proportional to $g_{ij}=\delta_{ij}+h_{ij}$, the operation of $TT$ projection drops out the terms proportional to $\delta_{ij}$, the other part of the second term is of second order in the tensor perturbations and so negligible. Then, the operation of projection can be performed on the remaining part of the stress-energy tensor, so that
\begin{equation}\label{sorgentepre}	
\Pi_{ij}\left(\varphi,\chi\right)=\frac{1}{a^{2}}\left[\partial_{i}\chi\partial_{j}\chi+\partial_{i}\varphi\partial_{j}\varphi\right]^{\rm TT}\,.
\end{equation}
In order to make numerical calculations on a lattice, we write the stress-energy tensor for the GW as in eq.\eqref{energygw}
where the average is to be considered on a sufficiently large volume $V=L^{3}$ in lattice space \cite{phdfigueroa}. 
Then the GW energy-density in Fourier space is written as 
\begin{equation}\label{gwdensity}	
\rho_{\rm gw}=\frac{M_{\rm pl}^{2}}{32\pi a^{2}}\frac{1}{L^{3}}\int\frac{\mathrm{d}^{3}\textbf{k}}{\left(2\pi\right)^{3}}\,h'_{ij}\left(\textbf{k},\tau\right)h'^{\ast}_{ij}\left(\textbf{k},\tau\right)\,,
\end{equation}
where $\ast$ stands for the complex conjugate. Unlike the case in which GW are produced in the inflationary stage and then re-enter the horizon in the subsequent eras, here GW are sub-horizon at the time of production and remain sub-horizon until the present time. Then, what we need is the solution of the equation of motion in the presence of the source \eqref{sorgentepre}. We then have to take into account the modulation of the 
GW amplitude that takes place at later times.
As seen in eq.\eqref{radiation}, during radiation dominance the GW amplitude scales as $1/a$, so that the integral in eq.\eqref{gwdensity} evolves with time as $\propto 1/a^{2}$.
Then we have:
\begin{equation}\label{vederepicco}
	\left(\frac{\mathrm{d}\rho_{\rm gw}}{\mathrm{d\,ln} k}\right)\left(k,\tau\right)= \frac{k^{3}M_{\rm pl}^{2}}{a^{2}\left(4\pi L\right)^{3}}\int\frac{\mathrm{d}\Omega_{k}}{4\pi}\,{h}'_{ij}\left(\textbf{k},\tau\right)h'^{\ast}_{ij}\left(\textbf{k},\tau\right)\,,
\end{equation}
so that it is clear that the quantity which directly contributes to the total spectral energy-density is $h'_{ij}$. Then the spectrum at late times can be expressed as
\begin{equation}\label{esse1}
	\left(\frac{\mathrm{d}\rho_{\rm gw}}{\mathrm{d\,ln} k}\right)_{\tau>\tau_{\rm f}}=\frac{S_{k}\left(\tau_{\rm f}\right)}{a^{4}\left(\tau\right)}\,,
\end{equation}
where $\tau_{\rm f}$ is the time when the GW source becomes negligible, and $S_{k}$ is referred to $\bar{h}_{ij}=a h_{ij}$ (which gives a new factor $1/a^{2}$) and encodes the amount of GW produced at the 
time the source was present.

\subsubsection{Analytical estimation of the gravitational waves spectral energy-density features}\label{stimepre}
For each of the two cases analyzed before, we estimate the amplitude and the shape of the GW spectral energy-density that could be measured today, in order to understand their rough dependence on the parameters of the system; see tab.\ref{tabb} for a summary.\\
To estimate the present spectrum one needs to take into account the evolution of the expansion rate of the Universe, between the time of emission of the GW and the present time. At the end of inflation the equation of state moves from $w=0$ to a value $w\simeq 1/3$ during the preheating stage \cite{Podolsky:2005bw,Dufaux:2006ee}, and then to $w=1/3$ during the radiation dominance. We define $t_{\rm i}$ as the time corresponding to the end of inflation, $t_{\rm j}$ a moment after the jump of the equation of state, where we assume a mean value $w$, $t_{\ast}$ the time when thermal equilibrium is established, and $t_{\rm 0}$ the present time. Then the relation between the present value of the scale-factor and that corresponding to the moment when GW are produced is given by
\begin{equation}	
\frac{a_{\rm i}}{a_{0}}\simeq\frac{a_{\rm i}}{a_{\rm j}\rho_{\rm j}^{1/4}}\left(\frac{a_{\rm j}}{a_{\ast}}\right)^{1-\frac{3}{4}\left(1+w\right)}\left(\frac{g_{\ast}}{g_{0}}\right)^{-1/12}\rho_{\rm rad0}^{1/4}\,,
\end{equation}
where $\rho_{\rm j}$ is the total energy-density at $t=t_{\rm j}$, $\rho_{\rm rad0}$ is the present energy-density of radiation, and $g$ is the number of effectively massless degrees of freedom. From this relation, the following connection between the present value of a frequency $f$ and the value at the emission time of the GW, is obtained:
\begin{equation}	
f\equiv\frac{k_{0}}{2\pi}\simeq\frac{k}{a_{\rm j}\rho_{\rm j}^{1/4}}\left(\frac{a_{\rm j}}{a_{\ast}}\right)^{1-\frac{3}{4}\left(1+w\right)}4\times 10^{10}\,\mathrm{Hz}\,,
\end{equation}
where $f$ is the frequency evaluated today. We took $\Omega_{\rm rad}h^{2}=h^{2}\rho_{\rm rad0}/\rho_{\rm c}=4.3\times 10^{-5}$ as the abundance of radiation today, where $h$ is defined via the Hubble parameter as $H_{0}\equiv100h$ km s$^{-1}$ Mpc$^{-1}$, and $g_{\ast}/g_{0}=100$ \cite{Dufaux:2007pt}. Analogously, we can obtain the relation between the GW spectral energy-density evaluated at the time of production and at the present time. From eq.\eqref{esse1} we have \cite{Dufaux:2007pt}:
\begin{equation}
	h^{2}\Omega_{\rm GW}=\frac{S_{k}\left(\tau_{f}\right)}{a_{\rm j}^{4}\rho_{\rm j}}\left(\frac{a_{\rm j}}{a_{\ast}}\right)^{1-3w}\left(\frac{g_{\ast}}{g_{0}}\right)^{-1/3}\Omega_{\rm rad}h^{2}\,,
\end{equation}
at the present time.
For preheating following chaotic inflation or hybrid inflation, the mean value of $w$ in the intermediate stage reaches $w=1/3$ soon after the end of inflation \cite{Dufaux:2007pt,Dufaux:2008dn}, so that the factor $\left(a_{\rm j}/a_{\ast}\right)^{1-3w}$
can be neglected. In this case the previous relations read
\begin{equation}\label{omegabu}
	f=\frac{k}{a_{\rm j}\rho_{\rm j}^{1/4}}\,4\times 10^{10}\,\mathrm{Hz} \qquad\mbox{and}\qquad
	\Omega_{\rm GW}h^{2}=\frac{S_{k}\left(\tau_{\rm f}\right)}{a_{\rm j}^{4}\rho_{\rm j}}\,9.3\times 10^{-6}\,.
\end{equation}
Note that if the transition from the end of inflation to radiation dominance is sufficiently fast
and the scale-factor is normalized to one at $t=t_{\rm i}$, 
the quantity $a_{\rm j}^{4}\rho_{\rm j}$ represents the energy-density at the end of inflation.
From eq.\eqref{omegabu} we can conclude that if inflation ends at lower energies, the GW produced after that time will be less diluted by the expansion till the present time. At the same time, lowering the energy scale of inflation means having less efficient GW sources during the preheating stage. These two effects roughly cancel each other and in conclusion the present time GW spectral energy-density does not strongly depend on the energy scale at the end of inflation \cite{Easther:2006gt}.\\
It is easy to guess that the spectrum will present a peak strictly related to the $k_{\ast}$ values of the excited fluctuations. The translation from the peak of the excited inhomogeneities to that of the GW, is well understood looking at the relations \eqref{vederepicco}. 
Roughly the peak wavelength is estimated to be $1\sim 2$ orders of magnitude smaller than the Hubble radius at the time of GW production $H_{\rm i}$ \cite{Greene:1997ge,Greene:1997fu,Kofman:1997yn}. 
Clearly, the actual value of the peak wavelength will depend on the model parameters, through the value of $q$ \cite{Easther:2006gt,Bethke:2013vca}. The connection between the resonant mode $k_{\ast}$ and the present frequency, is given by eq.\eqref{omegabu}, where the denominator $a_{\rm j}\rho_{\rm j}^{1/4}\simeq\sqrt{H_{\rm j}}$ can be approximated with $\sqrt{H_{\rm i}}$, so that roughly the peak frequency results $f_{\ast}\sim \sqrt{H_{\rm i}}$. 
Then, on one hand, lowering the energy scale of inflation corresponds to reduce the power of the GW produced by the quantum mechanism during inflation, making them harder to detect. On the other hand, lowering the energy scale means reddening the peak power of the GW produced during the preheating stage, which makes their detection easier. This is because the strain sensitivity of GW detectors scales as $\propto 1/f^{3}$ \cite{Maggiore:2000gv}.
In any case, being the excited $\chi_{k}$ corresponding to sub-horizon modes at the time of preheating, the frequency of the peak would give also an upper bound on the horizon size at that time. Moreover, in general, the peak wavelength is expected to be constrained by the upper bound on the inflation energy scale provided by CMB and the lower bound required by baryogenesis and nucleosynthesis \cite{Easther:2006gt}.\\
Considering that the main power in GW is emitted during the bubbly stage \cite{GarciaBellido:2007af}, an estimate of the fraction of energy which is converted into gravitational radiation can be made. 
The typical size $R_{\ast}$ of the bubble inhomogeneities depends on the resonant modes $k_{\ast}$, more precisely $R_{\ast}\sim a/k_{\ast}$ \cite{Dufaux:2007pt}. The gravitational energy-density with respect to the total, results \cite{GarciaBellido:2007af}
\begin{equation}\label{bubble}
	\left(\frac{\rho_{\rm gw}}{\rho_{\rm tot}}\right)_{\rm p}\propto\left(R_{\ast}H\right)_{\rm p}^{2}\,,
\end{equation}
where $p$ means the time of production and the coefficient of proportionality can be calculated numerically. This estimate will be valid also for the bubbly period of the tachyonic preheating.

\paragraph{Analytical estimate for gravitational waves produced during parametric resonance after chaotic inflation}
In the case of chaotic inflation \eqref{vchaotic}, where the energy-density can be approximated by $a^{4}_{j}\rho_{\rm j}\simeq1.15\lambda_{\rm c}\varphi_{0}^{4}/4$ \cite{Dufaux:2007pt}, from eq.\eqref{omegabu} the present 
frequency of the peak reads
\begin{equation}\label{risonanza}
	f_{\ast}\simeq \frac{k_{\ast}}{a_{\rm j}\rho_{\rm j}^{1/4}}\,4\times 10^{10}\,\mathrm{Hz}\simeq\frac{k_{\ast}}{\varphi_{0}}\lambda_{\rm c}^{-1/4}\,5\times 10^{10}\,\mathrm{Hz}\,,
\end{equation}
where $\varphi_{0}\simeq 0.342\,M_{\rm pl}$ is the amplitude of the inflaton at the end of inflation and the scale-factor $a$ is normalized to unity at the end of inflation.
From eq.\eqref{bubble}, the amplitude of the spectral energy-density in correspondence to the peak at the time of production goes like
\begin{equation}\label{risonanzabis}
	\left(\Omega_{\rm GW}^{\ast}\right)_{\rm p}\sim\left(\frac{aH}{k_{\ast}}\right)^{2}_{\rm p}
\end{equation}
where $R_{\ast}=a/k_{\ast}$ and $a_{\rm p}$ is the value of the scale-factor referred to the time of production, and, as before, the dependence on the particular model is included in $k_{\ast}$.

\paragraph{Analytical estimate for gravitational waves produced during tachyonic preheating after hybrid inflation}
To get to analogous estimate in the case of tachyonic preheating, it is useful to write \eqref{omegabu} in terms of the size of the bubbles $R_{\ast}$, that is
\begin{equation}
	f_{\ast}\sim \frac{4\times 10^{10}}{R_{\ast}\rho_{\rm p}^{1/4}}\,\mathrm{Hz}\,\qquad
	h^{2}\Omega^{\ast}_{\rm GW}\sim 10^{-6}\left(R_{\ast}H_{\rm p}\right)^{2}\,,
\end{equation}
where $\rho_{p}=\lambda_{\rm t} v^{4}/4$ and $H_{p}^{2}=8\pi\rho_{p}/\left(3M_{\rm pl}^{2}\right)$.
From the potential \eqref{ibrido}, the Hubble parameter and the energy-density at the time of GW production can be estimated, so that, neglecting the expansion rate of the Universe, the previous relations read \cite{Dufaux:2008dn}
\begin{equation}\label{tac1}
	f_{\ast}\sim \frac{k_{\ast}}{\lambda_{\rm t}^{1/4}v}6\times 10^{10}\,\mathrm{Hz}\,\qquad
	h^{2}\Omega^{\ast}_{\rm GW}\sim 2\times 10^{-6}\frac{\lambda_{\rm t} v^{4}}{k_{\ast}M_{\rm pl}^{2}}\,.
\end{equation}
In order to be able to write $k_{\ast}$ as a function of the model parameters $v$ and $\lambda_{\rm t}$, one needs to consider different cases depending on the field that dominates the dynamics around the critical point. 
See \cite{Dufaux:2008dn}.

\subsection{Computational strategies for numerical simulations and current results}\label{equaz}
The classical equations of motions which describe the evolution of the fields obtained from the Lagrangian are the starting point. For a system described by \eqref{azionepre} and a FRW metric the equations for the fields result
\begin{equation}\label{1}
	\ddot{\chi}+3H\dot{\chi}-\frac{1}{a^{2}}\nabla^{2}\chi+\frac{\partial \mathcal{V}}{\partial\chi}=0\,\qquad
	\ddot{\varphi}+3H\dot{\varphi}-\frac{1}{a^{2}}\nabla^{2}\varphi+\frac{\partial\mathcal{V}}{\partial\varphi}=0\,.
\end{equation}
Via the FRW metric and the stress-energy tensor obtained from the Lagrangian, the equations for the background are obtained from Einstein equations \cite{GarciaBellido:2007af}:
\begin{equation}\label{3}
	-\frac{\dot{H}}{4\pi G}=\dot{\chi}^{2}+\frac{1}{3a^{2}}\left(\nabla\chi\right)^{2}+\dot{\varphi}^{2}+\frac{1}{33a^{2}}\left(\nabla\varphi\right)^{2}\,,
\end{equation}
\begin{equation}\label{4}
	\frac{3H^{2}}{4\pi G}=\dot{\chi}^{2}+\frac{1}{a^{2}}\left(\nabla\chi\right)^{2}+\dot{\varphi}^{2}+\frac{1}{a^{2}}\left(\nabla\varphi\right)^{2}+2\mathcal{V}\left(\chi,\varphi\right)\,.
\end{equation}
The equation of motion for GW is \eqref{gweq}.
Considering that the production of GW starts at a specific time $t_{e}$, the solution of the above equation can be written as
\begin{equation}
	h_{ij}\left(\textbf{k},t\right)=\frac{16\pi}{M_{\rm pl}^{2}}\int_{t_{\rm e}}^{t}\mathrm{d}t'\,G\left(t,t'\right)\Pi_{ij}\left(\textbf{k},t\right)
\end{equation}
so that
\begin{equation}
	\dot{h}_{ij}\left(\textbf{k},t\right)=\frac{16\pi k}{M_{\rm pl}^{2}}\int^{t}_{t_{\rm e}}\mathrm{d}t'\,G\left(k\left(t-t'\right)\right)\Pi_{ij}\left(\textbf{k},t'\right)\,,
\end{equation}
where $G\left(t,t'\right)$ is the Green function relative to the differential equation \eqref{gweq}.
In order to follow the dynamics of reheating relative of a given inflationary model, one has to evolve in a lattice eq.\eqref{1} simultaneously to eqs.\eqref{3}-\eqref{4}, while GW are obtained from eq.\eqref{gweq}, which clearly is coupled to the previous ones.\\
In the last decade several methods have been developed to solve the previous system of equations. The evolution of the fields on a space-time lattice is commonly analyzed by the publicly available code \textsc{LatticeEasy} \cite{Felder:2000hq}. The issue of lattice simulations has been addressed and discussed specifically by \cite{Figueroa:2011ye} and \cite{Huang:2011gf}.\\
For a detailed comparison of the results of different methods and strategies see \cite{Dufaux:2007pt,Easther:2007vj,Price:2008hq}.

\subsubsection{Current results for preheating after chaotic inflation}
Numerical simulations performed by \cite{Price:2008hq}, \cite{Easther:2007vj} and \cite{Dufaux:2007pt} are in good agreement for the case of a massless preheating \cite{Price:2008hq}, while the first simulations performed by \cite{Easther:2006gt} present slightly different behaviors. This agreement constitutes a strong check on the reliability of numerical simulations. Moreover the behavior shown by eqs.\eqref{bubble}-\eqref{risonanza} and eq.\eqref{risonanzabis} results confirmed.
Let us present here the main results about chaotic inflation, following the analysis of \cite{Dufaux:2007pt}, which chooses as reference model $q=g^{2}/\lambda_{\rm c}=2$, fixing then $k_{\ast}$, and $\lambda_{\rm c}$ is taken to be $\lambda_{\rm c}=10^{-14}$.
The exponential rate of GW production is maximal during parametric resonance but most of final amount of GW is produced during the bubbly stage \cite{Dufaux:2007pt}. Assuming the previous parameters the peak amplitude at present results of $h^{2}\Omega_{\rm GW}^{\ast}\sim 3\times10^{-11}$ and the frequency of $f^{\ast}\sim 7\times 10^{7}$ Hz. Running simulations with different values of the coupling constant $g$, that is changing the resonance band, it is found that the peak amplitude mildly depends on $q$, roughly it decrease with its value but not in a perfect monotonic way, while its frequency depends on the value of $k_{\ast}$. 
For example, for the case $q=1.2$, the amplitude peak at present time is $h^{2}\Omega^{\ast}_{\rm GW}\sim 5\times 10^{-10}$, at a frequency of order $5\times 10^{6}$ Hz. The peak frequency results in any case higher than the accessible band of current and planned experiments, being not lower than $10^{6}$ Hz, even though with an amplitude of about $10^{-9}-10^{-11}$; see section \ref{exp}.

\subsubsection{Current results for preheating after hybrid inflation}
\begin{figure}[ht] 
    \centering
    \includegraphics[width=0.7\textwidth]{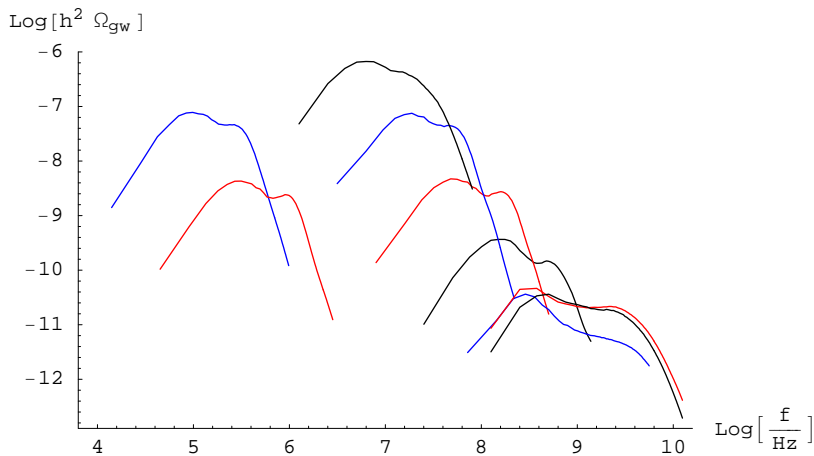}
    \caption{Gravitational waves spectral energy density from numerical simulations generated after hybrid inflation. On the left $\lambda_{\rm t} = 10^{-14}$, on the right $\lambda_{\rm t}=10^{-5}$. The spectra for $\lambda_{\rm t} = 10^{-5}$ are, from top to bottom, for $\lambda_{\rm t}/g^{2} = 20000,\, 5000,\, 500,\, 50,\, 0.5,\, 0.005,\, 0.0005$ respectively. The spectra for $\lambda_{\rm t} = 10^{-14}$ are for $\lambda_{\rm t}/g^{2} = 5000,\,500$ respectively. All the spectra are for $v = 10^{-3}M_{\rm pl}$, although for $g^{2}\ll \lambda_{\rm t}$ a lower value of $v$ may be necessary to consistently neglect expansion of the Universe. The figure is taken from \cite{Dufaux:2008dn} (\copyright$\,\,$IOP Publishing. Reproduced with permission. All rights reserved.).}
  
    \label{preheatingplot}
\end{figure}

Testing the estimations \eqref{tac1} with numerical simulations, the peak frequency and amplitude may cover a wide range of values, depending on the parameters of the model $\lambda_{\rm t}$, $g^{2}$, $v$ and on the potential evaluated at the critical point \cite{Dufaux:2008dn}. Compared with the parametric resonance mechanism, in this case the GW can cover a larger range of frequencies. The peak wavelength depends essentially on the coupling constant $\lambda_{\rm t}$ and is independent of the $\sigma$ vev $v$. In particular to lower the peak frequency a small coupling constant $\lambda_{\rm t}$ is needed. More precisely, imposing the maximum value $v$ compatible with the success of the model, a peak frequency $f_{\ast}< 10^{3}$ Hz can be obtained varying $\lambda_{\rm t}$ \cite{Dufaux:2008dn}.
The amplitude of the peak results $h^{2}\Omega_{\rm GW}^{\ast}\propto v^{2}$, so that to obtain a higher peak only increasing the value of the vacuum $v$ is necessary. This is true up to $h^{2}\Omega_{\rm GW}^{\ast}<10^{-6}$, a bound which mainly follows from requiring that the GW production takes place on sub-horizon scales (for further details, see \cite{Dufaux:2008dn}). For several examples of numerical simulations see fig.\ref{preheatingplot}.
In principle, for some extreme range of values of the model parameters, hybrid inflation leads to a GW spectrum with a peak accessible to future detection experiments \cite{Dufaux:2008dn}, such as Big Bang Observer (BBO) \cite{Crowder:2005nr}.

 \paragraph{Modifications to the basic models of preheating} 
Besides chaotic and hybrid inflation, the GW produced can be considered also in slightly different models of preheating.
In particular, the number of fields involved in the system \cite{Giblin:2010sp}, their nature (bosonic or fermionic) \cite{Enqvist:2012im} and the possible presence of self-interaction of the light field coupled to the inflaton \cite{Hyde:2015gwa} 
\cite{Figueroa:2013vif}, influence the GW production. The gravitational signal due to the Higgs field decay \cite{Figueroa:2014aya}, considering such a field negligible during the inflationary dynamics, could be another interesting phenomenon, representing a source of information on high-energy particle physics. However, planned experiments will not be able to detect such a high-frequency spectrum.
GW production on scales which are super-horizon at the time of formation during the preheating stage were studied too \cite{Fenu:2009qf}. 
Such gravitational radiation is expected from the self-ordering of randomly oriented scalar fields which can be present during preheating after hybrid inflation.

\subsection{Anisotropies in the gravitational wave background}
All the previous modeling considers the role of the secondary light scalar field $\chi$, only starting from the end of inflation. Actually this assumption is an approximation, being the field $\chi$ acting also during the inflationary dynamics. As a consequence, at the end of the accelerated expansion, also $\chi$ is left with amplified perturbations frozen on super-horizon scales, just like the inflaton. Then, at the beginning of preheating each Hubble region is characterized by a homogeneous background value of the field $\chi_{\rm i}$, superimposed with sub-horizon vacuum fluctuations.
If the GW spectral energy-density depends on the value $\chi_{\rm i}$, then a different amount of GW  is expected for different regions of the sky we observe today, in correspondence with the Hubble regions of the reheating stage \cite{Bethke:2013vca,Bethke:2013aba,Lozanov:2016pac}.\\
A crucial point in order to get anisotropies is the growth of long-wavelengths modes. The value $\chi_{\rm i}$ corresponds to the mode $k=0$ in each horizon volume; if it is amplified at the beginning of the preheating then it influences the whole dynamics of that region. When non linearities become important, long-wavelengths modes transfer energy to the short ones, so that the dynamics of the latter is eventually influenced by $\chi_{\rm i}$, via the long-wavelength modes. In such a way the spatial distribution of the field $\chi$, and so the amount of produced GW, is affected by $\chi_{\rm i}$. Therefore, if the mode $k=0$ belongs to the resonant band, a different amount of produced GW could be expected in correspondence with different values of $\chi_{\rm i}$. The homogeneous mode $k=0$ lies in the resonant band for quite a narrow region of parameter space for the massless case, while for the other cases $k=0$ is often amplified for a wider range of the coupling constant \cite{Bethke:2013aba}.
Clearly, in order to have anisotropies, a further fundamental requirement is the presence of a light scalar field, in addition to the inflaton during the accelerated expansion.\\
The case in which the inflationary field is massless is usually considered to simplify calculations, providing a useful starting point to understand the phenomenon. Consider a scenario of parametric resonance due to the potential
\begin{equation}
	\mathcal{V}\left(\varphi,\chi\right)=\frac{\lambda}{4}\varphi^{4}+\frac{1}{2}g^{2}\varphi^{2}\chi^{2}\,,
\end{equation}
where $\chi$ is light during inflation, that is $m_{\chi}=g\varphi$ is less than the Hubble rate $H$. To guarantee the lightness of $\chi$ during inflation, the coupling constant is taken to be $g^{2}/\lambda=2$, fixing at the same time the resonant band \cite{Bethke:2013vca} . With this choice a large amplification of long wavelengths during the non-linear evolution is ensured.
As for the inflaton, perturbations of the $\chi$ field at the end of inflation are described by a nearly scale-invariant power-spectrum $P_{\chi}\simeq H^{2}/4\pi^{2}$, because of its lightness.
The differential equations for describing the reheating phase are the same as those considered in section \ref{equaz}, but the initial conditions at the end of inflation now are different for each horizon volume, in correspondence with the value $\chi_{\rm i}$ (usually this value is set to zero).  
For a reasonable range of $\chi_{i}$ values \cite{Bethke:2013vca}, and employing the ``separate Universe" approach, the peaks of the GW spectral energy-density of two different simulations result located at the same frequency determined by the scale at which fluctuations are amplified, while the amplitudes of the peaks result not comparable \cite{Bethke:2013vca}. 
This means that actually $\chi_{\rm i}$ influences the gradient of $\chi$ and then the efficiency of GW production \cite{Bethke:2013vca}.
More precisely $\Omega_{\rm GW}$ varies by as much as a factor of five between nearby values of $\chi_{\rm i}$ \cite{Bethke:2013aba}.
The level of anisotropy of the produced GW is estimated by the angular spectrum of the relative GW spectral energy-density fluctuations as a function of the $\chi_{i}$ value. A general formula applicable to all the scenarios characterized by a light spectator field during inflation is \cite{Bethke:2013vca}:
\begin{equation}
	l\left(l+1\right)C_{l}=\frac{H_{\ast}^{2}}{8\pi}\frac{\langle \delta\chi_{\rm i}\Omega_{\rm GW}\left(\chi_{\rm i}\right)\rangle^{2}}{\sigma_{\chi}^{4}\langle\Omega_{\rm GW}\rangle^{2}}\,.
\end{equation}
For the considered massless case with $g^{2}/\lambda=2$, the relative amplitude of GW power on large scales relative to different $\chi_{\rm i}$ values results above the one percent level \cite{Bethke:2013vca} (to get a comparison, the relative amplitude of the CMB fluctuations is of the order of $10^{-5}$).\\
Furthermore, we can expect that the anisotropy of the GW spectral energy-density could be correlated with the non-Gaussianity of CMB anisotropies, being originated by the presence of a light field during inflation in both cases.
The details of this anisotropy, if observed, could provide a way to distinguish between different microscopic theories and also help in breaking degeneracies of the inflationary sector \cite{Dufaux:2007pt}.

\begin{table}[H]
\centering
\resizebox{15.5cm}{!} { %article
%\resizebox{9cm}{!} { %cimento
\begin{tabular}{|c|c|c|c|c|c|}
\hline
 & \textbf{Model} & \textbf{Peak frequency} & \textbf{Peack amplitude}  & \textbf{Peak frequency} & \textbf{Peack amplitude} \\ \hline \hline
\multirow{2}{*}{\textbf{Burst}} 
& Preheating - chaotic infl. & $f_{\ast}\simeq\frac{k_{\ast}}{\varphi_{0}}\lambda_{\rm c}^{-1/4}\,5\times 10^{10}$ Hz & $h^{2}\Omega_{\rm GW}^{\ast}\sim\left(\frac{aH}{k_{\ast}}\right)^{2}$& $f_{\ast}\sim 5\cdot10^{6}$ Hz& $h^{2}\Omega^{\ast}_{\rm gw}\sim 5\cdot10^{-10}$\\ \cline{2-6}
& Preheating - hybrid infl. & $f_{\ast}\sim \frac{k_{\ast}}{\lambda_{\rm t}^{1/4}v}6\times 10^{10}$ Hz & $h^{2}\Omega^{\ast}_{\rm gw}\sim 2\times 10^{-6}\frac{\lambda_{\rm t} v^{4}}{k_{\ast}M_{\rm pl}^{2}}$ & $f_{\ast}\sim 10^{5}$ Hz& $h^{2}\Omega^{\ast}_{\rm gw}\sim 10^{-7}$ \\ \hline
\end{tabular}
}
\caption{{\footnotesize Features of the GW spectral energy-density produced during the preheating phase. $k_{\ast}$ are resonant modes, $\lambda_{\rm c}$ and $\lambda_{\rm t}$ are the inflaton couplings in each model (see eqs.\eqref{vchaotic}-\eqref{ibrido}), and $v$ the vacuum value of the waterfall field; for more details see sections \ref{sezchaotic}-\ref{stimepre}. The first two columns refer to the time of GW production, while the last two refer to the present time. Values for chaotic inflation are obtained for: $\varphi_{0}=0.342 M_{\rm pl}$, $\lambda_{\rm c}=10^{-14}$ and $q=g^{2}/\lambda_{\rm c}=1.2$. Values for hybrid inflation are obtained for: $\lambda_{\rm t}=10^{-14}$, $v_{\rm c}=10^{-3}$, $v=10^{-3} M_{\rm pl}$ and $\lambda/g^{2}=5000$. Notice that actually the values of the frequency and amplitude of the peak can vary largely from those reported here in each case. For comparison, the expected GW spectral energy density for a \textit{standard inflation} and $r\simeq 0.1$, $n_{\rm T}=0$, is $h^{2}\Omega_{\rm GW}\simeq 10^{-16}$.}}
\label{tabb}
\end{table}

 %cambiare dim tabella
\section{Gravitational waves from inflation in Modified Gravity}
\label{capmodified}

\subsection{Why Modified Gravity?}

The interest in Modified Gravity (MG) theories, as an alternative to General Relativity (GR), is due to both fundamental physics issues and cosmological observations. From the point of view of fundamental physics, it is well known that a full quantum description of the space-time and of the gravitational interaction is not possible considering General Relativity as the theory of gravity \cite{Weinberg:100595}. A quantum theory of gravity lacks. Moreover when attempts to unify fundamental interactions are made (such as superstrings, supergravity etc.), effective actions with non-minimal coupling with the geometry or higher-order terms involving curvature invariants appear. From the cosmological point of view, there are at least two main reasons leading to the introduction of MG \cite{Clifton:2011jh}: on one side GR combined with the Standard Model of particle physics cannot solve some internal inconsistencies of the Standard Big Bang cosmology, such as the horizon and flatness problems. To get the primordial accelerated expansion of the Universe, while remaining in GR, one needs to introduce at least a scalar field beyond the known Standard Model of elementary particles. On the other hand, recent cosmological observations tell us that the Universe is presently undergoing a period of accelerated expansion. 
The most hasty way to explain such a dynamics is that of introducing the well-know cosmological constant, representing the dark energy content of the Universe. However this way of proceeding, actually does not give a motivation for the current comparable amounts of matter and dark energy density (the so-called ``coincidence problem"). In a rough summary, it seems then that GR is not able to explain successfully the gravitational accelerated dynamics in its extreme regimes.\\
Motivated by these considerations in recent years several attempts to face these issues have been made. The most natural way to proceed consists in extending Einstein theory with corrections and extensions, which are required to reproduce GR in the regimes in which it is well tested. 
The most immediate modification consists in adding higher-order curvature invariants and minimally, or non-minimally, coupled scalar fields. 
On the other hand, MG scenarios seem to have the chance to address the early Universe issues, since they may naturally provide a period of accelerated expansion in the early Universe, originating from the gravity sector only.\\
In scenarios of MG, GW are produced by the same \textit{quantum} mechanism we have presented in section \ref{sezionegw} for the model single-field slow-roll inflation in GR, that is by vacuum fluctuations of the gravitational field. Then the significant point here, is to investigate how features of the produced GW changes, building the inflationary models on theory of gravity different from GR.

\subsection{Signatures of primordial tensor modes generated in modified gravity theories}
When inflation is built on a theory of MG, usually primordial GW are expected to be produced by the same quantum mechanism of the models related to GR, but in general it is possible that they get new features, such as a speed of sound $c_{\rm T}\neq {\rm c}$ and a non vanishing mass $m_{\rm T}$, and a modified equation of motion with respect to the eq.\eqref{mototens}. Then in general a tensor power-spectrum different from \eqref{tens} is expected.\\
Considering this fact, several works have considered the effects on the CMB of tensor modes with non-standard features, such as with a generic speed of graviton propagation, $c_{\rm T}$ \cite{Amendola:2014wma,Pettorino:2014bka,Raveri:2014eea,Xu:2014uba,Saltas:2014dha,Cai:2015dta} (see also \cite{Cai:2016ldn}, not for CMB), a non-vanishing mass \cite{Dubovsky:2009xk,Fasiello:2015csa,Malsawmtluangi:2016agy} or a non-standard friction term \cite{Pettorino:2014bka,Xu:2014uba} in their equations of motion. \cite{Noumi:2014zqa} studied the relation between $c_{\rm T}$ and the non-Gaussianity of primordial tensor perturbations in the Effective Field Theory (EFT) approach to inflation. For a recent discussion about the propagation speed of GW during inflation, see \cite{Creminelli:2014wna,Burrage:2016myt}.\\
Exploiting the detection of the GW signal by the aLIGO observatory \cite{Abbott:2016blz}, new constraints, for such a kind of GW, have been obtained for $c_{\rm T}$ \cite{Blas:2016qmn,Ellis:2016rrr,Collett:2016dey,Bicudo:2016pps} and $m_{\rm T}$ \cite{TheLIGOScientific:2016src}.

\subsection{Overview of the main models}
There is a great variety of MG models which work well with respect to our request of an inflationary period. We can organize them into three main categories: theories in which extra-fields are involved, theories in which 
higher-order derivatives are introduced and theories built on higher-dimensional spaces. In particular, the first and the second categories actually intertwine and are not clearly separated. For a review see \cite{Copeland:2006wr,amendola2010dark,Clifton:2011jh}.

\subsubsection{Gravity with extra fields}
In GR the gravitational degrees of freedom are described by the metric field, that is by a rank$-2$ tensor. There are no strong reasons to suppose that there could not be in addition other fields to describe the coupling between the matter fields and the gravity sector. The simplest way to implement such an idea consists in adding a scalar field to the usual metric tensor, whose effects clearly have to disappear on scales where GR is well tested. In alternative, one can add also vector or tensor fields. Among these theories the most significant ones, being the most natural and 
essential, are the scalar-tensor theories, which we will consider in more detail later.

\subsubsection{Gravity with higher-order derivatives}
GR is the most general theory based on a metric tensor which provides field equations of at most second order. A way to extend Einstein theory consists in admitting higher-order derivative in the field equations. Clearly this way of proceeding is not guaranteed to be preserved from the appearance of instabilities.
Anyhow, starting from a Lagrangian with higher-order derivative terms, there are several theories which actually lead to field equations of second order, providing only stable dynamics. 
The simplest and most immediate way to implement such an idea consists in replacing the usual curvature invariants with a function of them. This operation characterizes the so-called $f\left(R\right)-$theories \cite{Sotiriou:2008rp}. For example, considering an $R^{2}$ term generally leads to field equations of fourth order. 
Actually, what happens is that putting corrections to the curvature term coincides with adding new degrees of freedom to the system. In particular, theories in which $R$ is replaced by $R+R^{2}$ coincide with scalar-tensor theories.\\
For the primordial GW power-spectrum estimated in higher-order spatial derivatives theories, see, for example, \cite{DeFelice:2014bma,Cai:2015ipa}.

\subsection{Primordial gravitational waves in Scalar-Tensor theories of gravity}
In these theories the basic idea consists in adding a scalar degree of freedom to the gravitational sector. The simplest and most generic Lagrangian which includes such an extra degree of freedom reads \cite{Bergmann:1968ve,Nordtvedt:1970uv,Wagoner:1970vr}
\begin{equation}	
\mathscr{L}=\frac{1}{16\pi G}\sqrt{-g}\left[f\left(\Phi\right)R-g\left(\Phi\right)\nabla_{\mu}\Phi\nabla^{\mu}\Phi-2\Lambda\left(\Phi\right)\right]\,,
\end{equation}
where $f,g,h$ and $\Lambda$ are arbitrary functions of the scalar field $\Phi$. Clearly, this Lagrangian includes a large number of models, such as the Brans-Dicke theories \cite{Brans:1961sx}, which constitutes the first MG theory of the gravitational interaction alternative to GR.\\
In the last few years more general models have been considered starting from considerations about the symmetries of the system: \cite{Nicolis:2008in} built a Lagrangian with higher-derivative terms but leding to equations of motion of second order. Their model was based on the so-called Galilean symmetry, that is the invariance on the Minkowski spacetime under the \textit{Galilean} field transformation $\Phi\,\rightarrow\,\Phi+b_{\mu}x^{\mu}+c$, where $c$ is a constant and $b_{\mu}$ is a constant vector. In \cite{Nicolis:2008in} the theory is built on a flat and non-dynamical background; then later works \cite{Deffayet:2009mn,Deffayet:2009wt,Deffayet:2011gz} extended the theory on a dynamical background, obtaining to the so-called \textit{Covariant Galileon Inflationary} model \cite{Deffayet:2009wt}. To do so, new terms to the Lagrangian are added which lead to the breakdown of the Galilean symmetry \cite{Deffayet:2009mn,Deffayet:2009wt,deRham:2010eu}. Then \cite{Kobayashi:2011nu} generalized such a theory to the most general scalar-tensor theory which leads to second-order field equations on flat and curved space-time, calling it Generalized G-inflation. Such a model turns out to be equivalent to a Lagrangian proposed by Horndeski \cite{Horndeski:1974wa} as the most general scalar-tensor theory leading to second-order equations of motion starting from higher-derivative terms in the Lagrangian.\\
In the following subsections we will consider some of these models \cite{Burrage:2010cu,Mizuno:2010ag,Kobayashi:2010cm,Kobayashi:2011pc,Kobayashi:2011nu}, representing a natural way to extend GR and providing the possibility to get accurate predictions about GW.

\subsubsection{Generalized G-inflation}\label{sezginfl}
In 1974 Horndeski presented a work \cite{Horndeski:1974wa} in which he calculated the most general Lagrangian of a scalar-tensor theory in a four-dimensional space-time, which leads to second-order 
field equations, albeit starting from higher-order derivative terms. The latter generally lead to equations of motion at least of the fourth order, which can introduce gradient and ghost instabilities. Horndeski Lagrangians are built in order to avoid these problems. 
The Horndeski action reads \cite{Horndeski:1974wa,Deffayet:2010qz,Kobayashi:2010cm,Deffayet:2011gz}:
\begin{align}\label{azionegal}
	&S=\sum_{i=2}^{5}\int\mathrm{d}^{4}x\sqrt{-g}\mathscr{L}_{i}\,,\quad \mbox{with}\\ \nonumber
	&\mathscr{L}_{2}=K\left(\Phi,X\right)\,,\,\,\, \mathscr{L}_{3}=-G_{3}\left(\Phi,X\right)\Box\Phi\,,\\	\nonumber
	&\mathscr{L}_{4}=G_{4}\left(\Phi,X\right)R+G_{4X}\left[\left(\Box\Phi\right)^{2}-\left(\nabla_{\mu}\nabla_{\nu}\Phi\right)^{2}\right]\,,\\ \nonumber
	&\mathscr{L}_{5}=G_{5}\left(\Phi,X\right)G_{\mu\nu}\nabla^{\mu}\nabla^{\nu}\Phi-\frac{G_{5X}}{6}\left[\left(\Box\Phi\right)^{3}-3\left(\Box\Phi\right)\left(\nabla_{\mu}\nabla_{\nu}\Phi\right)^{2}+2\left(\nabla_{\mu}\nabla_{\nu}\Phi\right)^{3}\right]\,, \nonumber
\end{align}
where $G_{\mu\nu}$ is the Einstein tensor, $G_{iX}=\partial G_{i}/\partial X$, and $K,G_{i}$ are generic functions of $\Phi$ and $X\equiv -\partial_{\mu}\Phi\partial^{\mu}\Phi/2$. The system is described by four independent arbitrary functions of $\Phi$ and $X$. Adding separately the Hilbert Einstein action is not required. Notice that we are describing the system only as a gravitational one, since we are not introducing any matter fields.
The action \eqref{azionegal} includes a large class of models, such as single-field slow-roll inflation, the k-inflation \cite{ArmendarizPicon:1999rj}, Higgs G-inflation \cite{Kamada:2010qe} and Galileon inflation \cite{Burrage:2010cu} (the parameters of the latter model entering into the modified consistency relation for the tensor-to-scalar ratio have been constrained in the {\it Planck} analysis \cite{Ade:2015ava,Ade:2015lrj}, by combining the constraints from the power spectrum and  those on primordial non-Gaussianity). 
It can be seen \cite{Kobayashi:2011nu} that \eqref{azionegal} may lead to the sought stage of accelerated expansion in the early Universe and it offers reasonable ways to end such a period to get the usual hot Big Bang.

\paragraph{Background evolution}
To find the background equations of motion we can use the so-called unitary gauge, substituting $\Phi=\Phi\left(t\right)$ and the metric $\mathrm{d}s^{2}=-\mathrm{d}t^{2}+a^{2}\left(t\right)\mathrm{d}x^{2}$ into the action. There are several possibilities to get an accelerated expansion depending on the form of the functions $K$ and $G_{i}$. \cite{Kobayashi:2011nu} considers the case of a \textit{kinetically driven G-inflation} and a \textit{potential-driven slow-roll inflation}. 

\paragraph{Gravitational wave power-spectrum}
Let us consider the perturbed metric in the unitary gauge $\Phi=\Phi\left(t\right)$:
\begin{equation}\label{metricapert}	
\mathrm{d}s^{2}=-N^{2}\mathrm{d}t^{2}+\gamma_{ij}\left(\mathrm{d}x^{i}+N^{i}\mathrm{d}t\right)\left(\mathrm{d}x^{j}+N^{j}\mathrm{d}t\right)\,,
\end{equation}
where 
\begin{equation}
	N=1+\alpha\,,\qquad N_{i}=\partial_{i}\beta\,,\qquad \gamma_{ij}=a^{2}\left(t\right)e^{2\zeta}\left(\delta_{ij}+h_{ij}+\frac{1}{2}h_{ik}h_{kj}\right)\,,
\end{equation}
so that $\alpha$, $\beta$ and $\zeta$ are the scalar perturbations, while $h_{ij}$ are the tensor ones, which are traceless and divergence-free.\\
Perturbing the action \eqref{azionegal} at second order, two terms describing tensor perturbations are found \cite{Kobayashi:2011nu}:
\begin{equation}\label{azionetens}
S_{\rm T}^{\left(2\right)}=\frac{M_{\rm pl}^{2}}{8}\int\mathrm{d}t\mathrm{d}^{3}x\,a^{3}\left(t\right)\left[\mathscr{G}_{\rm T}\dot{h}_{ij}\dot{h}_{ij}-\frac{\mathscr{F}_{\rm T}}{a^{2}}\left(\nabla h_{ij}\right)^{2}\right]\,,
\end{equation}
with
\begin{align}\label{fg}
	&\mathscr{F}_{\rm T}\equiv\frac{2}{M_{\rm pl}^{2}}\left[G_{4}-X\left(\ddot{\Phi}G_{5X}+G_{5\Phi}\right)\right]\,,\\
	&\mathscr{G}_{\rm T}\equiv\frac{2}{M_{\rm pl}^{2}}\left[G_{4}-2XG_{4X}-X\left(H\dot{\Phi}G_{5X}-G_{5\Phi}\right)\right]\,.
\end{align}
The action \eqref{azionetens} shows the same structure of eq.\eqref{azionestandard}: it presents the same terms of an action describing wave propagation, but with unusual coefficients.
In particular, these new factors lead to a propagation speed of GW which generally differs from the speed of light, $c_{\rm T}^{2}=\mathscr{F}_{\rm T}/\mathscr{G}_{\rm T}$, and modify the GW amplitude with respect to the standard one.\\
Before calculating the power-spectrum, let us introduce some restrictions. To avoid ghosts and gradients instabilities, we require $\mathscr{F}_{\rm T}>0$ and $\mathscr{G}_{\rm T}>0$ \cite{Kobayashi:2011nu},
while, in order to simplify calculations, we assume
\begin{equation}
	\epsilon\equiv-\frac{\dot{H}}{H^{2}}\simeq \mathrm{const}\,,\qquad f_{\rm T}\equiv\frac{\dot{\mathscr{F}_{\rm T}}}{H\mathscr{F}_{\rm T}}\simeq \mathrm{const}\,,\qquad g_{\rm T}\equiv\frac{\dot{\mathscr{G}_{\rm T}}}{H\mathscr{G}_{\rm T}}\simeq \mathrm{const}\,.
\end{equation}
We now move to Fourier space and rescale the functions $h_{ij}$ and time $t$ in order to get an equation of motion completely analogous to eq.\eqref{mototens} \cite{Kobayashi:2011nu}
\begin{equation}
	\mathrm{d}y_{\rm T}\equiv\frac{c_{\rm T}}{a}\mathrm{d}t\,,\qquad z_{\rm T}\equiv\frac{a}{2}M_{\rm pl}\left(\mathscr{F}_{\rm T}\mathscr{G}_{\rm T}\right)^{1/4}\,,\qquad v_{ij}\equiv z_{\rm T}h_{ij}\,.
\end{equation}
These operations are analogous to those made in section \ref{modistandard}. However, besides the amplitude rescaling, here we have to perform also a time rescaling because of the generic propagation speed, in order to get a frame where $c_{\rm T}=c$. Performing these transformations the action \eqref{azionetens} becomes
\begin{equation}	
S_{\rm T}^{\left(2\right)}=\frac{M_{\rm pl}^{2}}{2}\int\mathrm{d}y_{\rm T}\mathrm{d}^{3}x\left[\left(v'_{ij}\right)^{2}+\frac{z''_{\rm T}}{z_{\rm T}}v_{ij}^{2}-\left(\nabla v_{ij}\right)^{2}\right]\,,
\end{equation}
where here the prime denotes differentiation w.r.t. $y_{\rm T}$. This action leads to the standard equation of motion, so that we can get the solutions, by proceeding as in section \ref{modistandard}. On super-horizon scales there are two independent solutions, one which corresponds to decaying $h_{ij}$ modes and the other which corresponds to constant $h_{ij}$. Requiring the canonical normalization which determines the behavior on the sub-horizon regime, we get the exact solution, analogous to \eqref{solstandard} \cite{Kobayashi:2011nu}:
\begin{equation}\label{nut}
	v_{ij}=\frac{\sqrt{\pi}}{2}\sqrt{-y}H_{\nu_{\rm T}}^{\left(1\right)}\left(-ky_{\rm T}\right)e_{ij}\,,\quad \mbox{where}\quad
	\nu_{\rm T}\equiv\frac{3-\epsilon+g_{\rm T}}{2-2\epsilon-f_{\rm T}+g_{\rm T}}\,,
\end{equation}
and $e_{ij}$ the polarization tensor. 
The latter is the generalization of $\nu\simeq3/2+\epsilon$ found in section \ref{sezionegw}. We are interested in the constant modes, being those that are still non-negligible at the re-entrance into the causal region. 
Exploiting the asymptotic behavior of the Hankel functions and moving back from $v_{ij}$ to $h_{ij}$, we get the tensor amplitudes on super-horizon scales,
\begin{equation}
	k^{3/2}h_{ij}\approx 2^{\nu_{\rm T}-2}\frac{\Gamma\left(\nu_{\rm T}\right)}{\Gamma\left(3/2\right)}\frac{\left(-y_{\rm T}\right)^{1/2-\nu_{\rm T}}}{z_{\rm T}}k^{3/2-\nu_{\rm T}}e_{ij}\,.
\end{equation}
Replacing our results in eq.\eqref{spettrogw}, we get the sought expression for the GW power-spectrum \cite{Kobayashi:2011nu}:
\begin{equation}\label{spettrogal}
P_{\rm T}=\frac{8\gamma_{\rm T}}{M_{\rm pl}^{2}}\frac{\mathscr{G}_{\rm T}^{1/2}}{\mathscr{F}_{\rm T}^{3/2}}\,\left(\frac{H}{2\pi}\right)^{2}\lvert_{ky_{\rm T}=1},
\quad \mbox{where}\quad\gamma_{\rm T}=2^{2\nu_{\rm T}-3}\left|\frac{\Gamma\left(\nu_{\rm T}\right)}{\Gamma\left(3/2\right)}\right|^{2}\left(1-\epsilon-\frac{f_{\rm T}}{2}+\frac{g_{\rm T}}{2}\right).
\end{equation}
The spectral index reads $n_{\rm T}=3-2\nu_{\rm T}$.
Let us make a few comments about these results. First, compare eq.\eqref{spettrogal} with the power-spectrum obtained in the single-field slow roll-inflation case eq.\eqref{tens}.
The power-spectrum \eqref{spettrogal} presents a new factor resulting from the unusual Lagrangian for the scalar field, so that the GW amplitude is modified with respect to the standard one. 
This difference is due to the fact that here GW are determined not only by the geometrical degrees of freedom alone but also by the coupling with the scalar one.
Even more interesting is the result concerning the spectral index: Generalized G-inflation admits a blue index, without NEC violation, contrary to what happens in GR. To get a positive spectral index one only needs to satisfy the condition $4\epsilon+3f_{\rm T}-g_{\rm T}<0$, so that we can have a blue index, while maintaining $\epsilon>0$, which means $\dot{H}<0$. See also \cite{Kunimitsu:2015faa}.

\subsubsection{Potential-driven G-inflation}\label{sezdriven}
Let us now investigate a particular case of the previous class of models, where only the first two terms of eq.\eqref{azionegal} are considered \cite{Kobayashi:2010cm,Kamada:2010qe}. We are still dealing with 
higher-derivative terms in the Lagrangian and with second-order equations of motion, both for the scalar field and the metric.
This model is interesting since it provides the possibility to write a consistency relation between the tensor-to-scalar ratio and the tensor spectral index 
that can be compared with that obtained in the standard model of inflation.
Actually GW do not present unusual features in this case.\\
The action of the system reads \cite{Kobayashi:2010cm}:
\begin{equation}\label{higgs}
	S=\int\mathrm{d}^{4}x\sqrt{-g}\left[\frac{1}{2}M_{\rm pl}^{2}R+\mathscr{L}_{\Phi}\right]\,,
\quad \mbox{with}\quad
	\mathscr{L}_{\Phi}=K\left(\Phi,X\right)-G\left(\Phi,X\right)\Box\Phi\,,
\end{equation}
where $K$ and $G$ are general functions of the scalar field $\Phi$ and $X=-\nabla_{\mu}\Phi\nabla^{\mu}\Phi/2$. Notice that the Einstein-Hilbert Lagrangian is obtained from $\mathscr{L}_{\Phi}$, 
with a particular choice of the function $G_{4}$.
The second term comes from the generalization of the interaction $X\Box\Phi$ which satisfies the Galilean symmetry. In particular we are concerned in the model identified by the following choice
\begin{equation}\label{choice}
	K\left(\Phi,X\right)=X-V\left(\Phi\right)\,,\qquad G\left(\Phi,X\right)=-g\left(\Phi\right)X\,,
\end{equation}
since, besides satisfying the basic requests, it provides the possibility to write the consistency relation. One of the most significant models included in the choice \eqref{choice} is the \textit{Higgs G-inflation} \cite{Kamada:2010qe} which represents an attempt to explain the inflationary accelerated expansion employing only fields of the Standard Model.

\paragraph{Background evolution}
As in the most general case, a slow-roll evolution of the scalar field leading to a stage of accelerated cosmic expansion can be obtained.\\
To get the sought dynamics we define the following quantities \cite{Kobayashi:2010cm}:
\begin{equation}
	\epsilon\equiv -\frac{\dot{H}}{H^{2}}\,,
\qquad
	\eta\equiv-\frac{\ddot{\Phi}}{H\dot{\Phi}}\,,
\qquad
	\alpha\equiv\frac{g_{\Phi}\dot{\Phi}}{gH}\,,
\qquad
\beta\equiv\frac{g_{\Phi\Phi}X^{2}}{V_{\Phi}}\,,
\end{equation}
where the subscript $\Phi$ means the derivation w.r.t. such a field and we where impose $\epsilon\,,\left|\eta\right|\,,\left|\alpha\right|\,,\left|\beta\right|\ll1$.
In particular, the first inequality includes the fundamental requirement of accelerated expansion, while the third corresponds to asking for the domain of the potential $V\left(\Phi\right)$ with respect to $X$ in the expression of $K\left(\Phi,X\right)$.
Choosing the regime in which the Galileon effect, represented by $g\left(\Phi\right)$, assumes a relevant role, that is $\left|gH\dot{\Phi}\right|\gg1$, the background equation of motion becomes $gH^{2}\dot{\Phi}^{2}+V_{\Phi}\simeq0$ \cite{Kobayashi:2010cm}.
Then, the slow-roll equation of motion is solved by a $\Phi\left(t\right)$, such that \cite{Kobayashi:2010cm}
\begin{equation}
	\dot{\Phi}\simeq-\mathrm{sgn}\left(g\right)M_{\rm pl}\left(\frac{V_{\Phi}}{3gV}\right)^{1/2}\,,
\end{equation}
requiring that the field slowly rolls down its potential. From this expression, one can realize that, compared to the standard slow-roll inflation, 
here the field velocity is suppressed by a factor $1/\sqrt{gV_{\Phi}}$, so that we can have the sought evolution, $H\simeq\mathrm{const}$, even if the potential is rather steep.

\paragraph{Gravitational wave power-spectrum}
Exploiting the results \eqref{spettrogal} of the general case investigated above, we can soon conclude that, in this model, GW do not show any unusual feature,
\begin{equation}\label{tensoriginflation}
	P_{\rm T}\left(k\right)=\frac{8}{M_{\rm pl}^{2}}\left(\frac{H}{2\pi}\right)^{2}\\,\qquad
	n_{\rm T}=-2\epsilon\,.
\end{equation}
This clearly leads to a degeneracy between this inflationary model and the standard one, so that GW do not seem the best way of getting information on the inflationary period and the underlying theory of gravity.
However, once again this result conceals a relevant difference with respect to the single-field slow-roll inflation power-spectrum: 
the tensor spectral index can be blue having the speed sound of scalar perturbations $c_{\rm S}^{2}>0$ at the same time since, as before, the NEC violation $\dot{H}>0$ does not imply $c_{\rm S}^{2}<0$.

\paragraph{Scalar perturbations}
In order to get the tensor-to-scalar ratio we give also the scalar power-spectrum.
Proceeding as in section \ref{scalaristandard} and adding assumptions needed to avoid ghosts \cite{Kobayashi:2010cm}, the scalar power-spectrum at horizon crossing results \cite{Kamada:2010qe}
\begin{equation}\label{scalariginflation}
P_{\zeta}=
\left(\frac{3\sqrt{6}}{64\pi^{2}}\right)\frac{H^{2}}{M_{\rm pl}^{2}\epsilon}\lvert_{\tau=1/c_{\rm S}k}\,.
\end{equation}
We obtained that the amplitude presents a different numerical factor with respect to that obtained in the single-field slow-roll inflation. 
As we will see, in a subsequent dedicated section, that this fact leads to a violation of the standard consistency relation \eqref{consistency}, and so to the possibility of distinguishing these inflationary models from the standard one.

\newpage

\section{Summary}\label{overview}
In the following table we provide an overview of the main mechanisms of GW production which have been investigated in the previous sections, that can take place during inflation and the preheating period. In the first three columns we put in evidence the different physical origin of such mechanisms. As illustrated up to here, the two ways for generating inflationary GW are vacuum oscillations of the gravitational field and the presence of a source term in the GW equation of motion that leads to a classical mechanism of GW production. In the first case the assumption that leads to different predictions for the features of the GW power spectrum is the theory of gravity underlying the inflationary model. In the second case, the form and the efficiency of the source terms are the discriminants for the generated tensor modes. In the last two columns, following such a scheme, we organize the main models we have investigated in the previous sections.

%\begin{landscape}

\begin{table}[H]
\centering
\resizebox{15cm}{!} { %article
%\resizebox{8cm}{!} { %cimento
\begin{tabular}{|c|c|c|c|c|}
\hline 
 \textbf{GW PRODUCTION} & \textbf{Discriminant}& \textbf{Specific discriminant} & \textbf{Examples of specific models} & \textbf{Produced GW} \\ \hline \hline \hline

\multirow{5}{3.6cm}{\textbf{Vacuum oscillations}\\ \vspace{0.3cm}
\begin{center}
 quantum 
fluctuations of the gravitational field stretched by the accelerated expansion
\end{center}
}
 & 
\multirow{5}{*}{\textbf{theory of gravity}} & \multirow{2}{*}{General 
Relativity} & \textit{single-field slow-roll} & \textit{broad spectrum}\\ \cline{4-5}

& & & all other models in GR & broad spectrum \\ \cline{3-5}

&  & \multirow{3}{*}{MG$/$EFT approach} & G-Inflation & broad spectrum\\ \cline{4-5}

&  &  & Potential-driven G-Inflation & broad spectrum\\ \cline{4-5}

&  &  & EFT approach  & broad spectrum\\ \hline \hline \hline

\multirow{8}{3.6cm}{\textbf{Classical production}\\ \vspace{0.3cm} 
\begin{center}
second-order GW generated by the presence of a source term in GW equation of motion
\end{center}
}& \multirow{8}{*}{\textbf{source term}} &  vacuum inflaton fluctuations & \textit{all models} & \textit{broad spectrum} \\ \cline{3-5}

& & \multirow{2}{4.5cm}{
\begin{center}
fluctuations of extra scalar fields
\end{center}
} & inflaton$+$spectator fields & broad spectrum\\ \cline{4-5}
& & & curvaton & broad spectrum\\ \cline{3-5}

& & \multirow{2}{*}{gauge particle production} & pseudoscalar inflaton$+$gauge field & broad spectrum\\ \cline{4-5}
& & & scalar infl.$+$pseudoscalar$+$gauge & broad spectrum \\ \cline{3-5}

& & scalar particle production & scalar inflaton$+$ scalar field & peaked \\ \cline{3-5}

& & \multirow{2}{4.5cm}{
\begin{center}
particle production during preheating
\end{center}
} & chaotic inflation & peaked \\ \cline{4-5}
& & & hybrid inflation & peaked \\ \cline{1-5}

\end{tabular}
}
\caption{\footnotesize {Summary of the main mechanisms of GW production during inflation and the preheating phase. In the fourth column, the scenarios mainly investigated in the present work are reported as examples for each mentioned case. They are discussed in the following sections respectively: \textit{``single-field slow-roll"} section \ref{sezionegw}, \textit{``G-Inflation"} section \ref{sezginfl}, \textit{``Potential-driven G-Inflation"} section \ref{sezdriven}, \textit{``EFT approach"} section \ref{eft}, \textit{``all models"} section \ref{secondordinarie}, \textit{``spectator fields"} section \ref{sezionespectator}, \textit{``curvaton"} section \ref{curvatone}, \textit{``pseudoscalar inflaton$+$gauge field"} section \ref{sezgauge}, \textit{``scalar infl.$+$pseudoscalar$+$gauge"} section \ref{pseudoparticelle}, \textit{``scalar inflaton$+$scalar field"} section \ref{particellescalari}, \textit{``chaotic inflation"} section \ref{sezchaotic}, \textit{``hybrid inflation"} section \ref{sezhybrid}.
To clarify the notation: \textit{``EFT approach"} refers to all models encoded in the generic action used in the EFT approach to inflation. \textit{``Broad spectrum"} means that a power spectrum, broad on a large range of scales is expected, while \textit{``peaked"} indicates a signal peaked on a narrow range of frequencies.}}
\end{table}

%\end{landscape}

\section{The issue of the quantum to classical transition for inflationary perturbations}\label{quantum}

According to the inflationary model, the seeds of perturbations present at last scattering are quantum fluctuations of the scalar field that has driven the accelerated expansion and of the gravitational field. Up to now, this is the only physical model where theoretical predictions coming from a simultaneous use of General Relativity and quantum mechanics, are testable, in principle, by observations. Therefore inflationary physics reveals itself as a framework where fundamental questions about quantum mechanics and cosmology arise too.
In facing such basic issues, inflationary GW play a significant role.\\
Up to now the most prominent and unsolved issue, in this framework, is the following: the CMB radiation is an observable and then, according to quantum mechanics, it corresponds to a quantum operator. 
Thus when we look at a CMB map we are considering  the results of a measurement corresponding to a specific observable. Following the Copenhagen interpretation, in obtaining CMB maps we are making a measurement which leads the quantum state of the CMB radiation falling into an eigenvalue of the related observable. The issue is that, in light of this interpretation, CMB perturbations get the value of the eigenvalue only at the present time when we are making the measurements, since no observers existed before us. Moreover, perturbations which lead to the CMB anisotropies are the same that give rise to the large-scale structures of the Universe. 
Then the fact that these perturbations get their determined value only today is in contrast with our understanding of the evolution of structures, in particular with the fact that they started growing at early times. 
Then, in a more impressive way with respect to laboratory systems, in cosmology we have to face with the single outcome problem.\\
Several steps have been made in order to face this issue. 
In \cite{Grishchuk:1990bj}, firstly, it has been pointed out that inflationary perturbations evolve into highly squeezed quantum states on super-horizon scales because of the accelerated expansion, resulting into highly non-classical states. On the other hand such a kind of quantum perturbations can be described as a realization of a classical stochastic process in virtue of their large occupation number \cite{Grishchuk:1990bj}, so that the usual approach of considering these quantities as classical is justified.\\
A further step in understanding this issue has been made introducing the phenomenon of ``decoherence", which selects the amplitude basis as the pointer basis of the system. At the same time this opens a discussion about the possibility of considering the Universe as a closed system \cite{Grishchuk:1990bj}.\\
Nevertheless, this new approach does not solve the problem of the single outcome, that still remains, not only in cosmology, but also for laboratory systems.
How a single outcome is produced still continues to be an open question, that in cosmology takes the name of \textit{macro-objectivation} problem. CMB maps bring information about a measurement that had to take place in the early Universe, but at that time no conscious observers were there. So, how does the realization we measure today is obtained?\\
Different solutions have been proposed to the issue. In particular, a way of solving the problem which presents discriminatory predictions comparable, in principle, with experimental data, has been proposed: the collapse models \cite{Pearle:1988uh,Bassi:2012bg}. The latter were employed for the first time in the framework of the early Universe by \cite{Perez:2005gh,DeUnanue:2008fw,DiezTejedor:2011jq}, and later they have been developed exploiting the Continuous Spontaneous Localization (CSL) approach \cite{Martin:2012pea,Das:2013qwa}. More recently, the CSL model was used also with a novel conceptual approach \cite{Leon:2015hwa}, which consists in facing the quantum-to-classical transition and the production of primordial perturbations at the same time.

\subsection{Observational predictions for CSL single field dynamics}
In the collapse model presented in \cite{Das:2013qwa}, the dynamics of the inflationary fields is investigated in the Schrodinger picture and the standard scalar and tensor perturbations are described by wave-functional \cite{Polarski:1995jg}. Then the corresponding Schrodinger equations are parametrized by several quantities which lead to non-standard scalar and tensor power-spectra.
In particular, both scalar and tensor perturbations present the same modification to the amplitude of the power-spectrum so that the tensor-to-scalar ratio is the same as the standard approach, that is $r=16\epsilon$. On the other hand, both scalar and tensor spectral index present a deviation with respect to eqs.\eqref{tiltscalari}-\eqref{tilttensori}:
\begin{align}
&n_{\rm S}-1=\delta+2\eta-4\epsilon\\
&n_{\rm T}=\delta-2\epsilon\,,
\end{align}
where $\delta$ is a number collecting the parametrization of the Schrodinger equation. Moreover, the fact that the tensor amplitude depends, besides the Hubble parameter, on the the new parameters of the model, implies that the GW amplitude is no longer a direct indication of the energy scale of inflation. Finally, the most interesting fact is the violation of the consistency relation between the tensor-to-scalar ratio and the tensor spectral index: $r=-8n_{\rm T}+8\delta$.
Clearly, in order to test these relations, the detection of tensor modes is a fundamental prerequisite.\\
In this framework, recently, interesting investigations have been made in order to find out if there are any signatures of the quantum origin of primordial perturbations in CMB anisotropies; see \cite{Martin:2015qta,Leon:2016mdn} and refs. therein.\\
In summary, inflationary perturbations represent a very interesting and particular area where fundamental questions about quantum mechanics and cosmology come into play, in particular for what concerns the single outcome problem. In this respect primordial GW play an important role in order to discriminate among different ways of solving this puzzle.

\section{Consistency relations and possible violations}\label{sezioneconsistency}
A detection of primordial GW would open the possibility of testing a powerful inflationary consistency relation, thus improving our understanding of the physics of the Early Universe.\\
In section \ref{sezconsistency} we have shown that for single-field slow-roll inflation a strict consistency relation holds between the tensor-to-scalar ratio $r$ and the GW spectral index $n_{\rm T}$, at first order in the slow-roll parameters, for each perturbation scale $k$:
\begin{equation}\label{consistencybb}
	r=-8n_{\rm T}\,,
\end{equation}
where, in general, $r$ and $n_{\rm T}$ are scale-dependent. 
An analogous relation which connects tensor features to the scalar ones, exists also for the running of the spectral index \eqref{tilttensori} \cite{Lidsey:1995np}:
\begin{equation}
	\frac{\mathrm{d\,}n_{\rm T}}{\mathrm{d\,ln}k}\simeq \frac{r}{8}\left[\frac{r}{8}+\left(n_{\rm S}-1\right)\right]\,.
\end{equation}
However, the running is slow-roll suppressed and then the tensor power-spectrum is usually described only by the spectral index $n_{\rm T}$. For higher-order extensions of the previous relations, see \cite{Lidsey:1995np}.
Relation \eqref{consistencybb} is a particular prediction of single-field slow-roll inflation, which is violated in several other inflationary scenarios. Therefore, verifying this equality would constitute a powerful test of single-field slow-roll inflationary models \cite{Camerini:2008mj,Dodelson:2014exa,Caligiuri:2014sla,Cai:2014uka,Biagetti:2015tja}.
Remember that the relation \eqref{consistencybb} comes from the possibility of expressing both sides in terms only of the slow-roll parameter $\epsilon$. In particular, this means that $r$ and $n_{\rm T}$ are two quantities directly related to the energy scale of inflation, therefore violating the relation \eqref{consistencybb} would imply loosing also the direct connection between $r$ and/or $n_{\rm T}$ and the energy scale of inflation.
A violation of the consistency relation, in the form of non-standard scalar power-spectrum amplitude, and/or non-standard GW power-spectrum, can arise for different reasons. The most explicit cause of a violation is represented by a blue tensor power-spectrum, that is by $n_{\rm T}>0$, which is obviously incompatible with eq.\eqref{consistencybb}. In general, inflationary models predict a red tilt of the GW power-spectrum, whose value depends on the model details. Notice that this fact, for most standard scenarios, is a direct consequence of the connection between $n_{\rm T}$ and $\epsilon$, which holds only for GW generated by vacuum fluctuations of the field that drives inflation. Then a blue tensor spectral index can arise when an extra amount of GW generated not from vacuum fluctuations is present and moreover when modifications in fundamental physics are assumed or in models built in the framework of modified gravity theories\footnote{See \cite{Ashoorioon:2015hya} for an investigation on how the trans-Planckian physics can modify the consistency relation too.}. In the first case the direct link between GW amplitude and energy scale of inflation is broken. 
However we will show that a violation of the standard consistency relation could come from a non-standard red GW spectral index too. Finally, the violation can be due to an unusual value of the scalar power-spectrum amplitude, which would lead to an anomalous expression for the tensor-to-scalar ratio.\\
Let us list here the main inflationary models in which, for different reasons, a violation of eq.\eqref{consistencybb} arises; the most significant, with respect to the role of GW, are collected in tab.\ref{taba}. Let us also mention some alternative scenarios where a violation of the relation \eqref{consistencybb} happens, but which are far from the goals of this review: string gas cosmology \cite{Brandenberger:2006xi} and matter bounce cosmology \cite{Cai:2012va,Cai:2013kja}.

\subsection{Inflationary models deviating from the single-field slow-roll scenario}\label{single}
The violation of the standard consistency relation can be the signature of a departure from the standard single-field slow roll inflation scenario.
The simplest situation leading to scalar perturbations or tensor power-spectra different from the usual ones, is the presence of extra sources for scalar or tensor perturbations, which enhance the amplitude of the respective power-spectra. Another reason could be a modification of some aspects in the physics of the inflaton, for example its speed of sound, which it imprints features in the power-spectrum.
Also the case of \textit{General} slow-roll Inflation \cite{Gong:2014qga} has been examined, where the slow-roll conditions are relaxed and in principle a blue tensor spectral index can be obtained, but leaving open the problem of how to get the end of inflation.

\subsubsection{Inflationary models with extra-sources of gravitational waves}
In several inflationary models extra mechanisms of GW production (w.r.t. the standard mechanism) can be implemented in several ways (see section \ref{secondo}). In the presence of additional contributions of tensor modes, the tensor-to-scalar ratio $r$ deviates from the standard relation $r=16\epsilon$ and also the tensor spectral index can be different from the standard value $n_{\rm T}=-2\epsilon$.
We have shown that second-order GW are generated in the presence of a transverse and traceless anisotropic stress tensor. The crucial point is in which cases and on which scales, the additional contribution so generated is significant, with respect to the standard one, providing a deviation from $r=16\epsilon$. Therefore the modifications to the consistency relation 
in these cases is quite model dependent, and we refer the reader to the previous sections for more details about the expression of the tensor-to-scalar ratio $r$. However an interesting point is that the consistency relation might receive very small (unmeasurable) corrections on CMB scales, while being very different on much smaller scales.\\
For example, in the case we have discussed in section \ref{secondoordine}, where a spectator scalar field, characterized by a small speed of sound is present during inflation, scalar perturbations of the inflaton act as a source of tensor modes, thus yielding an extra amount of GW. In this situation, the tensor spectral index related to the sourced GW can also be blue \cite{Biagetti:2013kwa}.
As explained in section \ref{particelle}, also inflationary scenarios where particle production takes place, predict an extra amount of GW, due to the anisotropic stress tensor introduced by the produced particles. A large amount of extra GW can be generated on some ranges of scales, with non-standard scale-dependence. This additional contribution clearly leads to a violation of the standard consistency relation on the related scales, as it provides an additional contribution to the GW power-spectrum, which is not directly related to the energy scale of inflation.

\subsubsection{Inflation driven by multi-fields}
If Inflation is driven by more than one degree of freedom, for example by two scalar fields, not only adiabatic scalar perturbations are generated but also isocurvature perturbations. Conversely, GW still depend only on the energy scale of inflation. Notice the difference with the case of a spectator scalar field in which the secondary field does not drive the accelerated expansion and then no significant isocurvature perturbations are produced.
Non-adiabatic perturbations lead to an extra contribution in the denominator of the tensor-to-scalar ratio $r$, and then a deviation from $r=16\epsilon$ is expected \cite{Wands:2002bn}, being $\epsilon$ related only to the adiabatic contribution of scalar perturbations. An analogous situation arises in the case of more than two scalar fields driving the inflationary dynamics \cite{Wands:2002bn} (in \cite{Price:2014ufa} the issue is faced with a {\it statistical} approach by the calculation of the probability distribution function of $n_{\rm T}/r$).
In the former scenario, the standard consistency relation involves a new factor \cite{Wands:2002bn}:
\begin{equation}
	r=-8n_{\rm T}\sin^{2}\Delta\,,
\end{equation}
where $\cos\Delta$ parametrizes the correlation between adiabatic and isocurvature perturbations at horizon exit.
In the case in which inflation is driven by more than two scalar fields, an additional \textit{uncorrelated} isocurvature mode could contribute to the non-adiabatic perturbations, 
giving an additional contribution to the primordial scalar curvature. The relation \eqref{consistencybb} then becomes \cite{Wands:2002bn}:
\begin{equation}
	r\leq -8n_{\rm T}\sin^{2}\Delta\,.
\end{equation}

\subsubsection{General single field inflation} 
By ``general single field inflation" we denote those scenarios in which the accelerated expansion is driven by scalar fields with a Lagrangian of the form $P(X,\varphi)$
where $X$ is the canonical kinetic term, and $P$ a generic function.
This model includes, for instance, DBI inflation and k-inflation \cite{ArmendarizPicon:1999rj,Chen:2006nt}. In these scenarios the violation of the standard consistency relation is due to the non-canonical dynamics of scalar perturbations. In those models, the inflaton sound speed reads $c_{\rm S}^{2} = dP/d\rho = P,_{X}/\left(P,_{X}+2XP,_{XX}\right)$, where $\rho$ is the energy-density of the inflaton; $c_S$ generally differs from the standard unitary value \cite{ArmendarizPicon:1999rj}. The scalar power-spectrum at leading order in the slow-roll parameters results enhanced by a factor $1/c_{\rm S}$. On the other hand, tensor perturbations have the usual behavior. In this case eq.\eqref{consistencybb} becomes
\begin{equation}
	r=-8c_{\rm S}n_{\rm T}\,.
\end{equation}
The Planck Collaboration obtained constraints on $c_{\rm S}$, for several specific scenarios belonging to this class of models \cite{Ade:2015lrj}, exploiting the scalar power-spectrum and non-Gaussianity estimations 
(since, in general, small sound speed of the scalar field means large non-Gaussianity of scalar perturbations \cite{Chen:2006nt}).

\subsubsection{Inflationary models with spatial and time variation of the inflaton decay rate}
A deviation from the standard consistency relation can be provided also by the reheating physics, when the inflaton decays into ordinary particles $\chi$ to which it can be coupled \cite{Matarrese:2003tk}. The inflaton decay rate $\Gamma$ depends on the vacuum expectation value of the field to which it is coupled. Usually $\Gamma$ is supposed to be constant, but $\chi$ can fluctuate during inflation and leave imprinted these variations on super-horizon scales, so that the inflaton decay rate $\Gamma$ cannot be considered constant. When this happens, the variation of $\Gamma$ leads to a shift of the curvature perturbations $\zeta$ on super-horizon scales, that can be parametrized by $\Delta\equiv\left(\zeta_{\rm f}-\zeta_{\rm i}\right)/\zeta_{\rm i}$, where $i$ and $f$ indicated the initial and final times of the reheating period. This variation introduces a new factor in the denominator of $r$ and then a deviation of the standard consistency relation of the form \cite{Matarrese:2003tk}:
\begin{equation}
	r=-8\left(1-2\Delta\right)n_{\rm T}\,.
\end{equation}

\subsection{Inflationary models with modifications in fundamental physics}
Also modification of specific features related to more fundamental physics can lead to a violation of the consistency relation. 

\subsubsection{Inflation with collapse model for quantum fluctuations}
As seen in section \ref{quantum}, introducing modifications to the Schrodinger equation which governs the evolution of primordial perturbations, in order to solve the quantum to classical transition issue, leads to non-standard scalar and tensor power-spectra and then to a violation of the consistency relation of the form \cite{Das:2014ada}:
\begin{equation}
	r=-8n_{\rm T}+8\delta\,,
\end{equation}
where $\delta$ includes the parameters which modify the Schrodinger equation.

\subsubsection{Inflation with general initial conditions}
There are several motivations to question about the initial vacuum conditions from which quantum fluctuations grow up. One of these is the Trans-Planckian problem. Inflationary perturbations arise in UV-completion scales, where we are not sure that quantum fluctuations of the fields were in their lowest energy state, that is we are not sure we can impose Bunch-Davies initial conditions. If general initial conditions are imposed, additional factors appear in the field power-spectrum \cite{Wang:2014kqa,Ashoorioon:2014nta}. In particular if non-Bunch-Davies initial conditions are applied to the gravitational field, in general the analog of eq.\eqref{solstandard} becomes \cite{Wang:2014kqa}:
\begin{equation}
	v_{k}=C_{+}\left(k\right)\frac{H}{\sqrt{2k^{3}}}\left(1+ik\tau\right)e^{-ik\tau}+C_{-}\left(k\right)\frac{H}{\sqrt{2k^{3}}}\left(1-ik\tau\right)e^{ik\tau}\,,
\end{equation}
where $\left|C_{+}\left(k\right)\right|^{2}-\left|C_{-}\left(k\right)\right|^{2}=1$.
These new factors introduce arbitrary quantities in the tensor power amplitude and then, in general, a deviation from the standard consistency relation \cite{Wang:2014kqa}. 
Moreover, the $k$ dependence of $C_{\pm}\left(k\right)$ can lead to a blue tensor spectrum \cite{Wang:2014kqa,Ashoorioon:2014nta}.

\subsubsection{Inflation with non-commutative phase space}
In the context of quantum gravity, and in the framework of FRW Universe models, the case of a single-field inflationary model with non-commutative phase-space, that is with a non-standard commutative relations for quantum fields has been considered \cite{Cai:2014hja}. Modifying the canonical commutation relation for the conjugate momenta leads to a non-standard equation of motion for tensor modes. In particular, the tensor spectral index vanishes for small scales and becomes blue at large scales. Scalar perturbations are not influenced by the new physics. The whole behavior is a deviation form the standard consistency relation at least on large scales, due to the blue tensor spectral index.

\subsection{Inflation in EFT approach}
As seen in section \ref{eft}, in the particular case of admitting the breaking of spatial diffeomorphisms by adding effective mass terms and derivative operators, it is found that tensor perturbations can get an effective mass and a speed of sound $c_{\rm T}$ different from unity \cite{Cannone:2014uqa}. The sound speed influences the amplitude of the tensor power-spectrum by a factor $1/c_{\rm T}$, while the mass leads to a deviation of the tensor spectral index from the standard value $n_{\rm T}\simeq -2\epsilon$.
Moreover a blue spectral index is possible, when preserving the NEC \cite{Cannone:2014uqa}. Then, in general, the standard consistency condition is violated. See also the case presented in \cite{Cai:2016ldn}.
\cite{Graef:2015ova} found the conditions under which a blue GW spectral index appears, while the scalar spectral index keeps red, as required by CMB observations.

\subsection{Inflationary models in Modified Gravity}
Another framework in which violations of the consistency relation arise are models built on modified gravity theories; see section \ref{capmodified}.
Here the deviation of the scalar and tensor power-spectra from the standard form has a simple reason: we are modifying the canonical dynamics of perturbations.
One of the most interesting scenarios considered in this context so far appear those related to Galileon Inflation \cite{Burrage:2010cu} (see \cite{Ade:2015lrj} for observational constraints on these models).

\subsubsection{G-Inflation and Generalized G-Inflation}
In G-Inflation model \cite{Kobayashi:2010cm} (see section \ref{sezdriven}), the inflationary dynamics is described by a Lagrangian of the form \eqref{higgs}. The GW dynamics is not altered by the new terms of the Lagrangian and then tensor perturbations present the same power-spectrum as the standard case. Moreover, in general the NEC is not guaranteed to be satisfied \cite{Kobayashi:2010cm} also in the case when ghost are avoided and $c_{\rm s}^{2}>0$, so that the tensor spectral index reads $n_{\rm T}\simeq -2\epsilon$, but it can be blue due to the possible negative values of the slow-roll parameter. However, in the general case of G-Inflation, due to the non-standard scalar power-spectrum produced, neither the standard consistency relation nor a simple expression between the tensor-to-scalar ratio and the tensor spectral index can be obtained \cite{Kobayashi:2010cm}, but in the particular case of potential-driven G-Inflation, a consistency relation holds \cite{Kamada:2010qe}. 
In fact, from \eqref{tensoriginflation}-\eqref{scalariginflation}, we have:
\begin{equation}
	r\simeq -\frac{32\sqrt{6}}{9}n_{\rm T}\,.
\end{equation}
As seen in section \ref{sezginfl}, the G-inflation model can be generalized by adding new terms in the Lagrangian, eq.\eqref{azionegal}, in order to get the most general equation of motion of second order \cite{Kobayashi:2011nu}. In this case the GW dynamics is influenced by the 
new terms and a deviation from the usual tensor power-spectrum is obtained, both in terms of amplitude and spectral index, and then in general a violation of the consistency relation can be expected. Scenarios of Generalized G-Inflation are investigated in connection with the consistency relation in \cite{Unnikrishnan:2013rka}.

\begin{table}[H]
\centering
\resizebox{15.5cm}{!} { %article
%\resizebox{9cm}{!} { %cimento
\begin{tabular}{|c|c|l|l|c|c|}
\hline 
 &\textbf{Model} & \multicolumn{1}{|c|}{\textbf{Tensor power-spectrum}} &  \multicolumn{2}{|c|}{\textbf{Tensor spectral index}} &\vtop{\hbox{\strut \textbf{Consistency}}\hbox{\strut \textbf{$\,\,\,\,\,\,$relation}}} \\ \hline \hline
\multirow{4}{*}{\textbf{Background}}& Standard infl. & $P_{\rm T}=\frac{8}{M_{\rm pl}^{2}}\left(\frac{H}{2\pi}\right)^{2}$ & $n_{\rm T}=-2\epsilon$ & red &$r=-8 n_{\rm T}$ \\ \cline{2-6}
 &EFT inflation$^{\left({\rm a}\right)}$ & $P_{\rm T}=\frac{8}{c_{\rm T}M_{\rm pl}^{2}}\left(\frac{H}{2\pi}\right)^{2}$ & $n_{\rm T}=-2\epsilon+\frac{2}{3}\frac{m_{\rm T}^{2}}{\alpha H^{2}}\left(1+\frac{4}{3}\epsilon\right)$& r/b &- \\ \cline{2-6}
&EFT inflation$^{\left({\rm b}\right)}$ & $P_{\rm T}=\frac{8}{c_{\rm T}M_{\rm pl}^{2}}\frac{2^{\frac{-p}{1+p}}}{\pi}\Gamma^{2}\left(\frac{1}{2\left(1+p\right)}\right)\left(\frac{H}{2\pi}\right)^{2}$ & $n_{\rm T}=\frac{p}{1+p}$ & blue & violation \\ \cline{2-6}
 &Gen. G-Infl.& $P_{\rm T}=\frac{8}{M_{\rm pl}^{2}}\gamma_{\rm T}\frac{\mathscr{G}_{\rm T}^{1/2}}{\mathscr{F}_{\rm T}^{3/2}}\,\left(\frac{H}{2\pi}\right)^{2}$ & $n_{\rm T}=3-2\nu_{\rm T}$ & r$/$b &-  \\ \cline{2-6}
 &Pot.-driv. G-Infl. & $P_{\rm T}=\frac{8}{M_{\rm pl}^{2}}\left(\frac{H}{2\pi}\right)^{2}$ & $n_{\rm T}=-2\epsilon$ &r/b & $r\simeq -\frac{32\sqrt{6}}{9}n_{\rm T}$  \\ \cline{1-6}
 \multirow{2}{*}{\vtop{\hbox{\strut $\,\,\,\,\,\,\,$\textbf{Extra}}\hbox{\strut \textbf{background}}}}&Particle prod.& $P_{\rm T}^{+}=8.6\times 10^{-7}\frac{4H^{2}}{M_{\rm pl}^{2}}\left(\frac{H}{2\pi}\right)^{2}\frac{e^{4\pi\xi}}{\xi^{6}}$ & -& blue & violation \\ \cline{2-6}
 &Spectator field & $P_{\rm T}\simeq 3\frac{H^{4}}{c_{\rm S}^{18/5}M_{\rm pl}^{4}}$ & $n_{\rm T}\simeq 2\left(\frac{2m^{2}}{3H^{2}}-2\epsilon\right)-\frac{18}{5}\frac{\dot{c}_{\rm S}}{Hc_{\rm S}}$& r$/$b& violation \\ \hline
 
\end{tabular}
}
\caption{{\footnotesize GW features for selected inflationary models. We show the prediction for the amplitude of the tensor power-spectrum at the horizon crossing and the related spectral index, as functions of the model parameters. In the next column we indicate if the tensor spectral index is expected to be red, $n_{\rm T}<0$, or blue $n_{\rm T}>0$, or if both possibilities are admitted \textit{r$/$b}. In last column we point out the consistency relation, where it is significant, and denote \textit{violation} the cases in which, due to an extra background of GW, a violation of the standard consistency relation can be expected on some ranges of scales (see discussion in section \ref{single}). \textit{Standard Inflation}: Lagrangian of eq.\eqref{azione}; see section \ref{sezionegw}. \textit{EFT inflation}$^{\left({\rm a}\right)}$: Lagrangian of eq.\eqref{left}, $c_{\rm T}$ GW propagation speed, $m_{\rm T}$ graviton mass; see section \ref{eft}. \textit{EFT inflation}$^{\left({\rm b}\right)}$: Lagrangian of the same form of eq.\eqref{left} with $\alpha=c_{\rm T}^{-2}/2$, $m=0$, $c_{\rm T}$ a time-dependent parameter and $p\equiv-\dot{c}_{\rm T}/c_{\rm T}H_{\ast}$ a positive quantity; for more details see \cite{Cai:2016ldn}.
\textit{Generalized G-Inflation}: Lagrangian of eq.\eqref{azionegal}, $\gamma_{\rm T}$, $\mathscr{G}_{\rm T}$, $\mathscr{G}_{\rm T}$ and $\nu_{\rm T}$ defined in \eqref{fg}, \eqref{spettrogal} and \eqref{nut} respectively; see section \ref{sezginfl}. \textit{Potential-driven G-Inflation}: Lagrangian of eq.\eqref{higgs}; see section \ref{sezdriven}. \textit{Particle production}: Lagrangian of eq.\eqref{modellogauge}, $\xi$ defined in \eqref{gauge}, and $\delta_{\xi}$ defined in section \ref{bluepart}; see section \ref{sezgauge}. \textit{Spectator field}: Lagrangian of eq.\eqref{azionespect}, $c_{\rm S}$ and $m$ the speed of sound and the mass of the spectator field; see section \ref{sezionespectator}. }}
\label{taba}
\end{table}

\subsection{Observational prospects}
In light of the power of the consistency relation \eqref{consistencybb}, constraining the tensor amplitude and spectral index would represent a powerful test for the single-field inflationary model or it would provide hints for a departure from that physics. 
In order to test the validity of the consistency relation, one has to obtain an estimate of the scalar and tensor perturbation amplitudes and of the spectral index of tensor perturbations. 
Clearly the most difficult task is that of observing features concerning the tensor sector.
The largest difficulty is, of course, estimating the GW spectral index, which requires a measurement of the GW amplitudes on different scales. CMB data alone cannot provide strong constraints on $n_{\rm T}$, but the advantage of those measurements is that they provide data \textit{directly} on the tensor power-spectrum. It is clear that measurements of GW on smaller scales, 
such as those related to the direct detection by laser interferometer experiments, could provide stronger constraints on tensor features \cite{Meerburg:2015zua,Cabass:2015jwe,Lasky:2015lej}. 
GW direct detection experiments are planned to work on range of scales $18-20$ orders of magnitudes smaller that those of the CMB. Up to now, on these small scales we have only upper bounds on the cosmological GW energy-density due to a non-detection of the primordial signal. Of course, a remarkable help in this direction, would come in the case of blue-tilted tensor spectra.\\
Note that measurements on small scales provide a GW energy-density affected by all the history of the Universe, see \cite{Boyle:2014kba} and refs. therein. 
Therefore in order to extract the primordial parameters $A_{\rm T}$ and $n_{\rm T}$ one has to clean the signal from all late-time effects.
Moreover it has been found that also the choice of the pivot scale influences the parameters estimation.\\
Up to now, the data leave open the possibility of a violation of the consistency relation, as arising from a blue tensor power-spectrum \cite{Camerini:2008mj}. Available data provide only an upper bound on the tensor-to-scalar ratio of $r_{0.05}<0.09$ at $95\%$ C.L. \cite{Array:2015xqh} and an estimate of the tensor spectral index 
of $n_{\rm T}=0.06\pm^{+0.63}_{-0.89}$ at $95\%$ C.L. \cite{Meerburg:2015zua} (for more details and further estimations, see section \ref{exp}).\\
It is also interesting to consider that, usually, the consistency relation is assumed by default in data analysis. A few papers \cite{Camerini:2008mj,Wang:2014kqa} showed how admitting a prior for the tensor spectral index that also allows negative values, leads to deviations in the cosmological parameter estimation, in particular for the scalar spectral index, but also for the baryonic and cold dark matter energy-densities \cite{Camerini:2008mj}.\\
Forecasts on the possibility of testing the consistency relation, both from CMB observations and direct detection experiments are presented in \cite{Errard:2015cxa,Huang:2015gca}.
 %cambiare dim tabella
\section{Gravitational waves as a source of information for the thermal history of the Universe}\label{sezstoria}

When primordial tensor modes enter the horizon after the accelerated expansion phase they start evolving, more precisely their amplitude is damped by a factor inversely proportional to the scale factor, so that the present GW spectrum reflects the expansion history of the Universe \cite{Watanabe:2006qe}.
If we could detect inflationary GW, knowing with a certain accuracy their primordial properties and being able to disentangle them from overlapping later effects, such GW would represent a possibility to trace the thermal evolution of the Universe, including the reheating phase; see \cite{Watanabe:2006qe,Kuroyanagi:2014qza,Kuroyanagi:2011fy,Kuroyanagi:2010mm,Boyle:2005se,Seto:2003kc,Kuroyanagi:2014qaa,Nakayama:2008wy,Kuroyanagi:2014nba,Nakayama:2008ip,Dai:2014jja}.

\subsection{Gravitational wave transfer function}
\begin{figure}[ht]
    \centering
    \includegraphics[width=0.5\textwidth]{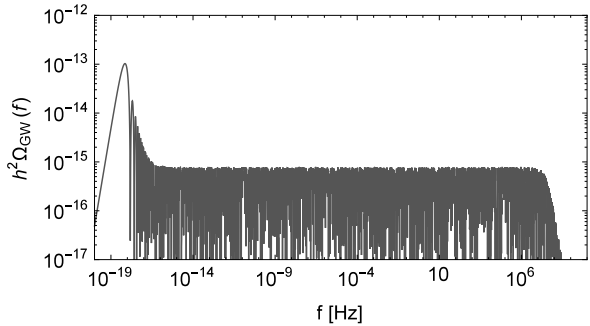}
    \caption{GW spectral energy-density at the present time, obtained from eqs.\eqref{trana}-\eqref{tranb}, for $r_{0.05}=0.07$, $g_{\ast s}=106.75$, $T_{\rm R}=10^{15}$ GeV and $\Omega_{\rm m}=0.3089$. }
  
    \label{plot_spettro}
\end{figure}

In a FRW spacetime, tensor modes obey eq.\eqref{mototens} if there are no sources. During inflation, GW wavelengths are stretched and moved to super-horizon scales. Solving the mentioned equation one finds that the GW amplitude $h_{k,\mbox{prim}}$ remains constant on super-horizon scales. Therefore, the general solution of eq.\eqref{mototens} can be written as
\begin{equation}\label{trasporto}
	h_{k}\left(\tau\right)\equiv h_{k,{\rm prim}}T_{\rm h}\left(\tau,k\right)\,,
\end{equation}
where the transfer function $T_{\rm h}\left(\tau,k\right)$ describes the evolution of the GW mode when they enter the horizon during later stages after inflation. The transfer function is normalized such that $T_{\rm h}\left(\tau,k\right)\rightarrow 1$ as $k\rightarrow 0$.\\
Defining 
\begin{equation}
	\Delta^{2}_{{\rm h},{\rm prim}}\left(k\right)\equiv\frac{\mathrm{d}\langle h_{ij} h^{ij}\rangle}{\mathrm{d\,ln} k}\,,
\end{equation}
from eq.\eqref{definizioneomega1} and eq.\eqref{trasporto}, we have
\begin{equation}
\Omega_{\rm GW}\left(k,\tau\right)=\frac{1}{12}\left(\frac{1}{aH}\right)^{2}\Delta^{2}_{{\rm h},{\rm prim}}\left(k\right)T'^{2}_{\rm h}\left(k,\tau\right)\,.
\end{equation}
For modes well inside the horizon, the previous formula can be approximated as \cite{Watanabe:2006qe}
\begin{equation}\label{trana}
\Omega_{\rm GW}\left(k,\tau\right)=\frac{1}{12}\left(\frac{k}{aH}\right)^{2}\Delta^{2}_{h,prim}\left(k\right)T^{2}_{\rm h}\left(k,\tau\right)\,.
\end{equation}
In single-field slow-roll inflation, the primordial power-spectrum is usually written in terms of the slow-roll parameters and of the Hubble scale during inflation \cite{liddle2000cosmological}:
\begin{equation}
	\Delta^{2}_{{\rm h},{\rm prim}}\left(k\right)=64\pi G\left(\frac{H}{2\pi}\right)^{2}\left[1-2\epsilon \mathrm{ln}\frac{k}{k_{\ast}}+2\epsilon\left(\eta-\epsilon\right)\left(\mathrm{ln}\frac{k}{k_{\ast}}\right)^{2}\right]\,.
\end{equation}
Solving \eqref{mototens} for radiation or matter dominated eras, one finds that in both cases the amplitude depends on the wavenumber and is modulated by the inverse of the scale-factor with the corresponding time-dependence (while the oscillatory behavior is described by Bessel functions). This damping factor is what we are interested in. Notice that the present GW amount is constituted by tensor modes that re-entered the horizon in different epochs of the history of the Universe, so that we have to take into account that each mode $k$ undergoes a different damping, depending on the time it evolves sub horizon and on the specific time dependence of the scale-factor during such an evolution. 
Well inside the matter dominated epoch, the solution of eq.\eqref{mototens} for all modes is
\begin{equation}
	h_{k}\left(\tau\right)=h_{k,{\rm prim}}\left(\frac{3j_{1}\left(k\tau\right)}{k\tau}\right)\,,
\end{equation}
with $j_{\ell}$ the $\ell$-th Bessel spherical function, given by $j_{1}\left(k\tau_{0}\right)\rightarrow1/\left(\sqrt{2}k\tau_{0}\right)$ in the limit $k\tau_{0}\rightarrow 0$. The subscript $0$ denotes the present time. 
Averaging over time the previous solution, to extract the amplitude behavior, one finds the factor $1/a$ mentioned above. Then the GW spectrum today attains the following form
\begin{equation}	
\Omega_{\rm GW}\left(k,\tau_{0}\right)=\frac{1}{12}\left(\frac{k}{aH}\right)^{2}\Delta^{2}_{{\rm h},{\rm prim}}\left(k\right)\left(\overline{\frac{3j_{1}\left(k\tau_{0}\right)}{k\tau_{0}}}\right)^{2}\left(\ldots\right)\,,
\end{equation}
where the last factor embodies all terms arising from the change of the scale-factor from the horizon re-entry to the present time, for a given mode $k$, and it will be specified now.\\
A first damping factor comes from the change of the relativistic degrees of freedom \cite{Watanabe:2006qe,Zhao:2006mm}:
\begin{equation}
	\left(\frac{g_{\ast}\left(T_{\rm in}\right)}{g_{\ast 0}}\right)\left(\frac{g_{\ast {\rm s} 0}}{g_{\ast {\rm s}}\left(T_{\rm in}\right)}\right)^{4/3}\,,
\end{equation}
where $g_{\ast}$ are the relativistic degrees of freedom, $g_{\ast {\rm s}}$ its counterpart for the entropy, \textit{in} denotes the time when a considered mode enters the horizon, and $T_{\rm in}$ is given by \cite{Nakayama:2008wy}
\begin{equation}
	T_{\rm in}\left(k\right)\simeq 5.8\times 10^{6}\mathrm{GeV}\left(\frac{g_{\ast {\rm s}}\left(T_{\rm in}\right)}{106.75}\right)^{-1/6}\left(\frac{k}{10^{14}\mathrm{Mpc}^{-1}}\right)\,.
\end{equation}
Another function is needed in order to connect the GW that enter the horizon before and after matter-radiation equality at $t=t_{\rm eq}$ \cite{Turner:1993vb}:
\begin{equation}
	T_{1}^{2}\left(x_{\rm eq}\right)=\left(1+1.57 x_{\rm eq}+3.42x_{\rm eq}^{2}\right)\,,
\end{equation}
where $x_{\rm eq}=k/k_{\rm eq}$ and $k_{\rm eq}\equiv a\left(t_{\rm eq}\right)H\left(t_{\rm eq}\right)=7.1\times 10^{-2}\Omega_{\rm m}h^{2}\mathrm{Mpc}^{-1}$. 
Analogously, a transfer function is needed to describe the change in the expansion rate at the end of reheating  $t=t_{\rm R}$, when the Universe moves from being inflaton-dominated to radiation-dominated \cite{Nakayama:2008wy}:
\begin{equation}
	T_{2}^{2}\left(x_{\rm R}\right)=\left(1-0.32x_{\rm R}+0.99x_{\rm R}^{2}\right)^{-1}\,,
\end{equation}
where $x_{\rm R}=k/k_{\rm R}$ and $k_{\rm R}\simeq 1.7\times 10^{14}\mathrm{Mpc}^{-1}\left(g_{\ast {\rm s}}\left(T_{\rm R}\right)/106.75\right)^{1/6}\left(T_{\rm R}/10^{7}{\rm GeV}\right)$. In terms of frequency it corresponds to
\begin{equation}
	f_{\rm R}\simeq 0.026\,\mathrm{Hz}\left(\frac{g_{\ast {\rm s}}\left(T_{\rm R}\right)}{106.75}\right)^{1/6}\left(\frac{T_{\rm R}}{10^{6}\mathrm{GeV}}\right)\,,
\end{equation}
which is the frequency for which the change in the frequency dependence of the spectrum due to the reheating stage appears.
In summary, the whole transfer function $T_{\rm h}^{2}\left(k\right)$ reads
\begin{equation}\label{tranb}
	T^{2}_{\rm h}\left(k\right)=\Omega_{\rm m}^{2}\left(\frac{g_{\ast}\left(T_{\rm in}\right)}{g_{\ast 0}}\right)\left(\frac{g_{\ast {\rm s} 0}}{g_{\ast {\rm s}}\left(T_{\rm in}\right)}\right)^{4/3} \left(\overline{\frac{3j_{1}\left(k\tau_{0}\right)}{k\tau_{0}}}\right)^{2}T_{1}^{2}\left(x_{\rm eq}\right)T^{2}_{2}\left(x_{\rm R}\right)\,.
\end{equation}
This expression tells that, once the values for the degrees of freedom evaluated at various epochs and $\Omega_{\rm m}$ are given, the GW spectrum is a function of the tensor-to-scalar ratio $r$ and of the reheating temperature $T_{\rm R}$; see also \cite{Boyle:2005se}. The so-obtained GW spectral energy-density at the present time, is shown in fig.\ref{plot_spettro}.

\subsubsection{Equation of state of the Universe and spectral tilt}\label{knee}
\begin{figure}[ht]
    \centering
    \includegraphics[width=.7\textwidth]{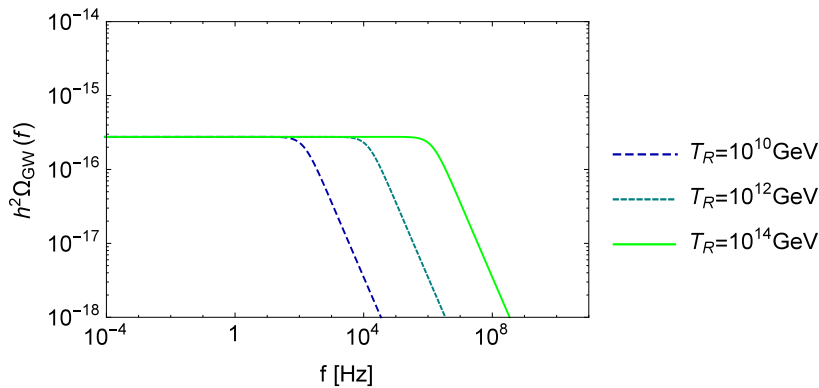}
    \caption{Change of slope of GW spectral energy-density in correspondence of the end of the reheating stage. Curves for different values of reheating temperature $T_{\rm R}$ are shown.}
  
    \label{plot-reh}
\end{figure}

In principle, from a direct detection of GW it could be also possible to extract information about the equation of state of the early Universe \cite{Nakayama:2008wy,Kuroyanagi:2011fy,Kuroyanagi:2014nba,Jinno:2014qka}. Assuming that the primordial GW power-spectrum has no tilt, the frequency dependence of the present power-spectrum is fully included in the transfer function, so that $\Omega_{\rm GW}\propto k^{2}T^{2}_{\rm T}\left(k\right)$. Since modes start evolving only when they enter the horizon and they are damped only by the factor $1/a$, the transfer function can be written as $T_{\rm T}\left(k\right)=\left|h_{\textbf{k},0}\right|/\left|h_{\textbf{k},{\rm prim}}\right|=\left(a_{0}/a_{\rm in}\right)^{-1}$, that is $\Omega_{\rm GW}\propto k^{2}a_{\rm in}^{2}$. If the equation of state at the time of the horizon crossing is given by $w=p/\rho$, that is the Hubble rate $H^{2}\propto a^{-3\left(1+w\right)}$, and we consider that at the horizon crossing of a mode k $k=aH$, we obtain $a_{\rm in}\propto k^{-2/\left(1+3w\right)}$. Then if the Universe evolves adiabatically, for a mode which enters the horizon when the Universe is described by $w$, we have
\begin{equation}\label{eqknee}
\Omega_{\rm GW}\left(f\right)=\Omega_{{\rm gw},F}\left(f/F\right)^{\left[2\left(3w-1\right)\right]/\left(1+3w\right)}\,,
\end{equation}
Then the spectrum behaves as $\propto f^{-2}$ for modes which enters the horizon during matter-dominance, and as $\propto f^{0}$ for the radiation dominated era. Then, a change in the Universe content dominance appears in $\Omega_{\rm GW}$ as a change in slope, in correspondence to the scale of Hubble horizon crossing at that time.
Therefore, extrapolating the tilt, in principle, one can find the equation of state that was governing the Universe at the time when the corresponding modes enter the horizon and identify the moments when the equation of state changed. Assuming that the reheating stage is a matter dominated stage, the transition towards the radiation dominated epoch, can be traced in the $\Omega_{\rm GW}$ change of slope. Exploiting the correspondence between cosmic time and temperature, for $10^{6}\mathrm{GeV}\lesssim T_{\rm R}\lesssim 10^{9}\mathrm{GeV}$ the frequency of the \textit{knee} varies respectively between $10^{-1}-10^{2}$ Hz; see fig.\ref{plot-reh} and tab.\ref{tabc}. For the possibility of constraining the reheating temperature assuming a GW detection, see \cite{Kuroyanagi:2011fy,Kuroyanagi:2014qza}. 

\subsubsection{Late-time entropy production}
It may happen that after the reheating stage a field other than the inflaton, dominates the energy-density and starts oscillating. In this case we have a phase of matter domination before the time of the equality, followed by the usual radiation domination when the field has decayed. In such a case another factor has to be considered in the transfer function \cite{Boyle:2005se,Nakayama:2008wy,Kuroyanagi:2014qza,Seto:2003kc,Kuroyanagi:2014nba}.

\subsection{Reheating parameters and gravitational waves}
The mechanism of reheating is still largely unknown. The difficulties in testing the physics of such a period 
are due essentially to the fact that all the features originating from that stage are washed out by the thermalization process. 
Moreover, reheating features are produced within the horizon so that they evolve during the whole history of the Universe without leaving imprints of their initial values.
In light of the this situation GW could represent a unique and powerful source of information about the reheating stage. At the same time, the dependence of GW evolution on the reheating process, means that the possibility of a detection of GW is influenced by the reheating temperature \cite{Kuroyanagi:2014nba,Jinno:2013xqa,Smith:2005mm,Smith:2008pf,Nakayama:2008wy}.
For example, an interesting aspect is the dependence of the tensor spectral index estimation on the reheating temperature assumed in the model with which data are fitted \cite{Kuroyanagi:2014qaa,Kuroyanagi:2014qza,Gong:2015qha}. Such a dependence can significantly relax the constraints on the tensor tilt $n_{\rm T}$, opening the possibility also to a blue-tilted spectrum on CMB scales \cite{Kuroyanagi:2014nba}.
On the other hand, given an inflationary potential, and then an energy scale of inflation, a detection of GW in a certain frequency band would represent the possibility to put a lower bound on the reheating temperature \cite{Kuroyanagi:2010mm,Kuroyanagi:2014qaa} and, for example, of breaking the degeneracy between inflationary models with a massive inflaton and those with a massless one, due to the different reheating stage that follows each of them \cite{Kuroyanagi:2014qaa}.

\subsubsection{CMB data and reheating parameters}
The constraints on the inflationary parameters $r$ and $n_{\rm S}$ usually extracted from CMB data actually have some dependence on the assumptions made about the reheating physics \cite{Martin:2014nya,Martin:2010kz,Creminelli:2014fca,Munoz:2014eqa,Cook:2015vqa,Cai:2015soa}. On the other hand this same fact means that CMB data encode information about the physics of the reheating stage \cite{Planck:2013jfk,Kinney:2005in,Liddle:2003as,Martin:2014nya}. Then it is interesting to study the possibility of extracting information on the thermal (reheating) history of the Universe from CMB data in order to improve the estimate of the inflationary parameters $r$ and $n_{\rm S}$.
The first possibility in order to extract constraints on inflationary parameters $r$ and $n_{\rm S}$ from CMB data, is to demand a reasonable number of e-folds between the time when the scale corresponding to the current Hubble radius left the inflationary horizon 
and the end of inflation $N_{k}$. The upper limit of $N_{k}$ is imposed by requiring that inflaton oscillations reheat the Universe instantaneously to a GUT scale-temperature, and the lower limit is obtained demanding that reheating is closer to the electroweak scale. 
But actually $N_{k}$ is related to the equation of state and the temperature of the reheating stage for each specific inflaton potential \cite{Dai:2014jja}. On the other hand $N_{k}$ can be written in terms of the inflationary parameters $n_{\rm S}$ and $r$ for a given inflationary model.
This connection between reheating physical quantities and inflationary parameters $n_{\rm S}$ and $r$, means that 
stronger constraints on one side would represent an improvement in understanding the physics of the other side, so that 
a measurement of $r$ would represents a source of information for the reheating physics.
This fact provides a way to break the degeneracy between inflationary models characterized by the same values of $n_{\rm S}$ and $r$. At the same time, for a given inflationary model, CMB data could be exploited in order to constrain the equation of state of reheating, through the estimation of $n_{\rm S}$ and $r$ and, moreover, if only a specific range of values of $w_{\rm re}$ is assumed as possible, some inflationary models can be excluded.\\
The other possibility in order to extract constraints on inflationary parameters is to consider a range for $w$, instead of a range for $N_{k}$ \cite{Munoz:2014eqa}. In this way slightly better constraints on inflationary parameters are obtained \cite{Munoz:2014eqa}. In particular, a higher lower limit for the tensor-to-scalar ratio is found and strong constraints in the plane $n_{\rm S}$-$r$ \cite{Creminelli:2014fca}. Besides the parameter $w$ also the reheating duration $N_{\rm reh}$ can be considered in order to model such a stage \cite{Cook:2015vqa,Cai:2015soa}.
In this case for some specific inflationary models, in the plane $n_{\rm S}$-$r$ a curve is identified, corresponding to a range of values of the equation of state of reheating, assuming constant $w$ value during the reheating stage, so that constraints on $n_{\rm S}$ and $r$ lead to an improvement of the comprehension of reheating physics. The reheating parameter space reveals that some inflationary models can be saved only for exotic equations of state of the reheating phase and others can be re-admitted with the help of the reheating parameters \cite{Cai:2015soa}.\\
A constraint on the reheating temperature, as a function of the energy scale of inflation, expressed by the tensor-to-scalar ratio $r$, stronger than that provided by BBN, can also be found \cite{Domcke:2015iaa}.

\subsection{Gravitational waves and neutrino free streaming}\label{neutrino}
The interaction of GW with matter and radiation is usually neglected, since it vanishes in the case of perfect fluids. However if a tensor part, that is a traceless and transverse term, is present in the anisotropic stress tensor, GW result coupled to matter and radiation too. During the history of the Universe, particle free streaming gives rise to anisotropic stress terms which modify the GW equation of motion.\\
Due to their relative energy-density and their relevance also in fundamental physics, the most studied particles in this context are neutrinos, which decouple from matter at $T\simeq 1 {\rm MeV}$, hence freely streaming afterwards. Calculations about their effects on GW have been made both at the first \cite{Weinberg:2003ur,Bashinsky:2005tv,Boyle:2005se,Dicus:2005rh,Watanabe:2006qe,Zhao:2009we,Jinno:2012xb,Shchedrin:2012sp,Stefanek:2012hj,Dent:2013asa} and the second order \cite{Mangilli:2008bw,Saga:2014jca} of the anisotropic stress tensor: the effect of the first-order term is a damping of the primordial GW amplitude, while the second-order contribution, besides providing a damping counterpart, acts as a source for second-order GW. It is found that their influence could be non-negligible, so that taking into account their presence could be relevant for the interpretation of possible data, but also for getting constraints on the neutrino physics itself.

\subsubsection{First-order neutrino anisotropic stress tensor}
A calculation of the effects of the first-order anisotropic stress tensor due to the neutrino free streaming has been done in \cite{Weinberg:2003ur} and then developed by several later 
works, e.g. \cite{Dent:2013asa} and references therein.\\
Defining a phase-space distribution function for neutrinos $n\left(\textbf{x},\textbf{p},t\right)$, its evolution is found by the Boltzmann equation in a perturbed FRW metric.
In the absence of metric perturbations the solution for $n\left(\textbf{x},\textbf{p},t\right)$ looks like that of an ideal gas, while metric perturbations give rise to a deviation $\delta n$ from such a behavior. The latter leads to a traceless and transverse contribution to the anisotropic stress tensor $\pi_{ij}$ of the Universe depending on the metric perturbations themselves, hence providing a source term in the GW equation of motion. 
Therefore, in order to find the solution for the equation of motion of tensor modes taking into account also for neutrino content of the Universe, the GW equation of motion and the Boltzmann equation have to be solved simultaneously. The results is an integro-differential equation for $h_{ij}\left(t\right)$ whose general solution is obtained numerically.\\
Of course, the distribution function $n$, and then the solution for the GW depends on the features assigned to the neutrinos. In light of recent developments in neutrino physics, it is interesting to generalize the calculations parameterizing the effects of neutrino masses and the possibility of additional degrees of freedom \cite{Dent:2013asa,Jinno:2012xb}.
The effect of massless neutrino on GW is a damping of the amplitude. In particular, for modes which enter the horizon during the radiation-dominated epoch, the squared amplitude is reduced by $35.6\%$ \cite{Weinberg:2003ur}, independently of any cosmological parameter, while for larger modes, which enter the horizon during the matter-dominated epoch, the damping of the amplitude ranges from $10.7\%$ to $9.0\%$ for different values of $\Omega_{\rm m}h^{2}$ \cite{Weinberg:2003ur}. 
In summary, the damping effect is most significant for modes with $k>k_{\rm eq}$, where $k_{\rm eq}$ is the horizon scale at the time of matter-radiation equality.\\
Considering neutrino masses, a further k-dependence of the damping arises \cite{Dent:2013asa}: for modes which enter the horizon before the Universe temperature reaches the neutrino mass, the damping is more efficient, while for GW that come into the horizon when the neutrino mass becomes significant, the damping effect is less strong. Furthermore, the scale dependence of the damping due to neutrino features means that a detection of a primordial GW signal would provide also constraints on the neutrino physics itself.\\
Furthermore, the beginning of the damping, corresponding to the neutrino decoupling era, affects those scales that correspond to present GW frequencies of $f\sim 10^{-11}$ Hz \cite{Lattanzi:2010gn}, that is close to the range of frequencies analyzed by pulsar timing array observatories; see tab.\ref{tabc}. Nevertheless such a feature is unlikely to be captured by planned experiments of that kind \cite{Lattanzi:2010gn}.

\subsubsection{Second-order neutrino anisotropic stress tensor}
We have shown how the presence of first-order scalar perturbations lead to a second-order source term in the GW equation of motion. But such a term could be given also by a pure second-order traceless and transverse anisotropic stress tensor. Free streaming particles give rise to such a kind of term. For the first time, \cite{Mangilli:2008bw} calculated such a contribution for cosmic neutrinos and photons solving the second-order Boltzmann equation for neutrino and photon distribution functions. Those calculations reveal that the usual approach of completely neglecting neutrino effects is a poor approximation in particular for studying the GW evolution, due to the large neutrino velocity dispersion during the radiation dominated era.\\
Numerically solving the full system of Einstein and Boltzmann equations at second order in tensor perturbations, using the tight-coupling approximation and taking into account both pure second-order terms and second-order terms coming from first-order scalar perturbations, the second-order GW contribution can be calculated \cite{Saga:2014jca}. Taking into account, besides neutrinos, also the effect of the photon stress-energy tensor, on large scales, both during the radiation and matter-dominated epochs, the role of the neutrino and photon second-order anisotropic stress is negligible with respect to the effect of the source term, due to first-order scalar perturbations. On the other hand, on small scales, the photon and neutrino free-streaming influence the GW evolution in a more efficient way with respect to the scalar-scalar source term. 
More precisely, the photon distribution function leads to an enhancement of the GW amplitude of about $150\%$, while the neutrino stress tensor provides a suppression of the amplitude of about $30\%$, so that the final result due to pure second-order sources is an amplification of $\sim 120\%$ of the GW amplitude with respect to the second-order GW calculated without taking into account the neutrino and photons anisotropic stress tensors \cite{Saga:2014jca}. This effect comes out to be of fundamental importance mainly for direct detection experiments if $r\lesssim 10^{-4}$ \cite{Saga:2014jca}, in which case the second-order GW background would become more important than that coming from the vacuum fluctuations.

\begin{table}[H]
\centering
\resizebox{15.5cm}{!} { %article
%\resizebox{9cm}{!} { %cimento
\begin{tabular}{|c|c|c|c|c|c|c|}
\hline
 & \textbf{Phenomenon} &  \textbf{Feature} & \textbf{Temperature} & \textbf{Frequency} & \textbf{Frequency}   \\ \hline \hline
\multirow{2}{*}{{\textbf{Thermal history}}}& Reheating & change of slope& $T_{\rm R}\in \left(10^{11};10^{16}\right)$ GeV  & \multirow{2}{*}{{$f_{\ast}\simeq 0.026{\rm Hz}\left(\frac{T_{\ast}}{10^{6}{\rm GeV}}\right)$}} 
 & $f_{\rm R}\in \left(10^{3};10^{8}\right)$ Hz\\ \cline{2-4} \cline{6-6}
& Neutrino decoupling & damping & $T_{\rm D}\simeq 1$ MeV & & $f_{\rm D}\sim 10^{-11}$ Hz\\ \hline
\end{tabular}
}
\caption{{\footnotesize Thermal history effects on the GW spectral energy-density. For the reheating effect see section \ref{knee}, for the role of neutrino decoupling see section \ref{neutrino}. In the column \textit{temperature}, we report the temperature of the Universe at the time of reheating phase and neutrino decoupling, respectively;
 next we report the estimation for the frequency at which one expects to have the feature pointed out.
}}
\label{tabc}
\end{table}

 %cambiare dim tabella
\section{Imprint of primordial gravitational waves on CMB and LSS}\label{cmb}

Primordial GW left several imprints on different physical observables along the history of the Universe. In this section we summarize which signature a primordial GW background left on the CMB and on LSS.

\subsection{Signature of primordial gravitational waves in the CMB}
The CMB forms when the Universe reaches a temperature of about $T\simeq 0.26$ eV and photons decouple from ordinary matter. 
After this moment photons propagate, nearly unperturbed, up to us. Therefore, the CMB provides a snapshot of the Universe at the time of recombination, and its polarization and temperature fluctuations carry a big amount of information about initial conditions at the end of Inflation, including imprints from primordial GW, e.g. \cite{Kamionkowski:1999qc,Hu:2001bc,liddle2000cosmological,Dodelson-Cosmology-2003}.\\
The presence of a GW background at the recombination epoch gives rise to both temperature and polarization anisotropies. However, the most important signature is clearly a ``curl-like" (B-mode) polarization pattern in CMB polarization. CMB radiation gets linearly polarized via Thomson scattering between photons and electrons at last scattering, in presence of a quadrupolar anisotropy in the intensity field of photons, see e.g. \cite{Hu:1997hv,1968ApJ,Kosowsky:1994cy}. 
Although primordial scalar, vector and tensor perturbations can all generate CMB polarization via this mechanism\cite{Hu:1997hv}, the specific signatures turn out to be different, in a way that we are going to quickly summarize in the following section, and which allows to single out contributions from tensor modes.\\
In general, polarization angular power-spectra have smaller amplitude than temperature ones, because only a few percent of the CMB photons gets  polarized by the aforementioned mechanism. This is due to the fact that, in the tight coupling regime between photons and electrons, Thomson scattering isotropizes the radiation field in the rest frame of the electron, thus erasing any incident quadrupole. Polarization is thus generated mostly around temperature dissipation scales, and close to recombination, in a photon-electron mild-coupling regime.
The polarization pattern is obtained by solving the Einstein-Boltzmann equations for the photon distribution, characterized by a generic polarization tensor, see e.g. \cite{Cabella:2004mk,Kamionkowski:2015yta}.

\subsubsection{Temperature and polarization angular power-spectra}
CMB temperature and polarization fluctuations are conveniently expanded in spherical harmonics. The Gaussianity of primordial CMB anisotropies then implies that all information is encoded in temperature and polarization angular power spectra.
The temperature field $T\left(\hat{n}\right)$ can be expanded on a spherical harmonics basis as
\begin{equation}
	T\left(\hat{n}\right)=\sum_{\ell,m}a_{\ell,m}^{\rm T}Y_{\ell m}\left(\hat{n}\right)\,,
\end{equation}
where $\hat{n}$ denotes a direction on the sky. The polarization field is described by a rank-2 tensor $I_{ij}\left(\hat{n}\right)$, where the usual Stokes parameters Q and U are given by $Q=1/4\left(I_{11}-I_{22}\right)$ and $U=I_{12}/2$. It is then useful to consider the combination Q$\pm i$U, which has definite transformation properties under rotations, and expand it on a spin-2 spherical harmonics $_{\pm 2}Y_{\ell m}$ basis \cite{Kamionkowski:1996ks,Zaldarriaga:1996xe,Newman:1961qr}:
\begin{equation}
	\left(Q\pm i U\right)=\sum_{\ell,m}a_{\ell m}^{\left(\pm 2\right)}\left[_{\pm 2}Y_{\ell m}\left(\hat{n}\right)\right]\,.
\end{equation}
For symmetry reasons, at this point it is convenient not to use directly the $a_{\ell m}^{\left(\pm 2\right)}$, but two independent combinations of them \cite{Newman:1961qr}:
\begin{equation}
	a_{\ell m}^{\rm E}=-\frac{1}{2}\left(a_{\ell m}^{\left(2\right)}+a_{\ell m}^{\left(-2\right)}\right)\,\qquad a_{\ell m}^{\rm B}=-\frac{1}{2i}\left(a_{\ell m}^{\left(2\right)}-a_{\ell m}^{\left(-2\right)}\right)\,,
\end{equation}
the so-called E- and B-modes. The former are invariant under parity transformation, while the latter have parity-odd properties.
Angular power-spectra are then defined as
\begin{equation}
	C_{\ell}^{\rm XY}\equiv\frac{1}{2\ell+1}\sum_{m}\langle a_{\ell m}^{\rm X}a_{\ell m}^{\rm Y}\rangle\,,
\end{equation}
where X,Y$=$T,E,B.\\
These spectra contain a large amount of information about cosmological parameters, and their accurate determination is a major experimental goal. In the following we briefly recall their most important features, in terms of GW signatures.
The temperature angular power-spectrum has been well measured by the Planck satellite \cite{Ade:2015xua}. The main source of temperature anisotropies are scalar perturbations. GW contribute to such anisotropies only at low multipoles $\ell \lesssim 60$, where the amount due to scalar perturbation is much larger and the cosmic variance prevents to extract unequivocally the information about tensor modes. More precisely, the amplitude of scalar perturbations is well constrained by current data but the present freedom on the scalar spectral index allows the tail of the angular power-spectrum to move, introducing then a degeneracy between $n_{\rm S}$ and $r$ \cite{Ade:2015lrj}. 
Moreover, the height and the slope of low-$\ell$ multipoles could be influenced also by a possible presence of isocurvature modes.
In summary the amount of primordial GW, expressed by $r$, cannot be extracted only by temperature measurements.
The most interesting observable in order to obtain information about GW is actually the CMB polarization.\\
The most interesting fact is that, since scalar perturbations locally produce only quadrupolar anisotropies of momentum $m=0$, at last scattering, they cannot generate B-mode patterns \cite{Hu:1997hv}. Therefore, since vector perturbations decay after inflation, {\em a measurement of primordial B-modes in the polarization pattern would provide unique evidence for the presence of a primordial GW background}. Their detection would allow to break the degeneracy between scalar and tensor perturbations, and provide an estimation for $r$. The primordial B-mode angular power-spectrum is thus the most significant one for what concerns primordial GW. At the same time, it is the most difficult to measure, due to its small amplitude. In general, GW affect low-multipole power-spectra, because they correspond to scales that  were super-horizon at the time of recombination (primordial GW on sub-horizon scales are damped by the expansion of the Universe). The contribution from primordial GW to the BB power spectrum essentially comes from multipoles $\ell<150$.  
A major issue when trying to measure primordial B-modes is that suitable cleaning  of the maps from the contamination due to galactic dust and astrophysical foregrounds  
is required \cite{Adam:2015wua}. In fact, interstellar dust grains cause thermal emission in the microwave band, so that, where galactic magnetic fields are present, such a radiation can be linearly polarized, generating a foreground contribution in E- and B-mode angular power-spectra. In a similar way, the interaction between cosmic rays and the galactic magnetic field is the origin of the synchrotron contribution in the polarization power-spectra. Foreground contamination at low-multipoles can give contributions to the BB spectrum up to two order of magnitude larger than the primordial one, therefore exquisitely accurate levels of cleaning are required.
Besides dust contamination, also gravitational lensing, due to the presence of clustered matter between the last scattering surface and the observer, affects the CMB polarization pattern \cite{Lewis:2001hp,Knox:2002pe,Kesden:2002ku}. In particular it can transform E-mode patterns into B-mode ones. Accurate de-lensing of B-modes is going to be an important task if we want to achieve the high-sensitivities in the measurement of $r$ which are promised by future experiments \cite{Knox:2002pe,Kesden:2002ku,Hu:2001kj}.
Besides TT, EE and BB auto-spectra, discussed above, we also have to consider the temperature-polarization cross-spectra. In a standard scenario, the only non-vanishing cross-spectrum is TE, since the other combination are parity breaking. The measurement of TE correlations confirms an important prediction of standard Cosmology and inflationary physics \cite{Peiris:2003ff}, allows tighter determination of cosmological parameters \cite{Galli:2014kla,Ade:2015xua}, and carries additional GW information at low-$\ell$ \cite{Polnarev:2007cv,Ade:2015lrj}. The study of TB and EB spectra is also interesting, since those can be generated, for example, in scenarios with primordial magnetic fields (see, e.g. \cite{Ade:2015cva} and refs. therein), or in presence of parity-breaking Physics in the Early Universe (see, e.g. \cite{Lue:1998mq,Feng:2004mq,Li:2008tma,Sorbo:2011rz}, and \cite{Bartolo:2014hwa} with refs. therein).

\subsubsection{Current constraints on tensor modes from the CMB}
Up to now, temperature and E-mode angular power-spectra have been measured with very high accuracy \cite{Ade:2015xua} on a wide range of multipoles. A certain amount of B-modes polarization has been detected too \cite{Ade:2014xna} for $30<\ell<150$, but its amplitude is compatible with foreground contamination and lensing of E-modes \cite{Ade:2015tva}. 
Current data actually provide only an upper bound on $r$.
The most strict constraint, that does not assume the consistency relation \eqref{consistency} but a scale-invariant GW power-spectrum, comes from the joint analysis of BICEP2 and Keck Array data, Planck polarization and WMAP9 23 GHZ and 33 GHZ maps: $r_{0.05}<0.09$ at $95\%$ C.L. \cite{Array:2015xqh}.
The Background Imaging of Cosmic Extragalactic Polarization 2 (BICEP2) and Keck Array are ground-based experiments for the detection of CMB polarization in a limited region of the sky, with high sensitivity. On the other hand their disadvantage is that they work on $150$ GHZ and $95$ GHz photon frequency channels. High frequencies, which are crucial to assess dust contributions, are not accessible from the ground. For this reason, data collected by the Planck satellite are essential to allow proper  component separation in the current analysis. The joint analysis of Planck, BICEP2 and Keck Array data has also produced a high-significance detection of gravitational lensing from LSS in B-mode polarization. Allowing the lensing amplitude $A_{\rm L}$ to vary in the data analysis, they found $A_{\rm L}=1.13\pm 0.18$ at $7.0\sigma$ of significance \cite{Ade:2015tva}.\\
For what concerns the tensor spectral index, CMB data alone do not have the possibility to provide strong bounds on it, also in the case of a B-mode detection, since they target only a narrow range of GW frequencies around $f\simeq 10^{-17}$ Hz. On the other hand, a measurement in the range of frequencies accessible via direct detection experiments, combined with CMB data, 
would provide very strict constraints on the tensor spectral index, and then on the consistency relation \eqref{consistency}; see section \ref{sezioneconsistency}.

\subsubsection{Further possible signatures of primordial gravitational waves}\label{cmbchirali}
The generation of non-standard GW during inflation or the production of primordial GW due to the presence of extra fields, would introduce additional characteristic signatures in the statistic of CMB anisotropies \cite{Kamionkowski:2010rb,Shiraishi:2010kd,Shiraishi:2010sm,Shiraishi:2011st,Barnaby:2010vf,Anber:2012du,Bartolo:2014hwa,Domcke:2016bkh,Meerburg:2016ecv}. 
In particular, if primordial GW are chiral or non-Gaussian, these features are expected to be encoded in temperature and polarization angular power-spectra and bispectra. A bound on tensor non-Gaussianity has been provided by the Planck Collaboration \cite{Ade:2015ava}.
For example, in the model described by eq. \eqref{modpseudo}, an extra production of chiral and non-Gaussian GW is expected. These specific features lead to non-vanishing temperature and polarization bispectra with parity-violation and non-vanishing TB and EB angular cross-spectra \cite{Shiraishi:2013kxa,Bartolo:2014hwa}. The latter, in general, would be a clear signature of parity-violation.
Then the search for features beyond CMB angular power-spectra, could show up interesting new insights in the inflationary mechanism by which GW have been produced.

\subsection{Imprint of long-wavelength tensors on cosmic structures}\label{lss}

Besides the imprints in CMB, GW affect the cosmic mass distribution too. Early and late time effects can be identified.  
The presence of tensor modes during the early epochs of the Universe is found to modify the power-spectrum of primordial scalar perturbations \cite{Jeong:2012df}. At late times the presence of a GW background leads to several effects: a tidal effect during structure formation due to the presence of long-wavelength tensor modes \cite{Masui:2010cz,Jeong:2012nu,Schmidt:2013gwa,Dai:2013kra}, a correlation of galaxy 
ellipticities (the shear) \cite{Schmidt:2012nw,Schmidt:2012ne}, and projection effects due to the perturbation of the space-time by the GW on the galaxy distribution \cite{Dodelson:2003bv,Jeong:2012nu,Dai:2012bc,Schmidt:2012nw}, the CMB \cite{Cooray:2005hm,Li:2006si,Dodelson:2003bv,Book:2011na} and the $21$-cm background \cite{Book:2011dz,Lu:2007pk,Pen:2003yv}.
Here we consider the imprint of long-wavelength tensor modes on the primordial power-spectrum of scalar perturbations and consequently on the late time matter power-spectrum \cite{Dai:2013kra,Pajer:2013ana}.\\
The basic idea is to indirectly trace the presence of degrees of freedom coupled to the source of curvature perturbations during the primordial stages of the Universe and uncoupled, or weakly coupled, during late-times cosmic evolution \cite{Jeong:2012df}. The coupling between curvature perturbations and other degrees of freedom leaves specific imprints in the primordial curvature power-spectrum. The interesting fact is that the information related to the primordial coupling contained in such features is left imprinted in the late-time matter perturbation power-spectrum, and then, in principle, it could be captured by galaxy surveys \cite{Ando:2008zza} and CMB experiments \cite{Pullen:2007tu}. The presence of such features in power-spectra of quantities evaluated at epochs when the coupling is off, suggests to refer to this extra degree of freedom as \textit{fossil field}.
Primordial GW produced during inflation can be understood as a fossil field \cite{Maldacena:2002vr,Maldacena:2011nz,Seery:2008ax,Giddings:2010nc,Giddings:2011zd}.
In view of these considerations, it is interesting to identify which is the imprint in the scalar power-spectrum of GW produced during the inflationary stage. In general, assuming statistically isotropic and Gaussian primordial curvature perturbations $\Phi_{\rm p}$, the coupling with GW implies that the scalar power-spectrum violates these properties \cite{Jeong:2012df}. There are two main effects due to the presence of GW. 
The first is the presence of non-vanishing off-diagonal terms in the correlation function $\langle\Phi_{\rm p}\Phi_{\rm p}\rangle_{h}$, which means that measurements related to two wavelengths $k_{1}$ and $k_{2}$ provide information also about a tensor mode $K$. \cite{Jeong:2012df} estimates the volume $V$ of a galaxy survey needed to detect such a departure from the Gaussianity of curvature perturbations, and finds that for $A_{\rm T}\simeq 10^{-9}$ a survey with $k_{\rm max}/k_{\rm min}\simeq 5000$ is required, where $k_{\rm max}$ and $k_{\rm min}$ are the modes related to the detection capability of the survey. Surveys of this size are unfortunately far form planned capabilities.\\
A second imprint left by fossil fields is a quadrupolar anisotropy in the local galaxy power-spectrum. Here we deepen this second aspect.

\subsubsection{Gravitational waves imprints on the primordial curvature power-spectrum}
Requiring global statistical isotropy, the primordial correlation function $\langle\Phi_{\rm p}\Phi_{\rm p}\rangle_{h}$ evaluated in the presence of a mode $h_{ij}\left(K\right)$, is modulated by the bispectrum $\langle\Phi_{\rm p}\Phi_{\rm p}h_{\rm p}\rangle$, as a direct consequence of the coupling between the two fields, and by the isotropic power-spectra $P_{\Phi}\left(k\right)$ and $P_{h}\left(k\right)$, where $\Phi_{\rm p}$ are the primordial scalar perturbations (including all orders) \cite{Jeong:2012df}. 
The interesting fact is that for several inflationary models, including single-field slow-roll inflation, a consistency relation holds between the isotropic component of the power-spectra of the scalar field $P_{\Phi}\left(k\right)$, and of the tensor field $P_{h}\left(K\right)$ and the 
bispectrum $\langle\Phi_{\rm p}\left(\textbf{k}_{1}\right)\Phi_{\rm p}\left(\textbf{k}_{2}\right)h_{{\rm p},\lambda}\left(\textbf{K}\right)\rangle$, evaluated in the squeezed limit $K\rightarrow 0$ \cite{Maldacena:2002vr,Creminelli:2004yq}:
\begin{align}\label{ccfossile}
	\langle\Phi_{\rm p}\left(\textbf{k}\right)\Phi_{\rm p}\left(\textbf{k}\right)h_
{{\rm p},\lambda}\left(\textbf{K}\right)\rangle\rightarrow &\left(2\pi\right)^{3}
\delta_{\rm D}\left(\textbf{k}_{1}+\textbf{k}_{2}+\textbf{K}\right)
\times\nonumber\\
&\times\frac{1}{2}\frac{{\rm d}\mathrm{ln}P_{\Phi}}{{\rm d}\mathrm{ln}k}\epsilon^{ij}_{s}
\left(\textbf{K}\right)\hat{k}_{1i}\hat{k}_{2j}P_{h}\left(K\right)P_{
\Phi}\left(k\right)\,,
\end{align}
where $\lambda$ indicates the polarization state of the GW and $p$ stands for primordial. Notice that this relation is valid up to corrections of order $\mathscr{O}\left(K^{2}/k^{2}\right)$, and that we are indicating the 
wave-numbers of scalar perturbations with $k$ and those of tensors with $K$. For more details about the origin and the meaning of this relation, see \cite{Dimastrogiovanni:2014ina}.
The limit \eqref{ccfossile} gives the possibility of specifying the power-spectrum $\langle\Phi_{\rm p}\Phi_{\rm p}\rangle_{h}$ in terms of the power-spectra of the scalar and of the tensor perturbation fields themselves. 
In the presence of the squeezed modes $K$ the scalar power-spectrum then reads
\begin{align}	
\langle\Phi_{\rm p}\left(\textbf{k}_{1}\right)\Phi_{\rm p}\left(\textbf{k}_{2}\right)\rangle_{h}=&\left(2\pi\right)^{3}\delta_{D}\left(\textbf{k}_{1}+\textbf{k}_{2}\right)P_{
\Phi}\left(k\right)+\int\frac{d^{3}\textbf{K}}{\left(2\pi\right)^{3}}\sum_{\lambda}\left(2\pi\right)^{3}\delta_{\rm D}\left(\textbf{k}_{1}+\textbf{k}_{2}+\textbf{K}\right)\nonumber\\
&\times\frac{1}{2}\frac{{\rm d}\mathrm{ln}P_{\Phi}}{{\rm d}\mathrm{ln}k}P_{\Phi}\left(k\right)h_{{\rm p},\lambda}\left(\textbf{K}\right)\epsilon^{ij}_{\lambda}\left(\textbf{K}\right)\hat{k}_{1i}\hat{k}_{2j}+\mathscr{O}\left(K^{2}/k^{2}\right)\,.
\end{align}
Notice that the correction to the statistically isotropic power-spectrum is of first order in the perturbation $h_{ij}$. Transforming to configuration space and considering a local region, that is associated to a Universe patch 
smaller than the tensor modes wavelength, a quadrupolar modulation asymmetry in the observed local power-spectrum for matter and galaxies is found.

\subsubsection{Local matter power-spectrum} 
In order to get a physical quantity that can be compared with the data, we need to consider the post-inflationary evolution and the projection effects due to the observation.
Let us indicate the observable matter power-spectrum including the signature of the fossil GW by $P_{\delta_{\rm g}}\left(\textbf{k};\textbf{x}_{c}\right)$, where $\textbf{x}_{\rm c}=\left(\textbf{x}_{1}+\textbf{x}_{2}\right)/2$ is the mean of the two points considered in the space where $\delta_{\rm g}$ is evaluated. With the $\textbf{x}_{\rm c}$ dependence, we mean that the power-spectrum is to be understood as a local quantity referred to a neighbourhood of 
$\textbf{x}_{\rm c}$.\\
First, one has to consider that second-order matter perturbations are coupled to scalar and tensor perturbations of the first order. This effect is negligible at early times but becomes significant at late times during the matter dominated epoch \cite{tomita,Matarrese:1997ay}. 
Actually, GW perturb the space-time during matter clustering leading to overdensity modes growing with a quadrupolar dependence. The most efficient modulation is due to long-wavelength tensor modes, in particular to those which are entering the horizon at the considered time.
Furthermore, in order to get to the \textit{observed} quadrupole of the $\delta_{\rm g}$ power-spectrum, the space-time distortion due to metric perturbations has to be taken into account. More precisely, scalar and tensor perturbations modify the geodesic curves leading to a gap between the observed space-time position and that in the comoving coordinates of the Universe.
Considering all these effects, for $K\ll k$, leads to an expression of $P_{\delta_{\rm g}}\left(\textbf{k};\textbf{x}_{\rm c}\right)$, which is a function of the primordial metric perturbations, of the primordial power-spectrum of $\Phi$ modulated by the fossil field, and of the isotropic power-spectrum $P_{\Phi}\left(k\right)$. 
Specifying $\langle\Phi_{\rm p}\Phi_{\rm p}\rangle_{h}$ by the consistency condition eq.\eqref{ccfossile}, $P_{\delta_{\rm g}}\left(\textbf{k};\textbf{x}_{\rm c}\right)$ in the squeezed limit is obtained in terms of $P_{\Phi}$ and $P_{h}$.

\subsubsection{Quadrupole anisotropy}
The obtained local matter power-spectrum presents a quadrupolar anisotropy \cite{Pullen:2007tu,Hanson:2009gu}. As anticipated, this fact is interesting in order to look for observable quantities that can probe the presence of a GW background originated during the primordial phases of the Universe evolution. 
Quadrupole moments are given by
\begin{equation}
	\mathnormal{Q}_{2m}\left(\textbf{x}_{\rm c}\right)\equiv \frac{\int d^{2}\hat{
\textbf{k}}P_{\rm g}\left(\textbf{k};\textbf{x}_{\rm c}\right)Y^{\ast}_{\left(2m
\right)}(\hat{\textbf{k}})}{\int d^{2}\hat{\textbf{k}}P_{\rm g}\left(\textbf{k};
\textbf{x}_{\rm c}\right)Y^{\ast}_{\left(00\right)}(\hat{\textbf{k}})}\,,
\end{equation}
for $m=\pm2,\pm1,0$, with $Y_{\left(\ell m\right)}(\hat{\textbf{k}})$, defined with respect to some chosen coordinate axes. 
It is useful to define a symmetric and traceless tensor ${Q}_{ij}$ that encodes the contribution due to a single large mode $K$:
\begin{equation}	
\mathnormal{Q}_{2m}\left(\textbf{x}_{\rm c}\right)=\int d^{2}\hat{\textbf{k}}
\,\mathnormal{Q}_{ij}\left(\textbf{x}_{\rm c}\right)\left(\hat{k}^{i}\hat{k}^{j}-
\frac{1}{3}\delta^{ij}\right)Y^{\ast}_{\left(2m\right)}(\hat{\textbf{k}})\,.
\end{equation}
From the expression of $P_{\delta_{\rm g}}\left(\textbf{k};\textbf{x}_{\rm c}\right)$, the contribution due to a single tensor mode $K$ is obtained as a sum of the following terms:
\begin{align}
	\mathnormal{Q}_{ij}&=\left(\text{\textit{terms from primordial stages
}}\right)+\nonumber\\
	&+\left(\text{\textit{terms from late time non linear coupling modes}}\right)+
\nonumber\\
	&+\left(\text{\textit{terms due to the projection effects}}\right)\,.
\end{align}
The first term is the quadrupole imprinted at early times by the scalar-scalar-tensor bispectrum in the squeezed limit. The second term represents the contribution due to the fact that matter clusters in an anisotropic space-time, because of the long-wavelength tensor modes. 
Notice that $Q_{ij}$ continues to be a local quantity.\\
Assuming that the tensor field is a realization of an underlying statistically homogeneous and isotropic Gaussian random field, one can find that the average quantity $\overline{Q^2(z)} \equiv \left(8\pi/15\right)\langle Q_{ij} \left(\textbf{x}_{\rm c},z\right) Q^{\ast ij}\left(\textbf{x}_{\rm c},z\right) \rangle$ is a function only of the redshift $z$ \cite{Dai:2013kra}. 
For $K\rightarrow 0$, at the present time, the contribution to the observed quadrupole vanishes, while for smaller scales there is a residual effect due to modes that are entering the horizon now \cite{Dai:2013kra}.
The consistency condition eq.\eqref{ccfossile} ensures that super-horizon modes do not affect sub-horizon observables.\\
Given the current constraints on tensor modes, the quadrupole amplitude is predicted to be very small, so that considering forecasts about future galaxies surveys, a quadrupole power detection is out of their detection capabilities \cite{Dai:2013ikl}.
In summary, the observation of a power-spectrum quadrupole by planned experiments, would rule out all that inflationary models which satisfy eq.\eqref{ccfossile}. It is then interesting to investigate the inflationary models which violate that condition; in this case a detection of the quadrupolar asymmetry would represent also the possibility to constrain inflationary parameters.
For example, in an inflationary scenario with a non-attractor phase, eq.\eqref{ccfossile} is satisfied but sub-leading terms contain information about the dynamics which are not included into the isotropic power-spectra, such as the time of the switch from the non-attractor to the attractor phase. 
The quadrupole term is found to be very small and undetectable by planned experiments, but an upper bound on its power would lead to an upper limit on the transition time \cite{Dimastrogiovanni:2014ina}. 
Another interesting model in this context is Solid Inflation, in which case eq.\eqref{ccfossile} is violated and a larger quadrupole in the galaxy clustering, still compatible with current constraints, is predicted \cite{Dimastrogiovanni:2014ina,Akhshik:2014bla,Endlich:2013jia}. Instead for a quasi-single-field model of inflation, the constraints on the quadrupole induced by super-horizon tensor modes do not significantly restrict the parameter space of the inflationary theory, while the departure from the statistical isotropy of the power-spectrum could represent in principle a powerful probe for the amplitude of primordial tensor modes \cite{Dimastrogiovanni:2015pla}. 
Moreover in \cite{Dimastrogiovanni:2015pla} the size of a galaxy survey necessary to probe a given tensor amplitude is estimated. \cite{Brahma:2013rua} investigated also inflation with non-Bunch-Davis initial conditions. More recently, \cite{Bartolo:2015qvr} examined the effects of fossil fields within the EFT approach of inflation, considering scenarios that simultaneously break time re-parameterization and spatial diffeomorphisms during inflation.
\section{Current constraints and observational prospects}\label{exp}

Primordial GW have never been detected directly and not even we have an unequivocal indirect measurement of them. Several efforts are underway on the two fronts. Assuming an inflationary period in the early history of the Universe, at the present time the space-time is expected to be filled with a GW spectral energy-density $\Omega_{\textrm{gw}}$ with a specific amplitude for each frequency $f=c2\pi a/k$. On the other hand, during the evolution of the Universe, GW could have left imprints on different physical observables, providing the possibility of indirect measurements.
Most experiments which try to detect them, directly or indirectly, have access to a specific range of frequencies, and then has to face with the GW evolution related to the frequency band in exam.

\subsection{Imprints of primordial gravitational waves on physical observables}
Starting from the early stages, a primordial GW background affects the BBN process: being GW relativistic degrees of freedom, they constitute a contribution to the radiation energy density of the Universe \cite{Mangano:2001iu,Smith:2006nka}, which results in a faster expansion rate of the background. In particular the latter means that neutrons have less time to decay before the freeze out of the weak interactions, and then in presence of a fixed amount of GW, a certain over-production of Helium during primordial nucleosynthesis is expected \cite{weinberg2008cosmology,Hou:2011ec}. An estimation of the Deuterium abundance combined with Planck data and BAO, then provides an integral upper bound $h^{2}\Omega_{\rm GW}<1.7\cdot10^{-6}$ at $95\%$ C.L. for $f\gtrsim 10^{-15}$ Hz \cite{Pagano:2015hma}. Phenomena due to the GW effects on the cosmic expansion have also been considered in \cite{Boyle:2007zx}.\\
Recently it has been noted that the current constraints on the abundance of primordial black holes leads to an upper bound on primordial tensor modes on very small scales \cite{Nakama:2015nea,Nakama:2016enz}. In fact, bounds on primordial black holes constitute a limit on the amplitude of scalar perturbation; as a consequence, since tensor modes play the role of source for second-order scalar perturbations, an upper bound on primordial gravitational waves can be found. It results: $\Omega_{\rm GW}<10^{-5}-10^{-5}$ for the frequency range $10^{-12}-10^{4}$ Hz \cite{Nakama:2016enz} (see figure $5$ of \cite{Nakama:2016enz} to find out the accurate scale-dependence).
Going on along the history of the Universe, the GW background left its imprints on CMB photons, both on their temperature and polarization distributions. In particular, the presence of GW at the time of recombination leads to the formation of a B-mode pattern in the polarization \cite{polnarev}, which is then modified by late time phenomena; see section \ref{cmb}. The measured CMB power-spectra include information about GW at frequencies $f\sim 10^{-17}$ Hz. Moreover the CMB energy spectrum too contains information about GW: the integrated tensor power in the frequency range $10^{-12}-10^{-9}${\rm Hz} leads to $\mu$-distortions of the CMB spectrum \cite{Ota:2014hha,Chluba:2014qia}. Furthermore, CMB power-spectra are affected by the GW contribution to the radiation energy-density through the time of matter-radiation equality and the expansion rate of the Universe \cite{Bashinsky:2003tk,Smith:2006nka}.
As explained in section \ref{lss}, the presence of a GW background also affects the mass distribution of the Universe, modifying the statistics of primordial curvature perturbations and perturbing the space-time when matters clusters during the matter-dominated epoch (by tidal effects) \cite{Masui:2010cz,Schmidt:2012nw,Dai:2013kra,Schmidt:2013gwa}.
Another interesting phenomenon that has to be considered for our purposes is the gravitational lensing effect due to the presence of GW in the space-time in which light signals propagate.
The observed mass distribution at high redshift is affected by this phenomenon (projection effects) \cite{Jeong:2012nu}, and a distortion of galaxy shapes (shear) \cite{Dodelson:2003bv,Dodelson:2010qu,Schmidt:2012nw,Jeong:2012nu} is expected to be there too.
Clearly, the presence of a GW background affects also light signals that are propagating to us from closer objects with respect to galaxies. This effect could be captured by pulsar timing array observations which, combining the perturbations in the signals coming form different ultra-stable millisecond pulsars, in principle, will be able to trace the presence of GW in the space-time in which the signal is propagating \cite{Joshi:2013at} (and refs. therein). This kind of observations are particularly sensitive to GW of frequencies $f\sim 10^{-9}-10^{-7}$ Hz \cite{Moore:2014lga}.
Finally, primordial GW are expected to permeate the present-time Universe and then a direct detection is in principle possible. 
For this purpose the laser interferometer experiments have been constructed and others are planned for the future.
The goal of these experiments is the detection of GW at frequencies spanning from $f\sim 10^{-4}$ Hz for space-based interferometers, to $f\sim10^{2}$ Hz for ground-based observatories \cite{Moore:2014lga}.

\subsection{Current constraints}
Parameterizing the primordial GW power-spectrum as in eq.\eqref{spettrotensori}, current data provide bounds on its amplitude and spectral tilt. From CMB data an upper bound on the GW amplitude at frequencies $f\sim 10^{-17}$ Hz is obtained. 
More precisely, as previously mentioned, the joint analysis of Planck and external data (named as \textit{Planck TT+lowP+lensing+ext} in \cite{Ade:2015tva}), BICEP2 and Keck Array data (including the 2014 observing run with the $95$ GHz channel), provided an upper bound of $r_{0.05}<0.07$ at $95\%$ C.L. \cite{Array:2015xqh}, assuming the consistency relation \eqref{consistency}. 
By employing BICEP2 and Keck Array data, Planck data only for polarization and WMAP9 23 GHZ and 33 GHZ maps, the bound becomes $r_{0.05}<0.09$ at $95\%$ C.L. \cite{Array:2015xqh}, assuming a scale-invariant power-spectrum. Notice that the constraints obtained in the first way are  more model-dependent than those obtained in the second one.
Other works \cite{Meerburg:2015zua,Cabass:2015jwe,Lasky:2015lej,Huang:2015gka} extended this analysis taking into account a non-vanishing spectral index and the measurements by LIGO \cite{Abbott:2007kv}. 
Considering current data, coming from CMB, LIGO and the Parker Pulsar Timing Array (PPTA) \cite{Manchester:2012za}, and fitting data allowing the tensor spectral index to vary, \cite{Meerburg:2015zua} found that a blue power-spectrum and $r>0.12$ are preferred (this value of $r$ refers to the previous constraint on such a quantity provided by \cite{Ade:2015tva}). It is also shown that Planck data are those which lead to such a high value of $r$, rather than BICEP data. At the same time, the LIGO bound is found to be fundamental in order to obtain such stringent limits.
Constraints on $n_{\rm T}$ have been obtained combining BICEP2/Keck Array \cite{Ade:2015tva}, temperature Planck data 2013 \cite{Ade:2013zuv}, WMAP \cite{Page:2006hz} low $\ell$ polarization, a prior on $H_{0}$ from HST \cite{Scoville:2006vr} data, BAO measurements from SDSS \cite{York:2000gk} and the upper limit on the intensity of a stochastic GW background from LIGO: $n_{\rm T, 0.01}=0.06^{+0.63}_{-0.89}$ at $95\%$ C.L. \cite{Meerburg:2015zua}, in correspondence of a best-fit for the tensor-to-scalar ratio of $r_{0.01}=0.02$. Clearly, admitting the spectral index to vary leads to a weaker bound on the tensor-to-scalar ratio $r$. 
A more recent paper \cite{Cabass:2015jwe} provides a further analysis which takes into account the latest data release of Keck Array at $95$ GHz \cite{Array:2015xqh} and one by one the bounds coming from the Helium abundance, $\mu$-distortions of the CMB and the LIGO-Virgo \cite{Acernese:2004jn} experiment. The limits resulting from their analysis with LIGO-Virgo bounds are $n_{\rm T}=0.04^{+0.61}_{-0.85}$ at $95\%$ C.L. with $r<0.085$, found putting a prior lower bound on the tensor-to-scalar ratio of $r>0.001$. \cite{Lasky:2015lej} makes a further analysis which takes into account CMB data form Planck, BICEP and SPTpol, current bounds form PPTA, LIGO-Virgo and BAO and BBN indirect constraints, providing an upper bound on the tensor spectral index of $n_{\rm T}<0.36$ at $95\%$ C.L. in correspondence of $r_{0.05}=0.11$.
In \cite{Meerburg:2015zua} and \cite{Cabass:2015jwe}, the same limits are obtained considering also a possible contribution of the GW background to the relativistic degrees of freedom $N_{\rm eff}$.\\
Notice that in the mentioned works the primordial power-spectrum of GW is always parametrized as a power law. This assumption could be not appropriate for frequency bands extending over several orders of magnitudes. In particular, at the end of inflation, the slow-roll conditions are no longer satisfied and then the GW power-spectrum is expected to deviate form a pure power law. 
Therefore, a more detailed parametrization could be a significant improvement of the outlined data analysis \cite{Boyle:2014kba}. In this direction, \cite{Meerburg:2015zua} parametrized the power-spectrum also taking into account a scale-dependence of the tensor spectral index, concluding that with available data no significant constraints can be obtained.
Also, the choice of the scale of the UV cutoff is still an open issue, which can significantly influence the results of data analysis \cite{Meerburg:2015zua} and is then worth being explored.

\subsection{Observational prospects}
\begin{figure}[ht]
    \centering
    \includegraphics[width=0.7\textwidth]{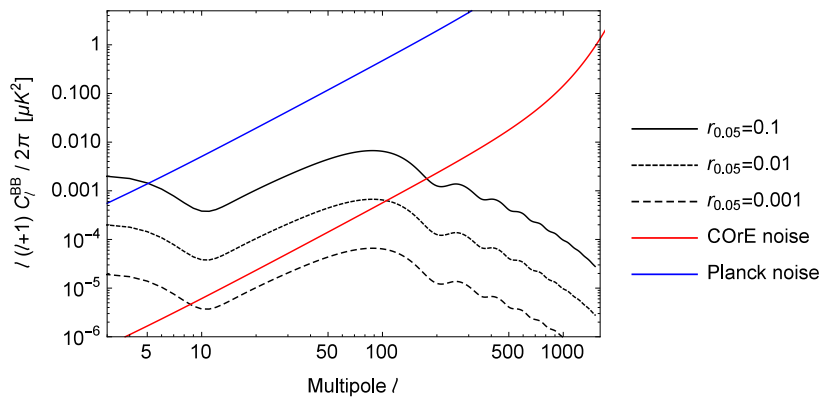}
    \caption{B-mode power-spectra obtained from Planck best-fit cosmological parameters \cite{Ade:2015xua} for different values of $r$, compared with idealized noise for the Planck satellite (blue curve) \cite{Adam:2015rua} and for a COrE-like experiment (red curve) \cite{Bouchet:2011ck}. From the top downwards (black curves): $r_{0.05}=0.1$, $r_{0.05}=0.01$ and $r_{0.05}=0.001$. Notice that the black curves do not include the lensing contribution, which contaminates the signal. As visible, the realization of a COrE-like CMB satellite would enhance considerably the possibilities of B-modes detection.}
  
    \label{plotB}
\end{figure}

Several future experiments are planned to improve the mentioned bounds and hopefully detect the cosmological GW background. For what concerns CMB polarization experiments, several ground-based, balloon and space-borne experiments are under construction or have been proposed.
Ground-based and balloon experiments, such as the Atacama Cosmology Telescope Polarization Experiment (ACTPol) \cite{Calabrese:2014gwa} and Polarbear \cite{Ade:2014afa} (which are already underway), the Cosmology Large Angular Scale Surveyor (CLASS) \cite{Eimer:2012ny}, the Primordial Inflation Polarization ExploreR (Piper) \cite{piper} and Spider \cite{Crill:2008rd}, 
are designed to improve the sensitivity over a restricted range of multipoles of the polarization power-spectra related to one or two frequency channels. 
On the other hand space-borne experiments have been proposed in order to span a larger multipoles range and to get data related to several frequency channels to improve the control of systematic errors and the component separation analysis. We mention, for example, the Cosmic Origins Explorer mission (COrE) \cite{Bouchet:2011ck}, the Polarized Radiation Imaging ans Spectroscopy Mission (PRISM) \cite{Andre:2013nfa}, LiteBIRD \cite{Matsumura:2013aja}, the Primordial Inflation Explorer (PIXIE) \cite{Kogut:2011xw} (that are planned to improve the $\mu$-distortions estimation too); see fig.\ref{plotB}. 
About detailed B-modes observational prospects, taking into account planned experiments for the next future and also for plausible far future experimental capabilities, see \cite{Errard:2015cxa,Verde:2005ff}.

\begin{figure}[ht]
    \centering
    \includegraphics[width=1\textwidth]{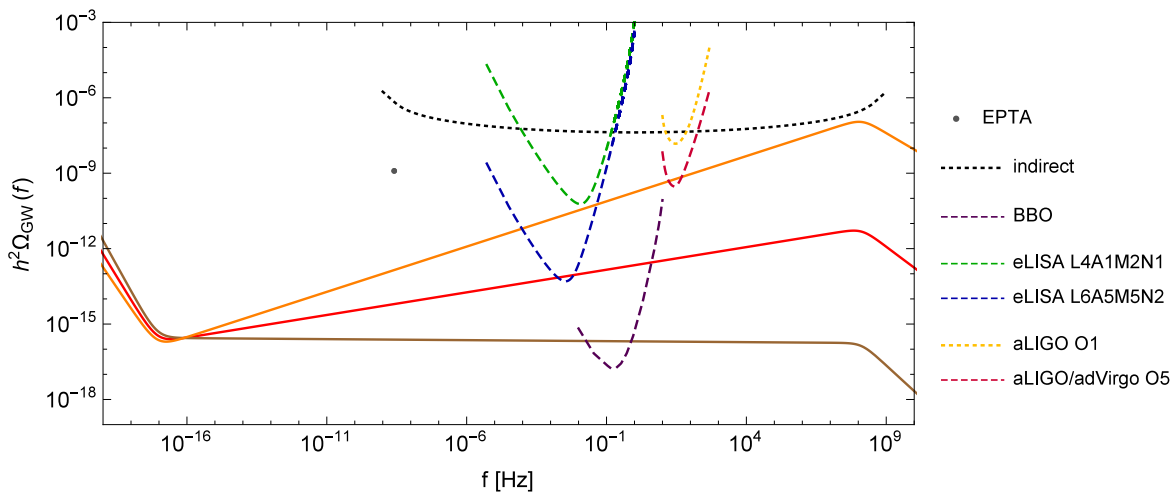}
    \caption{GW spectral energy-density for different values of $n_{\rm T}$ are shown with solid lines: $n_{\rm T}=-r/8$ (brown), $n_{\rm T}=0.18$ (red) and $n_{\rm T}=0.36$ (orange). The $r$ value is fixed at $r_{0.05}=0.07$. It is assumed also $T_{\rm R}=10^{16}$ GeV.
		Short-dashed lines are current bounds related to: aLIGO data, O$1\mbox{:}2015\mbox{-}16$ observing run (yellow) \cite{TheLIGOScientific:2016wyq}, combined analysis of Planck data, BAO and BBN measurements which provides an integral bound $\Omega_{\rm GW}<3.8\times 10^{-6}$ (black) \cite{Pagano:2015hma}; see \cite{Lasky:2015lej} for the manner of employing this limit. 
		The gray dot corresponds to the bound provided by EPTA \cite{Lentati:2015qwp}, which assumes $n_{T}=0$; see \cite{Lasky:2015lej} for comments about this choice. 
		Long-dashed lines are expected power-law integrated sensitivity curves for the following experiments: BBO (violet) \cite{Crowder:2005nr,Thrane:2013oya}, eLISA configuration L6A5M5N2 (blue) \cite{pet}, eLISA configuration L4A1M2N1 (green) \cite{pet}, 
		aLIGO-adVirgo, O$5\mbox{:}2020\mbox{-}22$ observing run (magenta) \cite{TheLIGOScientific:2016wyq}. Plotted upper bounds and expected sensitivity curves are obtained by the method provided by \cite{Thrane:2013oya} (see also \cite{Caprini:2015zlo}), assuming a power-law signal. The mentioned eLISA configurations are described in \cite{Klein:2015hvg}. 
		}
  
    \label{plot_exp}
\end{figure}

\noindent
Also pulsar timing array experiments are underway, such as PPTA \cite{Manchester:2012za}, the European Pulsar Timing Array (EPTA) \cite{Lentati:2015qwp}, and the North American Nanohertz Observatory for Gravitational Waves (NANOGrav) \cite{McLaughlin:2013ira}, and others are planned, such as the Square Kilometre Array (SKA) \cite{Janssen:2014dka} (as a secondary aim). The upper limit provided by
EPTA is $\Omega_{\rm GW}<1.2\times 10^{-9}$ at $95\%$ C.L. and for $f=2.8\times 10^{-9}$ Hz \cite{Lentati:2015qwp}.\\
Also for what concerns direct GW detection several efforts are underway. LIGO, Virgo and GEO$600$ \cite{Abadie:2010mt} have already collected data. The joint analysis of LIGO and Virgo provides an upper limit of $\Omega_{\rm GW}<5.6\times10^{-6}$ at $95\%$ C.L. for $f\in\left(41.4,169.25\right)$ Hz \cite{Aasi:2014zwg}.
The updated LIGO, that is aLIGO \cite{TheLIGOScientific:2014jea}, has collected data too for a few months at the end of 2015. Further upgrades are planned for aLIGO and for Virgo, that might become adVirgo in summer 2016, and a sequence of observing runs are expected for the more and more improved configurations of such laser interferometers (see for example \cite{TheLIGOScientific:2016wyq}, table I). A number of ground-based experiments have also been proposed for the next future, such as LIGO India (IndIGO) \cite{Unnikrishnan:2013qwa} (that will be included in the network aLIGO-adVirgo), the Kamioka Gravitational Wave Detector (KAGRA) \cite{Aso:2013eba} and the Einstein Telescope (ET) \cite{Punturo:2010zz}.
Moreover, the space-born experiment eLISA \cite{elisaweb,Klein:2015hvg} has been planned.
However, taking into account current bounds on $r$ related to CMB scales, if the primordial GW power-spectrum is scale-invariant, planned experiments will be not able to detect them \cite{Moore:2014lga}.
For next future planned experiments, such as upgraded aLIGO, a direct detection of primordial GW might be possible only in the case of a blue inflationary power-spectrum, that is, in the case of non-single-field slow-roll inflation; see fig.\ref{plot_exp}. For a possible detection of a scale-invariant inflationary power-spectrum, bold experiments, such as the DECI-Hertz Interferometer Gravitational wave Observatory (DECIGO) \cite{Kawamura:2011zz} and BBO \cite{Crowder:2005nr}, are required. These observatories might be useful also in order to get information about the reheating temperature after inflation, as shown in section \ref{sezstoria}.
Anyway, also a non-detection of the primordial tensor modes by next future experiments, combined with other data, would represent a powerful way to put limits on the tensor spectral index and then to test the consistency relation \eqref{consistency}. In fact experiments of direct detection would provide information about frequencies of more than $20$ orders of magnitude larger than those related to CMB data.
A detection of primordial GW by such a kind of experiments would give evidence for a blue tensor power-spectrum and then rule out the single-field slow-roll inflation. For example, inflationary models in which particle production takes place, admit the production of GW with a blue tensor power-spectrum, in principle, detectable by future experiments such as eLISA.
On the other hand a non-detection by those experiments would put a bound on the spectral index and then a limit on the violation of the consistency relation. For example, a non-detection by an experiment with upgraded aLIGO capabilities would put a constraint on the tensor spectral index of $n_{\rm T}<0.34$ at the $95\%$ C.L. for $r=0.11$ on CMB scales \cite{Cabass:2015jwe}, and analogously eLISA would put an upper limit of $n_{\rm T}\simeq 0.2$, depending on the configuration.
Experiments which investigate frequencies in the $f\sim 10^{-2}-10^{3}$ Hz range are then extremely significant in order to put constraints on the tensor spectral index, exploiting a combined analysis with CMB data \cite{Meerburg:2015zua,Cabass:2015jwe,Lasky:2015lej}.\\
To get the comparison between the sensitivity curves of planned experiments for GW detection and primordial signals parametrized by a power-law, see \cite{Kuroda:2015owv,Lasky:2015lej}. \cite{Cabass:2015jwe} presents also forecasts for a joint analysis of CMB data from a COrE-like experiment and a detection from the ground-based laser interferometer aLIGO.

\noindent
For the far future, it could be possible to obtain information about the inflationary physics also from the features of the GW, such as their level of non-Gaussianity and chirality. The latter would provide interesting bounds on parameters of those inflationary models which present events of particle production, such as the parameter $\xi$ defined in \eqref{gauge} \cite{Sorbo:2011rz,Cook:2011hg}.
For what concerns non-Gaussianity, from CMB data strict bounds on the scalar bispectrum have been obtained, but also the tensor three-point function has been, more weakly, constrained by Planck measurements \cite{Ade:2015ava}. Non-Gaussianities represent a clear example where information coming from features of scalar perturbations can provide constraints on tensor perturbation properties, and viceversa. As an example, the Planck Collaboration \cite{Ade:2015ava}, considering Galileon inflationary models, from the bounds of the scalar non-Gaussianity parameter $f_{\rm NL}^{\rm scal}$, constrained the sound speed $c_{\rm S}$ of the Galileon scalar field and another parameter $\bar{c}_{s}$ strictly linked to the tensor spectral index $n_{\rm T}$ in the modified consistency relation, finding that the constraints on $f_{\rm NL}^{\rm scal}$ leaves open the possibility of a blue tensor power-spectrum for that model \cite{Cook:2013xea,Shiraishi:2013kxa}. On the other hand, also the tensor bispectrum could provide constraints on inflationary physics. An example is given by the inflationary scenario associated with a pseudo-scalar coupling to a gauge field, where $f_{\rm NL}^{\rm tens}$ can provide upper bounds for the model parameter $\xi$, complementary to those coming from the scalar bispectrum \cite{Cook:2013xea,Shiraishi:2013kxa}. In the context of primordial non-Gaussianity from tensor modes, it might be interesting to consider also the most general three-point function for tensor modes obtained by \cite{Maldacena:2011nz}, taking into account the isometries of the inflationary space-time and the cross-bispectra between scalar and tensor modes introduced in \cite{Maldacena:2002vr}.

\section{Conclusions}\label{conclusioni}

The inflationary model of the Early Universe predicts the production of a stochastic GW background by quantum fluctuations of the gravitational field. 
Such a radiation encodes a unique probe of the physics of the Early Universe and fundamental physics theories.
In addition, during the primordial inflationary and reheating phases, further mechanisms of GW production can take place as a consequence of non-basic inflationary scenarios. 
Interestingly, each of them introduces peculiar contributions and features in the primordial GW power-spectrum. 
Precisely the shown multiplicity of predictions makes these GW significant information messengers able to discriminate among the variety of inflationary models. In this direction, testing the validity or the violation of the 
so-called consistency relation between the tensor-to-scalar ratio $r$ and the tensor spectral index $n_{\rm T}$ plays a substantial role. 
However, there are still some inflationary scenarios for which it could be significant to examine more deeply of what already done, the aspect of GW production, especially in light of the forthcoming experimental capabilities of detection. Not secondarily, the coming up of new ideas about inflationary scenarios in which a further GW production takes place would be a stimulating progress.
Besides the physics of the Early Universe, we have also shown that the present day inflationary GW spectral energy-density, would in principle provide the intriguing possibility of tracing the thermal history of the Universe.\\
In light of all this crucial information encoded in the primordial GW background, several efforts are underway and planned, to detect them directly and indirectly. 
Up to now, the most promising way to detect primordial GW seems to be the search for B-modes in CMB polarization anisotropies. In the more distant future, also the imprint of inflationary GW on the energy distribution of the CMB and on large-scale structure of the Universe, might provide interesting signatures of these GW. 
Current data put only upper bounds on the tensor-to-scalar ratio $r$ and the tensor spectral index $n_{\rm T}$, leaving open the possibility of several inflationary mechanisms of GW production, 
besides the standard (single-field, slow-roll) one. The forthcoming experimental capabilities concerning GW detection, then represent a promising direction
for improving these constraints and better understand the physics of the Early Universe.

\section*{Acknowledgments} 
The authors are involved in the eLISA cosmology working group and in the preparation of a related report about cosmological GW with respect to the eLISA experiment (https://www.elisascience.org); we are grateful to all people involved in this activity for stimulating discussions. We also thank Emanuela Dimastrogiovanni, Raul Jimenez, Marco Peloso and Yun-Song Piao for useful comments and Alvise Raccanelli for checking the draft.
M.C.G. thanks financial support from a Cariparo foundation grant. N.B., M.L. and S.M. acknowledge partial financial support by the ASI/INAF Agreement I/072/09/0 for the Planck LFI Activity of Phase E2.

\bibliographystyle{h-elsevier}
\bibliography{main}

\begin{thebibliography}{100}

\bibitem{Huber:2015znp}
S.J. Huber et~al.,
\newblock JCAP 1603 (2016) 036, 1512.06357.

\bibitem{Caprini:2015zlo}
C. Caprini et~al.,
\newblock JCAP 1604 (2016) 001, 1512.06239.

\bibitem{Dev:2016feu}
P.S.B. Dev and A. Mazumdar,
\newblock Phys. Rev. D93 (2016) 104001, 1602.04203.

\bibitem{Sanidas:2012ee}
S.A. Sanidas, R.A. Battye and B.W. Stappers,
\newblock Phys. Rev. D85 (2012) 122003, 1201.2419.

\bibitem{Figueroa:2012kw}
D.G. Figueroa, M. Hindmarsh and J. Urrestilla,
\newblock Phys. Rev. Lett. 110 (2013) 101302, 1212.5458.

\bibitem{Array:2015xqh}
BICEP2, Keck Array, P.A.R. Ade et~al.,
\newblock Phys. Rev. Lett. 116 (2016) 031302, 1510.09217.

\bibitem{Sorbo:2011rz}
L. Sorbo,
\newblock JCAP 1106 (2011) 003, 1101.1525.

\bibitem{Barnaby:2012xt}
N. Barnaby et~al.,
\newblock Phys. Rev. D86 (2012) 103508, 1206.6117.

\bibitem{Senatore:2011sp}
L. Senatore, E. Silverstein and M. Zaldarriaga,
\newblock JCAP 1408 (2014) 016, 1109.0542.

\bibitem{Binetruy:2012ze}
P. Binetruy et~al.,
\newblock JCAP 1206 (2012) 027, 1201.0983.

\bibitem{Biagetti:2013kwa}
M. Biagetti, M. Fasiello and A. Riotto,
\newblock Phys. Rev. D88 (2013) 103518, 1305.7241.

\bibitem{Khlebnikov:1996mc}
S.{\relax Yu}. Khlebnikov and I.I. Tkachev,
\newblock Phys. Rev. Lett. 77 (1996) 219, hep-ph/9603378.

\bibitem{Calabrese:2014gwa}
E. Calabrese et~al.,
\newblock JCAP 1408 (2014) 010, 1406.4794.

\bibitem{Ade:2014afa}
POLARBEAR, P.A.R. Ade et~al.,
\newblock Astrophys. J. 794 (2014) 171, 1403.2369.

\bibitem{Eimer:2012ny}
J.R. Eimer et~al.,
\newblock Proc. SPIE Int. Soc. Opt. Eng. 8452 (2012) 845220, 1211.0041.

\bibitem{piper}
PIPER, N.N. Gandilo et~al.,
\newblock 2016, 1607.06172.

\bibitem{Crill:2008rd}
B.P. Crill et~al.,
\newblock Proc. SPIE Int. Soc. Opt. Eng. 7010 (2008) 2P, 0807.1548.

\bibitem{Bouchet:2011ck}
COrE, F.R. Bouchet et~al.,
\newblock (2011), 1102.2181.

\bibitem{Andre:2013nfa}
PRISM, P. Andr\`e et~al.,
\newblock JCAP 1402 (2014) 006, 1310.1554.

\bibitem{Matsumura:2013aja}
T. Matsumura et~al.,
\newblock (2013), 1311.2847,
\newblock [astro-ph.IM] [J. Low. Temp. Phys.176,733(2014)].

\bibitem{Kogut:2011xw}
A. Kogut et~al.,
\newblock JCAP 1107 (2011) 025, 1105.2044.

\bibitem{Dodelson:2003bv}
S. Dodelson, E. Rozo and A. Stebbins,
\newblock Phys. Rev. Lett. 91 (2003) 021301, astro-ph/0301177.

\bibitem{Masui:2010cz}
K.W. Masui and U.L. Pen,
\newblock Phys. Rev. Lett. 105 (2010) 161302, 1006.4181.

\bibitem{Jeong:2012nu}
D. Jeong and F. Schmidt,
\newblock Phys. Rev. D86 (2012) 083512, 1205.1512.

\bibitem{Smith:2006nka}
T.L. Smith, E. Pierpaoli and M. Kamionkowski,
\newblock Phys. Rev. Lett. 97 (2006) 021301, astro-ph/0603144.

\bibitem{TheLIGOScientific:2014jea}
LIGO Scientific, J. Aasi et~al.,
\newblock Class. Quant. Grav. 32 (2015) 074001, 1411.4547.

\bibitem{elisaweb}
https://www.elisascience.org.

\bibitem{Klein:2015hvg}
A. Klein et~al.,
\newblock Phys. Rev. D93 (2016) 024003, 1511.05581.

\bibitem{Bird:2016dcv}
S. Bird et~al.,
\newblock Phys. Rev. Lett. 116 (2016) 201301, 1603.00464.

\bibitem{Sasaki:2016jop}
M. Sasaki et~al.,
\newblock Phys. Rev. Lett. 117 (2016) 061101, 1603.08338.

\bibitem{Clesse:2016vqa}
S. Clesse and J. García-Bellido,
\newblock (2016), 1603.05234.

\bibitem{Abbott:2016blz}
Virgo, LIGO Scientific, B.P. Abbott et~al.,
\newblock Phys. Rev. Lett. 116 (2016) 061102, 1602.03837.

\bibitem{Blair:2016idv}
D. Blair et~al.,
\newblock Sci. China Phys. Mech. Astron. 58 (2015) 120402, 1602.02872.

\bibitem{Brout:1977ix}
R. Brout, F. Englert and E. Gunzig,
\newblock Annals Phys. 115 (1978) 78.

\bibitem{Starobinsky:1980te}
A.A. Starobinsky,
\newblock Phys. Lett. B91 (1980) 99.

\bibitem{Kazanas:1980tx}
D. Kazanas,
\newblock Astrophys. J. 241 (1980) L59.

\bibitem{Sato:1980yn}
K. Sato,
\newblock Mon. Not. Roy. Astron. Soc. 195 (1981) 467.

\bibitem{Guth:1980zm}
A.H. Guth,
\newblock Phys. Rev. D23 (1981) 347.

\bibitem{Linde:1981mu}
A.D. Linde,
\newblock Phys. Lett. B108 (1982) 389.

\bibitem{Albrecht:1982mp}
A. Albrecht et~al.,
\newblock Phys. Rev. Lett. 48 (1982) 1437.

\bibitem{Abbott:1982hn}
L.F. Abbott, E. Farhi and M.B. Wise,
\newblock Phys. Lett. B117 (1982) 29.

\bibitem{Mukhanov:1981xt}
V.F. Mukhanov and G.V. Chibisov,
\newblock JETP Lett. 33 (1981) 532,
\newblock [Pisma Zh. Eksp. Teor. Fiz.33,549(1981)].

\bibitem{Hawking:1982cz}
S.W. Hawking,
\newblock Phys. Lett. B115 (1982) 295.

\bibitem{Guth:1982ec}
A.H. Guth and S.Y. Pi,
\newblock Phys. Rev. Lett. 49 (1982) 1110.

\bibitem{Starobinsky:1982ee}
A.A. Starobinsky,
\newblock Phys. Lett. B117 (1982) 175.

\bibitem{Abbott:1984fp}
L.F. Abbott and M.B. Wise,
\newblock Nucl. Phys. B244 (1984) 541.

\bibitem{Mukhanov:1985rz}
V.F. Mukhanov,
\newblock JETP Lett. 41 (1985) 493,
\newblock [Pisma Zh. Eksp. Teor. Fiz.41,402(1985)].

\bibitem{Ade:2015xua}
Planck, P.A.R. Ade et~al.,
\newblock (2015), 1502.01589.

\bibitem{Lidsey:1995np}
J.E. Lidsey et~al.,
\newblock Rev. Mod. Phys. 69 (1997) 373, astro-ph/9508078.

\bibitem{Liddle:1994dx}
A.R. Liddle, P. Parsons and J.D. Barrow,
\newblock Phys. Rev. D50 (1994) 7222, astro-ph/9408015.

\bibitem{Lyth:1998xn}
D.H. Lyth and A. Riotto,
\newblock Phys. Rept. 314 (1999) 1, hep-ph/9807278.

\bibitem{Ade:2015lrj}
Planck, P.A.R. Ade et~al.,
\newblock (2015), 1502.02114.

\bibitem{Lucchin:1985wy}
F. Lucchin and S. Matarrese,
\newblock Phys. Lett. B164 (1985) 282.

\bibitem{Dolgov:1982th}
A.D. Dolgov and A.D. Linde,
\newblock Phys. Lett. B116 (1982) 329.

\bibitem{Kofman:1994rk}
L. Kofman, A.D. Linde and A.A. Starobinsky,
\newblock Phys. Rev. Lett. 73 (1994) 3195, hep-th/9405187.

\bibitem{Linde:1982uu}
A.D. Linde,
\newblock Phys. Lett. B116 (1982) 335.

\bibitem{Ashoorioon:2012kh}
A. Ashoorioon, P.S. Bhupal~Dev and A. Mazumdar,
\newblock Mod. Phys. Lett. A29 (2014) 1450163, 1211.4678.

\bibitem{Krauss:2013pha}
L.M. Krauss and F. Wilczek,
\newblock Phys. Rev. D89 (2014) 047501, 1309.5343.

\bibitem{Bruni:1996im}
M. Bruni et~al.,
\newblock Class. Quant. Grav. 14 (1997) 2585, gr-qc/9609040.

\bibitem{Matarrese:1997ay}
S. Matarrese, S. Mollerach and M. Bruni,
\newblock Phys. Rev. D58 (1998) 043504, astro-ph/9707278.

\bibitem{Acquaviva:2002ud}
V. Acquaviva et~al.,
\newblock Nucl. Phys. B667 (2003) 119, astro-ph/0209156.

\bibitem{Bardeen:1980kt}
J.M. Bardeen,
\newblock Phys. Rev. D22 (1980) 1882.

\bibitem{Kodama:1985bj}
H. Kodama and M. Sasaki,
\newblock Prog. Theor. Phys. Suppl. 78 (1984) 1.

\bibitem{Mukhanov:1990me}
V.F. Mukhanov, H.A. Feldman and R.H. Brandenberger,
\newblock Phys. Rept. 215 (1992) 203.

\bibitem{liddle2000cosmological}
A.R. Liddle and D.H. Lyth,
\newblock {Cosmological inflation and large scale structure} (Cambridge, UK:
  Univ. Pr., 2000).

\bibitem{Sasaki:1986hm}
M. Sasaki,
\newblock Prog. Theor. Phys. 76 (1986) 1036.

\bibitem{Taruya:1997iv}
A. Taruya and Y. Nambu,
\newblock Phys. Lett. B428 (1998) 37, gr-qc/9709035.

\bibitem{Lukash:1980iv}
V.N. Lukash,
\newblock Sov. Phys. JETP 52 (1980) 807,
\newblock [Zh. Eksp. Teor. Fiz.79,1601(1980)].

\bibitem{Lyth:1984gv}
D.H. Lyth,
\newblock Phys. Rev. D31 (1985) 1792.

\bibitem{Liddle:1994cr}
A.R. Liddle and M.S. Turner,
\newblock Phys. Rev. D50 (1994) 758, astro-ph/9402021,
\newblock [Erratum: Phys. Rev.D54,2980(1996)].

\bibitem{Ade:2015ava}
Planck, P.A.R. Ade et~al.,
\newblock (2015), 1502.01592.

\bibitem{Grishchuk:1974ny}
L.P. Grishchuk,
\newblock Sov. Phys. JETP 40 (1975) 409,
\newblock [Zh. Eksp. Teor. Fiz.67,825(1974)].

\bibitem{Starobinsky:1979ty}
A.A. Starobinsky,
\newblock JETP Lett. 30 (1979) 682,
\newblock [Pisma Zh. Eksp. Teor. Fiz.30,719(1979)].

\bibitem{Rubakov:1982df}
V.A. Rubakov, M.V. Sazhin and A.V. Veryaskin,
\newblock Phys. Lett. B115 (1982) 189.

\bibitem{Fabbri:1983us}
R. Fabbri and M.d. Pollock,
\newblock Phys. Lett. B125 (1983) 445.

\bibitem{Grishchuk:1974jy}
L.P. Grishchuk,
\newblock Zh. Eksp. Teor. Fiz. 66 (1974) 833.

\bibitem{Grishchuk:1977zz}
L.P. Grishchuk,
\newblock Annals N. Y. Acad. Sci. 302 (1977) 439.

\bibitem{Misner:1974qy}
C.W. Misner, K.S. Thorne and J.A. Wheeler,
\newblock {Gravitation} (W. H. Freeman, San Francisco, 1973).

\bibitem{Bunch:1978yq}
T.S. Bunch and P.C.W. Davies,
\newblock Proc. Roy. Soc. Lond. A360 (1978) 117.

\bibitem{Mollerach:1993sy}
S. Mollerach, S. Matarrese and F. Lucchin,
\newblock Phys. Rev. D50 (1994) 4835, astro-ph/9309054.

\bibitem{Alba:2015cms}
V. Alba and J. Maldacena,
\newblock JHEP 03 (2016) 115, 1512.01531.

\bibitem{Adamek:2015mna}
J. Adamek, R. Durrer and V. Tansella,
\newblock JCAP 1601 (2016) 024, 1510.01566.

\bibitem{Maggiore:1900zz}
M. Maggiore,
\newblock {Gravitational Waves. Vol. 1: Theory and Experiments}Oxford Master
  Series in Physics (Oxford University Press, 2007).

\bibitem{Watanabe:2006qe}
Y. Watanabe and E. Komatsu,
\newblock Phys. Rev. D73 (2006) 123515, astro-ph/0604176.

\bibitem{Lyth:1984yz}
D.H. Lyth,
\newblock Phys. Lett. B147 (1984) 403,
\newblock [Erratum: Phys. Lett.B150,465(1985)].

\bibitem{Lyth:1996im}
D.H. Lyth,
\newblock Phys. Rev. Lett. 78 (1997) 1861, hep-ph/9606387.

\bibitem{Boubekeur:2005zm}
L. Boubekeur and D. Lyth,
\newblock JCAP 0507 (2005) 010, hep-ph/0502047.

\bibitem{Baumann:2006cd}
D. Baumann and L. McAllister,
\newblock Phys. Rev. D75 (2007) 123508, hep-th/0610285.

\bibitem{Boubekeur:2012xn}
L. Boubekeur,
\newblock Phys. Rev. D87 (2013) 061301, 1208.0210.

\bibitem{Efstathiou:2005tq}
G. Efstathiou and K.J. Mack,
\newblock JCAP 0505 (2005) 008, astro-ph/0503360.

\bibitem{Garcia-Bellido:2014eva}
J. Garcia-Bellido et~al.,
\newblock JCAP 1409 (2014) 006, 1405.7399.

\bibitem{Garcia-Bellido:2014wfa}
J. Garcia-Bellido et~al.,
\newblock Phys. Rev. D90 (2014) 123539, 1408.6839.

\bibitem{Easther:2006qu}
R. Easther, W.H. Kinney and B.A. Powell,
\newblock JCAP 0608 (2006) 004, astro-ph/0601276.

\bibitem{Endlich:2012pz}
S. Endlich, A. Nicolis and J. Wang,
\newblock JCAP 1310 (2013) 011, 1210.0569.

\bibitem{Gruzinov:2004ty}
A. Gruzinov,
\newblock Phys. Rev. D70 (2004) 063518, astro-ph/0404548.

\bibitem{Bartrum:2013fia}
S. Bartrum et~al.,
\newblock Phys. Lett. B732 (2014) 116, 1307.5868.

\bibitem{Bastero-Gil:2014jsa}
M. Bastero-Gil et~al.,
\newblock JCAP 1405 (2014) 004, 1401.1149.

\bibitem{Bastero-Gil:2016qru}
M. Bastero-Gil et~al.,
\newblock (2016), 1604.08838.

\bibitem{Cheung:2007st}
C. Cheung et~al.,
\newblock JHEP 03 (2008) 014, 0709.0293.

\bibitem{Tsujikawa:2014mba}
S. Tsujikawa,
\newblock Lect. Notes Phys. 892 (2015) 97, 1404.2684.

\bibitem{Cannone:2014uqa}
D. Cannone, G. Tasinato and D. Wands,
\newblock JCAP 1501 (2015) 029, 1409.6568.

\bibitem{Cannone:2015rra}
D. Cannone, J.O. Gong and G. Tasinato,
\newblock JCAP 1508 (2015) 003, 1505.05773.

\bibitem{Graef:2015ova}
L. Graef and R. Brandenberger,
\newblock JCAP 1510 (2015) 009, 1506.00896.

\bibitem{Lin:2015cqa}
C. Lin and L.Z. Labun,
\newblock JHEP 03 (2016) 128, 1501.07160.

\bibitem{Abolhasani:2015cve}
A.A. Abolhasani et~al.,
\newblock JCAP 1603 (2016) 020, 1511.03218.

\bibitem{Bartolo:2015qvr}
N. Bartolo et~al.,
\newblock JCAP 1603 (2016) 044, 1511.07414.

\bibitem{Cai:2016ldn}
Y. Cai, Y.T. Wang and Y.S. Piao,
\newblock Phys. Rev. D94 (2016) 043002, 1602.05431.

\bibitem{Brandenberger:2006xi}
R.H. Brandenberger et~al.,
\newblock Phys. Rev. Lett. 98 (2007) 231302, hep-th/0604126.

\bibitem{Gasperini:2002bn}
M. Gasperini and G. Veneziano,
\newblock Phys. Rept. 373 (2003) 1, hep-th/0207130.

\bibitem{Khoury:2001wf}
J. Khoury et~al.,
\newblock Phys. Rev. D64 (2001) 123522, hep-th/0103239.

\bibitem{tomita}
K. Tomita,
\newblock Prog. Theor. Phys. 37 (1967) 831.

\bibitem{Carrilho:2015cma}
P. Carrilho and K.A. Malik,
\newblock JCAP 1602 (2016) 021, 1507.06922.

\bibitem{Matarrese:1992rp}
S. Matarrese, O. Pantano and D. Saez,
\newblock Phys. Rev. D47 (1993) 1311.

\bibitem{Matarrese:1993zf}
S. Matarrese, O. Pantano and D. Saez,
\newblock Phys. Rev. Lett. 72 (1994) 320, astro-ph/9310036.

\bibitem{Nakamura:2003wk}
K. Nakamura,
\newblock Prog. Theor. Phys. 110 (2003) 723, gr-qc/0303090.

\bibitem{Nakamura:2004rm}
K. Nakamura,
\newblock Prog. Theor. Phys. 117 (2007) 17, gr-qc/0605108.

\bibitem{Carbone:2004iv}
C. Carbone and S. Matarrese,
\newblock Phys. Rev. D71 (2005) 043508, astro-ph/0407611.

\bibitem{Noh:2003yg}
H. Noh and J.c. Hwang,
\newblock (2003), astro-ph/0305123.

\bibitem{Carbone:2005nm}
C. Carbone, C. Baccigalupi and S. Matarrese,
\newblock Phys. Rev. D73 (2006) 063503, astro-ph/0509680.

\bibitem{Baumann:2007zm}
D. Baumann et~al.,
\newblock Phys. Rev. D76 (2007) 084019, hep-th/0703290.

\bibitem{Ananda:2006af}
K.N. Ananda, C. Clarkson and D. Wands,
\newblock Phys. Rev. D75 (2007) 123518, gr-qc/0612013.

\bibitem{Mollerach:2003nq}
S. Mollerach, D. Harari and S. Matarrese,
\newblock Phys. Rev. D69 (2004) 063002, astro-ph/0310711.

\bibitem{Fidler:2014oda}
C. Fidler et~al.,
\newblock JCAP 1407 (2014) 011, 1401.3296.

\bibitem{Assadullahi:2009nf}
H. Assadullahi and D. Wands,
\newblock Phys. Rev. D79 (2009) 083511, 0901.0989.

\bibitem{Cook:2011hg}
J.L. Cook and L. Sorbo,
\newblock Phys. Rev. D85 (2012) 023534, 1109.0022,
\newblock [Erratum: Phys. Rev.D86,069901(2012)].

\bibitem{Enqvist:2001zp}
K. Enqvist and M.S. Sloth,
\newblock Nucl. Phys. B626 (2002) 395, hep-ph/0109214.

\bibitem{Bartolo:2007vp}
N. Bartolo et~al.,
\newblock Phys. Rev. D76 (2007) 061302, 0705.4240.

\bibitem{Enqvist:2008be}
K. Enqvist, S. Nurmi and G.I. Rigopoulos,
\newblock JCAP 0810 (2008) 013, 0807.0382.

\bibitem{Suyama:2011pu}
T. Suyama and J. Yokoyama,
\newblock Phys. Rev. D84 (2011) 083511, 1106.5983.

\bibitem{Kawasaki:2013xsa}
M. Kawasaki, N. Kitajima and S. Yokoyama,
\newblock JCAP 1308 (2013) 042, 1305.4464.

\bibitem{Biagetti:2014asa}
M. Biagetti et~al.,
\newblock JCAP 1504 (2015) 011, 1411.3029.

\bibitem{Fujita:2014oba}
T. Fujita, J. Yokoyama and S. Yokoyama,
\newblock PTEP 2015 (2015) 043E01, 1411.3658.

\bibitem{Chung:1999ve}
D.J.H. Chung et~al.,
\newblock Phys. Rev. D62 (2000) 043508, hep-ph/9910437.

\bibitem{Barnaby:2010vf}
N. Barnaby and M. Peloso,
\newblock Phys. Rev. Lett. 106 (2011) 181301, 1011.1500.

\bibitem{Barnaby:2011vw}
N. Barnaby, R. Namba and M. Peloso,
\newblock JCAP 1104 (2011) 009, 1102.4333.

\bibitem{Barnaby:2011qe}
N. Barnaby, E. Pajer and M. Peloso,
\newblock Phys. Rev. D85 (2012) 023525, 1110.3327.

\bibitem{Anber:2012du}
M.M. Anber and L. Sorbo,
\newblock Phys. Rev. D85 (2012) 123537, 1203.5849.

\bibitem{Namba:2015gja}
R. Namba et~al.,
\newblock JCAP 1601 (2016) 041, 1509.07521.

\bibitem{Ben-Dayan:2016iks}
I. Ben-Dayan,
\newblock (2016), 1604.07899.

\bibitem{Garretson:1992vt}
W.D. Garretson, G.B. Field and S.M. Carroll,
\newblock Phys. Rev. D46 (1992) 5346, hep-ph/9209238.

\bibitem{Cook:2013xea}
J.L. Cook and L. Sorbo,
\newblock JCAP 1311 (2013) 047, 1307.7077.

\bibitem{Kofman:1997yn}
L. Kofman, A.D. Linde and A.A. Starobinsky,
\newblock Phys. Rev. D56 (1997) 3258, hep-ph/9704452.

\bibitem{Pearce:2016qtn}
L. Pearce, M. Peloso and L. Sorbo,
\newblock (2016), 1603.08021.

\bibitem{Anber:2009ua}
M.M. Anber and L. Sorbo,
\newblock Phys. Rev. D81 (2010) 043534, 0908.4089.

\bibitem{Meerburg:2012id}
P.D. Meerburg and E. Pajer,
\newblock JCAP 1302 (2013) 017, 1203.6076.

\bibitem{Domcke:2016bkh}
V. Domcke, M. Pieroni and P. Binétruy,
\newblock JCAP 1606 (2016) 031, 1603.01287.

\bibitem{Anber:2006xt}
M.M. Anber and L. Sorbo,
\newblock JCAP 0610 (2006) 018, astro-ph/0606534.

\bibitem{Lue:1998mq}
A. Lue, L.M. Wang and M. Kamionkowski,
\newblock Phys. Rev. Lett. 83 (1999) 1506, astro-ph/9812088.

\bibitem{Vachaspati:2001nb}
T. Vachaspati,
\newblock Phys. Rev. Lett. 87 (2001) 251302, astro-ph/0101261.

\bibitem{Shiraishi:2013kxa}
M. Shiraishi, A. Ricciardone and S. Saga,
\newblock JCAP 1311 (2013) 051, 1308.6769.

\bibitem{Bartolo:2014hwa}
N. Bartolo et~al.,
\newblock JCAP 1501 (2015) 027, 1411.2521.

\bibitem{Acernese:2004jn}
VIRGO, F. Acernese et~al.,
\newblock Class. Quant. Grav. 22 (2005) S869, gr-qc/0406123.

\bibitem{Gluscevic:2010vv}
V. Gluscevic and M. Kamionkowski,
\newblock Phys. Rev. D81 (2010) 123529, 1002.1308.

\bibitem{Mirbabayi:2014jqa}
M. Mirbabayi et~al.,
\newblock Phys. Rev. D91 (2015) 063518, 1412.0665.

\bibitem{Ferreira:2014zia}
R.Z. Ferreira and M.S. Sloth,
\newblock JHEP 12 (2014) 139, 1409.5799.

\bibitem{Mukohyama:2014gba}
S. Mukohyama et~al.,
\newblock JCAP 1408 (2014) 036, 1405.0346.

\bibitem{Shiraishi:2016yun}
M. Shiraishi et~al.,
\newblock Phys. Rev. D94 (2016) 043506, 1606.06082.

\bibitem{Baumann:2008aq}
CMBPol Study Team, D. Baumann et~al.,
\newblock AIP Conf. Proc. 1141 (2009) 10, 0811.3919.

\bibitem{Seto:2007tn}
N. Seto and A. Taruya,
\newblock Phys. Rev. Lett. 99 (2007) 121101, 0707.0535.

\bibitem{Seto:2008sr}
N. Seto and A. Taruya,
\newblock Phys. Rev. D77 (2008) 103001, 0801.4185.

\bibitem{Crowder:2012ik}
S.G. Crowder et~al.,
\newblock Phys. Lett. B726 (2013) 66, 1212.4165.

\bibitem{Hyde:2015gwa}
J.M. Hyde,
\newblock Phys. Rev. D92 (2015) 044026, 1502.07660.

\bibitem{lyth2009primordial}
D.H. Lyth and A.R. Liddle,
\newblock {The primordial density perturbation: Cosmology, inflation and the
  origin of structure} (Cambridge, UK: Cambridge Univ. Pr., 2009).

\bibitem{Traschen:1990sw}
J.H. Traschen and R.H. Brandenberger,
\newblock Phys. Rev. D42 (1990) 2491.

\bibitem{Albrecht:1982wi}
A. Albrecht and P.J. Steinhardt,
\newblock Phys. Rev. Lett. 48 (1982) 1220.

\bibitem{Linde:1983gd}
A.D. Linde,
\newblock Phys. Lett. B129 (1983) 177.

\bibitem{Linde:1993cn}
A.D. Linde,
\newblock Phys. Rev. D49 (1994) 748, astro-ph/9307002.

\bibitem{GarciaBellido:1997wm}
J. Garcia-Bellido and A.D. Linde,
\newblock Phys. Rev. D57 (1998) 6075, hep-ph/9711360.

\bibitem{Felder:2000hj}
G.N. Felder et~al.,
\newblock Phys. Rev. Lett. 87 (2001) 011601, hep-ph/0012142.

\bibitem{GarciaBellido:2002aj}
J. Garcia-Bellido, M. Garcia~Perez and A. Gonzalez-Arroyo,
\newblock Phys. Rev. D67 (2003) 103501, hep-ph/0208228.

\bibitem{Easther:2006gt}
R. Easther and E.A. Lim,
\newblock JCAP 0604 (2006) 010, astro-ph/0601617.

\bibitem{Easther:2007vj}
R. Easther, J.T. Giblin and E.A. Lim,
\newblock Phys. Rev. D77 (2008) 103519, 0712.2991.

\bibitem{Greene:1997fu}
P.B. Greene et~al.,
\newblock Phys. Rev. D56 (1997) 6175, hep-ph/9705347.

\bibitem{Felder:2006cc}
G.N. Felder and L. Kofman,
\newblock Phys. Rev. D75 (2007) 043518, hep-ph/0606256.

\bibitem{Dufaux:2007pt}
J.F. Dufaux et~al.,
\newblock Phys. Rev. D76 (2007) 123517, 0707.0875.

\bibitem{Dufaux:2008dn}
J.F. Dufaux et~al.,
\newblock JCAP 0903 (2009) 001, 0812.2917.

\bibitem{Felder:2001kt}
G.N. Felder, L. Kofman and A.D. Linde,
\newblock Phys. Rev. D64 (2001) 123517, hep-th/0106179.

\bibitem{GarciaBellido:2007dg}
J. Garcia-Bellido and D.G. Figueroa,
\newblock Phys. Rev. Lett. 98 (2007) 061302, astro-ph/0701014.

\bibitem{Price:2008hq}
L.R. Price and X. Siemens,
\newblock Phys. Rev. D78 (2008) 063541, 0805.3570.

\bibitem{GarciaBellido:2007af}
J. Garcia-Bellido, D.G. Figueroa and A. Sastre,
\newblock Phys. Rev. D77 (2008) 043517, 0707.0839.

\bibitem{GarciaBellido:1998gm}
J. Garcia-Bellido,
\newblock {Proceedings, 33rd Recontres de Moriond fundamental parameters in
  cosmology}, pp. 29--34, 1998, hep-ph/9804205.

\bibitem{Easther:2006vd}
R. Easther, J.T. Giblin, Jr. and E.A. Lim,
\newblock Phys. Rev. Lett. 99 (2007) 221301, astro-ph/0612294.

\bibitem{Figueroa:2016ojl}
D.G. Figueroa, J. García-Bellido and F. Torrentí,
\newblock (2016), 1602.03085.

\bibitem{phdfigueroa}
D.G. Figueroa,
\newblock Aspects of reheating, 2010,
\newblock Madrid, PhD Thesis.

\bibitem{Podolsky:2005bw}
D.I. Podolsky et~al.,
\newblock Phys. Rev. D73 (2006) 023501, hep-ph/0507096.

\bibitem{Dufaux:2006ee}
J.F. Dufaux et~al.,
\newblock JCAP 0607 (2006) 006, hep-ph/0602144.

\bibitem{Greene:1997ge}
B.R. Greene, T. Prokopec and T.G. Roos,
\newblock Phys. Rev. D56 (1997) 6484, hep-ph/9705357.

\bibitem{Bethke:2013vca}
L. Bethke, D.G. Figueroa and A. Rajantie,
\newblock JCAP 1406 (2014) 047, 1309.1148.

\bibitem{Maggiore:2000gv}
M. Maggiore,
\newblock {Gravitational waves: A challenge to theoretical astrophysics.
  Proceedings, Trieste, Italy, June 6-9, 2000}, pp. 397--414, 2000,
  gr-qc/0008027.

\bibitem{Felder:2000hq}
G.N. Felder and I. Tkachev,
\newblock Comput. Phys. Commun. 178 (2008) 929, hep-ph/0011159.

\bibitem{Figueroa:2011ye}
D.G. Figueroa, J. Garcia-Bellido and A. Rajantie,
\newblock JCAP 1111 (2011) 015, 1110.0337.

\bibitem{Huang:2011gf}
Z. Huang,
\newblock Phys. Rev. D83 (2011) 123509, 1102.0227.

\bibitem{Crowder:2005nr}
J. Crowder and N.J. Cornish,
\newblock Phys. Rev. D72 (2005) 083005, gr-qc/0506015.

\bibitem{Giblin:2010sp}
J.T. Giblin, Jr, L.R. Price and X. Siemens,
\newblock JCAP 1008 (2010) 012, 1006.0935.

\bibitem{Enqvist:2012im}
K. Enqvist, D.G. Figueroa and T. Meriniemi,
\newblock Phys. Rev. D86 (2012) 061301, 1203.4943.

\bibitem{Figueroa:2013vif}
D.G. Figueroa and T. Meriniemi,
\newblock JHEP 10 (2013) 101, 1306.6911.

\bibitem{Figueroa:2014aya}
D.G. Figueroa,
\newblock JHEP 11 (2014) 145, 1402.1345.

\bibitem{Fenu:2009qf}
E. Fenu et~al.,
\newblock JCAP 0910 (2009) 005, 0908.0425.

\bibitem{Bethke:2013aba}
L. Bethke, D.G. Figueroa and A. Rajantie,
\newblock Phys. Rev. Lett. 111 (2013) 011301, 1304.2657.

\bibitem{Lozanov:2016pac}
K.D. Lozanov and M.A. Amin,
\newblock JCAP 1606 (2016) 032, 1603.05663.

\bibitem{Weinberg:100595}
S. Weinberg,
\newblock {Gravitation and Cosmology: Principles and Applications of the
  General Theory of Relativity} (Wiley, New York, NY, 1972).

\bibitem{Clifton:2011jh}
T. Clifton et~al.,
\newblock Phys. Rept. 513 (2012) 1, 1106.2476.

\bibitem{Amendola:2014wma}
L. Amendola, G. Ballesteros and V. Pettorino,
\newblock Phys. Rev. D90 (2014) 043009, 1405.7004.

\bibitem{Pettorino:2014bka}
V. Pettorino and L. Amendola,
\newblock Phys. Lett. B742 (2015) 353, 1408.2224.

\bibitem{Raveri:2014eea}
M. Raveri et~al.,
\newblock Phys. Rev. D91 (2015) 061501, 1405.7974.

\bibitem{Xu:2014uba}
L. Xu,
\newblock Phys. Rev. D91 (2015) 103520, 1410.6977.

\bibitem{Saltas:2014dha}
I.D. Saltas et~al.,
\newblock Phys. Rev. Lett. 113 (2014) 191101, 1406.7139.

\bibitem{Cai:2015dta}
Y. Cai, Y.T. Wang and Y.S. Piao,
\newblock Phys. Rev. D91 (2015) 103001, 1501.06345.

\bibitem{Dubovsky:2009xk}
S. Dubovsky et~al.,
\newblock Phys. Rev. D81 (2010) 023523, 0907.1658.

\bibitem{Fasiello:2015csa}
M. Fasiello and R.H. Ribeiro,
\newblock JCAP 1507 (2015) 027, 1505.00404.

\bibitem{Malsawmtluangi:2016agy}
N. Malsawmtluangi and P.K. Suresh,
\newblock (2016), 1603.05836.

\bibitem{Noumi:2014zqa}
T. Noumi and M. Yamaguchi,
\newblock (2014), 1403.6065.

\bibitem{Creminelli:2014wna}
P. Creminelli et~al.,
\newblock Phys. Rev. Lett. 113 (2014) 231301, 1407.8439.

\bibitem{Burrage:2016myt}
C. Burrage, S. Cespedes and A.C. Davis,
\newblock JCAP 1608 (2016) 024, 1604.08038.

\bibitem{Blas:2016qmn}
D. Blas et~al.,
\newblock Pisma Zh. Eksp. Teor. Fiz. 103 (2016) 708, 1602.04188,
\newblock [JETP Lett.103,no.10,624(2016)].

\bibitem{Ellis:2016rrr}
J. Ellis, N.E. Mavromatos and D.V. Nanopoulos,
\newblock Mod. Phys. Lett. A31 (2016) 1650155, 1602.04764.

\bibitem{Collett:2016dey}
T.E. Collett and D. Bacon,
\newblock (2016), 1602.05882.

\bibitem{Bicudo:2016pps}
P. Bicudo,
\newblock (2016), 1602.04337.

\bibitem{TheLIGOScientific:2016src}
Virgo, LIGO Scientific, B.P. Abbott et~al.,
\newblock Phys. Rev. Lett. 116 (2016) 221101, 1602.03841.

\bibitem{Copeland:2006wr}
E.J. Copeland, M. Sami and S. Tsujikawa,
\newblock Int. J. Mod. Phys. D15 (2006) 1753, hep-th/0603057.

\bibitem{amendola2010dark}
L. Amendola and S. Tsujikawa,
\newblock Dark energy: theory and observations (Cambridge University Press, New
  York, 2010).

\bibitem{Sotiriou:2008rp}
T.P. Sotiriou and V. Faraoni,
\newblock Rev. Mod. Phys. 82 (2010) 451, 0805.1726.

\bibitem{DeFelice:2014bma}
A. De~Felice and S. Tsujikawa,
\newblock Phys. Rev. D91 (2015) 103506, 1411.0736.

\bibitem{Cai:2015ipa}
Y. Cai, Y.T. Wang and Y.S. Piao,
\newblock JHEP 02 (2016) 059, 1508.07114.

\bibitem{Bergmann:1968ve}
P.G. Bergmann,
\newblock Int. J. Theor. Phys. 1 (1968) 25.

\bibitem{Nordtvedt:1970uv}
K. Nordtvedt, Jr.,
\newblock Astrophys. J. 161 (1970) 1059.

\bibitem{Wagoner:1970vr}
R.V. Wagoner,
\newblock Phys. Rev. D1 (1970) 3209.

\bibitem{Brans:1961sx}
C. Brans and R.H. Dicke,
\newblock Phys. Rev. 124 (1961) 925.

\bibitem{Nicolis:2008in}
A. Nicolis, R. Rattazzi and E. Trincherini,
\newblock Phys. Rev. D79 (2009) 064036, 0811.2197.

\bibitem{Deffayet:2009mn}
C. Deffayet, S. Deser and G. Esposito-Farese,
\newblock Phys. Rev. D80 (2009) 064015, 0906.1967.

\bibitem{Deffayet:2009wt}
C. Deffayet, G. Esposito-Farese and A. Vikman,
\newblock Phys. Rev. D79 (2009) 084003, 0901.1314.

\bibitem{Deffayet:2011gz}
C. Deffayet et~al.,
\newblock Phys. Rev. D84 (2011) 064039, 1103.3260.

\bibitem{deRham:2010eu}
C. de~Rham and A.J. Tolley,
\newblock JCAP 1005 (2010) 015, 1003.5917.

\bibitem{Kobayashi:2011nu}
T. Kobayashi, M. Yamaguchi and J. Yokoyama,
\newblock Prog. Theor. Phys. 126 (2011) 511, 1105.5723.

\bibitem{Horndeski:1974wa}
G.W. Horndeski,
\newblock Int. J. Theor. Phys. 10 (1974) 363.

\bibitem{Burrage:2010cu}
C. Burrage et~al.,
\newblock JCAP 1101 (2011) 014, 1009.2497.

\bibitem{Mizuno:2010ag}
S. Mizuno and K. Koyama,
\newblock Phys. Rev. D82 (2010) 103518, 1009.0677.

\bibitem{Kobayashi:2010cm}
T. Kobayashi, M. Yamaguchi and J. Yokoyama,
\newblock Phys. Rev. Lett. 105 (2010) 231302, 1008.0603.

\bibitem{Kobayashi:2011pc}
T. Kobayashi, M. Yamaguchi and J. Yokoyama,
\newblock Phys. Rev. D83 (2011) 103524, 1103.1740.

\bibitem{Deffayet:2010qz}
C. Deffayet et~al.,
\newblock JCAP 1010 (2010) 026, 1008.0048.

\bibitem{ArmendarizPicon:1999rj}
C. Armendariz-Picon, T. Damour and V.F. Mukhanov,
\newblock Phys. Lett. B458 (1999) 209, hep-th/9904075.

\bibitem{Kamada:2010qe}
K. Kamada et~al.,
\newblock Phys. Rev. D83 (2011) 083515, 1012.4238.

\bibitem{Kunimitsu:2015faa}
T. Kunimitsu et~al.,
\newblock JCAP 1508 (2015) 044, 1504.06946.

\bibitem{Grishchuk:1990bj}
L.P. Grishchuk and {\relax Yu}.V. Sidorov,
\newblock Phys. Rev. D42 (1990) 3413.

\bibitem{Pearle:1988uh}
P.M. Pearle,
\newblock Phys. Rev. A39 (1989) 2277.

\bibitem{Bassi:2012bg}
A. Bassi et~al.,
\newblock Rev. Mod. Phys. 85 (2013) 471, 1204.4325.

\bibitem{Perez:2005gh}
A. Perez, H. Sahlmann and D. Sudarsky,
\newblock Class. Quant. Grav. 23 (2006) 2317, gr-qc/0508100.

\bibitem{DeUnanue:2008fw}
A. De~Unanue and D. Sudarsky,
\newblock Phys. Rev. D78 (2008) 043510, 0801.4702.

\bibitem{DiezTejedor:2011jq}
A. Diez-Tejedor, G. Leon and D. Sudarsky,
\newblock Gen. Rel. Grav. 44 (2012) 2965, 1106.1176.

\bibitem{Martin:2012pea}
J. Martin, V. Vennin and P. Peter,
\newblock Phys. Rev. D86 (2012) 103524, 1207.2086.

\bibitem{Das:2013qwa}
S. Das et~al.,
\newblock Phys. Rev. D88 (2013) 085020, 1304.5094,
\newblock [Erratum: Phys. Rev.D89,no.10,109902(2014)].

\bibitem{Leon:2015hwa}
G. León and G.R. Bengochea,
\newblock Eur. Phys. J. C76 (2016) 29, 1502.04907.

\bibitem{Polarski:1995jg}
D. Polarski and A.A. Starobinsky,
\newblock Class. Quant. Grav. 13 (1996) 377, gr-qc/9504030.

\bibitem{Martin:2015qta}
J. Martin and V. Vennin,
\newblock Phys. Rev. D93 (2016) 023505, 1510.04038.

\bibitem{Leon:2016mdn}
G. León, G.R. Bengochea and S.J. Landau,
\newblock Eur. Phys. J. C76 (2016) 407, 1605.03632.

\bibitem{Camerini:2008mj}
R. Camerini et~al.,
\newblock Phys. Rev. D77 (2008) 101301, 0802.1442.

\bibitem{Dodelson:2014exa}
S. Dodelson,
\newblock Phys. Rev. Lett. 112 (2014) 191301, 1403.6310.

\bibitem{Caligiuri:2014sla}
J. Caligiuri and A. Kosowsky,
\newblock Phys. Rev. Lett. 112 (2014) 191302, 1403.5324.

\bibitem{Cai:2014uka}
Y.F. Cai et~al.,
\newblock Nucl. Phys. B900 (2015) 517, 1412.7241.

\bibitem{Biagetti:2015tja}
M. Biagetti, A. Kehagias and A. Riotto,
\newblock Phys. Rev. D91 (2015) 103527, 1502.02289.

\bibitem{Ashoorioon:2015hya}
A. Ashoorioon,
\newblock Phys. Lett. B747 (2015) 446, 1502.00556.

\bibitem{Cai:2012va}
Y.F. Cai, D.A. Easson and R. Brandenberger,
\newblock JCAP 1208 (2012) 020, 1206.2382.

\bibitem{Cai:2013kja}
Y.F. Cai et~al.,
\newblock JCAP 1310 (2013) 024, 1305.5259.

\bibitem{Gong:2014qga}
J.O. Gong,
\newblock JCAP 1407 (2014) 022, 1403.5163.

\bibitem{Wands:2002bn}
D. Wands et~al.,
\newblock Phys. Rev. D66 (2002) 043520, astro-ph/0205253.

\bibitem{Price:2014ufa}
L.C. Price et~al.,
\newblock Phys. Rev. Lett. 114 (2015) 031301, 1409.2498.

\bibitem{Chen:2006nt}
X. Chen et~al.,
\newblock JCAP 0701 (2007) 002, hep-th/0605045.

\bibitem{Matarrese:2003tk}
S. Matarrese and A. Riotto,
\newblock JCAP 0308 (2003) 007, astro-ph/0306416.

\bibitem{Das:2014ada}
S. Das et~al.,
\newblock Phys. Rev. D90 (2014) 043503, 1404.5740.

\bibitem{Wang:2014kqa}
Y. Wang and W. Xue,
\newblock JCAP 1410 (2014) 075, 1403.5817.

\bibitem{Ashoorioon:2014nta}
A. Ashoorioon et~al.,
\newblock Phys. Lett. B737 (2014) 98, 1403.6099.

\bibitem{Cai:2014hja}
Y.F. Cai and Y. Wang,
\newblock Phys. Lett. B735 (2014) 108, 1404.6672.

\bibitem{Unnikrishnan:2013rka}
S. Unnikrishnan and S. Shankaranarayanan,
\newblock JCAP 1407 (2014) 003, 1311.0177.

\bibitem{Meerburg:2015zua}
P.D. Meerburg et~al.,
\newblock Phys. Rev. D91 (2015) 103505, 1502.00302.

\bibitem{Cabass:2015jwe}
G. Cabass et~al.,
\newblock Phys. Rev. D93 (2016) 063508, 1511.05146.

\bibitem{Lasky:2015lej}
P.D. Lasky et~al.,
\newblock Phys. Rev. X6 (2016) 011035, 1511.05994.

\bibitem{Boyle:2014kba}
L. Boyle et~al.,
\newblock Phys. Rev. D92 (2015) 043504, 1408.3129.

\bibitem{Errard:2015cxa}
J. Errard et~al.,
\newblock JCAP 1603 (2016) 052, 1509.06770.

\bibitem{Huang:2015gca}
Q.G. Huang, S. Wang and W. Zhao,
\newblock JCAP 1510 (2015) 035, 1509.02676.

\bibitem{Kuroyanagi:2014qza}
S. Kuroyanagi, K. Nakayama and J. Yokoyama,
\newblock PTEP 2015 (2015) 013E02, 1410.6618.

\bibitem{Kuroyanagi:2011fy}
S. Kuroyanagi, K. Nakayama and S. Saito,
\newblock Phys. Rev. D84 (2011) 123513, 1110.4169.

\bibitem{Kuroyanagi:2010mm}
S. Kuroyanagi, T. Chiba and N. Sugiyama,
\newblock Phys. Rev. D83 (2011) 043514, 1010.5246.

\bibitem{Boyle:2005se}
L.A. Boyle and P.J. Steinhardt,
\newblock Phys. Rev. D77 (2008) 063504, astro-ph/0512014.

\bibitem{Seto:2003kc}
N. Seto and J. Yokoyama,
\newblock J. Phys. Soc. Jap. 72 (2003) 3082, gr-qc/0305096.

\bibitem{Kuroyanagi:2014qaa}
S. Kuroyanagi et~al.,
\newblock Phys. Rev. D90 (2014) 063513, 1406.1369.

\bibitem{Nakayama:2008wy}
K. Nakayama et~al.,
\newblock JCAP 0806 (2008) 020, 0804.1827.

\bibitem{Kuroyanagi:2014nba}
S. Kuroyanagi, T. Takahashi and S. Yokoyama,
\newblock JCAP 1502 (2015) 003, 1407.4785.

\bibitem{Nakayama:2008ip}
K. Nakayama et~al.,
\newblock Phys. Rev. D77 (2008) 124001, 0802.2452.

\bibitem{Dai:2014jja}
L. Dai, M. Kamionkowski and J. Wang,
\newblock Phys. Rev. Lett. 113 (2014) 041302, 1404.6704.

\bibitem{Zhao:2006mm}
W. Zhao and Y. Zhang,
\newblock Phys. Rev. D74 (2006) 043503, astro-ph/0604458.

\bibitem{Turner:1993vb}
M.S. Turner, M.J. White and J.E. Lidsey,
\newblock Phys. Rev. D48 (1993) 4613, astro-ph/9306029.

\bibitem{Jinno:2014qka}
R. Jinno, T. Moroi and T. Takahashi,
\newblock JCAP 1412 (2014) 006, 1406.1666.

\bibitem{Jinno:2013xqa}
R. Jinno, T. Moroi and K. Nakayama,
\newblock JCAP 1401 (2014) 040, 1307.3010.

\bibitem{Smith:2005mm}
T.L. Smith, M. Kamionkowski and A. Cooray,
\newblock Phys. Rev. D73 (2006) 023504, astro-ph/0506422.

\bibitem{Smith:2008pf}
T.L. Smith, M. Kamionkowski and A. Cooray,
\newblock Phys. Rev. D78 (2008) 083525, 0802.1530.

\bibitem{Gong:2015qha}
J.O. Gong, S. Pi and G. Leung,
\newblock JCAP 1505 (2015) 027, 1501.03604.

\bibitem{Martin:2014nya}
J. Martin, C. Ringeval and V. Vennin,
\newblock Phys. Rev. Lett. 114 (2015) 081303, 1410.7958.

\bibitem{Martin:2010kz}
J. Martin and C. Ringeval,
\newblock Phys. Rev. D82 (2010) 023511, 1004.5525.

\bibitem{Creminelli:2014fca}
P. Creminelli et~al.,
\newblock Phys. Rev. D90 (2014) 083513, 1405.6264.

\bibitem{Munoz:2014eqa}
J.B. Munoz and M. Kamionkowski,
\newblock Phys. Rev. D91 (2015) 043521, 1412.0656.

\bibitem{Cook:2015vqa}
J.L. Cook et~al.,
\newblock JCAP 1504 (2015) 047, 1502.04673.

\bibitem{Cai:2015soa}
R.G. Cai, Z.K. Guo and S.J. Wang,
\newblock Phys. Rev. D92 (2015) 063506, 1501.07743.

\bibitem{Planck:2013jfk}
Planck, P.A.R. Ade et~al.,
\newblock Astron. Astrophys. 571 (2014) A22, 1303.5082.

\bibitem{Kinney:2005in}
W.H. Kinney and A. Riotto,
\newblock JCAP 0603 (2006) 011, astro-ph/0511127.

\bibitem{Liddle:2003as}
A.R. Liddle and S.M. Leach,
\newblock Phys. Rev. D68 (2003) 103503, astro-ph/0305263.

\bibitem{Domcke:2015iaa}
V. Domcke and J. Heisig,
\newblock Phys. Rev. D92 (2015) 103515, 1504.00345.

\bibitem{Weinberg:2003ur}
S. Weinberg,
\newblock Phys. Rev. D69 (2004) 023503, astro-ph/0306304.

\bibitem{Bashinsky:2005tv}
S. Bashinsky,
\newblock Submitted to: Phys. Rev. D  (2005), astro-ph/0505502.

\bibitem{Dicus:2005rh}
D.A. Dicus and W.W. Repko,
\newblock Phys. Rev. D72 (2005) 088302, astro-ph/0509096.

\bibitem{Zhao:2009we}
W. Zhao, Y. Zhang and T. Xia,
\newblock Phys. Lett. B677 (2009) 235, 0905.3223.

\bibitem{Jinno:2012xb}
R. Jinno, T. Moroi and K. Nakayama,
\newblock Phys. Rev. D86 (2012) 123502, 1208.0184.

\bibitem{Shchedrin:2012sp}
G. Shchedrin,
\newblock (2012), 1204.1384.

\bibitem{Stefanek:2012hj}
B.A. Stefanek and W.W. Repko,
\newblock Phys. Rev. D88 (2013) 083536, 1207.7285.

\bibitem{Dent:2013asa}
J.B. Dent et~al.,
\newblock Phys. Rev. D88 (2013) 084008, 1307.7571.

\bibitem{Mangilli:2008bw}
A. Mangilli et~al.,
\newblock Phys. Rev. D78 (2008) 083517, 0805.3234.

\bibitem{Saga:2014jca}
S. Saga, K. Ichiki and N. Sugiyama,
\newblock Phys. Rev. D91 (2015) 024030, 1412.1081.

\bibitem{Lattanzi:2010gn}
M. Lattanzi, R. Benini and G. Montani,
\newblock Class. Quant. Grav. 27 (2010) 194008, 1010.3849.

\bibitem{Kamionkowski:1999qc}
M. Kamionkowski and A. Kosowsky,
\newblock Ann. Rev. Nucl. Part. Sci. 49 (1999) 77, astro-ph/9904108.

\bibitem{Hu:2001bc}
W. Hu and S. Dodelson,
\newblock Ann. Rev. Astron. Astrophys. 40 (2002) 171, astro-ph/0110414.

\bibitem{Dodelson-Cosmology-2003}
S. Dodelson,
\newblock {Modern Cosmology} (Academic Press, Amsterdam, 2003).

\bibitem{Hu:1997hv}
W. Hu and M.J. White,
\newblock New Astron. 2 (1997) 323, astro-ph/9706147.

\bibitem{1968ApJ}
M.J. Rees,
\newblock ApJ 153 (1968) 1.

\bibitem{Kosowsky:1994cy}
A. Kosowsky,
\newblock Annals Phys. 246 (1996) 49, astro-ph/9501045.

\bibitem{Cabella:2004mk}
P. Cabella and M. Kamionkowski,
\newblock {International School of Gravitation and Cosmology: The Polarization
  of the Cosmic Microwave Background Rome, Italy, September 6-11, 2003}, 2004,
  astro-ph/0403392.

\bibitem{Kamionkowski:2015yta}
M. Kamionkowski and E.D. Kovetz,
\newblock (2015), 1510.06042.

\bibitem{Kamionkowski:1996ks}
M. Kamionkowski, A. Kosowsky and A. Stebbins,
\newblock Phys. Rev. D55 (1997) 7368, astro-ph/9611125.

\bibitem{Zaldarriaga:1996xe}
M. Zaldarriaga and U. Seljak,
\newblock Phys. Rev. D55 (1997) 1830, astro-ph/9609170.

\bibitem{Newman:1961qr}
E. Newman and R. Penrose,
\newblock J. Math. Phys. 3 (1962) 566.

\bibitem{Adam:2015wua}
Planck, R. Adam et~al.,
\newblock (2015), 1502.01588.

\bibitem{Lewis:2001hp}
A. Lewis, A. Challinor and N. Turok,
\newblock Phys. Rev. D65 (2002) 023505, astro-ph/0106536.

\bibitem{Knox:2002pe}
L. Knox and Y.S. Song,
\newblock Phys. Rev. Lett. 89 (2002) 011303, astro-ph/0202286.

\bibitem{Kesden:2002ku}
M. Kesden, A. Cooray and M. Kamionkowski,
\newblock Phys. Rev. Lett. 89 (2002) 011304, astro-ph/0202434.

\bibitem{Hu:2001kj}
W. Hu and T. Okamoto,
\newblock Astrophys. J. 574 (2002) 566, astro-ph/0111606.

\bibitem{Peiris:2003ff}
WMAP, H.V. Peiris et~al.,
\newblock Astrophys. J. Suppl. 148 (2003) 213, astro-ph/0302225.

\bibitem{Galli:2014kla}
S. Galli et~al.,
\newblock Phys. Rev. D90 (2014) 063504, 1403.5271.

\bibitem{Polnarev:2007cv}
A.G. Polnarev, N.J. Miller and B.G. Keating,
\newblock Mon. Not. Roy. Astron. Soc. 386 (2008) 1053, 0710.3649.

\bibitem{Ade:2015cva}
Planck, P.A.R. Ade et~al.,
\newblock (2015), 1502.01594.

\bibitem{Feng:2004mq}
B. Feng et~al.,
\newblock Phys. Lett. B620 (2005) 27, hep-ph/0406269.

\bibitem{Li:2008tma}
M. Li and X. Zhang,
\newblock Phys. Rev. D78 (2008) 103516, 0810.0403.

\bibitem{Ade:2014xna}
BICEP2, P.A.R. Ade et~al.,
\newblock Phys. Rev. Lett. 112 (2014) 241101, 1403.3985.

\bibitem{Ade:2015tva}
BICEP2, Planck, P. Ade et~al.,
\newblock Phys. Rev. Lett. 114 (2015) 101301, 1502.00612.

\bibitem{Kamionkowski:2010rb}
M. Kamionkowski and T. Souradeep,
\newblock Phys. Rev. D83 (2011) 027301, 1010.4304.

\bibitem{Shiraishi:2010kd}
M. Shiraishi et~al.,
\newblock Prog. Theor. Phys. 125 (2011) 795, 1012.1079.

\bibitem{Shiraishi:2010sm}
M. Shiraishi et~al.,
\newblock Phys. Rev. D82 (2010) 103505, 1003.2096.

\bibitem{Shiraishi:2011st}
M. Shiraishi, D. Nitta and S. Yokoyama,
\newblock Prog. Theor. Phys. 126 (2011) 937, 1108.0175.

\bibitem{Meerburg:2016ecv}
P.D. Meerburg et~al.,
\newblock Phys. Rev. D93 (2016) 123511, 1603.02243.

\bibitem{Jeong:2012df}
D. Jeong and M. Kamionkowski,
\newblock Phys. Rev. Lett. 108 (2012) 251301, 1203.0302.

\bibitem{Schmidt:2013gwa}
F. Schmidt, E. Pajer and M. Zaldarriaga,
\newblock Phys. Rev. D89 (2014) 083507, 1312.5616.

\bibitem{Dai:2013kra}
L. Dai, D. Jeong and M. Kamionkowski,
\newblock Phys. Rev. D88 (2013) 043507, 1306.3985.

\bibitem{Schmidt:2012nw}
F. Schmidt and D. Jeong,
\newblock Phys. Rev. D86 (2012) 083513, 1205.1514.

\bibitem{Schmidt:2012ne}
F. Schmidt and D. Jeong,
\newblock Phys. Rev. D86 (2012) 083527, 1204.3625.

\bibitem{Dai:2012bc}
L. Dai, M. Kamionkowski and D. Jeong,
\newblock Phys. Rev. D86 (2012) 125013, 1209.0761.

\bibitem{Cooray:2005hm}
A. Cooray, M. Kamionkowski and R.R. Caldwell,
\newblock Phys. Rev. D71 (2005) 123527, astro-ph/0503002.

\bibitem{Li:2006si}
C. Li and A. Cooray,
\newblock Phys. Rev. D74 (2006) 023521, astro-ph/0604179.

\bibitem{Book:2011na}
L.G. Book, M. Kamionkowski and T. Souradeep,
\newblock Phys. Rev. D85 (2012) 023010, 1109.2910.

\bibitem{Book:2011dz}
L. Book, M. Kamionkowski and F. Schmidt,
\newblock Phys. Rev. Lett. 108 (2012) 211301, 1112.0567.

\bibitem{Lu:2007pk}
T. Lu and U.L. Pen,
\newblock Mon. Not. Roy. Astron. Soc. 388 (2008) 1819, 0710.1108.

\bibitem{Pen:2003yv}
U.L. Pen,
\newblock New Astron. 9 (2004) 417, astro-ph/0305387.

\bibitem{Pajer:2013ana}
E. Pajer, F. Schmidt and M. Zaldarriaga,
\newblock Phys. Rev. D88 (2013) 083502, 1305.0824.

\bibitem{Ando:2008zza}
S. Ando and M. Kamionkowski,
\newblock Phys. Rev. Lett. 100 (2008) 071301, 0711.0779.

\bibitem{Pullen:2007tu}
A.R. Pullen and M. Kamionkowski,
\newblock Phys. Rev. D76 (2007) 103529, 0709.1144.

\bibitem{Maldacena:2002vr}
J.M. Maldacena,
\newblock JHEP 05 (2003) 013, astro-ph/0210603.

\bibitem{Maldacena:2011nz}
J.M. Maldacena and G.L. Pimentel,
\newblock JHEP 09 (2011) 045, 1104.2846.

\bibitem{Seery:2008ax}
D. Seery, M.S. Sloth and F. Vernizzi,
\newblock JCAP 0903 (2009) 018, 0811.3934.

\bibitem{Giddings:2010nc}
S.B. Giddings and M.S. Sloth,
\newblock JCAP 1101 (2011) 023, 1005.1056.

\bibitem{Giddings:2011zd}
S.B. Giddings and M.S. Sloth,
\newblock Phys. Rev. D84 (2011) 063528, 1104.0002.

\bibitem{Creminelli:2004yq}
P. Creminelli and M. Zaldarriaga,
\newblock JCAP 0410 (2004) 006, astro-ph/0407059.

\bibitem{Dimastrogiovanni:2014ina}
E. Dimastrogiovanni et~al.,
\newblock JCAP 1412 (2014) 050, 1407.8204.

\bibitem{Hanson:2009gu}
D. Hanson and A. Lewis,
\newblock Phys. Rev. D80 (2009) 063004, 0908.0963.

\bibitem{Dai:2013ikl}
L. Dai, D. Jeong and M. Kamionkowski,
\newblock Phys. Rev. D87 (2013) 103006, 1302.1868.

\bibitem{Akhshik:2014bla}
M. Akhshik,
\newblock JCAP 1505 (2015) 043, 1409.3004.

\bibitem{Endlich:2013jia}
S. Endlich et~al.,
\newblock Phys. Rev. D90 (2014) 063506, 1307.8114.

\bibitem{Dimastrogiovanni:2015pla}
E. Dimastrogiovanni, M. Fasiello and M. Kamionkowski,
\newblock (2015), 1504.05993.

\bibitem{Brahma:2013rua}
S. Brahma, E. Nelson and S. Shandera,
\newblock Phys. Rev. D89 (2014) 023507, 1310.0471.

\bibitem{Mangano:2001iu}
G. Mangano et~al.,
\newblock Phys. Lett. B534 (2002) 8, astro-ph/0111408.

\bibitem{weinberg2008cosmology}
S. Weinberg,
\newblock {Cosmology} (Oxford, UK: Oxford Univ. Pr., 2008).

\bibitem{Hou:2011ec}
Z. Hou et~al.,
\newblock Phys. Rev. D87 (2013) 083008, 1104.2333.

\bibitem{Pagano:2015hma}
L. Pagano, L. Salvati and A. Melchiorri,
\newblock Phys. Lett. B760 (2016) 823, 1508.02393.

\bibitem{Boyle:2007zx}
L.A. Boyle and A. Buonanno,
\newblock Phys. Rev. D78 (2008) 043531, 0708.2279.

\bibitem{Nakama:2015nea}
T. Nakama and T. Suyama,
\newblock Phys. Rev. D92 (2015) 121304, 1506.05228.

\bibitem{Nakama:2016enz}
T. Nakama and T. Suyama,
\newblock Phys. Rev. D94 (2016) 043507, 1605.04482.

\bibitem{polnarev}
A.G. Polnarev,
\newblock Sov. Astron. 29 (1985) 607.

\bibitem{Ota:2014hha}
A. Ota et~al.,
\newblock JCAP 1410 (2014) 029, 1406.0451.

\bibitem{Chluba:2014qia}
J. Chluba et~al.,
\newblock Mon. Not. Roy. Astron. Soc. 446 (2015) 2871, 1407.3653.

\bibitem{Bashinsky:2003tk}
S. Bashinsky and U. Seljak,
\newblock Phys. Rev. D69 (2004) 083002, astro-ph/0310198.

\bibitem{Dodelson:2010qu}
S. Dodelson,
\newblock Phys. Rev. D82 (2010) 023522, 1001.5012.

\bibitem{Joshi:2013at}
B.C. Joshi,
\newblock Int. J. Mod. Phys. D22 (2013) 1341008, 1301.5730.

\bibitem{Moore:2014lga}
C.J. Moore, R.H. Cole and C.P.L. Berry,
\newblock Class. Quant. Grav. 32 (2015) 015014, 1408.0740.

\bibitem{Huang:2015gka}
Q.G. Huang and S. Wang,
\newblock JCAP 1506 (2015) 021, 1502.02541.

\bibitem{Abbott:2007kv}
LIGO Scientific, B.P. Abbott et~al.,
\newblock Rept. Prog. Phys. 72 (2009) 076901, 0711.3041.

\bibitem{Manchester:2012za}
R.N. Manchester et~al.,
\newblock Publ. Astron. Soc. Austral. 30 (2013) 17, 1210.6130.

\bibitem{Ade:2013zuv}
Planck, P.A.R. Ade et~al.,
\newblock Astron. Astrophys. 571 (2014) A16, 1303.5076.

\bibitem{Page:2006hz}
WMAP, L. Page et~al.,
\newblock Astrophys. J. Suppl. 170 (2007) 335, astro-ph/0603450.

\bibitem{Scoville:2006vr}
N. Scoville et~al.,
\newblock Astrophys. J. Suppl. 172 (2007) 38, astro-ph/0612306.

\bibitem{York:2000gk}
SDSS, D.G. York et~al.,
\newblock Astron. J. 120 (2000) 1579, astro-ph/0006396.

\bibitem{Verde:2005ff}
L. Verde, H. Peiris and R. Jimenez,
\newblock JCAP 0601 (2006) 019, astro-ph/0506036.

\bibitem{Adam:2015rua}
Planck, R. Adam et~al.,
\newblock (2015), 1502.01582.

\bibitem{Lentati:2015qwp}
L. Lentati et~al.,
\newblock Mon. Not. Roy. Astron. Soc. 453 (2015) 2576, 1504.03692.

\bibitem{McLaughlin:2013ira}
M.A. McLaughlin,
\newblock Class. Quant. Grav. 30 (2013) 224008, 1310.0758.

\bibitem{Janssen:2014dka}
G. Janssen et~al.,
\newblock PoS AASKA14 (2015) 037, 1501.00127.

\bibitem{Abadie:2010mt}
VIRGO, LIGO, J. Abadie et~al.,
\newblock Phys. Rev. D81 (2010) 102001, 1002.1036.

\bibitem{Aasi:2014zwg}
VIRGO, LIGO Scientific, J. Aasi et~al.,
\newblock Phys. Rev. Lett. 113 (2014) 231101, 1406.4556.

\bibitem{TheLIGOScientific:2016wyq}
Virgo, LIGO Scientific, B.P. Abbott et~al.,
\newblock Phys. Rev. Lett. 116 (2016) 131102, 1602.03847.

\bibitem{Unnikrishnan:2013qwa}
C.S. Unnikrishnan,
\newblock Int. J. Mod. Phys. D22 (2013) 1341010, 1510.06059.

\bibitem{Aso:2013eba}
KAGRA, Y. Aso et~al.,
\newblock Phys. Rev. D88 (2013) 043007, 1306.6747.

\bibitem{Punturo:2010zz}
M. Punturo et~al.,
\newblock Class. Quant. Grav. 27 (2010) 194002.

\bibitem{Kawamura:2011zz}
S. Kawamura et~al.,
\newblock Class. Quant. Grav. 28 (2011) 094011.

\bibitem{Kuroda:2015owv}
K. Kuroda, W.T. Ni and W.P. Pan,
\newblock Int. J. Mod. Phys. D24 (2015) 1530031, 1511.00231.

\bibitem{Thrane:2013oya}
E. Thrane and J.D. Romano,
\newblock Phys. Rev. D88 (2013) 124032, 1310.5300.

\bibitem{pet}
A. Petiteau,
\newblock {to appear}.

\end{thebibliography}

\end{document}